\documentclass[trackchanges]{aastex701}
\usepackage{graphicx}
\usepackage{txfonts}
\usepackage{xcolor}
\usepackage{float}
\usepackage{comment}
\usepackage{pgf}   
\usepackage{enumitem}
\usepackage{eurosym}
\graphicspath{{./}{figures/}}
\def\degr{\hbox{$^\circ$\,}}

\def\arcsec{\hbox{$^{\prime\prime}$}}

\def\gsim{\mathrel{\hbox{\rlap{\lower.55ex \hbox {$\sim$}}
                   \kern-.3em \raise.4ex \hbox{$>$}}}}
\def\lsim{\mathrel{\hbox{\rlap{\lower.55ex \hbox {$\sim$}}
                   \kern-.3em \raise.4ex \hbox{$<$}}}}

\def\he2{\hbox{He\,{\sc ii} $\lambda$4686}}

\def\RL1{\hbox{{$R_{L_{1}}$}}}

\begin{document}

\title{Archival transients detected with the MeerLICHT telescope I: Supernovae}

    \author[0000-0002-2261-0382]{O.~Mogawana}
    \affiliation{Department of Astronomy, University of Cape Town,
    Private Bag X3, Rondebosch, 7701, South Africa}
    \affiliation{ South African Astronomical Observatory, P.O. Box 9,
    Observatory, 7935, South Africa}
    \affiliation{The Inter-University Institute for Data Intensive
    Astronomy, University of Cape Town, Private Bag X3, Rondebosch,
    7701, South Africa}
    \email[]{orapeleng@saao.ac.za}
    
    \author[0000-0002-4488-726X]{P.J.~Groot}
    \affiliation{Department of Astrophysics/IMAPP, Radboud University,
    P.O. Box 9010, 6500 GL, Nijmegen, The Netherlands}
    \affiliation{Department of Astronomy, University of Cape Town,
    Private Bag X3, Rondebosch, 7701, South Africa}
    \affiliation{ South African Astronomical Observatory, P.O. Box 9,
    Observatory, 7935, South Africa}
    \affiliation{The Inter-University Institute for Data Intensive
    Astronomy, University of Cape Town, Private Bag X3, Rondebosch,
    7701, South Africa}
    \email[]{p.groot@astro.ru.nl}

    \author[0000-0003-0901-1606]{N.~Blagorodnova}
    \affiliation{Departament de F\'isica Qu\'antica i Astrof\'isica (FQA), Universitat de Barcelona (UB), c. Martí i Franquès, 1, 08028 Barcelona, Spain}
    \affiliation{Institut de Ci\'encies del Cosmos (ICCUB), Universitat de Barcelona (UB), c. Martí i Franquès, 1, 08028 Barcelona, Spain}
    \affiliation{Institut d'Estudis Espacials de Catalunya (IEEC), Edifici RDIT, Campus UPC, 08860 Castelldefels (Barcelona), Spain}
    \affiliation{Department of Astrophysics/IMAPP, Radboud University,
    P.O. Box 9010, 6500 GL, Nijmegen, The Netherlands}
    \email[]{nblago@fqa.ub.edu}

    \author{P.M.~Vreeswijk}
    \affiliation{Department of Astrophysics/IMAPP, Radboud University,
    P.O. Box 9010, 6500 GL, Nijmegen, The Netherlands}
    \email[]{p.vreeswijk@astro.ru.nl}

    \author[0000-0003-3114-2733]{D.L.A.~Pieterse}
    \affiliation{Department of Astrophysics/IMAPP, Radboud University,
    P.O. Box 9010, 6500 GL, Nijmegen, The Netherlands}
    \email[]{d.pieterse@astro.ru.nl}

    \author[0000-0002-6636-921X]{S.~Bloemen}
    \affiliation{Department of Astrophysics/IMAPP, Radboud University,
    P.O. Box 9010, 6500 GL, Nijmegen, The Netherlands}
    \email[]{s.bloemen@astro.ru.nl}

    \author[0000-0002-8597-0756]{G.~Leloudas}
    \affiliation{DTU Space, National Space Institute,
    Technical University of Denmark, Elektrovej 327, 2800 Kgs. Lyngby,
    Denmark}
    \email[]{giorgos@space.dtu.dk}

    \author[0000-0002-3516-2152]{R.A.D.~Wijnands}
    \affiliation{Anton Pannekoek Institute for Astronomy, University
    of Amsterdam, P.O. Box 94249, 1090 GE Amsterdam, The Netherlands}
    \email[]{r.a.d.wijnands@uva.nl}

    \author[0000-0002-7004-9956]{D. A. H. ~Buckley}
    \affiliation{South African Astronomical Observatory, P.O. Box 9, Observatory 7935, Cape Town, South Africa}
    \affiliation{Department of Astronomy, University of Cape Town, Private Bag X3, Rondebosch 7701, South Africa}
    \affiliation{Department of Physics, University of the Free State, PO Box 339, Bloemfontein 9300, South Africa}
    \email[]{dibnob@saao.ac.za}

    \author[0000-0002-6896-1655]{P. A. ~Woudt}
    \affiliation{Department of Astronomy, University of Cape Town, Private Bag X3, Rondebosch 7701, South Africa}
    \email[]{patrick.woudt@uct.ac.za}

    \author[]{R. ~Fender}
    \affiliation{Astrophysics, The University of Oxford, Denys Wilkinson Building, Keble Road, Oxford OX1 3RH, UK}
    \email[]{rob.fender@astro.ox.ac.uk}
    
    \author[]{B. W. ~Stappers}
    \affiliation{Jodrell Bank Centre for Astrophysics, University of Manchester, Oxford Road, Manchester M13 9PL, UK}
    \email[]{ben.stappers@manchester.ac.uk}

\correspondingauthor{Orapeleng Mogawana}
\email[]{orapeleng@saao.ac.za}

%\collaboration{all}{(PASP Editorial Office)}

%% Use the \collaboration command to identify collaborations. This command
%% takes an optional argument that is either a number or the word "all"
%% which tells the compiler how many of the authors above the command to
%% show. For example "\collaboration[all]{(DELVE Collaboration)}" wil include
%% all the authors above this command.
%%
%% Mark off the abstract in the ``abstract'' environment. 
%\include{authors_PASP}

%\date{Received ...,; accepted ...}

% \abstract{}{}{}{}{} 
% 5 {} token are mandatory

\begin{abstract}
 {Wide-ﬁeld optical transient surveys collectively discover $\sim$50 new transients per night, making rapid, multi-filter and/or spectroscopic follow-up of all but the most exceptional events unfeasible. Consequently, the (early) light-curve properties of most supernovae (SNe), particularly sub-luminous events, are often poorly characterised. Here, we showcase recovery of key parameters by combining publicly available data from multiple, contemporaneous surveys. We built a catalogue of $125$ SNe by cross-matching the MeerLICHT (ML) transient survey with the Transient Name Server list of spectroscopically classified SNe detected between August 2017 and October 2022. For a subset of $14$ SNe with good multi-survey photometric coverage, we combined ML-$q$, ZTF-$g,\,r$, and ATLAS-$c,\,o$ data to construct composite early-time light curves. We then applied an expanding fireball ($t^2$) model to constrain their explosion dates and rest-frame rise times. We find that this simple model provides a reliable description for the early emission of most SNe in our sample and successfully constrains their rise time, a key parameter for characterising SN categories and probing progenitor origins. This multi-survey approach, only using standard survey products, provides a powerful method to characterise large numbers of SNe that would otherwise be lost, a technique that will be essential for maximising the scientific return from current and future surveys such as GOTO, BlackGEM, the Vera C. Rubin Observatory's LSST, and Global Open Transient Telescope Array (GOTTA).}
\end{abstract}

%% Keywords should appear after the \end{abstract} command. 
%% PASP uses Unified Astronomy Thesaurus (UAT) concepts:
%% https://astrothesaurus.org
%% You will be asked to selected these concepts during the submission process
%% but this old "keyword" functionality is maintained in case authors want
%% to include these concepts in their preprints.

%%
%% You can use the \uat command to link your UAT concepts back its source.
%\keywords{\uat{Galaxies}{573} --- \uat{Cosmology}{343} --- \uat{High Energy astrophysics}{739} --- \uat{Interstellar medium}{847} --- \uat{Stellar astronomy}{1583} --- \uat{Solar physics}{1476}}

%% From the front matter, we move on to the body of the paper.
%% Sections are demarcated by \section and \subsection, respectively.
%% Observe the use of the LaTeX \label
%% command after the \subsection to give a symbolic KEY to the
%% subsection for cross-referencing in a \ref command.
%% You can use LaTeX's \ref and \label commands to keep track of
%% cross-references to sections, equations, tables, and figures.
%% That way, if you change the order of any elements, LaTeX will
%% automatically renumber them.

%-------------------------------------------------%
%%%%%%%%%%%%%%%%%%%%%%%%%%%%%%%%%%%%%%%%%%%%%%%
\section{Introduction}\label{sec:introduction}
%%%%%%%%%%%%%%%%%%%%%%%%%%%%%%%%%%%%%%%%%%%%%%%
The systematic collection and categorization of transient data is crucial for statistical studies of both known and novel transient classes.
The field of time-domain astrophysics (TDA)\textemdash the study of variable phenomena, moving celestial objects, and transients \textemdash has grown significantly over the past two decades, though its origins trace back to pioneering efforts by Fritz Zwicky. 
Their survey with an $18-$inch Schmidt telescope at the Palomar Observatory in the late 1930s discovered $380$ supernovae (SNe) over $38$ years, an average rate of $10$ SNe per year \citep{1938ApJ....88..529Z, 1942ApJ....96...28Z}. 
The SNe discovery rate increased to approximately 30 per year between the 1960s and late 1990s, approaching $100$\textemdash $150$ per year by the early 2000s. Despite this, the detection rate of bright SNe (apparent magnitude $< 15$) remained roughly constant at about $10$ per year \citep{Cappellaro2003}. 
This plateau underscored the intrinsic rarity of SNe\textemdash roughly one event per century in a typical galaxy—and demonstrated that increasing the detection rate would require dedicated surveys capable of exploring larger volumes of space to much fainter magnitudes.
The shift to modern CCD detectors and advances in computational processing allowed for SN searches to exploit more robust analysis techniques such as accurate template image subtraction \citep{1986ApJ...304....1P,1998ApJ...503..325A,2015ascl.soft04004B}. With the adoption of these techniques, more precise photometry has become available from ongoing CCD transient survey facilities.
The use of type Ia SNe as cosmological distance calibrators \citep{Reiss_1998,1998Natur.391...51P,1999ApJ...517..565P, Riess_2007, 2009A&A...500L..17B, Suzuki_2012} and the evidence associating peculiar type Ib/c SNe with gamma-ray burst events \citep{1998Natur.395..670G,2003Natur.423..847H,2006Natur.442.1008C, doi:10.1146/annurev.astro.43.072103.150558, Woosley_2006} propelled the rise of optical wide-field TDA surveys. 
Through the combined effort of these TDA surveys, a vast amount of observational data and time-series forced-photometry has become available, increasing the discovery rate of classified SNe to over $2000$ per year \citep{2020arXiv200403511K}. Current state-of-the-art optical surveys\textemdash including Pan-STARRS \citep{2002SPIE.4836..154K, 2010SPIE.7733E..0EK, 2016arXiv161205560C}, ASAS-SN \citep{Shappee_2014, 10.1093/mnras/stw2273, 10.1093/mnras/stad355}, ATLAS \citep{Tonry_2011, Tonry_2018}, GOTO \citep{2022MNRAS.511.2405S}, ZTF \citep{2014htu..conf...27B, Bellm_2019}, and BlackGEM \citep{2015ASPC..496..254B,Groot24}\textemdash are already producing a high rate of transient discoveries. However, this rate will increase by about 100 times with the upcoming Vera C. Rubin Observatory's Legacy Survey of Space and Time (LSST; \citealt{2019ApJ...873..111I}), which is projected to generate around $\sim$1 million supernovae-related alerts per year \citep{2025A&A...698A.153S}, leading to the discovery of roughly 3500 brand-new supernovae per night. \\
\\
The study of SN light curves, particularly their early rising phase, provides crucial insights into the diverse physical mechanisms that power these stellar explosions and the characteristics of their plausible progenitor stars. Accurate determination of SN explosion and rise times is essential for a comprehensive understanding of the underlying population of these cosmic events \citep{2015MNRAS.446.3895F,2015A&A...574A..60T}. 
Most wide-field surveys currently operate with cadences ranging from several days to multiple observations per night, and in addition are often dual-band at most. They are designed to maximize discovery rates, but the sheer number of transients overwhelms dedicated follow-up resources, making follow-up with the same survey facility challenging or often unfeasible. 
Gaps in observational cadence, even for high-cadence surveys, can result in missed transients or large uncertainties in explosion time estimates, impacting the precision of rise time determination. These challenges underscore the need for a concerted approach, combining multiple survey products to fully unlock the potential of SN rise times.\\
\\
In this paper, we present a multi-band catalogue of classified supernovae (SNe) that were detected and followed up by MeerLICHT from 2017 to 2022. Section \ref{sec:Telescope} provides an overview of the MeerLICHT telescope design and characteristics. 
In Section \ref{sec:methods}, we describe the nightly survey observations and data processing techniques, followed by an explanation of how the sample of detected SNe was recovered in Section \ref{sec:supernovae_sample_selection}. 
Section \ref{sec:results} presents the detailed results for several SNe that warranted in-depth analysis, examining their early light curve evolution and parameter fits. 
Finally, Section \ref{sec:conclusion} discusses our findings and their astrophysical implications, and provides a conclusion to this study.
%%%%%%%%%%%%%%%%%%%%%%%%%%%%%%%%%%%%%%%%%%%%%%%

%--> Introduction
%---> Supernovae and Transient surveys
%--> Supernovae background and importance of their systematic surveys
%--> Introduction to all sections
%-------------------------------------------------%

%-------------------------------------------------%
%%%%%%%%%%%%%%%%%%%%%%%%%%%%%%%%%%%%%%%%%%%%%%%%%%%
\section{MeerLICHT Telescope Overview}\label{sec:Telescope}
%%%%%%%%%%%%%%%%%%%%%%%%%%%%%%%%%%%%%%%%%%%%%%%%%%%
%%%%%%%%%%%%%%%%%%%%%%%%%%%%%%%%%%%%%%%%%%%%%%%%%%%
\subsection{Telescope, Optics \& Operations}
%%%%%%%%%%%%%%%%%%%%%%%%%%%%%%%%%%%%%%%%%%%%%%%%%%%
MeerLICHT is a modified Dall-Kirkham optical wide-field telescope attached to a standard equatorial mount. It is hosted by the South African Astronomical Observatory (SAAO) located in Sutherland, South Africa. It is the prototype of the BlackGEM Array, and all specifications are provided in \cite{Groot24}.
%%%%%%%%%%%%%%%%%%%%%%%%%%%%%%%%%%%%%%%%%%%%%%%%%%%
%\subsubsection{Imaging Camera}
%%%%%%%%%%%%%%%%%%%%%%%%%%%%%%%%%%%%%%%%%%%%%%%%%%%
The MeerLICHT imaging camera is a thinned monolithic back-illuminated STA1600 CCD consisting of a $10560 \times 10560$ pixel array with a pixel size of $9\, \mathrm{\mu m}$. This offers a pixel scale of $0.56\, \mathrm{arcsec/pixel}$ at the focus of the MeerLICHT telescope. 
%%%%%%%%%%%%%%%%%%%%%%%%%%%%%%%%%%%%%%%%%%%%%%%%%%%
%\subsection{Optics and Filters}
%%%%%%%%%%%%%%%%%%%%%%%%%%%%%%%%%%%%%%%%%%%%%%%%%%%
The telescope optics achieve a low-distortion image quality over the entire $2.7\, \mathrm{square\, degree}$ footprint in six filters. 
MeerLICHT observes the sky through a set of six different filters, designated as $u,\, g,\, r,\, i,\, z$, which are similar to SDSS \citep{York_2000,2006AJ....131.2332G} filters and further extended with a sixth, wide passband filter designated $q$ \citep{Groot24}. The $q$ bandpass is significantly wide compared to other filters because it is specified by a FWHM of $1810.42\AA$ centered at an effective wavelength of $5745.80\AA$ with an RMS width of $783.34\AA$.
%%%%%%%%%%%%%%%%%%%%%%%%%%%%%%%%%%%%%%%%%%%%%%%%%%%
\subsection{MeerLICHT Local Transient Survey}\label{sec:MeerLICHT_LTP}
%%%%%%%%%%%%%%%%%%%%%%%%%%%%%%%%%%%%%%%%%%%%%%%%%%%

The MeerLICHT telescope is designed to follow the MeerKAT \citep{2009arXiv0910.2935B} radio array. However, this link was only fully established in September 2022, and in addition, the radio-wavelength observing strategies and the optical-wavelength observing strategies are sufficiently different that large gaps are present in the scheduler even when following MeerKAT. In between MeerKAT observations, and before September 2022, the MeerLICHT telescope concentrated on obtaining a six-filter Southern sky survey (at $\delta<30\degr$) and the Local Transient Program. Most of these Local Transient Survey observations have been obtained in the $u$-,\,$q$-, and $i$-band filters, which thereby make up the bulk of all MeerLICHT observations. The choice for these three filters was made as they span the full optical window at relatively high sensitivity. Excluding red-flagged observations, 68 percent of MeerLICHT q-band observations have the same-filter cadence of $\leq1$ hour, while 75 percent of observations, regardless of the filter, have a cadence of $\leq2$ minutes (Pieterse et al., in prep.)\\
\\
A full description of the MeerLICHT Local Transient Survey will be given in a forthcoming paper, but an outline is given here, as most supernova detections were obtained in Local Transient Survey data.
In between MeerKAT-triggered observations, the MeerLICHT telescope has been observing a set of Local Universe mass-overdensities out to $d=100\,\mathrm{Mpc}$. 
These fields were chosen to contain a significant fraction of the mass in the Local Universe and/or be of special interest to study stellar populations in the Southern hemisphere skies, such as the Magellanic Clouds and the Galactic Bulge region. In addition, a number of fields centered on sky areas intensely studied by several MeerKAT Large Survey Projects (LSPs) were defined, such as the LADUMA survey area \citep{2025ApJ...981..208K}, the ThunderKAT survey area \citep{2016mks..confE..13F,2022MNRAS.512.5037D}, as well as fields targeting the MHONGOOSE survey \citep{2024A&A...688A.109D} of nearby underluminous galaxies. The Local Transient Survey contains $139$ fields in the BlackGEM/MeerLICHT sky grid, covering $375$ square degrees in total.
%%%%%%%%%%%%%%%%%%%%%%%%%%%%%%%%%%%%%%%%%%%%%%%%%%%
\subsection{Image Processing \& Source Extraction}
%%%%%%%%%%%%%%%%%%%%%%%%%%%%%%%%%%%%%%%%%%%%%%%%%%%

All data from the MeerLICHT telescope is automatically processed by the \textsc{BlackBOX} data processing pipeline (\citealt{2021ascl.soft05011V}, Vreeswijk et al., in prep). In short, the pipeline performs standard de-biasing and flat-fielding steps. For this study, all data were processed with a previous build of the pipeline, where reduced images are astrometrically calibrated using Gaia DR2 stars. Typical astrometric residuals for the frame solution are at the $50\, \mu\mathrm{as}$ level in both coordinates. In the data described here, the photometric calibration was performed using Gaia DR2 $BP$ and $RP$ magnitudes, supplemented by {\sc Galex} \citep{2005ApJ...619L...1M,2009ApJ...694.1281B},  Pan-STARRS, SDSS, and 2MASS \citep{2006AJ....131.1163S} observations wherever available. Per frame, $50$-$150$ photometric standard stars were used to deduce a frame-averaged zeropoint. Typical zeropoint accuracies are $0.01\, \mathrm{mag}$ for the $g$, $q$, $r$, and $i$ bands and can reach up to $0.03\, \mathrm{mag}$ for the $u$ and $z$ bands. In particular, the $u$ band can show occasional larger zeropoint residuals due to a lack of {\sc Galex} \citep{2017ApJS..230...24B} data to anchor the spectrophotometric calibration of the reference stars. This holds in particular for low Galactic latitude fields, of which there are relatively few in this study. Uncertainties on the photometric measurement of each individual source are generally dominated by photon-counting noise, which is propagated through to the magnitude measurement. The final uncertainty is the statistical combination of the stochastic counting uncertainty and the systematic zeropoint uncertainty. All photometric measurements for MeerLICHT are in the AB system \citep{1990AJ.....99.1621O}. All astrometry is in the ICRS frame with epoch $2015.5$, derived from the Gaia DR2 astrometric frame. 

%%%%%%%%%%%%%%%%%%%%%%%%%%%%%%%%%%%%%%%%%%%%%%%%%%%
\subsection{Transient detections}
%%%%%%%%%%%%%%%%%%%%%%%%%%%%%%%%%%%%%%%%%%%%%%%%%%%

After a frame is fully astrophotometrically calibrated, it is compared to a reference image from the MeerLICHT All Sky Survey using the \textsc{ZOGY} image subtraction software (\citealt{2016ApJ...830...27Z}; Vreeswijk et al., in prep.). In the corrected-statistics, \texttt{Scorr} image, a transient candidate is identified when the \texttt{Scorr} pixel-value is $>$6, implying a 6-$\sigma$ detection if all noise contributions have been properly modeled. All 6-$\sigma$ detections are saved to a transient catalog for each frame, and submitted to the \textsc{MeerCRAB} real-bogus algorithm \citep{2021ascl.soft05009H,2021ExA....51..319H}. At this stage, a transient is assigned a real-bogus (RB) value between $0$ (bogus) and $1$ (real). Within the MeerLICHT consortium, RB-values $>$0.8 are generally considered to be real, although a small fraction of artefacts still find their way through the RB filter.\\
%%%%%%%%%%%%%%%%%%%%%%%%%%%%%%%%%%%%%%%%%%%%%%%%%%%
%\subsection{Forced photometry}
%%%%%%%%%%%%%%%%%%%%%%%%%%%%%%%%%%%%%%%%%%%%%%%%%%%

After the detection of a possible transient, a dedicated forced-photometry routine can be run on the sky coordinates of the transient. The advantage of the forced-photometry routine over the standard processing is that a user-defined signal-to-noise (SNR) limit can be set, and also that even when no significant source is present, a flux measurement is made at the location of the transient. In addition, the forced-photometry routine includes the measurement at the specified sky coordinates in the reference frame and the new reduced frame. For all transients described here, the forced-photometry routine was run on all available frames, with an SNR limit of 3. In some cases, the transient was present in (some of) the frames that went into making the reference frame, resulting in a positive detection of the transient in the reference frame and a negative detection in the template-subtracted frame. Hence applying a SNR cut allows for the flexibility to recover the transient flux (if significant) present in the reference frame. Thus, the resulting forced-photometry light curves have been corrected for the brightness of the source in the reference frame.\\
%%%%%%%%%%%%%%%%%%%%%%%%%%%%%%%%%%%%%%%%%%%%%%%%%%%
%\subsection{Subtraction of underlying emission}
%%%%%%%%%%%%%%%%%%%%%%%%%%%%%%%%%%%%%%%%%%%%%%%%%%%
Some supernovae are superimposed on detectable underlying emission, e.g., from H{\sc ii} regions or unresolved star clusters. In cases where there was significant emission more than $60$ days before the supernova explosion or a late-time $\left(\geq 365\;\mathrm{days}\right)$ emission consistent with pre-explosion emission, this baseline of emission has been subtracted from all intermediate detections. This baseline was calculated as a mean flux of no less than $3$ detections in each MeerLICHT band at $\geq60$ days before or $\geq365$ days after the SN discovery date.
%--> MeerLicht (ML) Telescope Overview
%--> (ML) telescope operations and technology
%-------------------------------------------------%

%%%%%%%%%%%%%%%%%%%%%%%%%%%%%%%%%%%%%%%%%%%%%%%%%%%
\section{Light curve fitting}\label{sec:methods}
%%%%%%%%%%%%%%%%%%%%%%%%%%%%%%%%%%%%%%%%%%%%%%%%%%%
We combined MeerLICHT data with ATLAS \citep{Tonry_2018,2020PASP..132h5002S} and ZTF \citep{Bellm_2019,2021AJ....161..141S} public data, which collectively offered sufficient early-time data to constrain the explosion time\footnote{Throughout this paper, we do not distinguish between explosion time and the first-light time (first photons escaping the SN ejecta), as has been proposed by some SNe Ia exhibiting a \textit{dark phase} between the explosion time and the time at which first photons emerge from the ejecta (e.g., \citealt{2013ApJ...769...67P}, \citealt{2014ApJ...784...85P}}, $t_0$, and derive the rest-frame rise time $t_{\mathrm{rise,z}}$ (rise time henceforth) of each SN presented in Section \ref{sec:well_sampled_sne}.
The phase, $\tau=t-t_{\mathrm{max}}$, in the light curves is defined as shifting every observation relative to the time of maximum brightness, $t_{\mathrm{max}}$, hence epochs before peak brightness have negative phases. 
Since we are primarily concerned with determining the light curve parameters $t_0$ and $t_\mathrm{rise,z} = t_{\mathrm{max}} - t_0$ (defined as the intrinsic time elapsed between the assumed time of maximum brightness and the estimated explosion epoch), from the early rising phase of the SNe, we have assumed that the photometric evolution of their early time data obeys a simple expanding fireball model \citep{1982ApJ...253..785A,1999AJ....118.2675R}. This fireball model is parameterised for the behaviour found in early-time SN light curves as,
\begin{equation}
\label{eq:1}
F_{\mathrm{opt}} = A \left(t - t_{0}\right)^2 + B\, ,
\end{equation}
for $t>t_0$, where all observed epochs are corrected for time-dilation ($1+z$) effects, and $A>0$, where the optical flux, $F_{\mathrm{opt}}$, is proportional to the square of the time elapsed ($t - t_{0}$) since the moment of the explosion. $B$ is a constant background term accounting for any residual flux present before the onset of the SN explosion. An additional parameter, $A$, is usually treated as a normalisation coefficient, which contains contributions from the mass, expansion velocity, $^{56}\mathrm{Ni}$ distribution and effective opacity of the ejecta \citep{1982ApJ...253..785A,2010ApJ...708.1025K,2013ApJ...769...67P,2000ApJ...530..744P}. 
This $t^2$ dependence (widely known as the "$t^2$-model") of optical flux has been used extensively for estimating the time of explosion in both thermonuclear and core-collapse SNe, with notable examples including; SN\,2010jn \citep{2013MNRAS.429.2228H}, SN\,2011fe \citep{2011Natur.480..344N,2012A&A...546A..12V,2013A&A...554A..27P,10.1093/mnras/stu077}, SN\,2012ht \citep{Yamanaka_2014}, and SN\,2018dz \citep{10.1093/mnras/staa2273}, SN\,1987A, SN\,2006aj, SN\,2008D \citep{COWEN201019}, respectively. 
With only three free parameters ($A$, $t_0$ and $B$) required for the $t^2$-model parametrisation, we fit equation \ref{eq:1} to time-dilation corrected data using a curve-fitting Python library \textsc{lmfit} \citep{newville.2025..15014437} within a specified set of constraints until the $\chi^2$ statistic is minimised. \\
The rising phase, during which the standard expanding fireball model is considered valid, is loosely defined. The $t^2$-model assumptions break down as the light curve approaches peak brightness and begins to roll over. Several SNe Ia studies have shown that the early-time fitting region where the $t^2$-model holds is within $\sim$$10-15\,\mathrm{days}$ before time of maximum light \citep{2015MNRAS.446.3895F,2024ApJ...974..164Y}, coincident with the time that the light curve reaches $40-50\%$ of its peak brightness value \citep{2015Natur.521..332O,2020ApJ...902...47M,2021ApJ...908...51F}. As demonstrated by \citet{2020ApJ...902...47M}, the choice of rise phase truncation threshold (e.g., 40\% and 50\% of the peak flux) impacts the inferred population-level parameters. Specifically, increasing the maximum flux fraction threshold generally leads to an increase in the inferred $t_{\mathrm{rise, z}}$.
Fitting was first performed in a temporal range of --30\,$\leq\tau\leq$\,--10 for each SN light curve, where $\tau$ covers the rise region of up to $10$ days before peak. This yielded the first $t_0$ estimate, which we then used to refit the light curve data in the range --30\,$\leq\tau\leq$\,--$\tau_{N}$\,while arbitrarily setting $\tau_{N}$ that ensures the latest ﬁtted epochs remain consistent with the best-ﬁt $t^2$-model. Varying $\tau_{N}$ for individual SNe allows for sufficient flexibility in fitting a maximum number of data points while defining a limit to epochs where the $t^2$ model diverges. Due to the varied observational cadence of our SN light curves, the maximum flux fraction included in our fits varied significantly on an object-by-object basis, ranging from 40.7\% to 97.4\%. We caution that the inferred $t_{\mathrm{rise, z}}$ for SNe with high maximum flux fraction values may be systematically influenced by the light curve's turnover. The maximum flux fraction included in the fit is denoted as $\left[f_{\mathrm{fit}}/f_{\mathrm{peak}}\right]$ in Table \ref{table:tsquared_fit_parameters} for each SN. In standard early-time light curve fitting of SNe, the recovered power-law index ($\sim$2 in the standard fireball model) depends on the specific filter being modeled, as the ejecta expands, cools, and its color evolves. As a result, the power-law rise across different optical bands is expected to exhibit systematic differences \citep{2020ApJ...902...47M,2024ApJ...974..164Y}. However, because individual transient surveys optimize for discovery rather than cadence, our early-time SN data is highly sparse and scattered across different passbands (MeerLICHT-$q$, ATLAS-$c, o$ and ZTF-$g, r$). To reliably reconstruct the early-time rise, we combine these detections into a single composite light curve. This combined dataset effectively acts as a broad, pseudo-filter comparable to the MeerLICHT $q$-band. By fixing the power-law index to 2, we enforced the classic expanding fireball approximation and imposed a single filter-independent parameter $B$ for the background flux. While this approach is a practical necessity to stabilize the fits of our under-sampled data, we acknowledge the inherent trade-off that a single $t^2$ rise assumes negligible early-time color evolution and prevents us from measuring filter-specific variations that trace the cooling of the ejecta. We caution that this approach may mask intrinsic population diversity if individual objects deviate from the $t^2$ rise (e.g., \citealt{Fausnaugh_2023}). \\
We used the \texttt{conf\_interval2d()} module in \texttt{lmfit} to calculate maps of $\chi^2$ values converted to probabilities around the minimum fit for a pair of key parameters $A$ and $t_0$, and visualised these as contour plots representing a confidence region. 
The uncertainty on $t_\mathrm{rise,z}$ is computed as the quadrature sum of errors for $t_0$ and $t_{\mathrm{max}}$, where the uncertainty on the explosion epoch largely dominates when the peak is well sampled. 
The date of peak brightness is measured by fitting a second-degree polynomial specifically to the data points near the peak region of the combined multi-survey light curve. For the majority of SNe, we contrast the absolute peak brightness and rise-time parameter with their average population class. To verify the robust determination of $t_0$, we performed a fit using night-binned data. The resulting parameters differed negligibly from the unbinned analysis (by less than 0.1 days), justifying our presentation of the model fits overlaid across both binned and unbinned data.
%%%%%%%%%%%%%%%%%%%%%%%%%%%%%%%%%%%%%%%%%%%%%%%%%%%
%--> Image Processing And Source Extraction
%--> Transient detections
%--> Forced photometry
%--> Subtraction of underlying emission
%--> Colour evolution
%--> fitting method to early SN light curves

%-------------------------------------------------%
%%%%%%%%%%%%%%%%%%%%%%%%%%%%%%%%%%%%%%%%%%%%%%%%%%%
\section{Supernovae Sample}\label{sec:supernovae_sample_selection}
\begin{figure}
    %\captionsetup{justification=centering}
     \centering
     \includegraphics[width=\columnwidth]{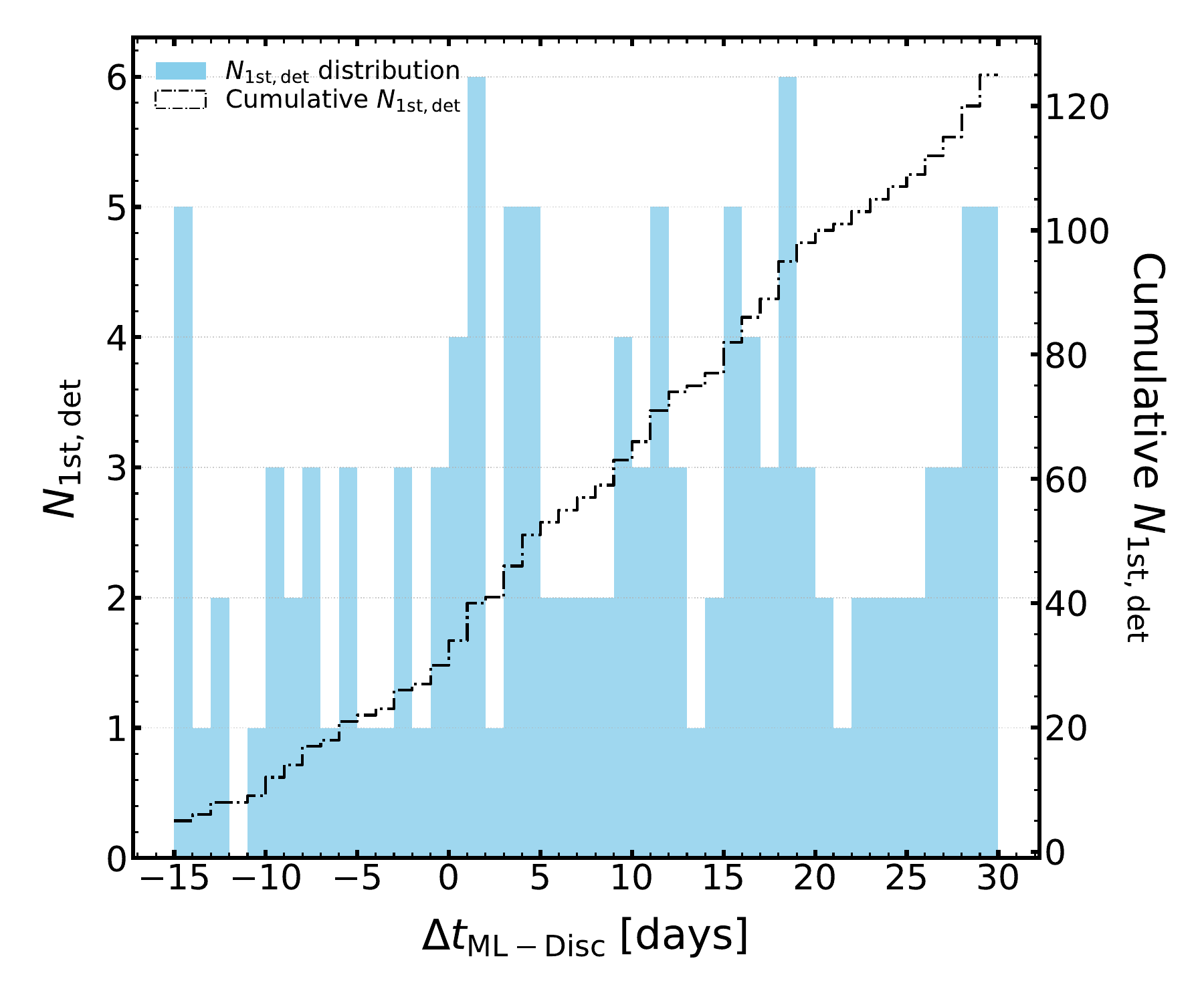}
     %--------------------------------------------%
     \caption{Distribution of SNe first (MeerLICHT) detections relative to the reported time of discovery on TNS. Blue bars show SN frequency of first detections within a bin-width equivalent to a day. The black "dash-dotted" line is a cumulative distribution showing the uniformity of the underlying distribution over the range of bins.}
     \label{fig:ML2TNSDiscoveryGap}
\end{figure}

\begin{figure}
    %\captionsetup{justification=centering}
     \centering
     \includegraphics[width=\columnwidth]{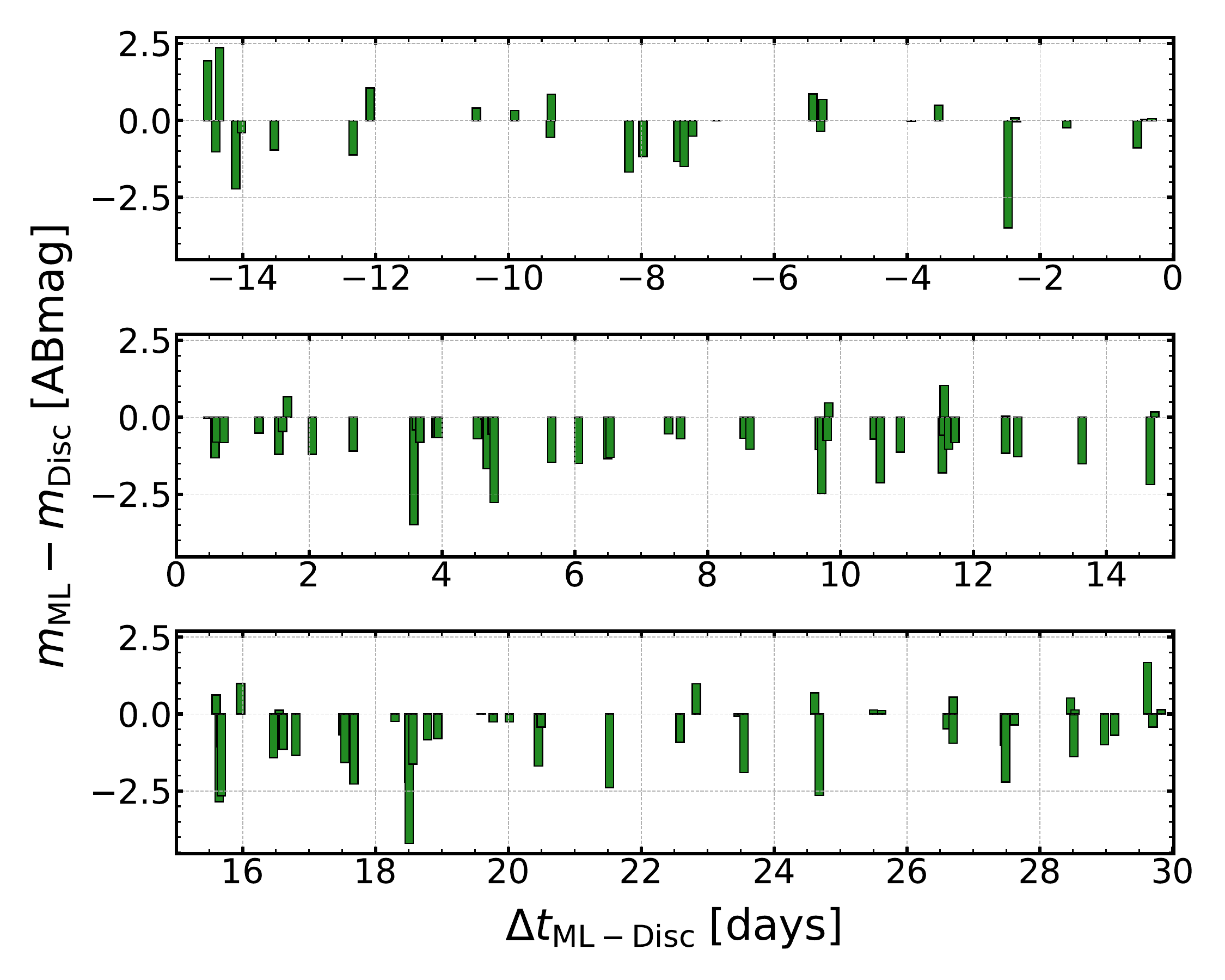}
     %--------------------------------------------%
     \caption{Distribution of the magnitude difference between MeerLICHT first $q$-band detection and reported discovery detection as a function of the gap to discovery date reported on TNS. 
     \textit{top}: First detections observed no more that $15$ days before TNS discovery date. 
     \textit{middle}: First detections observed between $0$ and $15$ days after TNS discovery date.
     \textit{bottom}: First detections observed between $15$ and $30$ days after TNS discovery date.}
     \label{fig:dist_DetMag_TNSMag_DiscoveryGap}
\end{figure}

%%%%%%%%%%%%%%%%%%%%%%%%%%%%%%%%%%%%%%%%%%%%%%%%%%%
To build our SN sample and simultaneously check SNe that MeerLICHT may have failed to observe, we first downloaded the entire catalog of spectroscopically classified transients from the Transient Name Server\footnote{\url{https://www.wis-tns.org/}} (TNS). Our initial selection included all transients reported within the first $5$-year survey period with a declination of $\delta < 30^{\circ}$. 
We further applied a temporal filter, retaining transients with MeerLICHT detections that fall within a time window of $-14$ to $+30$ days relative to the TNS-reported discovery date. Employing these criteria resulted in a parent sample of $2787$ unique TNS transients potentially observed by the Local Transient Survey. 
From this parent sample, we extracted MeerLICHT transient candidates with a positional offset of less than 5\arcsec\,from the reported TNS position and at least two detections within the specified time window of discovery. 
This cut removed two-thirds $(67.20\%)$ of transients from the parent sample. 
About $61.93\%$ of discarded transients can be attributed to unfavourable weather, observational coverage gaps, and MeerLICHT telescope downtime.  
This filtering process yielded a final sample of $914$ confirmed and independently observed transients. From this sample, we down-selected $139$ transients with spectroscopic classifications, of which $125$ were confirmed as supernovae (SNe). Thus, the final sample used in this work consists of $125$ archival MeerLICHT SNe, originally reported on the TNS between 2017-08-25 and 2022-10-15.\\
\\
The top section of Table \ref{table:ML_observed_transients} summarises our sample selection cuts in order of selection execution. Also presented in Table \ref{table:ML_observed_transients} is a breakdown of the $139$ spectroscopically classified transients co-discovered and observed during the survey period. 
A histogram of MeerLICHT $q$-band first detections of the $125$ confirmed SNe relative to their respective TNS-reported discovery date is shown in Figure \ref{fig:ML2TNSDiscoveryGap}. The overall distribution appears uniform, indicating that MeerLICHT detects SNe at any point in their evolution, regardless of their phase or SN-like time-scale within the specified discovery window. 
We also needed to know if MeerLICHT initially detected these SNe independent of their publicly reported discovery magnitude. To quantify our assessment for each SN, a bar chart of the difference in brightness between MeerLICHT $q$-band first detection and their respective discovery detection as a function of gap to discovery date is shown in Figure \ref{fig:dist_DetMag_TNSMag_DiscoveryGap}. This reveals that most SN events were preferentially discovered by MeerLICHT when they were already brighter than their respective discovery brightness in other extragalactic transient surveys. MeerLICHT's detection limit during its transient survey period was shallower than other transient surveys, but significant gaps between the occurrence of new SNe and their detection in any transient survey still existed at this time frame.\\
%%%%%%%%%%%%%%%%%%%%%%%%%%%%%%%%%%%%%%%%%%%%%%%%%%%
%---------------All transients by type------------%
%\setlength{\arrayrulewidth}{0.5mm}
%\setlength{\tabcolsep}{18pt}
%\setcounter{table}{0}
\begin{table}[ht!]
    \centering
    \caption{Transient category by main class.}
    \label{table:ML_observed_transients}
    % Use tabular* to force the table to fit the exact column width
    \begin{tabular*}{\columnwidth}{@{\extracolsep{\fill}} lr}
        \hline\hline
        \noalign{\smallskip}
        \textbf{Selection Cuts} & \textbf{Total} \\
        \noalign{\smallskip}
        \hline
        \noalign{\smallskip}
        Window to discovery time                       & 2787 \\
        At least 2 detections within 5$''$ radius      & 914  \\
        Spectroscopically classified                   & 139  \\
        Supernovae                                     & 125  \\
        \noalign{\smallskip}
        \hline\hline
        \noalign{\smallskip}
        \textbf{Class (Total = 139)}                   & \textbf{Observed} \\
        \noalign{\smallskip}
        \hline
        \noalign{\smallskip}
        SN Type Ia                  & 80 \\
        SN Type II                  & 23 \\
        SN Type Ia-91T-like         & 7  \\
        SN Type Ic                  & 5  \\
        Variable star               & 4  \\
        SN Type IIP                 & 3  \\
        Cataclysmic variable star   & 3  \\
        SN Type IIn                 & 2  \\
        Active Galactic Nucleus     & 2  \\
        Other                       & 2  \\
        SN Type Ia-91bg-like        & 1  \\
        SN Type Ib                  & 1  \\
        SN Type Ibn                 & 1  \\
        SLSN Type I                 & 1  \\
        SLSN Type II                & 1  \\
        Nova                        & 1  \\
        Luminous Red Nova           & 1  \\
        Light-Echo                  & 1  \\
        \noalign{\smallskip}
        \hline
    \end{tabular*}
    \tablecomments{\textit{\sl Top:} Table summarises the sample selection. \textit{\sl Bottom:} Table breaks down the observed transients by their classification. The ``Observed'' column represents the number of objects recovered from the MeerLICHT transient database in each category. The classification ``Other'' refers to a category of transients with extensive spectroscopic follow-up but cannot be associated with any of the established transient classes or subclasses and likely belong to a new category of transient events (e.g., AT\,2018cow).
    }
\end{table}
%---------------All transients by type------------%
Table \ref{table:ML1_sne} presents SNe information detected by MeerLICHT over the five-year transient survey period.
The supernova IAU names, coordinates, supernova type, redshift, discovery magnitude, reporting groups, discovery filter, and discovery dates were extracted from the TNS website. The discovery date for the SN is the date of first detection by the reporting group. The table also includes their discovery filter and discovery internal names, in addition to the presented IAU Designation names.
Additionally, we obtained host galaxy redshifts, which we used to estimate luminosity distances by adopting a local value of the Hubble constant as $H_0=73.04\, \mathrm{km}\,\mathrm{s^{-1}}\,\mathrm{Mpc^{-1}}$ \citep{Riess_2022} and thereby computing the observed absolute magnitudes light curves. 
Host names, redshifts, extinctions, and offsets were all collected from NED\footnote{\url{https://ned.ipac.caltech.edu/}}. For cases where host offset is not provided in NED, we manually use the Aladin Sky Atlas\footnote{\url{https://aladin.cds.unistra.fr/}} \citep{2000A&AS..143...33B} \texttt{dist} tool to measure the separation between the SN astrometric position and the assumed associated host galaxy nucleus.\\
\\ 
We prioritized spectroscopically measured host redshifts and only used the photometric host redshifts wherever the spectroscopic redshifts were unavailable in the NED catalogue.
We adopt the host redshifts from the extended version of the Galaxy List for the Advanced Detector Era (GLADE+, \citet{10.1093/mnras/sty1703,10.1093/mnras/stac1443}) using the VizieR catalogue access tool \citep{10.26093/cds/vizier,vizier2000} for SNe where the host redshifts are unavailable in NED.
For instances where redshift-independent host galaxy distance measurements exist, we adopt the mean of the distances cataloged in the NASA/IPAC Extragalactic Database (NED) and take one standard deviation ($1\sigma$) from this mean as the uncertainty in the host distance, which is used to derive the absolute peak brightness and its associated uncertainty where applicable. In cases where redshift-derived distances are used, they include corrections for the cosmic microwave background (CMB) reference frame, as well as the peculiar velocity field \citep{2020AJ....159...67K,2024NatAs...8.1610V} while accounting for a typical systematic uncertainty of $150\,\mathrm{km\,s^{-1}}$ introduced by peculiar velocity field dispersion. In instances where only photometric redshifts are available, their poor accuracy leads to large measurement errors in distance estimates, hence we derive distances from the redshift adopted from the SN classification spectrum. Since redshifts derived from SN classification spectra often lack reported uncertainties, we used the transient classification software \textsc{snid-sage} \citep{Stoppa2026} to derive the uncertainties from spectra published on TNS and propagate those uncertainties in accordance with distance and absolute magnitude calculations. 
%%%%%%%%%%%%%%%%%%%%%%%%%%%%%%%%%%%%%%%%%%%%%%%%%%%
%%%%%%%%%%%%%%%%%%%%%%%%%%%%%%%%%%%%%%%%%%%%%%%%%%%
%%%%%%%%%%%%%%%%%%%%%%%%%%%%%%%%%%%%%%%%%%%%%%%%%%%
%%%%%%%%%%%%%%%%%%%%%%%%%%%%%%%%%%%%%%%%%%%%%%%%%%%
%-------------------------------------------------%
% custom command to title images
%-------------------------------------------------%
\newcommand*{\figuretitle}[1]{%
    {\centering
    %   <--------  will only affect the title because of the grouping (by the
    \textbf{#1}
    %              braces before \centering and behind \medskip). If you remove
    \par\medskip}
    %            these braces the whole body of a {figure} env will be centred.
}
%%%%%%%%%%%%%%%%%%%%%%%%%%%%%%%%%%%%%%%%%%%%%%%%%%%
%--> Supernovae Sample Construction

%-------------------------------------------------%
%%%%%%%%%%%%%%%%%%%%%%%%%%%%%%%%%%%%%%%%%%%%%%%%%%%
\section{Supernovae Light Curves}\label{sec:results}
%%%%%%%%%%%%%%%%%%%%%%%%%%%%%%%%%%%%%%%%%%%%%%%%%%
\noindent The majority of light curves have a limited number of detections in the MeerLICHT transient database, owing to extremely sparse coverage. However, a small number of cases warrant special mention. 
Fourteen of the detected supernovae are located in the fields of the MeerLICHT Local Transient Survey and, therefore, have extensive coverage. Here, we summarise the key published results for each SN in this category and complement them with our MeerLICHT light curve data.
For all light curves presented in this work, the inverted triangles represent the transient upper limit at the position of the SN in the MeerLICHT full-frame image, and the associated colour of the inverted triangles corresponds to the filter in which the upper limit was determined.
The dashed vertical line specifies the earliest SN discovery date (UTC dates and timestamps are used throughout this paper) published on TNS.
Peak luminosity measurements are corrected for the Milky Way extinction $A_{\textrm{V,G}}$, but not host-galaxy extinction $A_{\textrm{V,h}}$, nor for the expansion of the universe ($K$-correction). Regarding extinction correction for the Milky Way, we adopted color excess $E(B-V)$ values from all-sky infrared dust maps \citep{1998ApJ...500..525S,2011ApJ...737..103S} along the SN line of sight, assuming a \citealt{2023ApJ...950...86G} extinction law with $R_V=3.1$.
The effect of the K-correction is negligible due to the low redshift range ($z<<0.1$) of our well-sampled SNe data \citep{2015ApJ...799..208S}. 
As MeerLICHT often subsequently observes the sample field in multiple filters, the color characteristics and temporal evolution of the transient source can be traced. The most common combination of filters is $u$, $q$, and $i$, allowing for the construction of the $u-q$ and $q-i$ colours. Occasionally, observations will also be taken in the $g$, $r$, and $z$ filters. As the $q$-band filter is most often used and is always present in any combination of the other filters, we have constructed pseudo-simultaneous colours by associating any filter with the nearest $q$ band observation, within a limit of no more than $0.1$ days time difference.\\
\\
ATLAS $o$- and $c$-band photometry was acquired using the ATLAS forced photometry server\footnote{\url{https://fallingstar-data.com/forcedphot/}} \citep{2020PASP..132h5002S,2021TNSAN...7....1S}, where we specified the SN data to be in the time window $\left(30\, \mathrm{days} \right)$ before and $\left(1000\, \mathrm{days} \right)$ after the SN discovery date. ZTF $g\;,r$ and $i$-band photometry was obtained from the ALeRCE ZTF explorer \citep{2021AJ....161..242F,2021AJ....161..141S} using the ALeRCE client\footnote{\url{https://alerce.readthedocs.io/en/latest/index.html}}\citep{2021AJ....162..231C}, provided the SN had a ZTF Object identifier ($\texttt{oid}$). 
To ensure that the apparent magnitudes from MeerLICHT forced photometry are consistent with the retrieved apparent magnitudes from ATLAS and ZTF queries, we performed ATLAS forced photometry on reduced ATLAS images and used the corrected apparent magnitudes computed by the ALeRCE pipeline. Since ZTF includes PSF-fit forced-photometry in the public stream up to 30 days before a given transient detection, we followed the prescription in \citet{2023arXiv230516279M} to calculate the magnitude upper limits for low SNR epochs when applicable.\\
\\
To ensure the statistical confidence of the photometric detections for each SN from public data, our primary requirement was that individual flux measurements be statistically robust rather than marginal. We calculated the signal-to-noise ratio (SNR) for each observation from the ratio of the measured flux $F_{\nu}$ to its photometric error $\delta F_{\nu}$ and retained only those points with $\mathrm{SNR}\geq3$. Additionally, to reduce isolated scatter in the SN light curves, we required detections to be brighter than the epoch's $5\sigma$ limiting magnitude. Together, these criteria ensure that the background-subtracted fluxes used in our analysis represent physically significant detections rather than background noise.
Due to the diverse nature of our multi-survey data, we had to minimise the different survey systematics (i.e., filter response, depth) that may contaminate the early-time light curve model fitting. Therefore, our model only employed ATLAS ($c$ and $o$) and ZTF ($g$ and $r$) broadband filters as they overlap in bandpass with MeerLICHT $q$-band. 
Furthermore, a well-observed consequence of the cooling SN ejecta is that bluer bands (e.g., $u$ or $g$) usually peak before redder bands (e.g., $r$ or $i$), which may introduce additional uncertainties in $t_{\mathrm{rise,z}}$ estimates. However, we emphasize that due to inadequate photometric coverage, many SNe light curves in this work have gaps around the rise, peak, and/or decline region, as is often the case for SNe without dedicated follow-up campaigns. Our approach to early-time model fitting combines observations from multiple surveys with similar bandpasses to recover the overall SN light curve shape and its characteristic parameters. Therefore, capturing the wavelength-dependent temporal evolution of the peak by isolating and fitting individual photometric bands is unfeasible for this study.
%-----------estimated-parameters-table------------%
% Force the next table to be numbered 3
%\setcounter{table}{2}
\begin{table}[ht!]
\centering
\caption{$t^2$-model fit parameters.}
\label{table:tsquared_fit_parameters}
% Ensures the table fits the narrow A&A column width (approx 88mm)
\resizebox{\columnwidth}{!}{
    \begin{tabular}{llcccccc}
    \hline\hline
    \noalign{\smallskip}
    Name & Type & $t_0$ & $t_{\mathrm{rise, z}}$ & DOF & $\chi^2_{\mathrm{R}}$ & $t_{\mathrm{peak}}$ & $\frac{f_{\mathrm{fit}}}{f_{\mathrm{peak}}}$ \\
     &  & (MJD) & (days)  &  &  & (MJD) \\
    \noalign{\smallskip}
    \hline
    \noalign{\smallskip}
    2019ano  & Ia  & $58504.35\pm0.69$ & $22.34\pm0.67$ & 55  & 3.57 & $58528.21\pm0.19$ & 77.5\% \\
    2020oi  & Ic  & $58853.40\pm0.40$ & $13.68\pm0.87$ & 51   & 20.78  & $58867.15\pm0.77$ & 67.1\% \\
    2020ftl  & Ia  & $58938.47\pm0.23$ & $18.40\pm0.37$ & 43  & 4.12  & $58957.00\pm0.29$ & 83.4\% \\
    2020jfo  & IIP & $58970.83\pm0.17$ & $10.92\pm0.20$ & 64  & 5.39 & $58981.81\pm0.11$ & 97.4\%\\
    2020noz  & IIn & $59025.17\pm0.63$ & $45.96\pm7.76$ & 128  & 3.63 & $59072.28\pm7.93$ & 52.1\% \\
    2021koq  & Ia  & $59326.42\pm0.32$ & $19.98\pm0.45$ & 121  & 3.74  & $59346.81\pm0.33$ & 65.1\% \\
    2021tkm  & Ia  & $59406.00\pm0.11$ & $17.80\pm0.12$ & 90  & 1.12  & $59423.92\pm0.05$ & 40.7\% \\
    2021aele & Ia  & $59516.17\pm0.71$ & $21.19\pm0.70$ & 261 & 1.50  & $59538.99\pm0.23$ & 92.0\% \\
    2022ame  & II  & $59602.43\pm0.24$ & $13.50\pm0.38$ & 54  & 3.95  & $59616.01\pm0.29$ & 61.3\% \\
    2022hrs  & Ia  & $59681.93\pm0.06$ & $17.49\pm0.16$ & 90  & 5.64  & $59699.50\pm0.15$ & 51.5\% \\
    2022vrr  & Ia  & $59838.55\pm1.39$ & $18.56\pm1.37$ & 16   & 2.04  & $59858.28\pm0.45$ & 88.3\% \\
    %2019lub  & IIn & \textemdash       & \textemdash & \textemdash & \textemdash & \textemdash \\
    %2022bll  & IIP & \textemdash       & \textemdash & \textemdash & \textemdash & \textemdash \\
    %2022phj  & II  & \textemdash       & \textemdash & \textemdash & \textemdash & \textemdash \\
    \noalign{\smallskip}
    \hline
    \end{tabular}
}
\tablecomments{SNe (2019lub, 2022bll, 2022phj) with no parameter estimates are not listed here because they lack early-time data to perform $t^2$ model fitting. The $t_0$ and $t_{\mathrm{peak}}$ are in Modified Julian Days (MJD) while $t_{\mathrm{rise, z}}$ is in SN rest-frame days.}
\end{table}
%-----------estimated-parameters-table------------%
%%%%%%%%%%%%%%%%%%%%%%%%%%%%%%%%%%%%%%%%%%%%%%%%%%%
\subsection{Well-sampled SNe}\label{sec:well_sampled_sne}
Following the cuts in section \ref{sec:supernovae_sample_selection} to limit the recovered transient sample to SNe, we focus on the $14$ well-sampled SNe with sufficient multi-survey coverage at epochs earlier than $10$ days or more, in the observer frame, before reaching peak brightness. 
In this section, we provide a brief overview of each SN and the consolidated light curve data from two contemporary transient surveys. The resulting fitted parameters from the $t^2$ modeling and their $1\sigma$ uncertainties enclosing $68.3\%$ confidence interval are provided in Table \ref{table:tsquared_fit_parameters}. 
SNe are discussed in chronological order, and nightly-binned rise-region data are displayed in plots with $t^2$-model fits for comparative purposes.
%%%%%%%%%%%%%%%%%%%%%%%%%%%%%%%%%%%%%%%%%%%%%%%%%%%
    %%%%%%%%%%%%%%%%%%%%%%%%%%%%%%%%%%%%%%%%%%%%%%%
    %%%%%%%%%%%%%%%%%%%%%%%%%%%%%%%%%%%%%%%%%%%%%%%
    \subsubsection{SN 2020oi (Ic)}
    %%%%%%%%%%%%%%%%%%%%%%%%%%%%%%%%%%%%%%%%%%%%%%%
    SN\,2020oi was discovered by ZTF on 2020 January 07, UT 13:00:54 at an apparent $r-$band magnitude of $17.28\,\mathrm{mag}$ and $4.58$ hours later reported to the TNS by the ALeRCE broker \citep{2020TNSTR..67....1F}. The discovery report by \citet{2020TNSTR..67....1F} also mentions a non-detection limiting magnitude of $r>20.52\,\mathrm{mag}$, obtained $\sim$$3\,\mathrm{days}$ before the first detection.
    It was then classified as a Type Ic following an optical spectrum obtained within $1.8$ days of discovery, using the Goodman spectrograph at Southern Astrophysical Research Telescope (SOAR) Observatory \citep{2020TNSCR..90....1S}. SN\,2020oi is located near the nucleus of the intermediate spiral galaxy M\,100, with a host galaxy redshift of $z = 0.005240$ \citep{2014MNRAS.440..696A}. 
    It was also discovered $\sim$10 days before it reached maximum brightness in the ZTF $g$-band, triggering a rapid-response follow-up with the Swift UltraViolet and Optical Telescope (UVOT), which revealed a rising ultraviolet source \citep{2020TNSAN...8....1H}. As reported on TNS, the SN was also subsequently detected by other optical imaging surveys: ATLAS, the Young Supernova Experiment (YSE using PanSTARRS, \citealt{2020TNSTR1009....1C}), and Gaia Alerts \citep{2021A&A...652A..76H}.\\
    %\footnote{\url{https://www.wis-tns.org/object/2020oi}}
    %%%%%%%%%%%%%%%%%%%%%%%%%%%%%%%%%%%%%%%%%%%%%%%%%%
    \begin{figure}
    \centering
    \begin{minipage}{0.491\textwidth}
    \centering
    \includegraphics[width=1.0\linewidth,trim={0.0cm 0.0cm 0.0cm 0.0cm},clip]{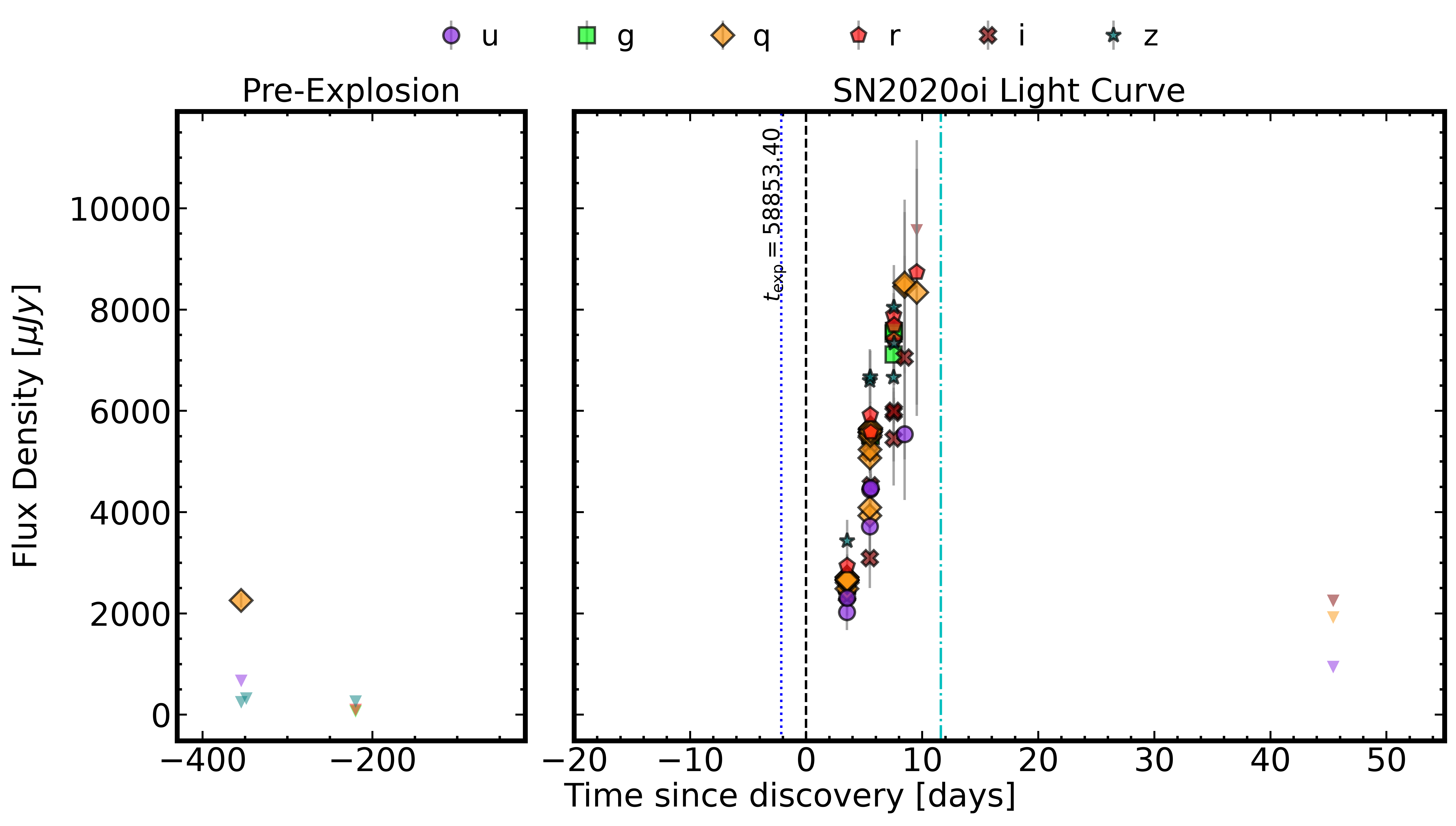}
    \includegraphics[width=1.0\linewidth,trim={0.0cm 0.0cm 0.0cm 0.0cm},clip]{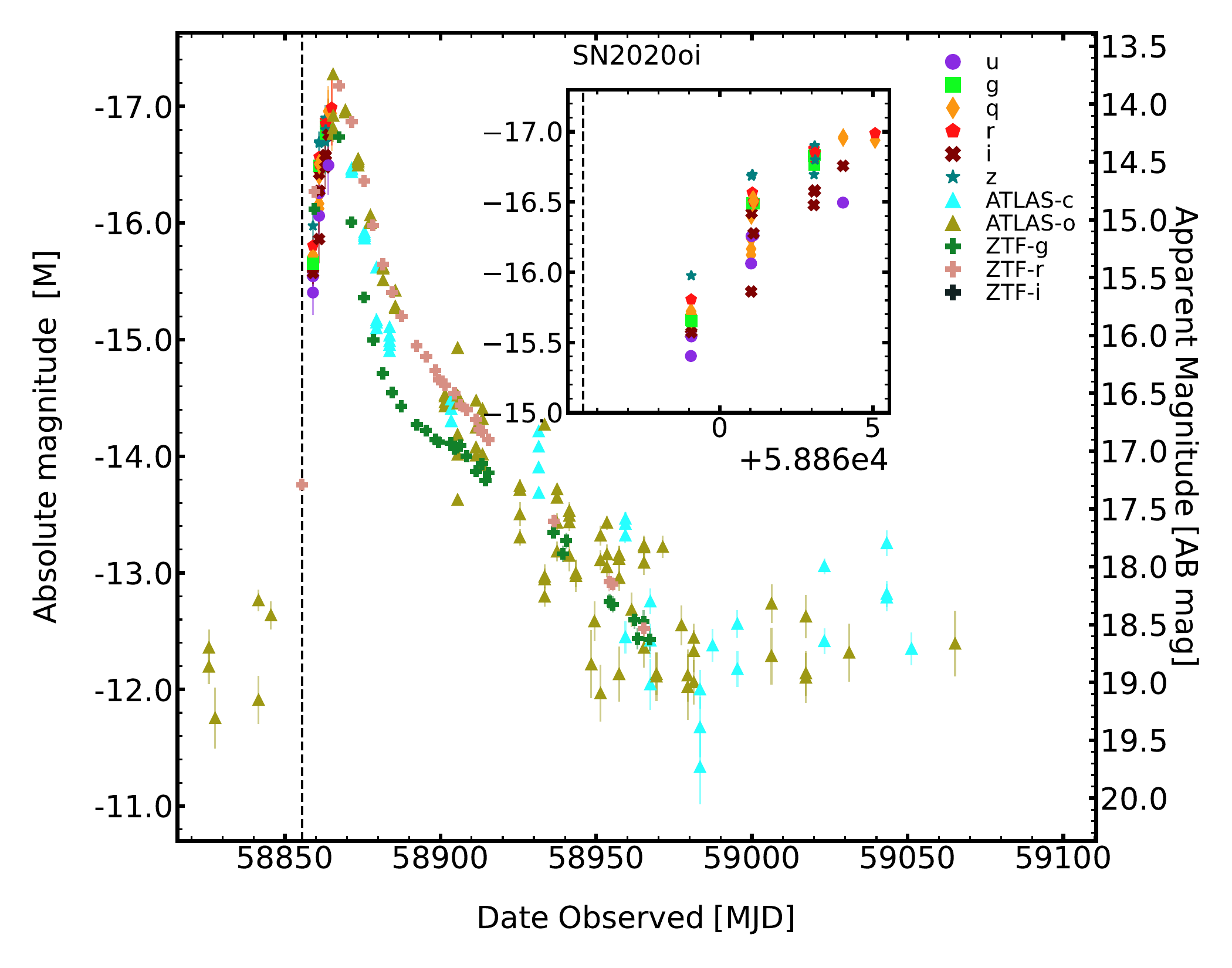}
    \end{minipage}
    \caption{
    Pre- and post-explosion light curves of SN\,2020oi from MeerLICHT.
    {\sl Top left}: Pre-explosion photometry at roughly 350 days before discovery. The shaded inverted triangles are upper limits, and all the other filled symbols are detections. 
    {\sl Top right}: Multi-band optical light curves with the underlying background emission subtracted. The phases are relative to the time of discovery as reported on TNS, marked by the black vertical dashed line. The estimated explosion epoch and peak brightness epoch are designated with the dotted blue line and dash-dotted cyan line, respectively.
    {\sl Bottom}: Absolute magnitudes light curves from MeerLICHT, ATLAS, and ZTF. Filled triangles represent ATLAS magnitudes, and ZTF corrected magnitudes are represented by filled `plus' symbols. A zoom into the early rise of SN\,2020oi is plotted in the insert window, showing the MeerLICHT observations of the relatively fast-rising phase of the SN. 
    }
    \label{fig:SN2020oi}
    \end{figure}
    %%%%%%%%%%%%%%%%%%%%%%%%%%%%%%%%%%%%%%%%%%%%%%%
    %--------------------color-curve--------------%
    %%%%%%%%%%%%%%%%%%%%%%%%%%%%%%%%%%%%%%%%%%%%%%%
    \begin{figure}
    \centering
    \begin{minipage}{0.491\textwidth}
    \includegraphics[width=1.0\linewidth,trim={0.0cm 0.0cm 0.0cm 0.0cm},clip]{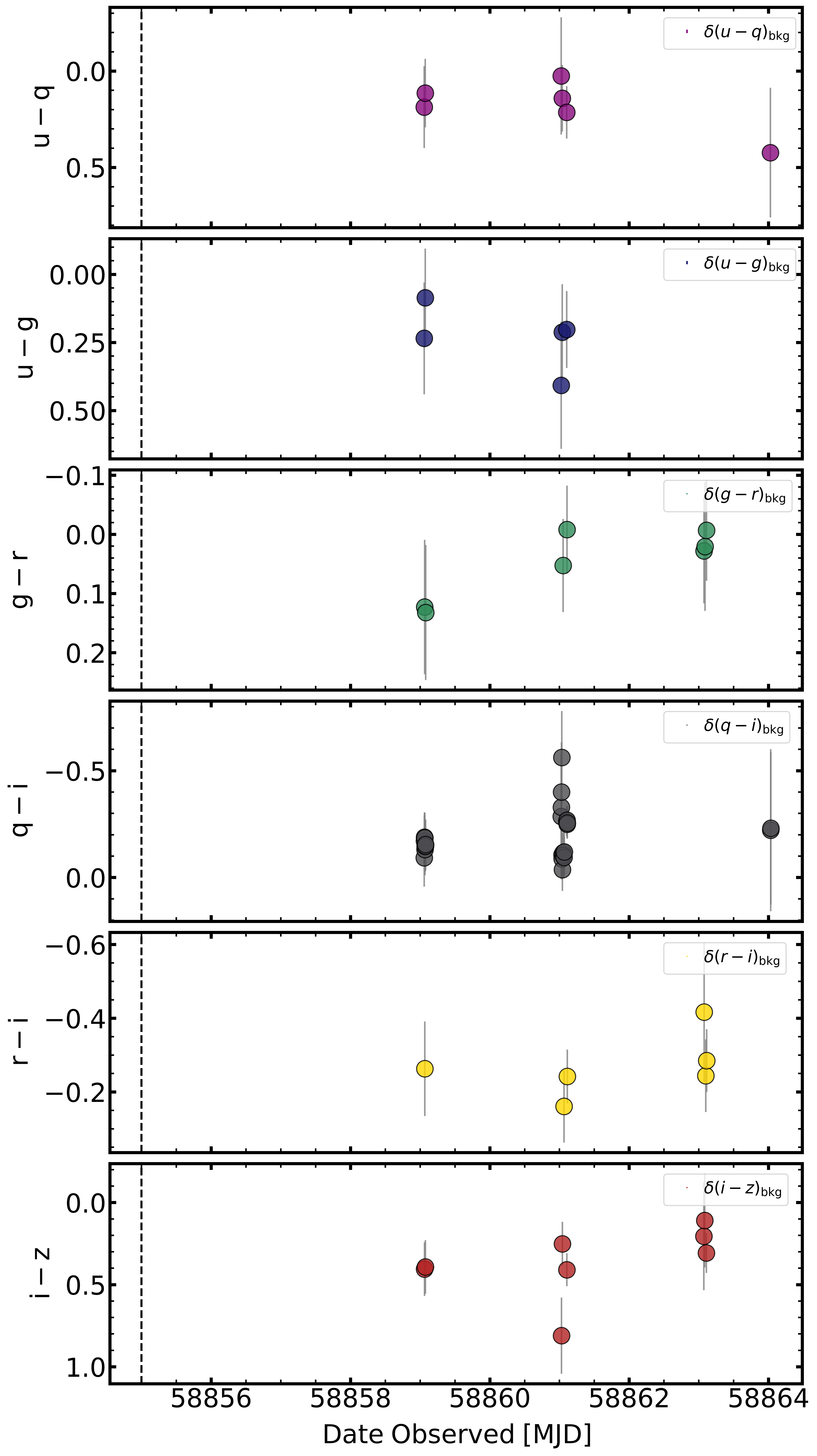}
    \end{minipage}
    \caption{
    Extinction corrected ($A_{V,\mathrm{G}}=0.073$) color evolution of SN\,2020oi from MeerLICHT observations. $\delta (color)_{\mathrm{bkg}}$ represents the magnitude of the systematic uncertainty in the computed color index. 
    }
    \label{fig:SN2020oi_color}
    \end{figure}
    %----------------------------------------------%
    MeerLICHT's earliest detection of SN\,2020oi was in the $u$-band, approximately $3.5$ days after the published TNS discovery date. The bottom panel of Figure\,\ref{fig:SN2020oi} shows the ensuing MeerLICHT multiband light curves of the early rising phase of SN\,2020oi, complemented by post-maximum data from ZTF and ATLAS. The earlier ATLAS detections $30-10\,\mathrm{days}$ before discovery are consistent with the background emission from the host. Adopting a redshift independent distance $d=16.117\pm3.032\,\mathrm{Mpc}$, the ATLAS $o$-band light curve reached an absolute peak magnitude at $-17.35\pm0.41\,\mathrm{mag}$, in agreement with the mean absolute peak magnitude of normal type Ic SNe ($M_B=-17.66\pm0.40\,\mathrm{mag}$, \citealt{2014AJ....147..118R}).
    While post-peak coverage by ATLAS and ZTF extends up to $\sim$$100$ days, early MeerLICHT data provides a rich multiband photometry in $u,\,g,\,r,\,i,\,z$ and $q$, spanning $6.02\,\mathrm{days}$ of the rise phase. 
    Benefiting from this extensive cadence at early times, we constructed color evolution light curves of the early rising region of the SN as illustrated in Figure\,\ref{fig:SN2020oi_color}. During this early rising phase, all colors appear to be relatively flat, with an average of $(u-q)=0.20$, $(u-g)=0.23$, $(g-r)=0.03$, $(r-i)=-0.27$, $(q-i)=-0.19$, and $(r-z)=0.29$.
    %%%%%%%%%%%%%%%%%%%%%%%%%%%%%%%%%%%%%%%%%%%%%%%
    %--------------------Modelling----------------%
    %%%%%%%%%%%%%%%%%%%%%%%%%%%%%%%%%%%%%%%%%%%%%%%
    \begin{figure}[!ht]
    %\captionsetup{justification=centering}
    %\figurenum{1} % Optional: specific figure number
     \centering
     \includegraphics[width=0.5\columnwidth]{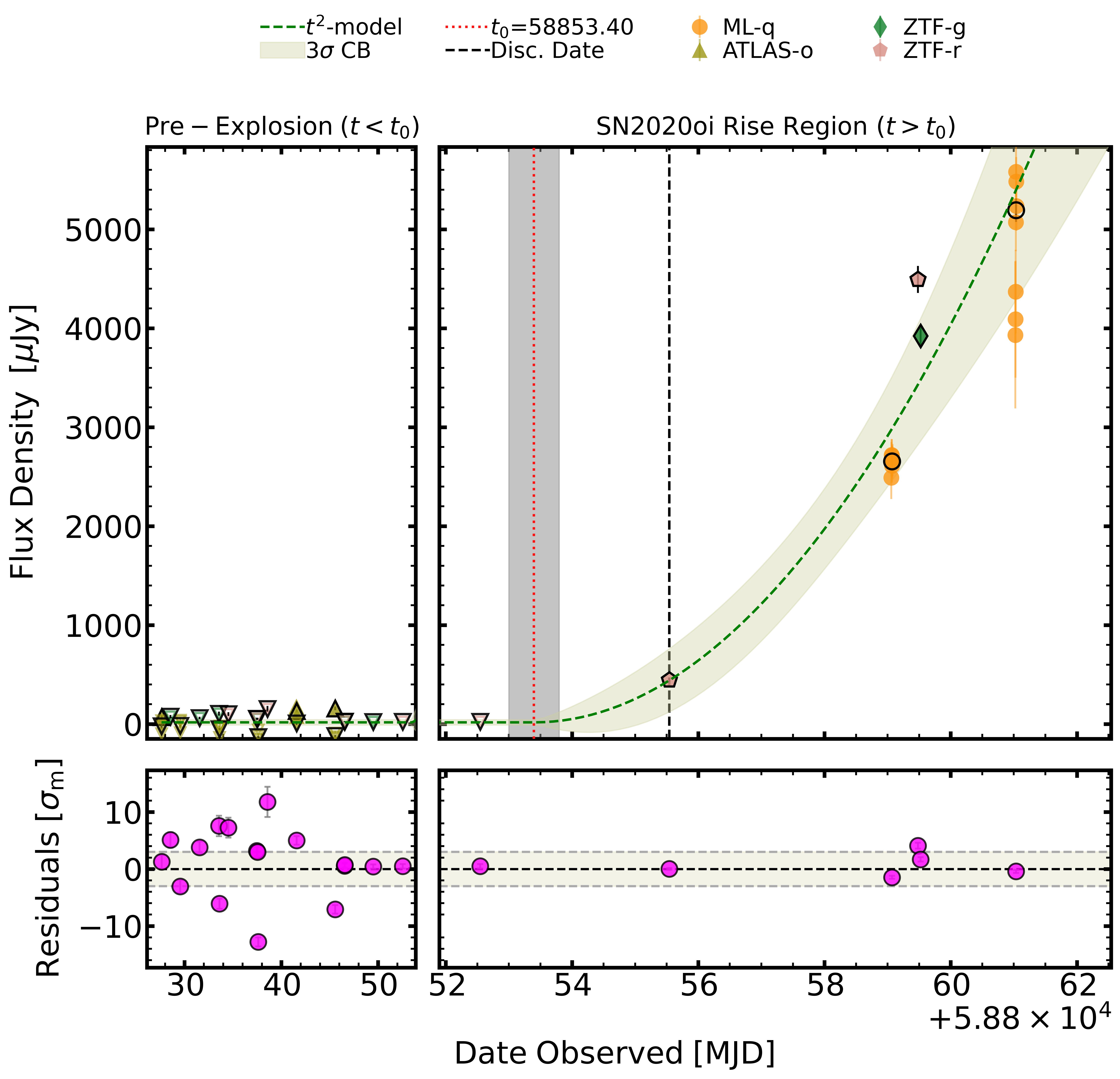}
     \includegraphics[width=0.5\columnwidth]{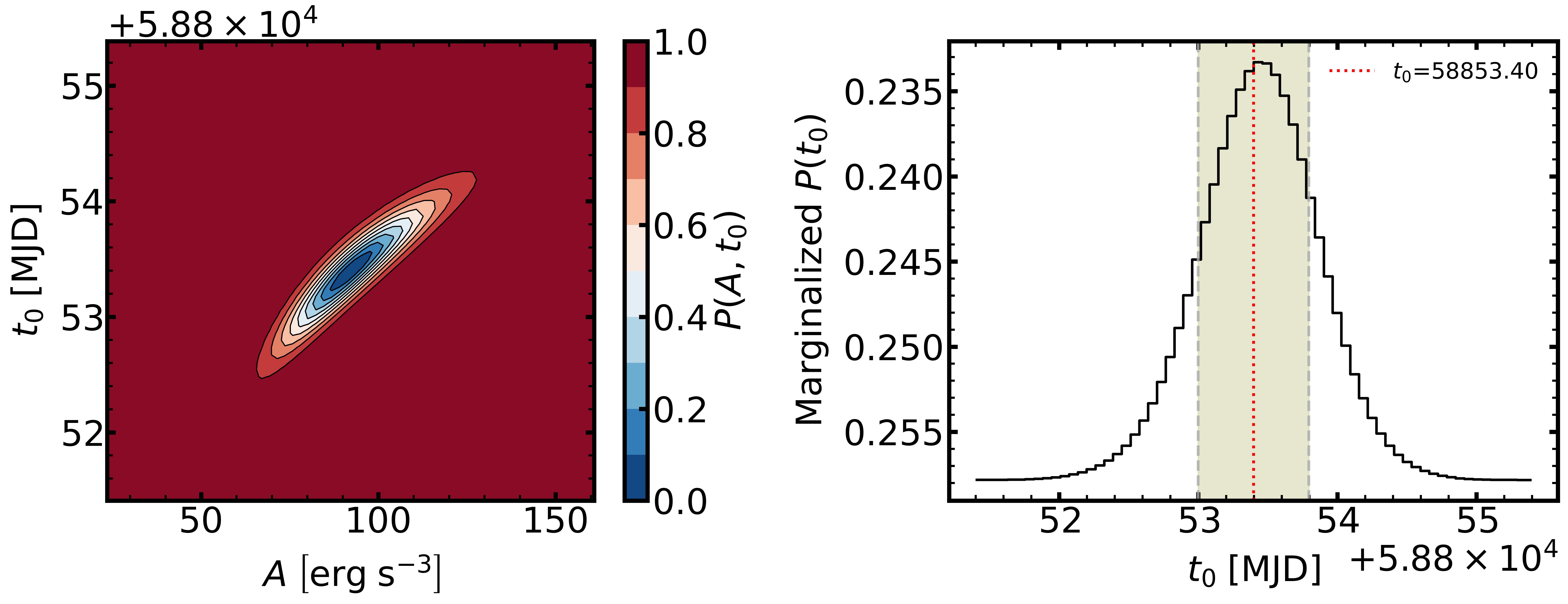}
     \caption{
     {\sl Top}: Early light curve data with best fitting $t^2$ model (dashed green line) for the rise time region of SN\,2020oi. The grey-shaded $3\sigma$ confidence interval represents the uncertainty in the model fit. The open black symbols indicate binned data where applicable. 
     {\sl Middle}: Residuals (normalized by model uncertainty) for the best fit computed from binned data. 
     {\sl Bottom left}: $\chi^2$ probability distribution contours from the best-fitting pair of parameters $A$ and $t_0$. 
     {\sl Bottom right}: A normalized probability distribution of $t_0$ marginalized along $A$. The red dotted vertical line represents the best-fitted $t_0$, and the filled region between grey lines on either side of $t_0$ represents its $1\sigma$ uncertainty.
     }
     \label{fig:SN2020oi_t2model}
    \end{figure}
    %%%%%%%%%%%%%%%%%%%%%%%%%%%%%%%%%%%%%%%%%%%%%%%%%%
    A composite MeerLICHT-$q$ and ZTF-$g, r$ band light curve used to model the initial early rise of SN\,2020oi is shown in the top panel of Figure\,\ref{fig:SN2020oi_t2model}, along with the $t^2$ model line of best fit extending up to $\tau=-6.11\,\mathrm{days}$. 
    Although this light curve is undersampled at the beginning of the rise, the model provides a reasonable fit, yielding an explosion date of $t_0=58853.40\pm0.40\,\mathrm{MJD}$ and a corresponding rise time of $t_{\text{rise, z}}=13.68\pm0.87\,\mathrm{days}$. This $t_{\text{rise, z}}$ value is consistent with the rise times of helium-poor (Ic and Ic-BL) SNe reported in other studies \citep{2009A&A...508..371H,2011MNRAS.416.3138V,2015A&A...574A..60T}, while $t_0$ agrees with the value used by \citet{Rho_2021}. 
    Even though the quality of the fit is statistically poor, with a reduced chi-squared of $\chi^2_R=20.78$, the $\chi^2$ contour map (bottom-left of Figure\,\ref{fig:SN2020oi_t2model}) reveals a stable confidence region with strong correlation between the fitted parameters. Furthermore, the marginalised probability distribution (bottom-right) of a range of fitted explosion times peaks at the best fit value, indicating a well-constrained $t_0$.
    %%%%%%%%%%%%%%%%%%%%%%%%%%%%%%%%%%%%%%%%%%%%%%%%%%
    \subsubsection{SN 2020ftl (Ia)}
    %%%%%%%%%%%%%%%%%%%%%%%%%%%%%%%%%%%%%%%%%%%%%%%%%%
    SN\,2020ftl was discovered by the CRTS on 2020 April 2, UT 06:23:46 at an apparent $V$-band magnitude of $18.5$ \citep{2020TNSTR.942....1C}, and subsequently detected and reported by other optical surveys including ATLAS, Pan-STARRS, ZTF with the AMPEL \citep{2019A&A...631A.147N} broker and Gaia Alerts. SN\,2020ftl is located in the galaxy NGC\,4277, with a host redshift of $z=0.007295$ \citep{2016A&A...595A.118V} corresponding to a luminosity distance of $d=33.38\pm2.13\,\mathrm{Mpc}$. 
    The proximity of its host galaxy prompted immediate photometric follow-up from many amateur astronomers, while professional surveys began spectroscopic monitoring shortly after the discovery was reported. 
    An initial optical spectrum, acquired on 2020 April 9 with the $0.2\,\mathrm{m}$ Newton telescope as part of the Italian Supernovae Search Project, was used to classify SN\,2020ftl as a Type Ia \citep{2020TNSCR1001....1B}.
    As shown by the multiband light curves in Figure\,\ref{fig:SN2020ftl}, MeerLICHT's first detection of SN\,2020ftl is $\sim$$9\,\mathrm{days}$ after the reported discovery date. Additionally, there appear to be MeerLICHT observations roughly $\sim$$18-40\,\mathrm{days}$ before the discovery date which are consistent with the underlying background emission. ATLAS data filled in the gap between the underlying background emission and the earliest MeerLICHT $q$-band detection of SN\,2020ftl. 
    The earliest ATLAS $o$-band detection at $\sim$$5.0\,\mathrm{days}$ before discovery is also consistent with background emission.
    Benefiting from the well-sampled MeerLICHT data around maximum light (as shown in the zoom-insert of Figure\,\ref{fig:SN2020ftl}), SN\,2020ftl reached an absolute maximum brightness of $-18.30\pm0.14\,\mathrm{mag}$ in the $q$-band, similar to $-18.45\pm0.14\,\mathrm{mag}$ in the ZTF $g$-band. 
    %\footnote{\url{https://www.wis-tns.org/object/2020ftl}}
    %%%%%%%%%%%%%%%%%%%%%%%%%%%%%%%%%%%%%%%%%%%%%%%
    %--------------------color-curve--------------%
    %%%%%%%%%%%%%%%%%%%%%%%%%%%%%%%%%%%%%%%%%%%%%%%
    Figure\,\ref{fig:SN2020ftl_color} shows the temporal colors derived from MeerLICHT data, despite the sparse photometric sampling before peak brightness. Post-peak, the color evolution is consistent with the reddening behavior observed in normal type Ia SNe during their decline phase \citep{1998AJ....115..234L, 2017ApJ...846...58H}. 
    Specifically, the $u-q$ color steadily reddens by $\sim$$1.26\,\mathrm{mag}$ over $\sim$$17\,\mathrm{days}$, corresponding to an average slope of $0.07\pm0.01\,\mathrm{mag}\,\mathrm{day}^{-1}$. 
    In contrast, the $q-i$ color first becomes bluer by $0.30\,\mathrm{mag}$ in $\sim$$4.0\,\mathrm{days}$, then remains constant at $-0.52\,\mathrm{mag}$ for $\sim$$6\,\mathrm{days}$, with a subsequent epoch showing a $0.8\,\mathrm{mag}$ reddening $\sim$$22\,\mathrm{days}$ later.
    For the remainder of the decline phase, the $u-q$ and $q-i$ colors slightly vary around $1.48\,\mathrm{mag}$ and $0.08\,\mathrm{mag}$, respectively, typical for the onset of the nebular phase in normal type Ia SNe \citep{2015MNRAS.448L..48T,2017ApJ...846...58H}.
    %%%%%%%%%%%%%%%%%%%%%%%%%%%%%%%%%%%%%%%%%%%%%%%%
    %--------------------Modelling-----------------%
    %%%%%%%%%%%%%%%%%%%%%%%%%%%%%%%%%%%%%%%%%%%%%%%%
    The combined MeerLICHT and predominantly ATLAS data reveal the early time region of the light curve with its characteristic rise shape.
    We fitted the $t^2$-model to this early rising phase using $q$ and $o$-band observations extending up to $\tau=-5.18$ days. 
    The top panel of Figure\,\ref{fig:SN2020ftl_t2model} presents the early rising light curve of SN\,2020ftl, along with the best-fit line. 
    We find an explosion date of $t_0=58938.47\pm0.23\, \mathrm{MJD}$ with a corresponding rise time of $t_{\mathrm{rise,\, z}}=18.40\pm0.37\; \mathrm{days}$. This rise time value is consistent with mean rise times of normal type Ia SNe \citep{10.1111/j.1365-2966.2011.19213.x,2015MNRAS.446.3895F}. In contrast, \citet{Fausnaugh_2023} independently derived $t_0=58943\pm0.07\, \mathrm{MJD}$ and $t_{\mathrm{rise,\, z}}=11\; \mathrm{days}$ using early-time light curve data from Transiting Exoplanet Survey Satellite (TESS) observations. While this is a large difference from our estimated parameters, \citet{Fausnaugh_2023} simultaneously fitted a model with a power-law index of $2.59\pm0.01$, which may have led to the reported shorter rise time.
    Despite the apparent satisfactory fit to our relatively sparse multi-survey data, the resulting quality of fit is unremarkable $(\chi^2_{R}=4.12)$.
    With no corrections made for host-galaxy extinction, an absolute $q$-band peak magnitude of $-18.30\pm0.14\,\mathrm{mag}$ is barely fainter than the $2\sigma$ uncertainty from the mean absolute peak brightness of normal type Ia SNe ($-18.25\,\mathrm{mag}$; \citealt{2014AJ....147..118R}). So
    we cannot conclusively place SN\,2020ftl into a category of rare "transitional" type Ia SNe events intermediate between normal and subluminous SN 1991bg–like objects, similar to iPTF\,13ebh \citep{2015A&A...578A...9H}, SN\,2015F \citep{2017MNRAS.464.4476C} and SN\,2015bh \citep{2021ApJ...914...57W}.
    %%%%%%%%%%%%%%%%%%%%%%%%%%%%%%%%%%%%%%%%%%%%%%%%%%
    \subsubsection{SN 2020jfo (IIP)}
    %%%%%%%%%%%%%%%%%%%%%%%%%%%%%%%%%%%%%%%%%%%%%%%%%%
    SN\,2020jfo was discovered on 2020 May 6, UT 04:51:41 by the ZTF survey 
    at $r_{\mathrm{ZTF}}=16.01\,\mathrm{mag}$,
    following 3.9 days of non-detections \citep{2020TNSAN..99....1S}. 
    SN 2020jfo occurred in a spiral arm of the nearby intermediate spiral galaxy M\,61, with a host redshift of $z=0.005224$ \citep{2014MNRAS.440..696A,2018MNRAS.474.2551S}. 
    A follow-up optical spectrum was acquired with the Liverpool (LT) and Nordic Optical (NOT) Telescope within $\sim$$17$ hours of the initial TNS discovery report. It was then used to classify SN\,2020jfo as a hydrogen-rich type II SN \citep{2020TNSCR1259....1P}. 
    An earlier spectrum obtained with the Gran Telescopio Canarias (GTC-OSIRIS:\citealt{2000SPIE.4008..623C,2003SPIE.4841.1739C}), nearly half an hour before the LT and NOT classification spectra, was also published on the TNS \citep{2022TNSCR3405....1P}.
    Other independent discoveries of SN\,2020jfo by ATLAS, Pan-STARRS1, MASTER, and Gaia were subsequently reported to the TNS.
    The MeerLICHT transient survey frequently observed the host galaxy field in the $u$, $q$, and $i$ bands before the reported discovery date. Observations taken $\sim$$7$ days before the discovery were consistent with the background emission level, confirming the SN was not yet optically visible to MeerLICHT at an upper limit of $20.5\,\mathrm{mag}$ in the $q$-band. 
    %\footnote{\url{https://www.wis-tns.org/object/2020jfo}}
    The bottom panel of Figure\,\ref{fig:SN2020jfo} presents the subsequent multi-band optical light curve of the SN event. Adopting a redshift-independent distance $d=14.623\pm7.322\,\mathrm{Mpc}$, the SN reached an absolute peak magnitude at $-16.45\pm1.09\,\mathrm{mag}$ in the ZTF $g$-band, consistent with the mean absolute peak magnitude of ordinary type IIP SNe ($M_B=-16.75\pm0.37\,\mathrm{mag}$, \citealt{2014AJ....147..118R}). 
    As depicted in the zoomed-in insert, following maximum brightness, the light curves for the different bands diverge. 
    The MeerLICHT $q$- and $i$-band light curves enter a plateau phase. In contrast, the $u$-band light curve shows no plateau and instead begins an immediate, rapid decline. This is a common behaviour in type IIP SNe, where iron-group nuclei synthesised in the SN explosion reabsorb some of the shorter wavelength light emitted at the photosphere \citep{2009ApJ...703.2205K}.
    %%%%%%%%%%%%%%%%%%%%%%%%%%%%%%%%%%%%%%%%%%%%%%
    %----------------color-curve-----------------%
    %%%%%%%%%%%%%%%%%%%%%%%%%%%%%%%%%%%%%%%%%%%%%%
    Figure\,\ref{fig:SN2020jfo_color} shows the $u-q$ and $q-i$ color evolution for SN\,2020jfo. In the $\sim$$35$ days following peak brightness, both colors gradually redden, though at different rates: the $u-q$ color reddens significantly by $2.01\,\textrm{mag}$, while the $q-i$ color increases by only $0.47\,\textrm{mag}$. 
    This corresponds to a steep slope of $0.06\pm0.004\,\mathrm{mag}\,\mathrm{day}^{-1}$ for $u-q$, in contrast to the much shallower slope of $0.01\pm0.002\,\mathrm{mag}\,\mathrm{day}^{-1}$ for $q-i$. 
    This reddening behaviour is characteristic of type IIP SNe, which generally show a similar trend for about $50$ days before their color evolution flattens for the remainder of the plateau phase \citep{10.1093/mnras/sty508}.\\
    %%%%%%%%%%%%%%%%%%%%%%%%%%%%%%%%%%%%%%%%%%%%%%%%
    %--------------------Modelling-----------------%
    %%%%%%%%%%%%%%%%%%%%%%%%%%%%%%%%%%%%%%%%%%%%%%%%
    Although MeerLICHT data provides extensive sampling for post-peak light curve evolution, it lacks early-time data. Here we relied on ZTF and ATLAS observations taken shortly after discovery to constrain $t_0$.  
    As illustrated by the top panel of Figure\,\ref{fig:SN2020jfo_t2model}, the early rising light curve from ZTF and ATLAS is comprised of only three epochs. Figure\,\ref{fig:SN2020jfo_t2model} also shows the fitted $t^2$-model up to $\tau=-3.56$ days.
    Despite the sparse photometric data during the rising phase of SN\,2020jfo, the model fits the joint ZTF-$g,\, r$ and ATLAS-$o$ bands reasonably well, albeit resulting in a statistically poor fit ($\chi^2_R=5.39$). 
    The explosion date, $t_0=58970.83\pm0.17\,\mathrm{MJD}$, is consistent with the explosion epochs computed by \citet{2021A&A...655A.105S}, \citet{10.1093/mnras/stac3234} and \citet{Teja_2022} using different methods. The derived rise time yields $t_{\mathrm{rise,z}}=10.92\pm0.20\,\mathrm{days}$, concurring with rise times of individual type II SNe (e.g., SN\,1987A and SN\,2004du, \citealt{2015A&A...582A...3G}). However, the elongated, banana-shaped $\chi^2$ probability distribution contour map in the bottom left panel of Figure\,\ref{fig:SN2020jfo_t2model} indicates the presence of degeneracy between fit parameters.
    %%%%%%%%%%%%%%%%%%%%%%%%%%%%%%%%%%%%%%%%%%%%%%%%%%
    \subsubsection{SN 2021tkm (Ia)}
    %%%%%%%%%%%%%%%%%%%%%%%%%%%%%%%%%%%%%%%%%%%%%%%%%%
    SN\,2021tkm was first discovered and reported by ASAS-SN on 2021 July 14, UT 17:45:36 at an apparent magnitude of $16.6\, \mathrm{mag}$ in the Sloan-$g$ filter \citep{2021TNSTR2446....1S}, and subsequently detected and reported by ATLAS and {\sl Gaia}\,Alerts as per TNS follow-up report. 
    SN\,2021tkm is located in a spiral arm of the intermediate spiral galaxy IC\,4441 with a host redshift of $z = 0.006535$ \citep{2006MNRAS.371.1855W}. As described in a classification report by \citet{2021TNSCR2542....1G}, an optical spectrum obtained with the Robert Stobie Spectrograph (RSS; using PG900 grating) attached to the Southern African Large Telescope (SALT) on 2021 July 19, 20:33:26UT was used to classify SN\,2021tkm as a Type Ia SN.
    Multi-survey light curves from MeerLICHT and ATLAS are presented in Figure\,\ref{fig:SN2021tkm}.
    %\footnote{\url{https://www.wis-tns.org/object/2021tkm}}
    MeerLICHT detected this SN $\sim$2 days after the reported TNS discovery date. Several ATLAS $o$- and $c$-band detections found before the discovery date were consistent with the background emission. MeerLICHT light curves show a $u,\, q\,$ and $i$-band photometric coverage of the early phase of the SN. 
    At a redshift-independent distance of $d=17.720\pm6.028\,\mathrm{Mpc}$, early data reveal a normal rise to peak brightness, reaching a maximum absolute magnitude of $-17.99\pm0.74\,\mathrm{mag}$ in the $q$-band. The ATLAS $o$-band light curve reached a similar absolute peak brightness level around the same epoch, placing this SN on the significantly fainter ($\sim$1.26$\,\mathrm{mag}$) tail of ordinary type Ia SNe \citep{2014AJ....147..118R}. 
    \\
    While early-time observations have sufficient temporal coverage and sampling, the light curve lacks decline-phase data, with post-peak ATLAS detections varying around the level of the background emission. 
    With the advantage of the well-sampled MeerLICHT multi-band light curves at early times, we constructed multi-color curves of the rising phase as depicted in Figure\,\ref{fig:SN2021tkm_color}. 
    %%%%%%%%%%%%%%%%%%%%%%%%%%%%%%%%%%%%%%%%%%%%%%%
    %--------------------color-curve--------------%
    %%%%%%%%%%%%%%%%%%%%%%%%%%%%%%%%%%%%%%%%%%%%%%%
    All the rising-phase temporal colors of SN\,2021tkm gradually become bluer shortly after the initial MeerLICHT observations. This is signified by the $u-q$, $q-i$, $u-g$, $g-r$, $r-i$ and $i-z$ colors decreasing by $0.22\,\mathrm{mag}$, $0.23\,\mathrm{mag}$, $0.09\,\mathrm{mag}$, $0.13\,\mathrm{mag}$, $0.22\,\mathrm{mag}$ and $0.19\,\mathrm{mag}$ over $10\,\mathrm{days}$, which corresponds to an average slope of $0.02\,\mathrm{mag}\,\mathrm{day}^{-1}$ for all colors except for $u-g$ and $g-r$ colors which exhibit a flat evolution.
    This bluing in type Ia SNe is characteristic of the initial heating phase by radioactive decay of $^{56}\mathrm{Ni}$ shortly after the explosion, where the photosphere grows in radius while receding through the expanding ejecta towards the heating source \citep{2017ApJ...846...58H}.
    %%%%%%%%%%%%%%%%%%%%%%%%%%%%%%%%%%%%%%%%%%%%%%%%
    %--------------------Modelling-----------------%
    %%%%%%%%%%%%%%%%%%%%%%%%%%%%%%%%%%%%%%%%%%%%%%%%
    A combined dataset from MeerLICHT-$q$ and ATLAS-$c, o$ bands showing the early-time light curve of SN\,2021tkm is presented in the top panel of Figure\,\ref{fig:SN2021tkm_t2model}. Also shown is the $t^2$ model best fit line to the early data extending up to $\tau=-10.60\,\mathrm{days}$. 
    Since no background emission was subtracted from the host, it is sufficient to assume that the earlier ATLAS $c$- and $o$-band detections are consistent with the background emission and do not contribute significantly to early SN emission.
    Therefore, the $q$-band data dominates the overall light curve shortly after discovery, and allowed for a strong constraining power on the early rise of SN\,2021tkm. Applying the $t^2$-model provided a statistically good fit ($\chi^2_R=1.12$), yielding an explosion time of $t_0=59406.00\pm0.11\,\mathrm{MJD}$ and a corresponding rise time of $t_{\mathrm{rise,z}}=17.80\pm0.12\,\mathrm{days}$. While this rise time value is shorter than the mean rise time of type Ia SNe, it is consistent with the range of rise times observed in a population of normal type Ia SNe \citep{2012ApJ...745...44G}. 
    To demonstrate the strong confidence in the explosion date and rise time, the $\chi^2$ contour map for a pair of model parameters is shown in the bottom left panel of Figure\,\ref{fig:SN2021tkm_t2model}, together with the probability distribution of a range of fitted explosion dates on the right panel. 
    %%%%%%%%%%%%%%%%%%%%%%%%%%%%%%%%%%%%%%%%%%%%%%%%%%
    \subsubsection{SN 2021aele (Ia)}\label{sec:SN2021aele}
    %%%%%%%%%%%%%%%%%%%%%%%%%%%%%%%%%%%%%%%%%%%%%%%%%%
    Only a brief discovery and classification report on TNS has been published so far for SN\,2021aele.
    It was discovered by the ATLAS survey on 2021 November 11, UT 10:03:22 at an apparent magnitude of $19.173$ in the ATLAS-$c$ filter \citep{2021TNSTR3894....1T}, and subsequently reported to the TNS by the YSE (using PanSTARRS; \citealt{Jones_2021, Aleo_2023}) following an ATLAS-$o$ band $19.17\, \mathrm{mag}$ upper limit $\sim$$1.98$ days earlier.
    An optical spectrum obtained $\sim$$3.91$ days later with the KAST spectrograph on the $3.0\, \textrm{m}$ Lick Shane Reflector telescope at Lick Observatory was used to classify SN\,2021aele as a type Ia SN \citep{2021TNSCR3967....1D}. It is located in the galaxy GALEXASC\,J033532.47--283050.7 at a redshift of $z=0.077$ \citep{2011MNRAS.413..249K}, corresponding to luminosity distance $d=315.66\pm4.94\,\mathrm{Mpc}$. 
    Additional data from ATLAS forced photometry provides a few detections before discovery but lacks regular coverage to discern the rise to peak and the decay phase of the light curve. 
    The MeerLICHT multi-band optical light curves of SN\,2021aele are depicted in Figure\,\ref{fig:SN2021aele}, showing a clear rise and fall consistent with type Ia SNe and a high cadence coverage in $u,\; q$ and $i$ around peak brightness. MeerLICHT data fills in the ATLAS light curve gaps predominantly in the $q$-band.
    The initial detection in the $q$-band was observed nearly $\sim$$7.56$ days earlier than the reported discovery date, with subsequent observations showing a rise towards peak brightness. 
    The MeerLICHT$-q$ band light curve reached an absolute peak magnitude of $-19.41\pm0.08\,\mathrm{mag}$, in agreement with peak absolute magnitudes of a typical type Ia SN \citep{2014AJ....147..118R}.
    %\footnote{\url{https://www.wis-tns.org/object/2021aele}}
    %%%%%%%%%%%%%%%%%%%%%%%%%%%%%%%%%%%%%%%%%%%%%%%%%%
    %--------------------color-curve-----------------%
    %%%%%%%%%%%%%%%%%%%%%%%%%%%%%%%%%%%%%%%%%%%%%%%%%%
    Using the good cadence multiband data from MeerLICHT, we derived color evolution curves for SN\,2021aele, shown in Figure\,\ref{fig:SN2021aele_color}.
    Due to SNR cuts, the $u-q$ and $q-i$ colors are only limited to $\sim$8 days and $\sim$52 days before and after the reported TNS discovery date.
    In the first $\sim$$7.15$ days after the reported discovery date, the $u-q$ color appears to steadily become bluer by $\sim$$0.14\,\mathrm{mag}$ during the rise to maximum brightness, and the $q-i$ color reddens by $0.27\,\mathrm{mag}$ in the same period. Indicative of a wide minimum seen in color evolution curves of normal type Ia SNe around peak brightness, their corresponding average slopes appear rather flat, with $0.02\pm0.06\,\mathrm{mag}\,\mathrm{day}^{-1}$ and $0.04\pm0.05\,\mathrm{mag}\,\mathrm{day}^{-1}$ for $u-q$ and $q-i$ respectively.
    However, it is important to take note that the computed color indices are spread over 1.0 magnitude in range, with large overlapping uncertainties in a single epoch of observation. During the SN decline phase (starting from $\sim$8.5 days after discovery), which is largely dominated by $q$ and $i$ band observations, the $q-i$ color significantly increases redward by $0.91\,\mathrm{mag}$ in $37.21$ days, corresponding to a slope of $0.03\pm0.01\,\mathrm{mag}\,\mathrm{day}^{-1}$.
    %%%%%%%%%%%%%%%%%%%%%%%%%%%%%%%%%%%%%%%%%%%%%%%%%%
    %--------------------Modelling-----------------%
    %%%%%%%%%%%%%%%%%%%%%%%%%%%%%%%%%%%%%%%%%%%%%%%%%%
    The top panel of Figure\,\ref{fig:SN2021aele_t2model} shows the early rise portion from MeerLICHT$-q$ and ATLAS$-c$ band light curves of SN\,2021aele, as well as the $t^2$ best fit line up to $\tau=-4.96\,\mathrm{days}$.  
    Applying the $t^2$-model fit to the joint early-time data provides a satisfactory fit ($\chi^2_{R}=1.5$), which yields $t_0=59516.17\pm0.71\,\mathrm{MJD}$ with a corresponding $t_{\mathrm{rise,\, z}}=21.19\pm0.70$ days.
    The derived $t_{\mathrm{rise,\, z}}$ value, which is $\sim$$2.21$ days longer than the mean rise time of normal type Ia SNe, is within the acceptable range of individual normal type Ia SNe \citep{10.1111/j.1365-2966.2011.19213.x, 2015MNRAS.446.3895F}.
    %%%%%%%%%%%%%%%%%%%%%%%%%%%%%%%%%%%%%%%%%%%%%%%%%%
    \subsubsection{SN 2022ame (II)}
    %%%%%%%%%%%%%%%%%%%%%%%%%%%%%%%%%%%%%%%%%%%%%%%%%%
    SN\,2022ame was discovered by amateur astronomer K\=oichi Itagaki on 2022 January 27, UT 12:08:18 at an unfiltered apparent magnitude of $17.3\,\mathrm{mag}$ \citep{2022TNSTR.219....1I} and was reported to the TNS $0.49$ hours later. 
    The ATLAS survey also detected and reported it following a non-detection obtained $2.64$ days before the discovery date, and later by ZTF with the ALeRCE broker. SN\,2022ame occurred in one of the arms of the barred spiral galaxy NGC\,1255, with a host redshift of $z=0.005624$ \citep{2004AJ....128...16K}. As described by \citet{2022TNSCR.233....1B}, an optical spectrum obtained on 2022 January 28 with the Gemini Multi-Object Spectrograph-South (GMOS-S) by the Gemini Observatory DLT40 project, showed a flat continuum (indicative of high extinction) with a broad $\textrm{H}\,\alpha$ emission feature, which classified SN\,2022ame as a type II SN. 
    %\footnote{\url{https://www.wis-tns.org/object/2022ame}}
    The host galaxy field was observed by MeerLICHT a dozen times in $u, \, q$ and $i$ filters before the discovery of SN\,2022ame. 
    As shown by MeerLICHT's $u,\,q$ and $i$-band light curves in Figure\,\ref{fig:SN2022ame}, subtracting background emission of the host from the extracted light curves reveals that MeerLICHT's initial detection of SN\,2022ame was $\sim$$0.61$ days before the discovery date. The remaining pre-SN detections were consistent with background emission. 
    In contrast to the $u$-band light curve that appears fainter and only limited to $15\,\mathrm{days}$, the SN peaks in the redder $i$-band, with both $q$ and $i$ light curves exhibiting a steady decline by more than $0.5\,\mathrm{mag}$ for $47$ days thereafter, similar to the linearly declining light curve shape observed in Type IIL SNe \citep{2014MNRAS.445..554F}. \\
    Adopting a redshift-independent distance of $d=17.739\pm4.135\,\mathrm{Mpc}$, SN\,2022ame reached a MeerLICHT-$q$ band absolute brightness of $-14.64\pm0.51\,\mathrm{mag}$ at peak ($\sim$$3.34\,\mathrm{mag}$ fainter than an average type IIL SN, \citealt{2014AJ....147..118R}), confirming that it is indeed a highly extinguished type II SN.
    %%%%%%%%%%%%%%%%%%%%%%%%%%%%%%%%%%%%%%%%%%%%%%%
    %--------------------color-curve--------------%
    %%%%%%%%%%%%%%%%%%%%%%%%%%%%%%%%%%%%%%%%%%%%%%%
    The overall color evolution of the event further indicates the substantial reddening seen in its light curve.
    Figure\,\ref{fig:SN2022ame_color} shows the evolution of $u-q$ and $q-i$ colors for SN\,2022ame. They exhibit a steady reddening in the $\sim$$10\,\mathrm{days}$ following discovery, with $u-q$ increasing by $0.48\,\textrm{mag}$ and $q-i$ increasing by $0.33\,\textrm{mag}$. This corresponds to an average slope of $0.05\pm0.03\,\mathrm{mag}\,\mathrm{day}^{-1}$ and $0.03\pm0.008\,\mathrm{mag}\,\mathrm{day}^{-1}$ for both $u-q$ and $q-i$ respectively. The $u-q$ color then moderately varies around $1.69\,\mathrm{mag}$ for $7.0$ days, while the $q-i$ color shows a flat progression around $0.71\,\mathrm{mag}$ for the remaining $\sim$$40$ day decline, with the last observation showing a significant redward jump of $0.1\,\mathrm{mag}$.\\
    %%%%%%%%%%%%%%%%%%%%%%%%%%%%%%%%%%%%%%%%%%%%%%%
    %--------------------Modelling----------------%
    %%%%%%%%%%%%%%%%%%%%%%%%%%%%%%%%%%%%%%%%%%%%%%%
    We combined MeerLICHT-$q$ and ATLAS-$c$ band data to construct the early-time light curve of SN\,2022ame (see the top panel of Figure\,\ref{fig:SN2022ame_t2model}). Even though the ATLAS $o$-band data is present in the post-peak phase of the light curve, it does not constitute the early ATLAS detections, hence we omitted them from the model. The early rising phase of the SN has two epochs, with MeerLICHT offering the earliest observation of the two. Also shown in Figure\,\ref{fig:SN2022ame_t2model} is the $t^2$ model line of best fit up to $\tau=-9.15$. Given the sparse data, the model provides a mediocre fit ($\chi^2_R=1.14$), yielding an explosion date of $t_0=595602.43\pm0.24\,\mathrm{MJD}$ and a rise time of $t_{\mathrm{rise,z}}=13.50\pm38\,\mathrm{days}$. This $t_{\mathrm{rise,z}}$ value is in agreement with the range of values found in a population of type II SNe  \citep{2015A&A...582A...3G}, with a notable example being the type IIP SN\,1987A.
    %%%%%%%%%%%%%%%%%%%%%%%%%%%%%%%%%%%%%%%%%%%%%%%
    \subsubsection{SN 2022hrs (Ia)}
    %%%%%%%%%%%%%%%%%%%%%%%%%%%%%%%%%%%%%%%%%%%%%%%
    SN\,2022hrs was discovered by amateur astronomer K\=oichi Itagaki on 2022 April 16, 14:50:40UT at an unfiltered apparent magnitude of $15.0\,\mathrm{mag}$ \citep{2022TNSTR.994....1I}, and as per TNS follow-up reports, it was subsequently detected and reported by ATLAS, ZTF, and MASTER. 
    SN\,2022hrs occurred in the outer spiral arm of the intermediate spiral galaxy NGC\,4647 with a host redshift of $z=0.004700$ \citep{2008AJ....136..713K}. 
    %\footnote{\url{https://www.wis-tns.org/object/2022hrs}}
    An optical spectrum obtained on 2022 April 16, UT 20:04:00 with the $0.2\, \textrm{m}$ Newton telescope in collaboration with the Italian Supernova Search Project was used to classify SN\,2022hrs as a type Ia SN \citep{2022TNSCR.997....1B}.
    The consolidated data from MeerLICHT and ATLAS of SN\,2022hrs are shown in the bottom panel of Figure\,\ref{fig:SN2022hrs}. 
    This SN was detected by MeerLICHT $1.18\,\mathrm{days}$ after the reported discovery date. Several $q$-band detections spanning $\sim$$400\,\mathrm{days}$ earlier than the discovery date were found to be consistent with the background emission. 
    The MeerLICHT light curve is reasonably sampled during the early rise and post-peak decline phase. However, it lacks critical data around the maximum, which is populated by the well-sampled ATLAS data.
    At a redshift-independent distance of $d=17.591\pm3.043\,\mathrm{Mpc}$, ATLAS light curves show an initial primary peak in both $o$- and $c$-band, with the $c$-band reaching an absolute peak magnitude at $-18.79\pm0.38\,\mathrm{mag}$. 
    Although this determined absolute peak magnitude is $\sim$0.48\,mag fainter than the average type Ia SN, it is consistent with the mean absolute peak magnitude of normal type Ia SNe \citep{2014AJ....147..118R}.
    As the $o$-band light curve continued to decay, it exhibited a shoulder at roughly $19\,\mathrm{days}$ post-peak. This characteristic shoulder in the redder bands is a common feature of normal type Ia SNe \citep{2009ApJ...697..380W}. 
    %%%%%%%%%%%%%%%%%%%%%%%%%%%%%%%%%%%%%%%%%%%%%%%
    %--------------------color-curve--------------%
    %%%%%%%%%%%%%%%%%%%%%%%%%%%%%%%%%%%%%%%%%%%%%%%
    Figure\,\ref{fig:SN2022hrs_color} shows the temporal color curves of SN\,2022hrs computed from MeerLICHT observations. 
    At early times, MeerLICHT obtained three epochs in $u,\,q$ and $i$ bands. 
    In the initial $2.17$ days, the $u-q$ color rapidly decreases by $0.17\,\mathrm{mag}$ during the early rise. Due to gaps in the light curve, the subsequent observation shows a redder ($u-q\sim0.98\,\mathrm{mag}$) color at $7.6\,\mathrm{days}$ after peak, which steadily becomes redder in a linear fashion thereafter. 
    In contrast, the $q-i$ color barely changes with a marginal increase of $0.07\,\mathrm{mag}$ during the early rise phase. The next observation is at $7.6\,\mathrm{days}$ after peak, showing a bluer ($q-i=-0.62\,\mathrm{mag}$) color, which progressively becomes red in a similar evolution to $u-q$. The $u-q$ and $q-i$ colors then settle on a plateau, at $1.78\,\mathrm{mag}$ and $0.14\,\mathrm{mag}$ respectively, indicative of SN\,2022hrs entering the nebular phase.\\
    %%%%%%%%%%%%%%%%%%%%%%%%%%%%%%%%%%%%%%%%%%%%%%%
    %--------------------Modelling----------------%
    %%%%%%%%%%%%%%%%%%%%%%%%%%%%%%%%%%%%%%%%%%%%%%%
    The rise phase portion of the light curve obtained by combining MeerLICHT-$q$ and ATLAS-$o$ band detections is presented in the top panel of Figure\,\ref{fig:SN2022hrs_t2model}, along with the $t^2$ model line of best fit extending up to $\tau-9.07\,\mathrm{days}$. The resulting best fit provided an explosion date of $t_0=59681.93\pm0.06\,\mathrm{MJD}$ with an unexceptional static ($\chi^2_R=3.26$) and a corresponding rise time of $t_{\mathrm{rise, z}}=17.49\pm0.16\,\mathrm{days}$. This derived rise time value is in line with the mean rise time determined from a population of normal type Ia SNe \citep{10.1111/j.1365-2966.2011.19213.x,2015MNRAS.446.3895F}. The $t^2$ model shows a statistically well-behaved confidence region around the pair of fitted parameters, as demonstrated by the $\chi^2$ contour map and the probability distribution around the explosion epoch.
    %%%%%%%%%%%%%%%%%%%%%%%%%%%%%%%%%%%%%%%%%%%%%%%
    %%%%%%%%%%%%%%%%%%%%%%%%%%%%%%%%%%%%%%%%%%%%%%%
    \subsubsection{SN 2022phj (II)}
    %%%%%%%%%%%%%%%%%%%%%%%%%%%%%%%%%%%%%%%%%%%%%%%
    SN\,2022phj was discovered on 2022 July 22, UT 11:00:50 by the ZTF imaging survey at $g=18.87\,\mathrm{mag}$ and reported to the TNS $10.36$ hours later by the ALeRCE broker \citep{2022TNSTR2056....1M}, and also subsequently detected and reported by ATLAS and Pan-STARRS. 
    SN\,2022phj is located near the galaxy LEDA\,907434, with no available redshift measurement in the NED. 
    An optical spectrum obtained on 2022 August 2 with the Supernova Integral Field Spectrograph (SNIFS) attached to the University of Hawaii $2.2\,\textrm{meter}$ telescope (UH\,88inch) revealed a broad $\textrm{H}\,\alpha$ feature, classifying SN\,2022phj as a hydrogen-rich type II SN \citep{2022TNSCR2185....1J}. 
    A redshift of $z=0.039$ was determined from the classification spectrum, which correlates with a luminosity distance of $160.08\pm6.67\,\mathrm{Mpc}$. No studies related to SN\,2022phj have been published since its discovery. 
    %\footnote{\url{https://www.wis-tns.org/object/2022phj}}
    MeerLICHT detected SN\,2022phj roughly $15.79\,\mathrm{hours}$ after the TNS discovery date, as shown by the multi-band light curve in the top panel of Figure\,\ref{fig:SN2022phj}. Also shown in the figure are the recovered light curves from ZTF and ATLAS, filling in the gaps where MeerLICHT was not observing. Background emission was observed in the MeerLICHT-$u$, $q$, and $i$ bands from the potential host galaxy, detected $71.70\,\mathrm{days}$ prior to the discovery date. This residual emission was subtracted from the main SN light curve. None of the recovered light curves has a clear rise to peak. As a result, the epoch of maximum brightness is unknown. 
    From the $100.78$ days of MeerLICHT $q$ and $i$ band coverage, SN\,2022phj resembles a steeply declining light curve shape usually observed in type IIL SN \citep{2014MNRAS.445..554F}. This is also confirmed by the ZTF-$g$ band steadily declining by $1.1\,\mathrm{mag}$ in $\sim$57 days and the ATLAS-$c$ band light curve exhibiting a similar behavior within the same timescale. 
    While the MeerLICHT $q$ and $i$ band data have good coverage, the $u$ band is sparse with only two epochs. Regardless of this limitation in temporal coverage, the color evolution of the SN is still salvageable.
    %%%%%%%%%%%%%%%%%%%%%%%%%%%%%%%%%%%%%%%%%%%%%%%
    %--------------------color-curve--------------%
    %%%%%%%%%%%%%%%%%%%%%%%%%%%%%%%%%%%%%%%%%%%%%%%
    The $u-q$ and $q-i$ color curves of SN\,2022phj are presented in Figure\,\ref{fig:SN2022phj_color}. Shortly after discovery, the $u-q$ color becomes bluer by $0.97\, \textrm{mag}$ in $\sim$$20\,\mathrm{days}$ while the $q-i$ color moderately vary around $-0.14\,\textrm{mag}$ in the last $3$ days of that period. 
    Whereas the $u$ band coverage is limited beyond the initial $\sim$$20\,\mathrm{days}$, the $q-i$ color gradually evolves redwards by $0.87\,\mathrm{mag}$ for the last $\sim$$55\,\mathrm{days}$ of the $100\,\mathrm{days}$ decline, which yields an average slope of $0.02\pm0.01\,\mathrm{mag}\,\mathrm{day}^{-1}$. 
    Without early time data from MeerLICHT and supplementary data from other surveys to estimate the time of peak brightness, we refrained from performing a $t^2$ model fit on SN\,2022phj.
    %%%%%%%%%%%%%%%%%%%%%%%%%%%%%%%%%%%%%%%%%%%%%%%
%%%%%%%%%%%%%%%%%%%%%%%%%%%%%%%%%%%%%%%%%%%%%%%%%%%
\subsection{Pre-alert detections}
In this section, we discuss SNe that show pre-alert detection in their MeerLICHT light curves relative to the reported time of discovery on the TNS. This group of SNe highlights cases where, although not publicly reported, an initial MeerLICHT detection was considerably earlier than other transient surveys, which further aids in better constraints for the explosion epoch.
%%%%%%%%%%%%%%%%%%%%%%%%%%%%%%%%%%%%%%%%%%%%%%%%%%%
    %%%%%%%%%%%%%%%%%%%%%%%%%%%%%%%%%%%%%%%%%%%%%%%
    %%%%%%%%%%%%%%%%%%%%%%%%%%%%%%%%%%%%%%%%%%%%%%%
    \subsubsection{SN 2019ano (Ia)}
    %%%%%%%%%%%%%%%%%%%%%%%%%%%%%%%%%%%%%%%%%%%%%%%
    SN\,2019ano was initially discovered and reported by ASAS-SN on 2019 February 6, UT 03:21:36 at an apparent magnitude of $18.1\,\mathrm{mag}$ in the SDSS-$g$ filter \citep{2019TNSTR.207....1S}, and as per follow-up TNS reports, subsequently detected and reported by ATLAS and ZTF with the AMPEL broker. The SN is located near the blue galaxy SDSS\,J084920.46+111152.7 with no available redshift measurement in the NED. A redshift of $z=0.068$ was determined from the classification spectrum, which correlates with a luminosity distance of $279.11\pm7.90\,\mathrm{Mpc}$.
    Based on a spectrum obtained on 2019 February 8, UT 07:07:46 by the $1.5\mathrm{m}$ Tillinghast Telescope at the Fred Lawrence Whipple Observatory (FLWO) using the FAst Spectrograph for the Tillinghast Telescope (FAST, \citealt{1998PASP..110...79F}), SN\,2019ano was classified as a type Ia SN \citep{2019TNSCR.227....1F}. 
    A subsequent spectrum obtained $17.8$ hours later with the Spectral Energy Distribution Machine (SEDM, \citealt{Blagorodnova_2018}) spectrograph mounted on the Palomar 60-inch telescope (P60) confirmed this classification and was also published on the TNS \citep{2019TNSCR2773....1F}. Only a single discovery and two classification reports have been published so far on this SN.
    %\footnote{\url{https://www.wis-tns.org/object/2019ano}}
    The combined multi-band light curves from MeerLICHT and ATLAS are shown in the bottom panel of Figure\,\ref{fig:SN2019ano}. No underlying background subtraction was applied for these light curves. There is significant scatter in MeerLICHT detections roughly $40$ days earlier than discovery, which are considered to be consistent with the background level. The initial MeerLICHT $q$-band detection was obtained $20.24$ days before the reported discovery date, with a subsequent $u$-band detection thereafter. 
    As shown by Figure\,\ref{fig:SN2019ano}, $q$-band observations leading up to the discovery exhibit a rise from $\sim$$20.30\pm0.23\,\mathrm{mag}$ to $\sim$$17.92\pm0.02\,\mathrm{mag}$ over $\sim$$20$ days, with a similar early rise behaviour seen on both the $u$ and $i$ band light curves, albeit sparsely sampled. 
    MeerLICHT data provided well-sampled early rise light curves. They terminated immediately before peak brightness, but the rest of the light curve was filled in by additional data from ATLAS. SN\,2019ano reached an absolute peak magnitude of $-19.65\pm0.07\,\mathrm{mag}$ in the ATLAS $c$-band, which is $0.40\,\mathrm{mag}$ brighter but consistent with the average peak of normal type Ia SNe \citep{2014AJ....147..118R}. 
    %%%%%%%%%%%%%%%%%%%%%%%%%%%%%%%%%%%%%%%%%%%%%%%
    %--------------------Color--------------------%
    %%%%%%%%%%%%%%%%%%%%%%%%%%%%%%%%%%%%%%%%%%%%%%%
    The $u-q$ and $q-i$ color curves displayed in Figure\,\ref{fig:SN2019ano_color} show the color behaviour during the rise phase. 
    As expected for early times, the SN is initially blue, with $u-q$ color increasing by $1.0\,\mathrm{mag}$ in $21$ days, compared with the $q-i$ color decreasing by $1.39\,\mathrm{mag}$ in $13$ days. 
    This corresponds to a gradual reddening slope of $0.05\,\mathrm{mag}\,\mathrm{day}^{-1}$ and a fast bluing slope of $0.1\,\mathrm{mag}\,\mathrm{day}^{-1}$ for $u-q$ and $q-i$ colors, respectively.\\
    \\
    %%%%%%%%%%%%%%%%%%%%%%%%%%%%%%%%%%%%%%%%%%%%%%%
    %--------------------Modelling----------------%
    %%%%%%%%%%%%%%%%%%%%%%%%%%%%%%%%%%%%%%%%%%%%%%%
    The early light curve of SN\,2019ano is shown in Figure\,\ref{fig:SN2019ano_t2model}, comprising detections from MeerLICHT-$q$ and both ATLAS bands. Also displayed in Figure\,\ref{fig:SN2019ano_t2model} is the $t^2$ model best-fit curve up to $\tau=-8.29\,\mathrm{day}$. 
    The fit provides an explosion time of $t_0=58504.35\pm0.69\,\mathrm{MJD}$ corresponding to a derived rise time of $t_{\mathrm{rise,z}}=22.34\pm0.67\,\mathrm{days}$. 
    With a rise time that is $\sim$$3.36\,\mathrm{days}$ longer than the mean rise time of normal type Ia SNe, this value is leaning towards a longer rise time threshold considered to be associated with overluminous type Ia SNe \citep{10.1111/j.1365-2966.2011.19213.x,2015MNRAS.446.3895F}. 
    Even with the mediocre statistical fit quality ($\chi^2_R=3.57$), the confidence region (Figure\,\ref{fig:SN2019ano_t2model}, bottom left) around the pair of best fitted parameters illustrates the well-behaved uncertainties in the model and justifies this with a bell-shaped marginalised distribution (Figure\,\ref{fig:SN2019ano_t2model}, bottom right) around the best fitted $t_0$. 
    %%%%%%%%%%%%%%%%%%%%%%%%%%%%%%%%%%%%%%%%%%%%%%%
    %%%%%%%%%%%%%%%%%%%%%%%%%%%%%%%%%%%%%%%%%%%%%%%
    \subsubsection{SN 2020noz (IIn)} 
    %%%%%%%%%%%%%%%%%%%%%%%%%%%%%%%%%%%%%%%%%%%%%%%
    The discovery of SN\,2020noz was reported by YSE (fulfilled on the PanSTARRS1 survey) on 2020 June 22, UT 06:33:07 at a magnitude of $21.03\,\mathrm{mag}$ in the $g_{\mathrm{P1}}$ filter \citep{2020TNSTR1958....1J}, following a $21.50\,\mathrm{mag}$ Gaia$-G$ band upper limit $1.86$ days before discovery. Located in a galaxy VCC\,1152 ($z=0.024847$, \citealt{2017ApJS..233...25A}) at a luminosity distance of $d=111.50\pm2.62\,\mathrm{Mpc}$, it was subsequently detected and reported by the ATLAS, GaiaAlerts, ZTF, and PanSTARRS1 surveys. 
    \citet{2020TNSCR2271....1S} classified SN\,2020noz as a type IIn SN, based on the spectrum obtained $32.73$ days after discovery with a Low Resolution Imaging Spectrograph (LRIS; \citealt{1995PASP..107..375O, 2010SPIE.7735E..0RR}) on the Keck I $10\, \mathrm{m}$ telescope of the W. M. Keck Observatory (WMKO). No dedicated studies on SN\,2020noz have been published thus far. 
    %\footnote{\url{https://www.wis-tns.org/object/2020noz}}\\
    A series of SN\,2020noz light curves taken from MeerLICHT, ATLAS, and ZTF observations are displayed in the bottom panel of Figure\,\ref{fig:SN2020noz}. Whereas the consolidated data from ATLAS and ZTF shows a good temporal coverage during the early rising phase of the SN, the MeerLICHT data provides the earliest detections at $70.45$ days before discovery as shown in Figure\,\ref{fig:SN2020noz}, indicating a probable SN precursor outburst, similar to the recent supernova SN\,2024xuo (Blagorodnova et al. in prep). It must be noted that these earlier detections suffered from bad seeing ($\geq3''$) conditions and cannot be taken as conclusive evidence for a real precursor event. 
    With ATLAS data points populating the peak region of the light curve, the SN reached an absolute peak brightness of $-18.88\pm0.06\,\mathrm{mag}$ in the ATLAS $c$-band, consistent with the mean absolute magnitude of ordinary type IIn SNe \citep{2014AJ....147..118R}. 
    %%%%%%%%%%%%%%%%%%%%%%%%%%%%%%%%%%%%%%%%%%%%%%%
    
    %%%%%%%%%%%%%%%%%%%%%%%%%%%%%%%%%%%%%%%%%%%%%%%
    %--------------------Color--------------------%
    %%%%%%%%%%%%%%%%%%%%%%%%%%%%%%%%%%%%%%%%%%%%%%%
    Figure\,\ref{fig:SN2020noz_color} shows the $u-q$ and $q-i$ temporal colors computed from MeerLICHT observations. Due to the sparse MeerLICHT data, the observed colors are comprised of two epochs $\sim$$21\,\mathrm{days}$ post-discovery. 
    The colors show that $u-q$ increased by $0.16\,\mathrm{mag}$ in $11$ days, compared to the decrease of $0.25\,\mathrm{mag}$ observed in $q-i$ for the same duration. Except for the $q-i$ color, this is characteristic of the early time color behaviour of type II SNe during the photospheric phase as the ejecta expands and cools \citep{10.1093/mnras/sty508}. 
    The observed rapid decrease in $q-i$ at such early times is uncommon, as redder colors are expected to increase more slowly during this phase \citep{2016AJ....151...33G}. 
    This can be attributed to a rapid drop in temperature and the progressively stronger line blanketing affecting this early phase, which causes the peak of the spectral energy distribution to quickly shift to redder wavelengths (see \citealt{2016AJ....151...33G}).
    %%%%%%%%%%%%%%%%%%%%%%%%%%%%%%%%%%%%%%%%%%%%%%%
    %--------------------Modelling----------------%
    %%%%%%%%%%%%%%%%%%%%%%%%%%%%%%%%%%%%%%%%%%%%%%%
    Even with the lack of MeerLICHT $q$-band data at early times, the additional ATLAS $o$-band data has sufficient cadence to highlight the properties of the early light curve behaviour.
    Figure\,\ref{fig:SN2020noz_t2model} shows the early rising light curve region of SN\,2020noz, as well as the best fitting $t^2$ model curve up to $\tau=-27.98\,\mathrm{days}$. The model yields a poor quality statistical fit with $\chi^2_R=3.63$. 
    Neverthelss, it provided an explosion date of $t_0=59025.17\pm0.63\,\mathrm{MJD}$ and a corresponding rise time of $t_{\mathrm{rise,z}}=45.96\pm7.76\,\mathrm{days}$. 
    This $t_{\mathrm{rise,z}}$ value is consistent with the rise time of "slow-rising" type IIn SNe found in literature (SN\,2006gy,\,\citealt{2007ApJ...666.1116S,2020A&A...637A..73N}). As demonstrated by the stable confidence region around the pair of fitted parameters in the bottom left panel of Figure\,\ref{fig:SN2020noz_t2model}, the fitted $t^2$ model is stable, which is also supported by the probability distribution of a range of fitted $t_0$ values centered on the best fit value (bottom right).
    %%%%%%%%%%%%%%%%%%%%%%%%%%%%%%%%%%%%%%%%%%%%%%%
    \subsubsection{SN 2021koq (Ia)}
    %%%%%%%%%%%%%%%%%%%%%%%%%%%%%%%%%%%%%%%%%%%%%%%
    SN\,2021koq was discovered by the ATLAS survey on 28 April 2021, UT 11:39:50 at a magnitude of $o$=17.84$\,\mathrm{mag}$, and succeeded by a limit of $18.95\,\mathrm{mag}$ non-detection $\sim$$2.93$ days before discovery \citep{2021TNSTR1370....1T}. 
    Located in the SB(s)b galaxy UGC\,11177 ($z=0.020534$, \citealt{1981RSA...C...0000S, 1999PASP..111..438F}) at a luminosity distance of $d=81.49\pm2.36\,\mathrm{Mpc}$, it was subsequently detected and reported by Pan-STARRS, ZTF, and GaiaAlerts. 
    Classification of SN\,2021koq as a type Ia SN was reported by \citet{2021TNSCR1460....1J}, based on an optical spectrum obtained $\sim$$6.0$ days post-discovery with the UH 88inch (SNIFS) telescope. Only a TNS discovery and classification report has been published thus far on this event. 
    %\footnote{\url{https://www.wis-tns.org/object/2021koq}}\\
    Depicted in Figure\,\ref{fig:SN2021koq} are light curves from MeerLICHT, ATLAS, and ZTF. ATLAS and ZTF showcase well-sampled data with substantial temporal coverage to discern the rise to peak brightness and the decline phase of the SN. 
    The ATLAS $o$-band light curve exhibits a similar behaviour seen in SN\,2022hrs, in that there is a shoulder approximately $20\,\mathrm{days}$ after peak.
    MeerLICHT light curve is inadequately sampled, with $u\;, q$ and $i$ band observations spread over only four epochs. However, it provided the earliest observation of the SN in the $q$-band detection at $5.40$ days before the discovery date. 
    These early observations are crucial data points towards reliably constraining the SN explosion date, especially in cases where survey data is obtained out of phase. 
    SN\,2021koq reached an absolute peak magnitude of $-18.74\pm0.07\,\mathrm{mag}$ in ATLAS $c$-band. Although marginally fainter ($0.51\,\mathrm{mag}$), it is consistent with the mean absolute peak magnitude of normal type Ia SNe \citep{2014AJ....147..118R}.
    %%%%%%%%%%%%%%%%%%%%%%%%%%%%%%%%%%%%%%%%%%%%%%%
    %--------------Color-Evolution----------------%
    %%%%%%%%%%%%%%%%%%%%%%%%%%%%%%%%%%%%%%%%%%%%%%%
    Temporal color curves constructed from MeerLICHT data are presented in Figure\,\ref{fig:SN2021koq_color}.
    Even with the limited MeerLICHT data, the colors evolve similarly to other type Ia SNe discussed in this work. The $u-q$ color increased by $0.42\,\mathrm{mag}$ in $16.98$ days, resulting in an average slope of $0.02\,\mathrm{mag}\,\mathrm{day}^{-1}$. 
    In contrast, the $q-i$ color deceased by $1.31\,\mathrm{mag}$ in $26$ days, with a resulting average slope of $0.05\,\mathrm{mag}\,\mathrm{day}^{-1}$. 
    Given that MeerLICHT data is largely sampling the peak for epochs after discovery (spanning $16.98$ days), both colors increase ($u-q$) and decrease ($q-i$) at similar a rate ($0.02\,\mathrm{mag}\,\mathrm{day}^{-1}$) because the increase of the photospheric radius compensates the increase in overall heating.
    %%%%%%%%%%%%%%%%%%%%%%%%%%%%%%%%%%%%%%%%%%%%%%%
    %--------------------Modelling----------------%
    %%%%%%%%%%%%%%%%%%%%%%%%%%%%%%%%%%%%%%%%%%%%%%%
    The joint MeerLICHT-$q$ and ATLAS-$o$ band data showing the early rise portion of the SN light curve is depicted in Figure\,\ref {fig:SN2021koq_t2model}. Also shown is the applied $t^2$ model line of best fit up to $\tau=-10.30\,\mathrm{days}$. The resulting fit is statistically underwhelming, with $\chi^2_R=3.74$. It yields an explosion time of $t_0=59326.42\pm0.32\,\mathrm{MJD}$ with a corresponding rise time $t_{\mathrm{rise,z}}=19.98\pm0.45\,\mathrm{days}$. Despite this $t_{\mathrm{rise,z}}$ value being $\sim 1.0\,\mathrm{days}$ longer than the expected rise time of a normal type Ia SN, it is consistent with the range of rise times displayed by individual type Ia SNe \citep{10.1111/j.1365-2966.2011.19213.x,2015MNRAS.446.3895F}. The model uncertainties are well-behaved, as shown by the ellipsoid confidence region around the pair of fitted parameters, as well as the bell-shaped probability distribution centred on the best fit value of $t_0$.
    %%%%%%%%%%%%%%%%%%%%%%%%%%%%%%%%%%%%%%%%%%%%%%%
    \subsubsection{SN 2022bll (IIP)} 
    %%%%%%%%%%%%%%%%%%%%%%%%%%%%%%%%%%%%%%%%%%%%%%%
    SN\,2022bll was discovered on 2022 February 4, UT 00:46:05 by the Gaia Space Telescope at an apparent G-band magnitude of $18.86\,\mathrm{mag}$ and reported to the TNS $1.45$ hours later by GaiaAlerts \citep{2022TNSTR.300....1H}. 
    Classification of SN\,2022bll as a type II SN was reported by \citet{2022TNSCR.373....1S}, following the evident broad P Cygni H$\alpha$ features in a spectrum obtained around peak brightness, using the Goodman spectrograph attached to the SOAR telescope. 
    SN\,2022bll is located in the galaxy 2MASX\,J05045678--6832059 behind the Large Magellanic Cloud (LMC) with a host redshift of $z=0.024482$ \citep{2014ApJS..210....9B} and a luminosity distance of $d=94.53\pm2.46\,\mathrm{Mpc}$.
    The MeerLICHT light curve of the SN (Figure\,\ref{fig:SN2022bll}, top panel) shows that it was detected $32$ days before the discovery date. Whereas most of the pre-SN detections are consistent with the average background emission from the host, predominantly in the $i$-band, some of the $q$-band detections exhibit an apparent rise with time (see Figure\,\ref{fig:SN2022bll} bottom panel insert), especially for observations leading up to the discovery date. Due to the lack of a characteristic rise to peak and fade in the light curve, the maximum brightness is approximated as the brightest $q$-band photometric point after discovery. This assumption yields an absolute peak magnitude of $\sim$$-16.53\pm0.16\,\mathrm{mag}$. Without knowing the uncertainties added by the host galaxy and LMC extinction, we consider this value a lower limit to the SN peak luminosity.
    %%%%%%%%%%%%%%%%%%%%%%%%%%%%%%%%%%%%%%%%%%%%%%%
    %%%%%%%%%%%%%%%%%%%%%%%%%%%%%%%%%%%%%%%%%%%%%%%
    %--------------------color-curve--------------%
    %%%%%%%%%%%%%%%%%%%%%%%%%%%%%%%%%%%%%%%%%%%%%%%
    Even with limited MeerLICHT multi-band data, the $u$, $q$, and $i$ temporal coverage is sufficient to compute post-discovery color light curves.
    The $u-q$ and $q-i$ color evolution curves of SN\,2022bll are illustrated in Figure \ref{fig:SN2022bll_color}. Due to the good observing coverage in the $q$ and $i$ band immediately after discovery, the $q-i$ color exhibits a $\sim$$71\,\mathrm{days}$ reddening of $\sim$$0.22\,\mathrm{mag}$ with an average shallow slope of $0.003\,\mathrm{mag}\,\mathrm{day}^{-1}$ while the short $9\,\mathrm{day}$ coverage of $u-q$ color indices are poorly determined to discern any color behaviour.
    SN\,2022bll may signify a highly extinguished SN from host-galaxy and LMC dust, as shown in some individual well-observed type IIP SNe; SN\,2001dc \citep{2004MNRAS.347...74P}, SN\,2013am \citep{2018MNRAS.475.1937T}, SN\,2016cok \citep{2017MNRAS.467.3347K}.
    %%%%%%%%%%%%%%%%%%%%%%%%%%%%%%%%%%%%%%%%%%%%%%%
    %%%%%%%%%%%%%%%%%%%%%%%%%%%%%%%%%%%%%%%%%%%%%%%
    %--------------------Modelling----------------%
    %%%%%%%%%%%%%%%%%%%%%%%%%%%%%%%%%%%%%%%%%%%%%%%
    %%%%%%%%%%%%%%%%%%%%%%%%%%%%%%%%%%%%%%%%%%%%%%%
    Without early time data from MeerLICHT, ATLAS, and ZTF, and due to the absence of a discernible peak, it was determined that the absolute magnitude at peak brightness in the $q$-band occurred more than $70\,\mathrm{days}$ after the initial MeerLICHT detection.
    We refrained from applying the $t^2$ model because of the fact that no clear maximum is seen.
    %%%%%%%%%%%%%%%%%%%%%%%%%%%%%%%%%%%%%%%%%%%%%%%%

%%%%%%%%%%%%%%%%%%%%%%%%%%%%%%%%%%%%%%%%%%%%%%%%%%%
\subsection{Hostless supernovae}\label{sec:hostless_sne}
In this section, we explore the two SNe (SN\,2019lub and SN\,2022vrr) that lack an apparent host. MeerLICHT cutouts of SN\,2019lub and SN\,2022vrr showing the location of these SNe with respect to other point/extended sources in the field can be found in Appendix \ref{app:sn_hostless}. The thumbnail images were generated by the forced photometry routine, and they subtend one arcminute from the central location of the SN. We also provide deep reference $r$-band Dark Energy Survey (DES; \citealt{DES_Survey_2018,DES_DR2_2021}) images to demonstrate the SN location with respect to the nearby galaxies in the field (see appendix \ref{app:sn_hostless}). These stacked calibrated images were obtained through the Astro Data Lab SIA (Simple Image Access) service \citep{NIKUTTA2020100411}. For each SN, we derive $5\sigma$ magnitude upper limits directly from the DES \texttt{FITS} images by placing a $2''$ aperture at the transient's location and sampling the surrounding local background noise via a circular annulus. By correcting for the distance modulus and Milky Way extinction along the SN line of sight, we quantified strict absolute magnitude upper limits to determine how intrinsically faint probable hosts could be. The Milky Way $r$-band extinction and color excess along the SN\,2019lub line of sight is $A_r=0.088\,\mathrm{mag}$ and $E(B-V)=0.03\,\mathrm{mag}$, respectively. Additionally, the Milky Way $r$-band extinction and color excess along the SN\,2020vrr line of sight is $A_r=0.701\,\mathrm{mag}$ and $E(B-V)=0.27\,\mathrm{mag}$. For both SNe, no strong qualitative claims can be made regarding the morphology and characteristics of the host, given the single low-resolution SN spectroscopic redshift with large measurement uncertainty.
%%%%%%%%%%%%%%%%%%%%%%%%%%%%%%%%%%%%%%%%%%%%%%%%%%%
    %%%%%%%%%%%%%%%%%%%%%%%%%%%%%%%%%%%%%%%%%%%%%%%
    \subsubsection{SN 2019lub (IIn)}
    %%%%%%%%%%%%%%%%%%%%%%%%%%%%%%%%%%%%%%%%%%%%%%%
    SN\,2019lub is a hostless SN discovered by the ATLAS survey on 2019 July 25, UT 09:34:33 at a magnitude of $18.80\,\mathrm{mag}$ in the ATLAS-orange filter, following an $18.52\,\mathrm{mag}$ non-detection $3.89$ days prior within the same filter \citep{2019TNSTR1327....1T}. 
    A follow-up spectroscopic observation $4.82$ days later by the ESO New Technology Telescope (NTT) revealed a blue continuum superimposed with narrow Balmer emission lines, at a redshift of $z=0.066$ \citep{2019TNSCR1379....1A}, classifying SN\,2019lub as a type IIn. 
    The spectroscopic redshift of this SN translates to a luminosity distance of $d=270.90\pm7.49\,\mathrm{Mpc}$. Regarding the invisible host galaxy, we established a faint absolute magnitude upper limit of $-14.97\,\mathrm{mag}$. The supplementary data from ATLAS is coarsely sampled and does not add any characteristic shape to the light curve as displayed on the top panel of Figure\,\ref{fig:SN2019lub}.
    MeerLICHT coverage of SN\,2019lub is minimal, with only two epochs in the $q$- and $g$-band and a single $r$-band detection at $16.49$ days after discovery. 
    Additional data from ATLAS is also limited to two epochs and has a very large scatter within each epoch. Course data like this does not provide any shape to the light curve. Thus, light curve features such as a clear rise to maximum and eventual epoch at maximum cannot be extracted.
    Based on the initial MeerLICHT epoch $\sim$$16$ days after discovery, we derived a $g-r$ color term of $0.01\,\mathrm{mag}$, implying that the SN may have been observed near maximum light.  
    The insufficient multi-survey coverage at early times renders our efforts to estimate the explosion time and rise time for this SN impossible.
    %%%%%%%%%%%%%%%%%%%%%%%%%%%%%%%%%%%%%%%%%%%%%%%
    \subsubsection{SN 2022vrr (Ia)}
    %%%%%%%%%%%%%%%%%%%%%%%%%%%%%%%%%%%%%%%%%%%%%%%
    SN\,2022vrr is another probable hostless SN in our sample. It is located between a bright 2MASS source and a relatively faint Gaia source. There is an apparent faint emission footprint in Pan-STARRS imaging, but it is consistent with the background level. 
    It was discovered by ZTF imaging survey at a ZTF-$r$ band magnitude of $19.65\,\mathrm{mag}$ on 2022 September 22, UT 03:34:34, preceded by a ZTF-g band limit of $20.37\, \mathrm{mag}$ non-detection $3.0$ days before discovery \citep{2022TNSTR2766....1F}. 
    SN\,2022vrr was classified as a type Ia SN with a redshift of $z=0.063$ based on a follow-up spectrum taken on 2022 September 28, UT 04:58:28 by SEDM \citep{2022TNSCR2811....1F}. 
    The spectroscopic redshift of SN\,2022vrr corresponds to a luminosity distance of $d=258.58\pm9.66\,\mathrm{Mpc}$. For the host galaxy that remains completely hidden, we established an absolute magnitude upper limit of $-18.59\,\mathrm{mag}$.
    ZTF-$r$ and $g$ band observations yield a moderately-sampled SN light curve characterized by a clear rise and fall, albeit with sparse coverage on the ZTF-$g$ band over the $\sim$$32$ days since discovery. 
    MeerLICHT only has a single $q$-band detection at $9.62$ days post-discovery, contributing an additional data point for the early-phase evolution of the SN, as shown in Figure\,\ref{fig:SN2022vrr}. The light curve reached an absolute peak magnitude of $-19.31\pm0.15\,\mathrm{mag}$ in the ZTF-$r$ band, which is consistent with the mean absolute peak magnitude of typical type Ia SNe. SN\,2022vrr exhibits a red color, as indicated by the ZTF-$r$ band light curve, which consistently shows it to be brighter than the $g$ band throughout its evolution. \\
    %%%%%%%%%%%%%%%%%%%%%%%%%%%%%%%%%%%%%%%%%%%%%%%
    %--------------------Modelling----------------%
    %%%%%%%%%%%%%%%%%%%%%%%%%%%%%%%%%%%%%%%%%%%%%%%
    The early rise region of SN\,2022vrr is shown in Figure\,\ref{fig:SN2022vrr_t2model}, predominantly composed of ZTF $g$ and $r$ band detections, as well as a single $q$ band detection from MeerLICHT. Also shown is the resulting $t^2$ model best-fit curve extending up to $\tau=-8.15$ days. The fit yields an explosion date of $t_0=59838.55\pm1.39\,\mathrm{MJD}$ and a corresponding rise time of $t_{\mathrm{rise, z}}=18.56\pm1.37$ days. This derived $t_{\mathrm{rise, z}}$ value is consistent with the mean rise time of normal type Ia SNe. While the resulting quality of fit is average ($\chi^2_R=2.04$), the confidence region displayed on the bottom left panel of Figure\,\ref{fig:SN2022vrr_t2model} shows a presence of degeneracy between the pair of fitted parameters. As illustrated by the slightly skewed probability distribution (right panel) of the fitted explosion epoch around the best value, the uncertainties in the fit parameters are large. This suggests that the $t^2$ model is not a reliable description of the early-time light curve of SN\,2022vrr.
    %%%%%%%%%%%%%%%%%%%%%%%%%%%%%%%%%%%%%%%%%%%%%%%
    
%-->Well-sampled SNe

%-------------------------------------------------%
%\input{section_6}
%--> Discussion

%-------------------------------------------------%
%\section{Conclusion}\label{sec:conclusion}

\section{Summary and Conclusions}\label{sec:conclusion}

\begin{figure}[!ht]
    %\captionsetup{justification=centering}
     \centering
     \includegraphics[width=0.5\columnwidth]{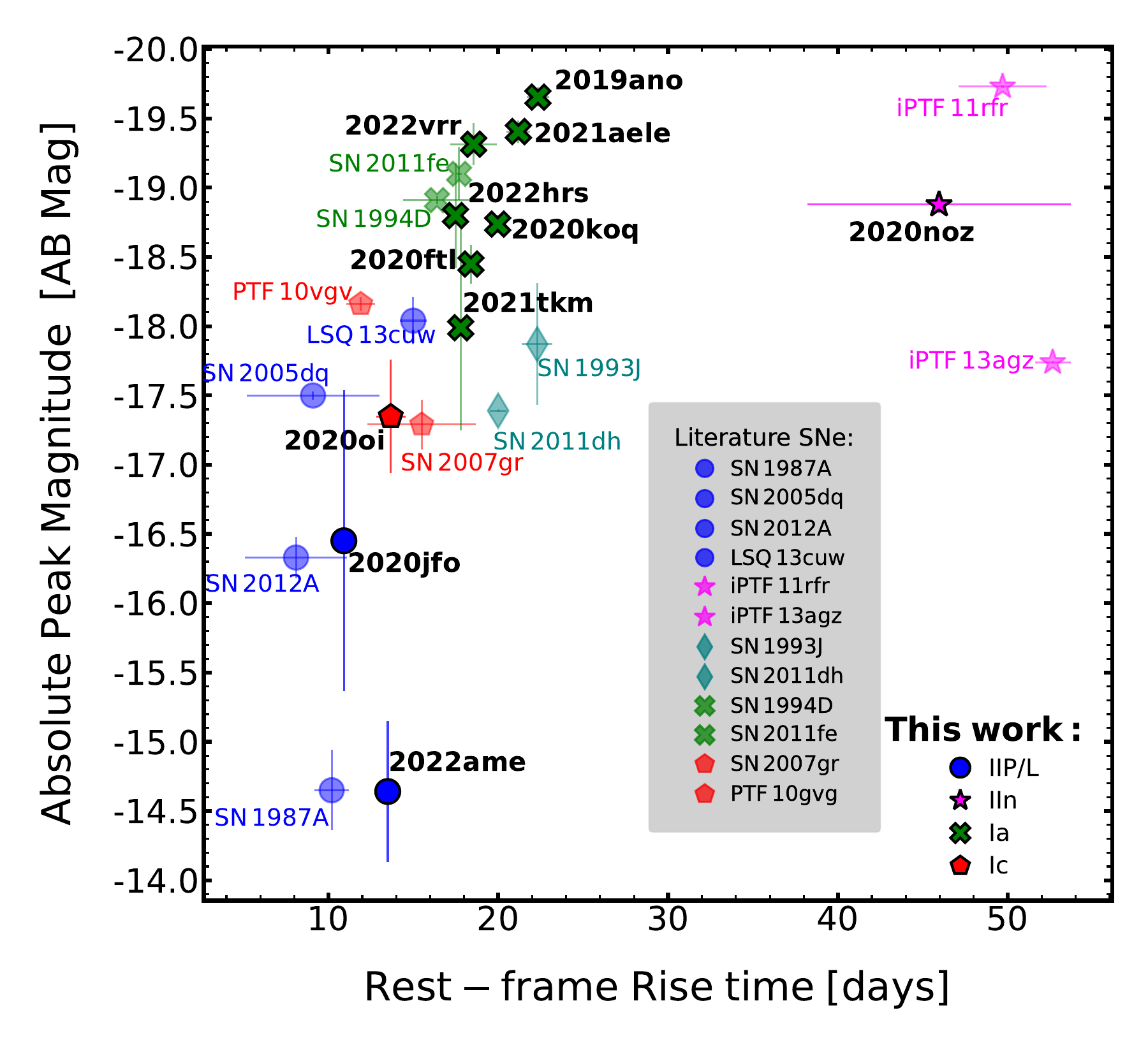}
     %--------------------------------------------%
     \caption{Heterogeneous sample of supernovae rise times as a function of absolute peak magnitude. Distinct symbols and colors highlight the different classes of SNe. The filled symbols represent SNe with determined redshift error, while the empty symbols are those with unknown redshift error. The individually labelled SNe of type: IIP/IIL/IIb (1987A, 2012A / 2005dq, LSQ\,13cuw / 1993J, 2011dh; \citealt{2015A&A...582A...3G}), IIn (iPTF\,11rfr, iPTF\,13agz; \citealt{2020A&A...637A..73N}), Ic (2007gr, PTF\,10vgv; \citealt{2009A&A...508..371H,2012ApJ...747L...5C,2015A&A...574A..60T}), and Ia (1994D, 2011f; \citealt{2000A&A...359..876C,2013A&A...554A..27P}) are displayed for comparison purposes as they provide ideal templates for their respective classes. SNe involved in this work are labelled in bold text.}
     \label{fig:luminosity_risetime_distribution}
\end{figure}

%\section{Discussion}\label{sec:discussion}
In this work, we used a sample of $14$ SNe with early-time multi-survey photometry to characterise their light curves and color evolution. A key goal was to test the effectiveness of the expanding fireball ($t^2$) model in constraining their explosion epochs and rise times. By applying this model to each supernova, we assessed its usefulness, identifying cases where it provides strong constraints and others where it is insufficient. Our findings show that the model is robust for the majority of `normal' SNe.
The $t^2$ model was found to be an excellent descriptor for the early-time evolution of eight SNe in our sample ( SN\,2019ano, SN\,2020oi, SN\,2020ftl, SN\,2020noz, SN\,2021koq, SN\,2021tkm, and SN\,2021aele). 
Similarly, for SN\,2022hrs, despite a gap in the early light curve, the model reliably captured the rising behaviour. 
In contrast, the model provided a poor description for a subset of our SNe, suggesting a deviation from the simple expanding fireball prescription. For instance, in the case of SN\,2020jfo, the poor statistical quality of the fit and degeneracy in the parameter space pointed towards a much steeper rise than the model can accommodate. Similarly, the early emission of SN\,2022ame and SN\,2022vrr was not well-captured by the model, leading to unreliable constraints on their explosion epochs. 
For three SNe, the available data were insufficient to conduct a meaningful model fit. Early-time coverage for SN\,2022phj and SN\,2019lub was too sparse to constrain the model parameters effectively. In the case of SN\,2022bll, the absence of a well-observed light-curve maximum made it impossible to apply the model, as it would lead to an unreliable estimation of $t_{z,\mathrm{rise}}$.\\
Figure\,\ref{fig:luminosity_risetime_distribution} shows a phase space diagram of SN rise times versus their absolute peak magnitudes, placing our results in the context of the broader SN population. The comparison SNe in the figure are from studies with well-constrained R-band rise times, with the exception of SN\,2011fe and SN\,1994D, for which the B-band rise times are quoted. Different classes of SNe occupy distinct regions in this phase space. Type Ia SNe occupy a compact region in the upper left corner of the phase diagram. This region demonstrates a clear correlation where brighter events tend towards longer rise times. This behaviour is fully consistent with the established width-luminosity relationship of type Ia SNe. Foundational early-time studies such as \citealt{Riess_1999} first established that longer rise times correlate with brighter peak luminosities, specifically noting that longer rise times are found in brighter SNe. This trend has been confirmed by modern high-cadence surveys, where \citealt{2020ApJ...902...47M} utilized ZTF data and found a clear correlation between the \textsc{SALT2} shape parameter ($x_1$) and $t_{\mathrm{rise}}$. Because $x_1$ correlates with peak luminosity, our phase diagram therefore independently recovers this classical relation.
In this context, SN\,2021tkm, SN\,2020ftl, and SN\,2021koq appear to be $\sim$$1\,\mathrm{mag}$ dimmer compared to the canonical SNe 2011fe and 1994D, albeit with rise times consistent with normal Type Ia events.\\
Determining whether these three SNe are intrinsically faint will require correcting for host extinction, but our sample lacks the high signal-to-noise spectra and complementary infrared photometry needed to disentangle intrinsic properties from host galaxy dust effects.
While methods such as measuring $\text{Na I D}$ equivalent widths or applying SALT light-curve fitting are standard estimators for host galaxy extinction, they were not viable for our dataset. $\text{Na I D}$ measurements are highly uncertain given our reliance on low-resolution classification spectra. Furthermore, when SALT fitting was attempted on a subset of the SNe Ia, sparse photometric coverage, specifically missing rise or decline phases, led to parameter degeneracy, which meant the resulting host $A_V$ could only serve as a lower limit.
The hydrogen-rich core-collapse SNe generally show a broader distribution \citep{2015A&A...582A...3G}. While SN\,2020jfo shows a shorter rise time and moderate peak brightness expected from an ordinary Type IIP event, the highly extinguished SN\,2022ame is much fainter with a rise time comparable to Type II events \citep{2015MNRAS.451.2212G}. Type IIn SNe are obvious outliers in this space, occupying the top-right of the plot, which is characterized by very long rise times and bright absolute peak magnitudes \citep{2020A&A...637A..73N}; our object SN\,2020noz occupies a parameter space between iPTF\,11rfr and iPTF\,13agz, indicating it was a normal, slow-rising Type IIn event. Finally, the only Type Ic in our sample, SN\,2020oi, falls within the compact region typical for helium-poor core-collapse SNe, which tend to have shorter rise times and are, on average, dimmer than Type Ia SNe \citep{2015A&A...574A..60T,2014AJ....147..118R}.
This analysis highlights a potential issue with current observational strategies. Follow-up facilities are often geared towards peculiar transients with extreme phase-space values, which risks overlooking the importance of placing better constraints on the rise times and peak magnitudes for the large majority of `normal' SNe. Such constraints, as shown here, offer valuable insight for subdividing traditional SN classifications. Therefore, our findings demonstrate a systematic way to constrain explosion epochs to better than one day using existing and future transient survey data, improving our understanding of the broader supernova population.
%%%%%%%%%%%%%%%%%%%%%%%%%%%%%%%%%%%%%%%%%%%%%%%%%%%
%%%%%%%%%%%%%%%%%%%%%%%%%%%%%%%%%%%%%%%%%%%%%%%%%%%
\\

From the $914$ transient events detected by MeerLICHT, $\sim$13.7\% were spectroscopically classified, and only $\sim$11.2\% of those were studied in detail in this work, which is only $\sim$1.5\% of the total sample. This highlights the fact that $\sim$99\% of newly discovered transients remain unstudied.
We presented an analysis of $14$ supernovae (SNe) detected by the MeerLICHT transient survey. 
Using the $t^2$ model, we estimated their explosion dates ($t_0$) and measured their rise times ($t_{\mathrm{rise,z}}$). 
We find that this model offers strong constraints on these parameters without relying on complex shock-cooling prescriptions, which often require high-quality, high-cadence, multi-wavelength (UV, optical, IR) photometric data. The high precision of the explosion epochs and rise times attained in this study was largely dependent on combining early photometric detections from multiple optical transient surveys to construct composite light curves. 
Although current high-cadence strategies provide well-sampled light curves for some events, most SNe discovered nightly do not receive dedicated follow-up once they receive a conventional spectroscopic classification. This highlights the importance of our work, which showcases the ability to effectively use multi-survey data, thereby substantially adding to the sample of rise time measurements for various SNe classes. 
Accurately determining SN rise times provides critical insights into differentiating between and within various SN subclasses, which could reveal a continuum of progenitor and environment properties rather than a single-parameter family. \\
\\
In combination with peak bolometric luminosity, the rise time provides a direct estimate of the ejected mass and the kinetic energy of the explosion.
Furthermore, the rise time versus peak absolute magnitude distribution subdivides spectroscopic classes into distinct clusters, aiding in the classification of different subtypes and providing insight into their underlying populations. Even with a small sample, our findings demonstrate that constraints on explosion epochs to better than $1$ day can be routinely extracted from multi-survey data.
The combined power of current and future surveys, such as PanSTARRS, GOTO, ZTF, ATLAS, BlackGEM, the Vera C. Rubin Observatory’s LSST, and the Global Open Transient Telescope Array (GOTTA, \citealt{2021AnABC..93..628L}), will generate high-quality, multi-band data ideal for this methodology. The added benefit of obtaining these multi-band, pre-maximum observations is the ability to probe the color evolution of SNe. This offers an extra dimension to explore the physical processes driving different SNe types (e.g., \citealt{2020ApJ...902...48B}), such as constraining ejecta mass and explosion energies. 
Applying our approach to these datasets will yield strong constraints on SNe explosion dates and consequently rise times by using an ensemble of early-time detections.  Motivated by our findings, individual SNe previously ignored due to sparse early data can now be analyzed to estimate their explosion dates and rise times. 
%--------------------------------------------------%
    \newpage
    \clearpage
    \pagebreak
%--------------------------------------------------%
    \newpage
    \newpage
    \clearpage
    \pagestyle{empty}
%--------------------------------------------------%
%--> Conclusion

%-------------------------------------------------%

%-------------------------------------------------%
\begin{acknowledgements}
\newline \small{
The National Research Foundation (NRF) fully supported this work under PJG's NRF SARChI grant 111692.
MeerLICHT is operated by a consortium consisting of Radboud University, the University of Cape Town, the South African Astronomical Observatory (SAAO), the University of Oxford, the University of Manchester and the University of Amsterdam, in association with, and partly supported by, the South African Radio Astronomy Observatory (SARAO), the European Research Council, the Netherlands Research School for Astronomy (NOVA) and the Netherlands Organization for Scientific Research
(NWO).
This work made use of the Inter-University Institute for Data Intensive Astronomy (IDIA) data-intensive research cloud for data processing. IDIA is a South African university partnership involving the University of Cape Town, the University of Pretoria, and the University of the Western Cape. We acknowledge the use of the ILIFU cloud computing facility – \url{www.ilifu.ac.za}, the ILIFU facility is supported by contributions from IDIA, the Computational Biology division at UCT, and the Data Intensive Research Initiative of South Africa
(DIRISA).
This work has made use of data from the Asteroid Terrestrial-impact Last Alert System (ATLAS) project. The ATLAS project is primarily funded to search for near earth asteroids through NASA grants NN12AR55G, 80NSSC18K0284, and 80NSSC18K1575; byproducts of the NEO search include images and catalogs from the survey area. The ATLAS science products have been made possible through the contributions of the University of Hawaii Institute for Astronomy, the Queen’s University Belfast, the Space Telescope Science Institute, the South African Astronomical Observatory, and The Millennium Institute of Astrophysics (MAS), Chile.
This work has made use of the ALeRCE broker and webclient. It is also based on observations obtained with the Samuel Oschin Telescope 48-inch and the 60-inch Telescope at the Palomar Observatory as part of the Zwicky Transient Facility (ZTF) project. ZTF is supported by the National Science Foundation under Grants No. AST-1440341 and AST-2034437 and a collaboration including current partners Caltech, IPAC, the Weizmann Institute for Science, the Oskar Klein Center at Stockholm University, the University of Maryland, Deutsches Elektronen-Synchrotron and Humboldt University, the TANC Consortium of Taiwan, the University of Wisconsin at Milwaukee, Trinity College Dublin, Lawrence Livermore National Laboratories, IN2P3, University of Warwick, Ruhr University Bochum, Northwestern University and former partners the University of Washington, Los Alamos National Laboratories, and Lawrence Berkeley National Laboratories. Operations are conducted by COO, IPAC, and UW.
This research has made use of "Aladin sky atlas" developed at Centre de Données astronomiques de Strasbourg (CDS), Strasbourg Observatory, France. This research also made use of the SIMBAD database and VizieR (\url{(DOI:10.26093/cds/vizier)}) services provided by the CDS, Strasbourg, France. This research used services and data provided by the Astro Data Lab at NSF's NOIRLab. NOIRLab is operated by the Association of Universities for Research in Astronomy (AURA), Inc. under a cooperative agreement with the National Science Foundation (NSF). 
This project used public archival data from the Dark Energy Survey (DES). Funding for the DES Projects has been provided by the U.S. Department of Energy, the U.S. National Science Foundation, the Ministry of Science and Education of Spain, the Science and Technology Facilities Council of the United Kingdom, the Higher Education Funding Council for England, the National Center for Supercomputing Applications at the University of Illinois at Urbana-Champaign, the Kavli Institute of Cosmological Physics at the University of Chicago, the Center for Cosmology and Astro-Particle Physics at the Ohio State University, the Mitchell Institute for Fundamental Physics and Astronomy at Texas A\&M University, Financiadora de Estudos e Projetos, Funda{\c c}{\~a}o Carlos Chagas Filho de Amparo {\`a} Pesquisa do Estado do Rio de Janeiro, Conselho Nacional de Desenvolvimento Cient{\'i}fico e Tecnol{\'o}gico and the Minist{\'e}rio da Ci{\^e}ncia, Tecnologia e Inova{\c c}{\~a}o, the Deutsche Forschungsgemeinschaft, and the Collaborating Institutions in the Dark Energy Survey.
The Collaborating Institutions are Argonne National Laboratory, the University of California at Santa Cruz, the University of Cambridge, Centro de Investigaciones Energ{\'e}ticas, Medioambientales y Tecnol{\'o}gicas-Madrid, the University of Chicago, University College London, the DES-Brazil Consortium, the University of Edinburgh, the Eidgen{\"o}ssische Technische Hochschule (ETH) Z{\"u}rich,  Fermi National Accelerator Laboratory, the University of Illinois at Urbana-Champaign, the Institut de Ci{\`e}ncies de l'Espai (IEEC/CSIC), the Institut de F{\'i}sica d'Altes Energies, Lawrence Berkeley National Laboratory, the Ludwig-Maximilians Universit{\"a}t M{\"u}nchen and the associated Excellence Cluster Universe, the University of Michigan, the National Optical Astronomy Observatory, the University of Nottingham, The Ohio State University, the OzDES Membership Consortium, the University of Pennsylvania, the University of Portsmouth, SLAC National Accelerator Laboratory, Stanford University, the University of Sussex, and Texas A\&M University.
Based in part on observations at Cerro Tololo Inter-American Observatory, National Optical Astronomy Observatory, which is operated by the Association of Universities for Research in Astronomy (AURA) under a cooperative agreement with the National Science Foundation.
NBM acknowledges being funded by the European Union (ERC, CET-3PO, 101042610). Views and opinions expressed are, however, those of the author(s) only and do not necessarily reflect those of the European Union or the European Research Council Executive Agency. Neither the European Union nor the granting authority can be held responsible for them. The publication is part of the research project PID2024-155585NA-I00 funded by MICIU/AEI /10.13039/501100011033. NBM acknowledges financial support from grant CEX2024-001451-M funded by MICIU/AEI/10.13039/501100011033.
\software{Analysis was performed using the Scientific Python stack, including \texttt{Astropy} \citep{2013A&A...558A..33A}, \texttt{matplotlib} \citep{2007CSE.....9...90H}, \texttt{NumPy} \citep{2020Natur.585..357H}, \texttt{SciPy} \citep{2020NatMe..17..261V}, \texttt{Synphot} \citep{pey_lian_lim_2020_3971036}, and \texttt{Pandas} \citep{the_pandas_development_team}. }
\facilities{ZTF, ATLAS, DES, Astro Data Lab}
}
\end{acknowledgements}

%% To help institutions obtain information on the effectiveness of their 
%% telescopes the AAS Journals has created a group of keywords for telescope 
%% facilities.
%
%% Following the acknowledgments section, use the following syntax and the
%% \facility{} or \facilities{} macros to list the keywords of facilities used 
%% in the research for the paper.  Each keyword is check against the master 
%% list during copy editing.  Individual instruments can be provided in 
%% parentheses, after the keyword, but they are not verified.
%\facilities{ML, ZTF, ATLAS, DES}
%% Similar to \facility{}, there is the optional \software command to allow 
%% authors a place to specify which programs were used during the creation of 
%% the manuscript. Authors should list each code and include either a
%% citation or url to the code inside ()s when available

%-------------------------------------------------%
%% Appendix material should be preceded with a single \appendix command.
%% There should be a \section command for each appendix. Mark appendix
%% subsections with the same markup you use in the main body of the paper.
%%
%% Each Appendix (indicated with \section) will be lettered A, B, C, etc.
%% The equation counter will reset when it encounters the \appendix
%% command and will number appendix equations (A1), (A2), etc. The
%% Figure and Table counter will not reset.
%-------------------------------------------------%
\section*{Data availability}
All data (e.g., a catalog of MeerLICHT supernova photometry tables) used as part of this work will be archived in VizieR at the Strasbourg Astronomical Data Center (CDS) once the article is accepted for publication, and can be accessed via \url{https://cdsarc.cds.unistra.fr/viz-bin/cat/J/PASP/138}. The analysis codes will be hosted in a private repository and can be shared upon request to the authors.
%-------------------------------------------------%

%\newpage
%\clearpage
%\pagestyle{empty}
\appendix
%%%%%%%%%%%%%%%%%%%%%%%%%%%%%%%%%%%%%%%%%%%%%%%%%%%%%%%%%%%%%%%%%%%%%%%%%%%%%%
% Table: MeerLICHT Supernovae
%%%%%%%%%%%%%%%%%%%%%%%%%%%%%%%%%%%%%%%%%%%%%%%%%%%%%%%%%%%%%%%%%%%%%%%%%%%%%%
\newpage
\clearpage
\pagestyle{empty}
\movetabledown=5cm
\begin{rotatetable}
\begin{deluxetable*}{lllclllcllllll} % 14 columns %
%\centering
\tablewidth{0pt} % Forces natural width centering
%% Set the font size to be small enough to fit all 14 columns %%
%\fontsize{6}{7.2}\selectfont
%\tabletypesize{\scriptsize} % Options: \tiny, \small, \footnotesize, \scriptsize
\tabletypesize{\tiny}
%% The AASTeX-specific \tablecaption{} command looks for \deluxetable{} parameters and will fail if using the standard \abular{} environment inside a rotatetable. %%
\tablecaption{Archival supernovae detected by the MeerLICHT telescope}
\label{table:ML1_sne}
%% Define column headers
\tablehead{
\colhead{SN} & \colhead{RA} & \colhead{DEC} & \colhead{SN} & \colhead{SN} & \colhead{Discovery} & \colhead{Reporting} & \colhead{Discovery} & \colhead{Discovery} & \colhead{Discovery} & \colhead{Host} & \colhead{Host} & \colhead{Host} & \colhead{Offset} \\
\colhead{Name} & \colhead{(h,m,s)} & \colhead{($^{\circ}$,$'$,$''$)} & \colhead{Type} & \colhead{Redshift} & \colhead{Mag} & \colhead{Group/s} & \colhead{Filter} & \colhead{Internal Name} & \colhead{Date (UT)} & \colhead{Name} & \colhead{Redshift} & \colhead{($A_V$)} & \colhead{(arcmin)}
}
%% Instead of using a simple dash -, the aastex package provides \nodata in the column for missing data. %%
\startdata
SN 2017bzb & 22:57:17.320 & -41:00:57.46 & II & 0.0031 & 13.0 & BOSS, GaiaAlerts, DLT40 & Clear- & Gaia17axu & 2017-03-07 05:06:55.000 & NGC7424 & 0.00310 & 0.029 & 3.283 \\ 
SN 2019qo & 12:18:54.894 & +05:14:56.98 & Ia & 0.076 & 19.46 & ZTF & g-ZTF & ZTF19aacilht & 2019-01-10 11:30:02.000 & SDSS J121855.23+051456.8 & 0.07614 & 0.053 & 0.084 \\ 
SN 2019ano & 08:49:20.655 & +11:11:56.71 & Ia & 0.068 & 18.1 & ASAS-SN, ATLAS, ZTF & g-Sloan & ASASSN-19ci & 2019-02-06 03:21:36.000 & SDSS J084920.46+111152.7 & - & 0.090 & 0.08 \\ 
SN 2019fcc & 13:51:30.490 & -52:55:36.62 & II & 0.0126 & 15.8 & MASTER, GaiaAlerts & Clear- & MASTER-OT-J135130.87-525534.4 & 2019-05-12 06:46:53.000 & 2MASX J13513072-5255237 & 0.01325 & 1.079 & 0.216 \\ 
SN 2019gak & 17:50:29.429 & -65:37:19.34 & Ia & 0.045 & 17.559 & ASAS-SN & g-Sloan & ASASSN-19nq & 2019-05-21 08:32:38.000 & GALEXASC J175029.43-653719.8 & - & 0.236 & 0.009 \\ 
SN 2019fte  & 12:56:49.013 & -30:17:20.38 & Ia & 0.054 & 18.214 & ATLAS & orange-ATLAS & ATLAS19lgg & 2019-05-22 09:11:31.000 & GALEXASC J125648.80-301723.8 & 0.05580 & 0.237 & 0.105 \\ 
SN 2019gkk & 17:04:49.609 & -08:20:55.42 & Ia & 0.0284 & 18.91 & Pan-STARRS, ZTF & w-P1 & PS19ayk & 2019-05-26 10:01:55.000 & 2MASX J17045008-0820548 & 0.02835 & 1.917 & 0.117 \\ 
SN 2019gvp & 16:46:35.521 & -31:35:31.28 & Ib & 0.015 & 19.419 & ATLAS & cyan-ATLAS & ATLAS19mco & 2019-06-03 10:04:48.000 & ESO 453- G 003 & 0.02216 & 1.184 & 0.162 \\ 
SN 2019gxa & 23:53:51.821 & -40:01:19.96 & Ia & 0.04 & 17.4 & ASAS-SN, ATLAS & g-Sloan & ASASSN-19oi & 2019-06-04 10:04:48.000 & WISEA J235351.27-400116.9 & - & 0.031 & 0.115 \\ 
SN 2019hcf & 19:45:32.611 & -15:30:57.45 & Ia & 0.081 & 19.65 & ZTF, ATLAS & g-ZTF & ZTF19aaxnlps & 2019-06-05 10:03:12.000 & 2MASX J19453205-1531038 & 0.08127 & 0.580 & 0.172 \\ 
SN 2019ibe & 13:21:07.804 & -10:54:58.10 & Ia & 0.047 & 18.2 & ZTF & r-ZTF & ZTF19abahsdc & 2019-06-21 04:10:33.000 & 2MASX J13210749-1054535 & 0.04727 & 0.143 & 0.104 \\ 
SN 2019icv & 22:18:46.109 & -36:48:04.86 & Ia & 0.02 & 16.0 & BOSS, ATLAS & orange-ATLAS & ATLAS19nsb & 2019-06-24 03:37:53.000 & IC 5186 & 0.01640 & 0.056 & 0.083 \\ 
SN 2019kau & 13:16:00.009 & -14:54:15.47 & Ia & 0.094 & 19.91 & Pan-STARRS, ZTF & w-P1 & PS19ddr & 2019-06-30 06:30:14.000 & J131600.02-145416.2 & - & - & 0.01467 \\ 
SN 2019lqn & 23:15:10.920 & -33:19:54.52 & Ia & 0.0756 & 18.76 & GaiaAlerts, ATLAS & G-Gaia & Gaia19dct & 2019-07-20 12:04:19.000 & 2MASS J23151015-3319530 & 0.07561 & 0.050 & 0.164 \\ 
SN 2019lrj & 20:18:37.850 & -54:00:27.97 & Ia & 0.051 & 17.9 & DLT40, GaiaAlerts & r-Sloan & DLT19g & 2019-07-24 04:45:07.000 & GALEXASC J201838.95-540025.1 & 0.04305 & 0.147 & 0.175 \\ 
SN 2019lub & 20:23:00.754 & -49:31:14.95 & IIn & 0.066 & 18.801 & ATLAS & orange-ATLAS & ATLAS19pvd & 2019-07-25 09:34:33.000 & - & - & - & - \\ 
SN 2019luz & 21:27:38.704 & -32:01:25.64 & II & 0.036 & 17.986 & ATLAS & orange-ATLAS & ATLAS19pwh & 2019-07-25 11:51:21.000 & MRSS 465-150838 & - & 0.204 & 0.011 \\ 
SN 2019ltx & 23:59:02.750 & -33:13:48.40 & Ia & 0.051 & 18.466 & ATLAS, GaiaAlerts & orange-ATLAS & ATLAS19puv & 2019-07-25 13:36:28.000 & DUKST 349-050 & 0.05909 & 0.040 & 0.148 \\ 
SN 2019mbq & 00:43:20.486 & -25:53:02.16 & II & 0.1008 & 19.04 & ZTF, Pan-STARRS, ATLAS & g-ZTF & ZTF19abkfmjp & 2019-07-30 11:01:19.000 & GALEXASC J004320.43-255300.6 & 0.10050 & 0.036 & 0.013 \\ 
SN 2019mkv & 22:52:29.940 & -46:41:43.76 & Ia & 0.054 & 17.6 & ASAS-SN & g-Sloan & ASASSN-19tc & 2019-08-02 00:47:31.000 & ESO 290- G 021 & 0.05170 & 0.028 & 0.135 \\ 
SN 2019mhm & 19:11:23.980 & -57:03:18.58 & IIP & 0.0106 & 16.6 & BOSS, GaiaAlerts & Clear- & Gaia19egp & 2019-08-02 13:13:21.000 & NGC 6753 & 0.01057 & 0.189 & 0.339 \\ 
SN 2019msz  & 20:32:09.437 & -59:11:58.92 & Ia & 0.063 & 17.6 & ASAS-SN & g-Sloan & ASASSN-19to & 2019-08-03 07:12:00.000 & GALEXASC J203208.46-591202.1 & 0.06050 & 0.178 & 0.117 \\ 
SN 2019mny & 21:47:09.805 & -45:13:26.49 & Ic & 0.038 & 17.948 & ATLAS & orange-ATLAS & ATLAS19rgs & 2019-08-04 09:46:04.000 & GALEXASC J214709.95-451327.3 & - & 0.058 & 0.031 \\ 
SN 2019nqr & 01:34:17.649 & -32:44:30.41 & II & 0.084 & 18.343 & DESGW & i-Sloan & desgw-190814d & 2019-08-16 06:31:40.000 & ESO 353- G 018 & 0.08324 & 0.074 & 0.026 \\ 
SN 2019nzx & 20:51:34.800 & -44:49:32.77 & Ia-91T & 0.065 & 17.409 & ATLAS, GaiaAlerts & orange-ATLAS & ATLAS19src & 2019-08-20 09:28:48.000 & GALEXASC J205135.34-444933.3 & - & 0.099 & 0.118 \\ 
SN 2019omk & 23:21:30.890 & -61:36:57.31 & Ia & 0.041 & 17.1 & ASAS-SN, GaiaAlerts & g-Sloan & ASASSN-19ui & 2019-08-23 08:09:36.000 & GALEXASC J232130.77-613657.8 & - & 0.050 & 0.007 \\ 
SN 2019ozz & 19:20:01.850 & -60:49:04.94 & Ia & 0.0362 & 17.6 & ASAS-SN, GaiaAlerts & g-Sloan & ASASSN-19vk & 2019-08-30 04:48:00.000 & FAIRALL 0056 & 0.03623 & 0.158 & 0.043 \\ 
SN 2019pcr & 01:04:26.890 & -64:07:32.34 & II & 0.0189 & 18.826 & OGLE, GaiaAlerts & I-Cousins & Gaia19dxe & 2019-08-30 06:06:58.000 & ESO 079- G 016 & 0.01888 & 0.069 & 0.341 \\ 
SN 2019phq & 00:35:30.420 & -52:10:57.00 & Ia-91T & 0.054 & 17.5 & ASAS-SN, GaiaAlerts & g-Sloan & ASASSN-19vo & 2019-09-02 06:14:24.000 & FAIRALL 0368 & 0.04203 & 0.040 & 0.254 \\ 
SN 2019qqh & 02:38:57.886 & -53:20:55.18 & Ia & 0.05 & 17.9 & ASAS-SN & g-Sloan & ASASSN-19xb & 2019-09-22 08:38:24.000 & GALEXASC J023858.68-532054.6 & - & 0.095 & 0.11 \\ 
SN 2019rwq & 02:18:57.869 & -66:23:29.08 & Ic & 0.0255 & 17.6 & ASAS-SN & g-Sloan & ASASSN-19yk & 2019-10-05 05:16:48.000 & GALEXASC J021857.57-662330.4 & - & 0.104 & 0.053 \\ 
SN 2019sss & 03:02:49.960 & -60:52:16.54 & Ia & 0.04 & 16.7 & ASAS-SN, GaiaAlerts & g-Sloan & ASASSN-19zh & 2019-10-18 07:12:00.000 & GALEXASC J030250.09-605212.5 & - & 0.064 & 0.0696 \\ 
SN 2019szt & 03:59:58.630 & -27:42:14.18 & Ia & 0.069 & 18.6077 & ALeRCE, GaiaAlerts & g-ZTF & ZTF19acgouje & 2019-10-21 11:33:28.000 & GALEXASC J035958.52-274213.5 & - & 0.044 & 0.027 \\ 
SN 2019urn & 00:57:49.292 & -17:38:26.92 & Ia & 0.06 & 18.51 & ZTF & g-ZTF & ZTF19acryaon & 2019-11-13 04:32:13.000 & 2MASX J00574929-1738270 & 0.05511 & 0.046 & 0.008 \\ 
SN 2019wyv & 08:34:03.161 & -15:52:43.22 & Ia & 0.08 & 18.53 & ZTF & r-ZTF & ZTF19aczjkxo & 2019-12-18 10:32:47.000 & PSO J128.5131-15.8786 & - & - & 0.00815 \\ 
SN 2019xit & 04:37:17.947 & -22:12:57.24 & Ia-91T & 0.064 & 18.0 & ASAS-SN, ZTF, Pan-STARRS & g-Sloan & ASASSN-19adq & 2019-12-22 06:57:36.000 & 2MASX J04371792-2212570 & 0.06826 & 0.140 & 0.003 \\ 
SN 2019xuj & 08:25:42.949 & -17:13:15.59 & Ia & 0.076 & 19.352 & ATLAS, ZTF & cyan-ATLAS & ATLAS19bebs & 2019-12-24 12:25:55.000 & GALEXASC J082542.88-171316.2 & - & 0.201 & 0.018 \\ 
SN 2019xvu & 01:53:08.500 & -36:14:09.92 & Ia & 0.063 & 18.92 & GaiaAlerts, ATLAS & G-Gaia & Gaia19fti & 2019-12-26 01:04:48.000 & J015308.25-361406.9 & - & - & 0.0696 \\ 
SN 2020oi & 12:22:54.930 & +15:49:24.96 & Ic & 0.0052 & 17.2799 & ALeRCE, ATLAS, Pan-STARRS1 & r-ZTF & ZTF20aaelulu & 2020-01-07 13:00:54.000 & M 100 & 0.00524 & 0.072 & 0.11 \\ 
SN 2020aqe & 11:43:29.760 & -16:47:53.16 & II & 0.0123 & 18.269 & ATLAS, Pan-STARRS, ZTF & orange-ATLAS & ATLAS20cvy & 2020-01-23 13:47:59.000 & 2MASX J11432985-1647473 & 0.01226 & 0.111 & 0.1 \\ 
SN 2020ckl & 05:38:22.170 & -20:35:24.47 & Ia & 0.1 & 19.2306 & ALeRCE, ZTF, ATLAS & r-ZTF & ZTF20aamurxn & 2020-02-13 03:29:42.000 & J053822.25-203524.5 & - & - & 0.0203 \\ 
SN 2020dny & 13:25:52.551 & -17:20:10.17 & Ia & 0.1 & 19.131 & ATLAS & orange-ATLAS & ATLAS20hai & 2020-02-24 14:24:00.000 & WISEA J132552.52-172009.4 & - & 0.241 & 0.014 \\ 
SN 2020dqw & 13:23:39.659 & -16:38:44.61 & Ia & 0.0235 & 17.4 & GOTO, ZTF, ATLAS & L-GOTO & GOTO20mn & 2020-02-28 02:49:43.104 & GALEXASC J132339.44-163839.0 & 0.02348 & 0.199 & 0.098 \\ 
SN 2020drh & 13:48:29.796 & -30:40:14.72 & Ia & 0.05 & 18.61 & ATLAS & orange-ATLAS & ATLAS20hgf & 2020-02-28 13:26:24.000 & J134829.90-304015.0 & - & - & 0.0238 \\ 
SN 2020ejm & 10:16:18.850 & -33:33:49.17 & Ia & 0.0098 & 16.0 & DLT40 & Clear- & DLT20h & 2020-03-11 05:42:35.424 & IC 2560 & 0.00976 & 0.263 & 0.029 \\ 
SN 2020enj & 10:20:57.071 & +00:34:25.55 & Ia & 0.104 & 19.235 & ATLAS, Pan-STARRS, ZTF & orange-ATLAS & ATLAS20hza & 2020-03-16 10:59:31.200 & SDSS J102057.07+003426.5 & - & 0.123 & 0.016 \\ 
SN 2020fqo & 08:01:23.078 & +06:08:26.57 & II & 0.0388 & 19.0444 & ALeRCE, ZTF & r-ZTF & ZTF20aatvupb & 2020-03-31 04:43:53.000 & CGCG 031-021 & 0.03878 & 0.070 & 0.164 \\ 
SN 2020ftl & 12:20:03.770 & +05:20:35.84 & Ia & 0.0073 & 18.5 & SNHunt, ATLAS, Pan-STARRS & V-crts-CRTS & ATLAS20jqt & 2020-04-02 06:23:46.000 & NGC 4277 & 0.00730 & 0.052 & 0.116 \\ 
SN 2020jfo & 12:21:50.480 & +04:28:54.05 & II & 0.0050 & 16.01 & ZTF, ATLAS, YSE & r-ZTF & ZTF20aaynrrh & 2020-05-06 04:51:41.184 & M 061 & 0.00522 & 0.061 & 1.201 \\ 
SN 2020mvk & 13:35:45.680 & -30:52:30.36 & II & 0.0150 & 18.151 & ATLAS, GaiaAlerts, DogsHeavenTS & cyan-ATLAS & ATLAS20pqr & 2020-06-17 07:35:02.400 & ESO 444- G 075 & 0.01504 & 0.143 & 0.318 \\
\enddata
\raggedright
\noindent 
\tablecomments{The inflated version of this table is available in machine-readable form in the online journal. The portion shown here is its reduced form due to space formatting constraints.\\
%$\texttt{DP}$ --- This table is arranged in descending order of the number of data points (DP) for each supernova in the MeerLich transient archive. \\
$\texttt{RA, DEC}$ --- Right ascension and declination are given in the J2000 epoch. \\
$\texttt{SN Type}$ --- Spectral classification of supernovae. For a subtype of Ia SNe designated the luminous Ia-1991T-like and subluminous Ia-1991bg-like, we use the notation Ia-91T and Ia-91bg respectively.\\
$\texttt{Disc. Mag.}$ --- Discovery magnitudes are broadband magnitudes corresponding to the reported discovery filter from TNS, depending on the camera used from the discovery data source. \\
$\texttt{Rep. Group/s}$ --- List of groups who published their independent photometric measurements of the discovery of a candidate supernova on TNS. \\
$\texttt{Disc. Filter}$ --- The photometric filter used for the discovery data source, ''Clear'' indicates discovery by any number of non-professional astronomers. \\
$\texttt{Disc. Int. Name}$ --- Name assigned to the candidate supernova by the discovery data source, which could be any of the reporting group/s (Here, we only take the earliest official name from the earliest reporting group). \\
$\texttt{Host}$ --- No host galaxy name, redshift and extinction listed for those galaxy footprints not detected in GALEX, SDSS, 2MASS, CatWISE, PanSTARRS and DES survey data. \\
$A_V$ --- Host Galactic extinction taken from \citet{2011ApJ...737..103S}. \\
$\texttt{Offset}$ Offset indicates the offset of the supernova in arcseconds from the coordinates of the host nucleus, taken from NED.}
\end{deluxetable*}
\end{rotatetable}
\pagebreak
%----------------------------------------------------------------------------%
\newpage
\clearpage
\pagestyle{empty}
\setcounter{table}{2}
\movetabledown=5cm
\begin{rotatetable}
\begin{deluxetable*}{lllclllcllllll} % 14 columns %
%\centering
\tablewidth{0pt} % Forces natural width centering
%% Set the font size to be small enough to fit all 14 columns %%
%\fontsize{6}{7.2}\selectfont
%\tabletypesize{\scriptsize} % Options: \tiny, \small, \footnotesize, \scriptsize
\tabletypesize{\tiny}
%% The AASTeX-specific \tablecaption{} command looks for \deluxetable{} parameters and will fail if using the standard \abular{} environment inside a rotatetable. %%
\tablecaption{,continued}
%\label{table:ML1_sne}
%% Define column headers
\tablehead{
\colhead{SN} & \colhead{RA} & \colhead{DEC} & \colhead{SN} & \colhead{SN} & \colhead{Discovery} & \colhead{Reporting} & \colhead{Discovery} & \colhead{Discovery} & \colhead{Discovery} & \colhead{Host} & \colhead{Host} & \colhead{Host} & \colhead{Offset} \\
\colhead{Name} & \colhead{(h,m,s)} & \colhead{($^{\circ}$,$'$,$''$)} & \colhead{Type} & \colhead{Redshift} & \colhead{Mag} & \colhead{Group/s} & \colhead{Filter} & \colhead{Internal Name} & \colhead{Date (UT)} & \colhead{Name} & \colhead{Redshift} & \colhead{($A_V$)} & \colhead{(arcmin)}
}
%% Instead of using a simple dash -, the aastex package provides \nodata in the column for missing data. %%
\startdata
SN 2020noz & 12:29:00.250 & +07:50:58.42 & IIn & 0.025 & 21.03 & YSE, ATLAS, GaiaAlerts & g-Sloan & PS20eml & 2020-06-22 06:33:07.200 & VCC 1152 & 0.02485 & 0.074 & 0.132 \\ 
SN 2020qey & 01:13:51.593 & -06:21:31.56 & Ia & 0.1 & 20.37 & ZTF, ATLAS, Pan-STARRS & r-ZTF & ZTF20abmsxxr & 2020-07-18 10:05:09.600 & GALEXASC J011351.54-062128.8 & - & 0.218 & 0.096 \\ 
SN 2020rod & 21:47:50.608 & -11:52:04.51 & Ia & 0.09 & 19.93 & ZTF, Pan-STARRS & g-ZTF & ZTF20abrhlfb & 2020-08-08 08:02:24.000 & 2MASX J21475037-1151586 & 0.09278 & 0.139 & 0.115 \\ 
SN 2020rfm & 00:56:15.851 & -13:18:02.55 & Ia & 0.06 & 18.257 & ALeRCE, ATLAS, ZTF & g-ZTF & ZTF20abrnklh & 2020-08-12 09:33:04.003 & 2MASS J00561587-1318018 & - & 0.065 & 0.016 \\ 
SN 2020rws & 03:39:59.250 & -04:03:59.94 & Ia & 0.052 & 17.7906 & ALeRCE, ATLAS, ZTF & g-ZTF & ZTF20abvykmy & 2020-08-23 10:45:55.999 & GALEXASC J033958.96-040405.0 & - & 0.149 & 0.1 \\ 
SN 2020sur & 01:28:59.576 & -11:29:29.29 & II & 0.063 & 19.66 & ZTF, ATLAS, Pan-STARRS & g-ZTF & ZTF20abywoaa & 2020-08-31 09:40:19.200 & GALEXASC J012859.58-112927.9 & - & 0.100 & 0.015 \\ 
SN 2020tap & 22:44:13.393 & -18:10:41.80 & Ia & 0.0537 & 18.435 & ATLAS, Pan-STARRS, ZTF  & orange-ATLAS & ATLAS20znm & 2020-09-11 11:15:21.600 & 2MASX J22441302-1810389 & 0.05374 & 0.082 & 0.098 \\ 
SN 2020tfb & 06:08:52.139 & -26:24:45.98 & IIP & 0.048 & 19.083 & ATLAS, YSE, ZTF & orange-ATLAS & ATLAS20babr & 2020-09-11 14:34:04.800 & WISEA J060852.23-262447.6 & - & 0.089 & 0.036 \\ 
SN 2020tld & 00:29:41.590 & -51:32:09.60 & Ia & 0.0112 & 17.0 & ASAS-SN, GaiaAlerts & g-Sloan & ASASSN-20ls & 2020-09-15 23:45:36.000 & ESO 194- G 021 & 0.01120 & 0.043 & 0.92 \\ 
SN 2020uli & 19:32:51.790 & -45:13:20.17 & Ia & 0.019 & 17.549 & ATLAS, GaiaAlerts & orange-ATLAS & ATLAS20bcev & 2020-09-29 05:54:14.400 & ESO 283- G 004 & 0.01656 & 0.228 & 0.315 \\ 
SN 2020uqm & 00:37:43.170 & -59:02:54.49 & II & 0.048 & 18.5 & GaiaAlerts & G-Gaia & Gaia20eos & 2020-10-01 19:29:16.800 & DES J003743.15-590254.2 & - & - & 0.0042 \\ 
SN 2020xoq & 20:03:00.630 & -41:33:02.23 & II & 0.032 & 18.166 & ATLAS, GaiaAlerts & orange-ATLAS & ATLAS20bdyf & 2020-10-17 05:29:45.600 & 2MASX J20030114-4133024 & 0.03066 & 0.266 & 0.101 \\ 
SN 2020xtq & 00:02:21.567 & -15:19:56.31 & Ia & 0.12 & 19.624 & ATLAS & cyan-ATLAS & ATLAS20beeq & 2020-10-20 09:00:00.000 & GALEXASC J000221.57-151955.3 & - & 0.091 & 0.009 \\ 
SN 2020abce & 23:19:19.097 & +08:05:59.99 & Ia & 0.06 & 19.7413 & ALeRCE, ZTF, Fink & g-ZTF & ZTF20actpbck & 2020-11-25 03:41:49.998 & 2MASX J23191904+0806045 & 0.06544 & 0.213 & 0.077 \\ 
SN 2020acqy & 03:24:24.540 & -43:29:39.98 & Ia & 0.073 & 18.356 & ATLAS & cyan-ATLAS & ATLAS20bihj & 2020-12-12 08:35:31.200 & GALEXASC J032424.73-432941.5 & - & 0.038 & 0.045 \\ 
SN 2021fmu & 10:31:56.240 & -15:57:19.62 & SLSN-II & 0.14 & 19.345 & ATLAS, ZTF, GaiaAlerts & orange-ATLAS & ATLAS21hra & 2021-03-14 08:32:38.400 & GALEXASC J103156.89-155721.2 & - & 0.194 & 0.161 \\ 
SN 2021fox & 09:57:36.360 & -13:41:45.42 & II & 0.032 & 19.045 & ATLAS, ALeRCE, Pan-STARRS & orange-ATLAS & ATLAS21hzm & 2021-03-15 10:04:48.000 & GALEXASC J095736.40-134149.4 & 0.03172 & 0.167 & 0.067 \\ 
SN 2021ftu & 12:17:16.360 & -13:30:46.85 & Ia & 0.08 & 19.7 & Pan-STARRS, ALeRCE, ATLAS & w-P1 & PS21btn & 2021-03-15 11:29:45.600 & GALEXASC J121716.65-133048.8 & - & 0.157 & 0.08 \\ 
SN 2021fxm & 13:54:18.313 & -25:54:30.45 & II & 0.02 & 18.89 & ZTF, ATLAS, Pan-STARRS & r-ZTF & ZTF21aapjphz & 2021-03-17 09:01:26.400 & GALEXASC J135418.24-255428.4 & - & 0.192 & 0.037 \\ 
SN 2021gwn & 10:47:11.930 & -29:23:10.21 & Ia & 0.048 & 18.031 & ATLAS, GaiaAlerts & orange-ATLAS & ATLAS21jcj & 2021-03-22 10:04:48.000 & 2MASX J10471209-2923112 & 0.05933 & 0.196 & 0.041 \\ 
SN 2021hjc & 09:25:33.310 & -24:45:53.93 & Ia & 0.0537 & 18.817 & ZTF, ATLAS, Fink & r-ZTF & ZTF21aaqyetj & 2021-03-30 05:32:38.400 & 2MASX J09253320-2445530 & 0.05369 & 0.246 & 0.028 \\ 
SN 2021hpv & 10:02:57.333 & -11:44:01.44 & Ic & 0.029 & 18.589 & ATLAS, Pan-STARRS & orange-ATLAS & ATLAS21jlk & 2021-04-02 08:52:48.000 & MCG -02-26-023 & 0.02926 & 0.181 & 0.089 \\ 
SN 2021ich & 10:19:29.620 & -03:19:57.94 & II & 0.05 & 18.976 & ATLAS, Pan-STARRS, GaiaAlerts & orange-ATLAS & ATLAS21jni & 2021-04-02 10:01:55.200 & 2MASX J10193041-0319563 & 0.04925 & 0.126 & 0.199 \\ 
SN 2021hvu & 10:13:38.390 & -24:31:14.81 & Ia-91T & 0.104 & 19.3 & ATLAS, ALeRCE, GaiaAlerts & cyan-ATLAS & ATLAS21jpb & 2021-04-03 08:25:26.400 & GALEXASC J101338.37-243115.9 & - & 0.134 & 0.02 \\ 
SN 2021iso & 11:57:21.460 & -20:13:58.56 & II & 0.031 & 18.39 & ALeRCE, ATLAS, Pan-STARRS & r-ZTF & ZTF21aatpsky & 2021-04-09 07:03:36.003 & GALEXASC J115721.58-201405.0 & - & 0.130 & 0.112 \\ 
SN 2021iyt & 10:18:44.410 & +00:03:19.43 & Ibn & 0.0712 & 19.97 & YSE, ALeRCE, Pan-STARRS & r-Sloan & PS21dau & 2021-04-09 08:39:50.400 & SDSS J101844.26+000315.3 & 0.07120 & 0.128 & 0.076 \\ 
SN 2021ahpl & 15:15:57.940 & -19:17:31.96 & SLSN-I & 0.051 & 21.03 & Pan-STARRS, ATLAS, ZTF & w-P1 & PS22bca & 2021-04-09 11:43:55.776 & GALEXASC J151557.90-191731.4 & - & 0.325 & 0.012 \\ 
SN 2021jbp & 13:24:57.974 & -24:10:50.77 & Ia-91bg & 0.042 & 18.553 & ATLAS, Pan-STARRS & cyan-ATLAS & ATLAS21mfg & 2021-04-11 09:36:00.000 & 2MASX J13245691-2410516  & 0.04226 & 0.276 & 0.243 \\ 
SN 2021juw & 19:47:05.210 & -20:12:11.52 & Ia & 0.039 & 18.59 & GaiaAlerts, AleRCE & G-Gaia & Gaia21bwh & 2021-04-12 14:58:33.600 & ESO 594- G 016 & 0.03924 & 0.239 & 0.144 \\ 
SN 2021jvi & 10:23:54.735 & -30:10:55.07 & Ia & 0.0672 & 19.256 & ATLAS & cyan-ATLAS & ATLAS21mwk & 2021-04-15 07:59:31.200 & GALEXASC J102354.71-301054.8 & - & 0.188 & 0.027 \\ 
SN 2021koq & 18:16:01.020 & +06:45:06.37 & Ia & 0.0205 & 17.84 & ATLAS, ZTF, Pan-STARRS & orange-ATLAS & ATLAS21nmi & 2021-04-28 11:39:50.400 & UGC 11177 & 0.02053 & 0.575 & 0.504 \\ 
SN 2021smj & 12:26:46.560 & +08:52:57.61 & Ia & 0.0042 & 17.025 & ATLAS, YSE, GaiaAlerts & cyan-ATLAS & ATLAS21baqj & 2021-07-08 06:25:55.200 & NGC 4411b & 0.00424 & 0.081 & 0.202 \\ 
SN 2021tkm & 14:31:41.590 & -43:24:51.66 & Ia & 0.0065 & 16.6 & ASAS-SN, ATLAS, GaiaAlerts & g-Sloan & ASASSN-21my & 2021-07-14 17:45:36.000 & IC 4441 & 0.00654 & 0.457 & 0.593 \\ 
SN 2021twb & 00:47:07.790 & -25:39:43.16 & Ia & 0.065 & 18.998 & ATLAS, GaiaAlerts, AleRCE & orange-ATLAS & ATLAS21bcev & 2021-07-20 12:47:31.200 & GALEXASC J004707.64-253938.2 & 0.06090 &  0.048 & 0.074 \\ 
SN 2021wjb & 20:00:49.920 & -38:34:38.03 & Ia & 0.02 & 16.5 & ASAS-SN, ATLAS, GaiaAlerts & g-Sloan & ASASSN-21qf & 2021-08-18 02:09:36.000 & IC 4931  & 0.02004 & 0.195 & 0.156 \\ 
SN 2021aele & 03:35:32.564 & -28:30:51.52 & Ia & 0.077 & 19.173 & ATLAS, YSE, Pan-STARRS & cyan-ATLAS & ATLAS21bmkw & 2021-11-11 10:03:21.600 & GALEXASC J033532.47-283050.7 & 0.07700 & 0.030 & 0.007 \\ 
SN 2021aexp & 01:38:11.964 & -29:42:04.38 & Ia & 0.07 & 18.23 & ATLAS, ZTF, Pan-STARRS & orange-ATLAS & ATLAS21bmox & 2021-11-19 08:16:47.000 & GALEXASC J013811.96-294205.1 & - & 0.049 & 0.01 \\ 
SN 2022da & 10:40:53.901 & -29:02:07.71 & Ia & 0.06 & 18.61 & ATLAS & cyan-ATLAS & ATLAS22apj & 2022-01-06 14:03:50.400 & GALEXASC J104053.61-290206.4 & - & 0.147 & 0.067 \\ 
SN 2022aff & 09:22:14.189 & -20:43:24.80 & Ia & 0.055 & 18.422 & ATLAS, ZTF & orange-ATLAS & ATLAS22cll & 2022-01-22 11:51:21.600 & GALEXASC J092214.10-204325.3 & 0.05464 & 0.187 & 0.039 \\ 
SN 2022aig & 04:48:10.845 & -17:50:27.38 & Ic & 0.04 & 19.277 & ATLAS, ZTF & orange-ATLAS & ATLAS22cpm & 2022-01-24 08:49:55.200 & PSO J072.0453-17.8410 & - & - & 0.003486 \\ 
SN 2022ame & 03:13:33.505 & -25:43:21.07 & II & 0.0056 & 17.3 & ATLAS, ALeRCE & Clear- & ATLAS22dap & 2022-01-27 12:08:18.000 & NGC 1255 & 0.00562 & 0.038 & 0.366 \\ 
SN 2022auf  & 05:21:31.042 & -16:05:18.03 & Ia-91T & 0.054 & 18.817 & ATLAS, Pan-STARRS & cyan-ATLAS & ATLAS22dpw & 2022-01-29 20:34:04.800 & 2MFGC 04381 & 0.05448 & 0.227 & 0.266 \\ 
SN 2022bll & 05:04:58.190 & -68:32:01.07 & IIP & 0.020 & 18.86 & GaiaAlerts & G-Gaia & Gaia22alq & 2022-02-04 00:46:04.800 & 2MASX J05045678-6832059 & 0.02448 & 0.206 & 0.147 \\ 
SN 2022blq & 12:28:33.442 & +08:52:14.67 & Ia & 0.089 & 21.1 & YSE, Pan-STARRS, AleRCE & g-Sloan & PS22aqv & 2022-02-04 12:30:14.400 & SDSS J122833.44+085213.2 & - & 0.053 & 0.024 \\ 
SN 2022bme & 04:41:43.320 & -25:13:19.60 & Ia & 0.071 & 18.621  & ATLAS, ZTF, GaiaAlerts, AMPEL & cyan-ATLAS & ATLAS22ewz & 2022-02-04 20:18:50.688 & GALEXASC J044143.31-251318.7 & 0.07648 & 0.084 & 0.005 \\ 
SN 2022btu & 12:54:14.440 & -24:26:02.22 & II & 0.01 & 18.84 & ATLAS, ALeRCE, GaiaAlerts & cyan-ATLAS & ATLAS22fmz & 2022-02-10 00:50:10.176 & WISEA J125414.14-242603.8 & - & 0.262 & 0.073 \\ 
SN 2022ctp & 04:58:57.857 & -15:58:18.03 & Ia & 0.06 & 19.3 & ZTF, ATLAS & r-ZTF & ZTF22aaazhly & 2022-02-17 03:33:07.200 & WISEA J045857.71-155815.1 & - & 0.196 & 0.059 \\ 
SN 2022ctu & 09:56:42.515 & -19:56:26.23 & Ia & 0.058 & 19.2469 & ALeRCE, ATLAS, Pan-STARRS & r-ZTF & ZTF22aabitze & 2022-02-19 08:09:52.001 & GALEXASC J095642.01-195630.8 & - & 0.105 & 0.135 \\ 
SN 2022eyk & 19:28:35.420 & -58:46:14.77 & II & 0.016 & 17.196 & ATLAS, GaiaAlerts & orange-ATLAS & ATLAS22jks & 2022-03-22 08:22:31.872 & IC 4857 & 0.01557 & 0.224 & 0.517 \\ 
SN 2022ffv & 03:36:29.715 & -35:17:21.59 & Ia & 0.006 & 17.0 & ASAS-SN, ATLAS, MASTER & g-Sloan & ASASSN-22dy & 2022-03-27 00:00:00.000 & NGC 1381 & 0.00575  & 0.036 & 0.533 \\ 
\enddata
\raggedright
\noindent 
\tablecomments{ The inflated version of this table is available in machine-readable form in the online journal. The portion shown here is its reduced form due to space formatting constraints.\\
%$\texttt{DP}$ --- This table is arranged in descending order of the number of data points (DP) for each supernova in the MeerLich transient archive. \\
$\texttt{RA, DEC}$ --- Right ascension and declination are given in the J2000 epoch. \\
$\texttt{SN Type}$ --- Spectral classification of supernovae. For a subtype of Ia SNe designated the luminous Ia-1991T-like and subluminous Ia-1991bg-like, we use the notation Ia-91T and Ia-91bg respectively.\\
$\texttt{Disc. Mag.}$ --- Discovery magnitudes are broadband magnitudes corresponding to the reported discovery filter from TNS, depending on the camera used from the discovery data source. \\
$\texttt{Rep. Group/s}$ --- List of groups who published their independent photometric measurements of the discovery of a candidate supernova on TNS. \\
$\texttt{Disc. Filter}$ --- The photometric filter used for the discovery data source, ''Clear'' indicates discovery by any number of non-professional astronomers. \\
$\texttt{Disc. Int. Name}$ --- Name assigned to the candidate supernova by the discovery data source, which could be any of the reporting group/s (Here we only take the earliest official name from the earliest reporting group). \\
$\texttt{Host}$ --- No host galaxy name, redshift and extinction listed for those galaxy footprints not detected in GALEX, SDSS, 2MASS, CatWISE, PanSTARRS and DES survey data. \\
$A_V$ --- Host Galactic extinction taken from \citet{2011ApJ...737..103S}. \\
$\texttt{Offset}$ Offset indicates the offset of the supernova in arcseconds from the coordinates of the host nucleus, taken from NED.}
\end{deluxetable*}
\end{rotatetable}
\pagebreak
%----------------------------------------------------------------------------%
\newpage
\clearpage
\pagestyle{empty}
\pagestyle{empty}
\setcounter{table}{2}
\movetabledown=5cm
\begin{rotatetable}
\begin{deluxetable*}{lllclllcllllll} % 14 columns %
%\centering
\tablewidth{0pt} % Forces natural width centering
%% Set the font size to be small enough to fit all 14 columns %%
%\fontsize{6}{7.2}\selectfont
%\tabletypesize{\scriptsize} % Options: \tiny, \small, \footnotesize, \scriptsize
\tabletypesize{\tiny}
%% The AASTeX-specific \tablecaption{} command looks for \deluxetable{} parameters and will fail if using the standard \abular{} environment inside a rotatetable. %%
\tablecaption{,continued}
%\label{table:ML1_sne}
%% Define column headers
\tablehead{
\colhead{SN} & \colhead{RA} & \colhead{DEC} & \colhead{SN} & \colhead{SN} & \colhead{Discovery} & \colhead{Reporting} & \colhead{Discovery} & \colhead{Discovery} & \colhead{Discovery} & \colhead{Host} & \colhead{Host} & \colhead{Host} & \colhead{Offset} \\
\colhead{Name} & \colhead{(h,m,s)} & \colhead{($^{\circ}$,$'$,$''$)} & \colhead{Type} & \colhead{Redshift} & \colhead{Mag} & \colhead{Group/s} & \colhead{Filter} & \colhead{Internal Name} & \colhead{Date (UT)} & \colhead{Name} & \colhead{Redshift} & \colhead{($A_V$)} & \colhead{(arcmin)}
}
%% Instead of using a simple dash -, the aastex package provides \nodata in the column for missing data. %%
\startdata
SN 2022fli & 13:42:41.752 & -34:05:30.17 & II & 0.025 & 17.677 & ATLAS & cyan-ATLAS & ATLAS22jyc & 2022-03-27 21:49:40.800 & GALEXASC J134241.68-340530.2 & - & 0.137 & 0.014 \\ 
SN 2022hrs & 12:43:34.335 & +11:34:35.87 & Ia & 0.0047 & 15.0 & ATLAS, ZTF, MASTER & Clear- & ATLAS22mip & 2022-04-16 14:50:40.000 & NGC 4647 & 0.00470 & 0.072 & 0.588 \\ 
SN 2022idx & 12:53:12.200 & -06:23:06.83 & Ia-91T & 0.0141 & 18.672 & ATLAS, ZTF, Pan-STARRS & cyan-ATLAS & ATLAS22mqi & 2022-04-23 09:54:26.784 & LCRS B125036.2-060655 & 0.01410 & 0.090 & 0.17 \\ 
SN 2022irr & 12:56:44.299 & -19:30:58.75 & Ia & 0.075 & 18.771 & ATLAS & orange-ATLAS & ATLAS22ncg & 2022-04-28 04:50:30.336 & 2MASS J12564407-1930570 & - & 0.199 & 0.059 \\ 
SN 2022jmp & 19:05:28.329 & -20:59:11.60 & Ia & 0.05 & 18.599 & ATLAS, ZTF & orange-ATLAS & ATLAS22nrq & 2022-05-07 08:04:12.000 & J190528.37-205910.7 & - & - & 0.0171 \\ 
SN 2022jnn & 14:06:55.443 & -25:01:10.48 & Ia & 0.049 & 19.4642 & ALeRCE, ATLAS, Pan-STARRS & r-ZTF & ZTF22aajidyk & 2022-05-08 06:32:42.999 & GALEXASC J140655.62-250111.7 & - & 0.171 & 0.047 \\ 
SN 2022kbn & 14:43:26.335 & -14:20:02.34 & II & 0.0411 & 18.77 & ZTF, Pan-STARRS, ATLAS & r-ZTF & ZTF22aajowwt & 2022-05-10 08:02:24.000 & GALEXASC J144326.40-142002.1 & 0.04281 & 0.350 & 0.009 \\ 
SN 2022kck & 12:33:07.540 & -14:29:24.72 & Ia & 0.05 & 19.78 & ZTF, ATLAS, Fink & g-ZTF & ZTF22aakfzqz & 2022-05-16 05:32:38.400 & GALEXASC J123307.86-142926.9   & 0.04700 & 0.130 & 0.052 \\ 
SN 2022lbj & 19:21:38.648 & -14:41:41.74 & Ia & 0.065 & 19.8726 & ALeRCE, ATLAS & r-ZTF & ZTF22aalrcmn & 2022-05-27 09:56:16.996 & WISEA J192138.80-144142.2 & - & 0.430 & 0.038 \\ 
SN 2022ljo & 14:46:06.566 & -24:47:24.94 & Ia & 0.07 & 19.24 & ZTF, ATLAS, Fink, Pan-STARRS & r-ZTF & ZTF22aalzeui & 2022-05-31 06:31:40.800 & PSO J221.5273-24.7905 & - & - & 0.012935 \\ 
SN 2022lmw & 21:31:04.516 & -11:57:32.47 & Ia & 0.087 & 19.38 & ZTF, Pan-STARRS, ATLAS & g-ZTF & ZTF22aamglev & 2022-06-01 09:33:07.200 & APMUKS(BJ) B212821.88-121048.8 & - & 0.149 & 0.047 \\ 
SN 2022mts & 18:42:43.388 & -22:04:42.73 & Ia & 0.0292 & 18.9767 & ZTF & g-ZTF & ZTF22aaobkbh & 2022-06-11 09:33:46.944 & 2MASX J18424365-2204447 & 0.02917 & 1.168 & 0.066 \\ 
SN 2022mxk & 14:14:20.657 & -19:25:22.48 & Ia & 0.07 & 18.9174 & AleRCE & r-ZTF & ZTF22aaogyhz & 2022-06-17 05:02:45.997 & GALEXASC J141420.52-192523.2 & - & 0.203 & 0.034 \\ 
SN 2022nah & 00:46:56.910 & -21:50:49.60 & II & 0.0214 & 17.81 & ATLAS, GaiaAlerts, ZTF & orange-ATLAS & ATLAS22qyk & 2022-06-19 07:37:19.776 & ESO 540- G 021 & 0.02143 & 0.049 & 0.067 \\ 
SN 2022pbw & 23:22:54.855 & -12:57:22.06 & Ia & 0.079 & 19.3571 & ALeRCE, ATLAS, Pan-STARRS & r-ZTF & ZTF22aaujqll & 2022-07-19 10:02:30.002 & GALEXASC J232254.48-125720.3 & 0.08150 & 0.089 & 0.105 \\ 
SN 2022phj & 23:59:48.777 & -15:42:54.40 & II & 0.039 & 18.8669 & AleRCE, ATLAS, Pan-STARRS & g-ZTF & ZTF22aavbcky & 2022-07-22 11:00:49.997 & GALEXASC J235948.85-154258.2 & - &  0.087 & 0.059 \\ 
SN 2022qio & 02:13:11.137 & -20:28:15.36 & Ia & 0.104 & 18.71 & ATLAS, ZTF, Pan-STARRS & cyan-ATLAS & ATLAS22xyo & 2022-08-01 02:36:46.368 & WISEA J021311.43-202814.9 & - & 0.044 & 0.032 \\ 
SN 2022tvz & 19:55:01.660 & -21:05:57.37 & Ia & 0.0233 & 17.976 & ATLAS, GaiaAlerts, Pan-STARRS & orange-ATLAS & ATLAS22beig & 2022-09-09 20:31:03.360 & ESO 595- G 003 & 0.02214 & 0.413 & 0.183 \\ 
SN 2022uax & 20:02:39.314 & -37:58:12.68 & Ia & 0.03 & 18.182 & ATLAS & orange-ATLAS & ATLAS22beju & 2022-09-12 02:04:12.864 & GALEXASC J200238.99-375815.0 & - & 0.245 & 0.074 \\ 
SN 2022ube & 21:51:54.530 & -03:53:54.46 & Ia & 0.06 & 19.27 & Pan-STARRS, ATLAS, ALeRCE & i-P1 & PS22ihp & 2022-09-12 09:59:51.648 & WISEA J215154.83-035354.7 & - & 0.111 & 0.077 \\ 
SN 2022udc & 04:23:51.146 & -10:18:30.19 & Ia & 0.05 & 18.218 & ATLAS, ALeRCE, Pan-STARRS & orange-ATLAS & ATLAS22bepz & 2022-09-14 14:31:36.192 & GALEXASC J042350.74-101826.9 & 0.04603 & 0.212 & 0.1 \\ 
SN 2022uoo & 22:09:58.397 & -07:33:59.00 & Ia & 0.0753 & 19.17 & ZTF, ATLAS & g-ZTF & ZTF22abfwthv & 2022-09-17 05:03:50.400 & SDSS J220958.42-073359.7 & 0.07526 & 0.152 & 0.015 \\ 
SN 2022vbc & 22:01:46.265 & -17:52:21.96 & Ia & 0.101 & 19.42 & ZTF, Fink & g-ZTF & ZTF22abfxwkm & 2022-09-17 07:03:21.600 & 2MASX J22014606-1752232 & 0.08839 & 0.095 & 0.052 \\ 
SN 2022vcv & 03:42:40.982 & -12:50:00.04 & Ia-91T & 0.071 & 19.69 & Pan-STARRS, ZTF, ATLAS, Fink & w-P1 & PS22jde & 2022-09-19 13:12:45.792 & GALEXASC J034241.21-125003.0 & - & 0.181 & 0.057 \\ 
SN 2022vrr & 19:17:44.071 & -07:06:52.61 & Ia & 0.063 & 19.65 & ZTF, Fink & r-ZTF & ZTF22abhencl & 2022-09-22 03:34:33.600 & - & - & - & - \\ 
SN 2022vqk & 01:51:25.990 & -25:10:54.67 & Ia & 0.05 & 17.3 & ASAS-SN, ATLAS & g-Sloan & ASASSN-22lz & 2022-09-23 02:52:48.000 & 2MASX J01512586-2510531 & 0.06759 & 0.030 & 0.035 \\ 
\enddata
\raggedright
\noindent 
\tablecomments{ The inflated version of this table is available in machine-readable form in the online journal. The portion shown here is its reduced form due to space formatting constraints.\\
%$\texttt{DP}$ --- This table is arranged in descending order of the number of data points (DP) for each supernova in the MeerLich transient archive. \\
$\texttt{RA, DEC}$ --- Right ascension and declination are given in the J2000 epoch. \\
$\texttt{SN Type}$ --- Spectral classification of supernovae. For a subtype of Ia SNe designated the luminous Ia-1991T-like and subluminous Ia-1991bg-like, we use the notation Ia-91T and Ia-91bg respectively.\\
$\texttt{Disc. Mag.}$ --- Discovery magnitudes are broadband-band magnitudes corresponding to the reported discovery filter from TNS, depending on the camera used from the discovery data source. \\
$\texttt{Rep. Group/s}$ --- List of groups who published their independent photometric measurements of the discovery of a candidate supernova on TNS. \\
$\texttt{Disc. Filter}$ --- The photometric filter used for the discovery data source, ''Clear'' indicates discovery by any number of non-professional astronomers. \\
$\texttt{Disc. Int. Name}$ --- Name assigned to the candidate supernova by the discovery data source, which could be any of the reporting group/s (Here we only take the earliest official name from the earliest reporting group). \\
$\texttt{Host}$ --- No host galaxy name, redshift and extinction listed for those galaxy footprints not detected in GALEX, SDSS, 2MASS, CatWISE, PanSTARRS and DES survey data. \\
$A_V$ --- Host Galactic extinction taken from \citet{2011ApJ...737..103S}. \\
% Redshifts are reported to five decimal places, if the SN host galaxy redshift $z_{host}$ is known. Otherwise, the SN redshift ($z_{SN}$) is reported to three decimal places.
$\texttt{Offset}$ Offset indicates the offset of the supernova in arcseconds from the coordinates of the host nucleus, taken from NED.}
\end{deluxetable*}
\end{rotatetable}
\pagebreak
%-------------------------------------------------%
%\input{appendix_binned_sne}
\section{Location of SN 2019lub and SN 2022vrr}\label{app:sn_hostless}
\begin{figure}[!ht]
    %\captionsetup{justification=centering}
     \centering
     \includegraphics[width=0.35\columnwidth]{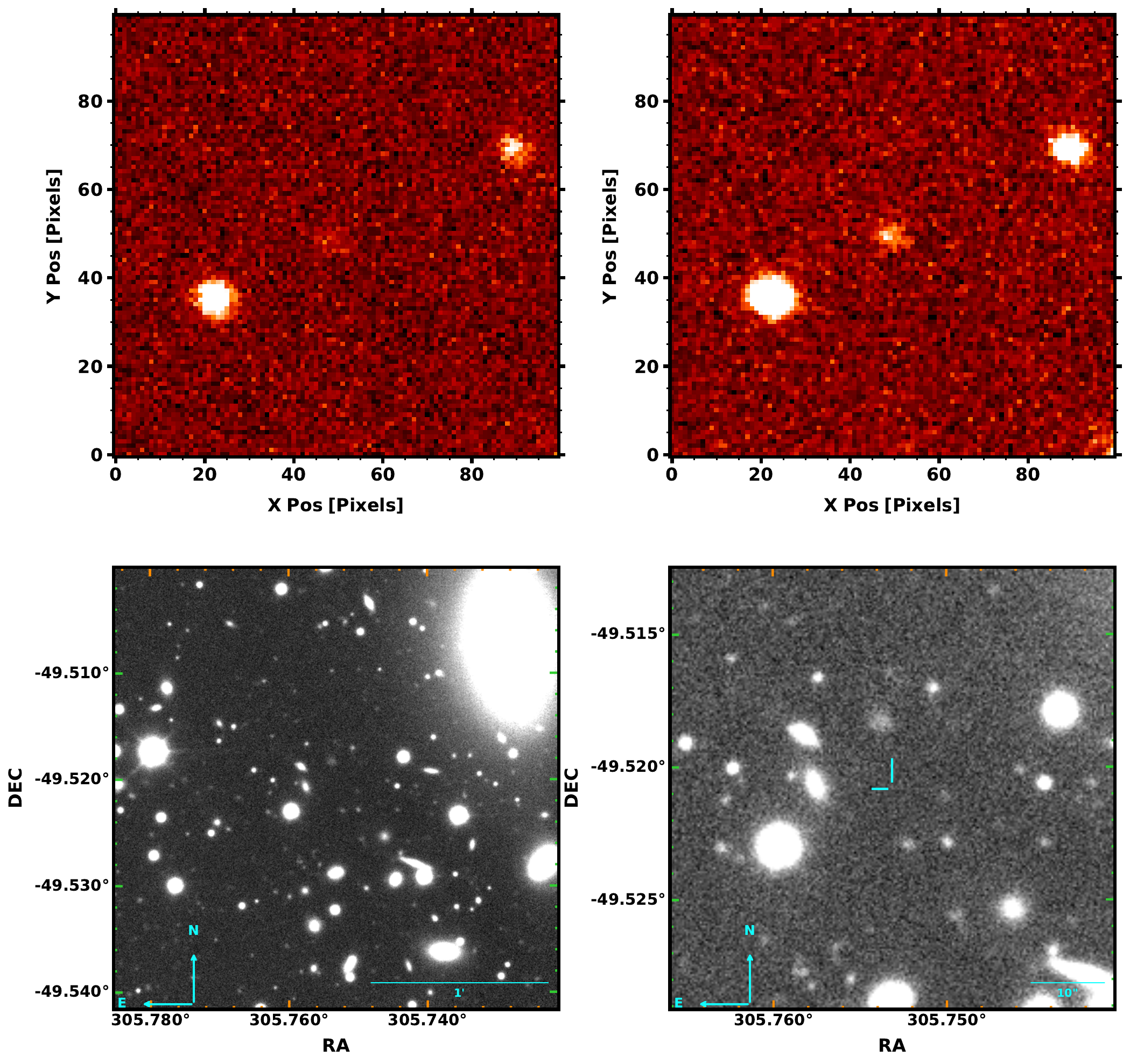}
     %--------------------------------------------%
     \caption{MeerLICHT postage stamps centered at the position of SN\,2019lub on 2019 August 17, UT 20:09:43.859. {\sl Top left}: $q$-band reduced image. {\sl Top right}: $q$-band reference science image. SN\,2019lub was discovered in the reference image, with the subsequent reduced images showing the SN in its fading phase. The cutouts are 1 arcminute across. {\sl Bottom left}: Deepest $2.5\times2.5$ arcminute $r$-band image from the DES Data Release 2 fully calibrated image products. {\sl Bottom right}: A zoomed-in central square cutout of the DES image exactly 1.0 arcminute across. The cyan cross markers show the location of the SN\,2019lub.}
     \label{fig:SN2019lub_hostless}
    \end{figure}

\begin{figure}[!ht]
    %\captionsetup{justification=centering}
     \centering
     \includegraphics[width=0.35\columnwidth]{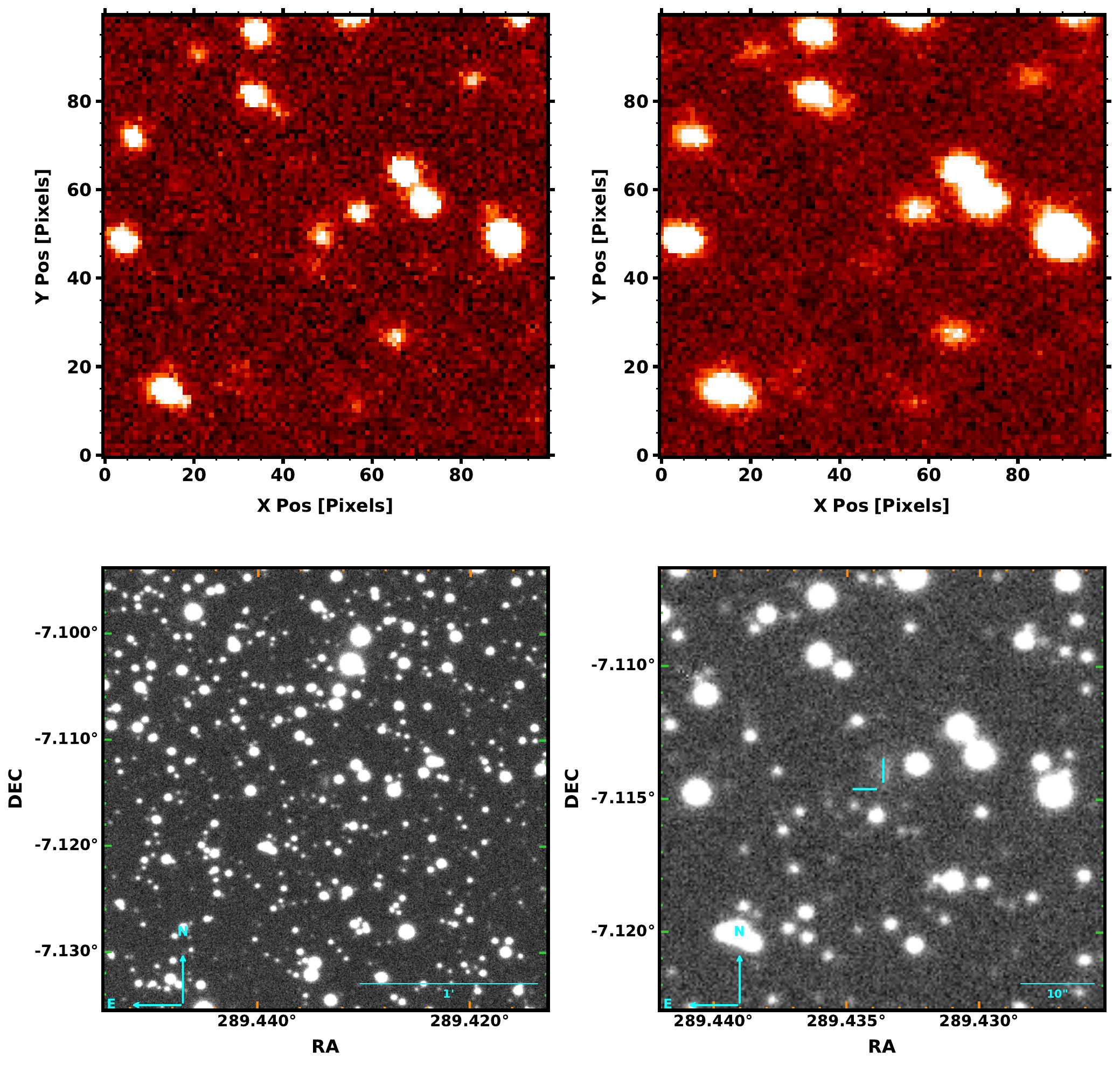}
     %--------------------------------------------%
     \caption{MeerLICHT postage stamps centered at the position of SN\,2022vrr on 2022 October 01, UT 18:27:57.380.  {\sl Top Left}: $q$-band reduced image. {\sl Top right}: $q$-band reference science image. {\sl Bottom left}: Deepest $2.5\times2.5$ arcminute $r$-band image from the DES Data Release 2 fully calibrated image products. {\sl Bottom right}: A zoomed-in central square cutout of the DES image exactly 1.0 arcminute across. The cyan cross markers show the location of the SN\,2022vrr.}
     \label{fig:SN2022vrr_hostless}
    \end{figure}

The location of SN\,2019lub and SN\,2022vrr, considered to be hostless, are displayed in Figure \ref{fig:SN2019lub_hostless} and Figure \ref{fig:SN2022vrr_hostless} respectively. The retrieved MeerLICHT cutouts are $1$ arcminute across, centered on the SN location. To confirm that these SNe are not associated with any galaxy in the field, we also inspected the deepest images from the Dark Energy Survey data release 2 (DES-DR2, \citealt{DES_DR2_2021}) due to their high photometric precision and depth. In the DES image of SN\,2022vrr field, there is an apparent extended background emission, but since it is not cataloged in NED and GLADE+, we cannot confirm it as the host galaxy. The DES image of SN\,2019lub field does not show any extended background emission, and the largest candidate host galaxy ESO\,234-G\,014 (north west of the SN location) is at $z=0.038683$, which is not consistent with the spectroscopic redshift of SN\,2019lub ($z=0.066$).
%%%%%%%%%%%%%%%%%%%%%%%%%%%%%%%%%%%%%%%%%%%%%%%%%%%%%%
    %%%%%%%%%%%%%%%%%%%%%%%%%%%%%%%%%%%%%%%%%%%%%%%
    %%%%%%%%%%%%%%%%%%%%%%%%%%%%%%%%%%%%%%%%%%%%%%%
    %%%%%%%%%%%%%%%%%%%%%%%%%%%%%%%%%%%%%%%%%%%%%%%
    %%%%%%%%%%%%%%%%%%%%%%%%%%%%%%%%%%%%%%%%%%%%%%%
    
    %%%%%%%%%%%%%%%%%%%%%%%%%%%%%%%%%%%%%%%%%%%%%%%
    %%%%%%%%%%%%%%%%%%%%%%%%%%%%%%%%%%%%%%%%%%%%%%%

\newpage
\clearpage
\pagestyle{empty}
\section*{Supplementary Material}\label{supp:figures}
%%%%%%%%%%%%%%%%%%%%%%%%%%%%%%%%%%%%%%%%%%%%%%%
%\subsubsection{SN 2020ftl (Ia)}
%%%%%%%%%%%%%%%%%%%%%%%%%%%%%%%%%%%%%%%%%%%%%%%
%%%%%%%%%%%%%%%%%%%%%%%%%%%%%%%%%%%%%%%%%%%%%%%
\begin{figure}[!ht]
\centering
\begin{minipage}{0.491\textwidth}
\centering
\includegraphics[width=0.85\linewidth,trim={0.0cm 0.0cm 0.0cm 0.0cm},clip]{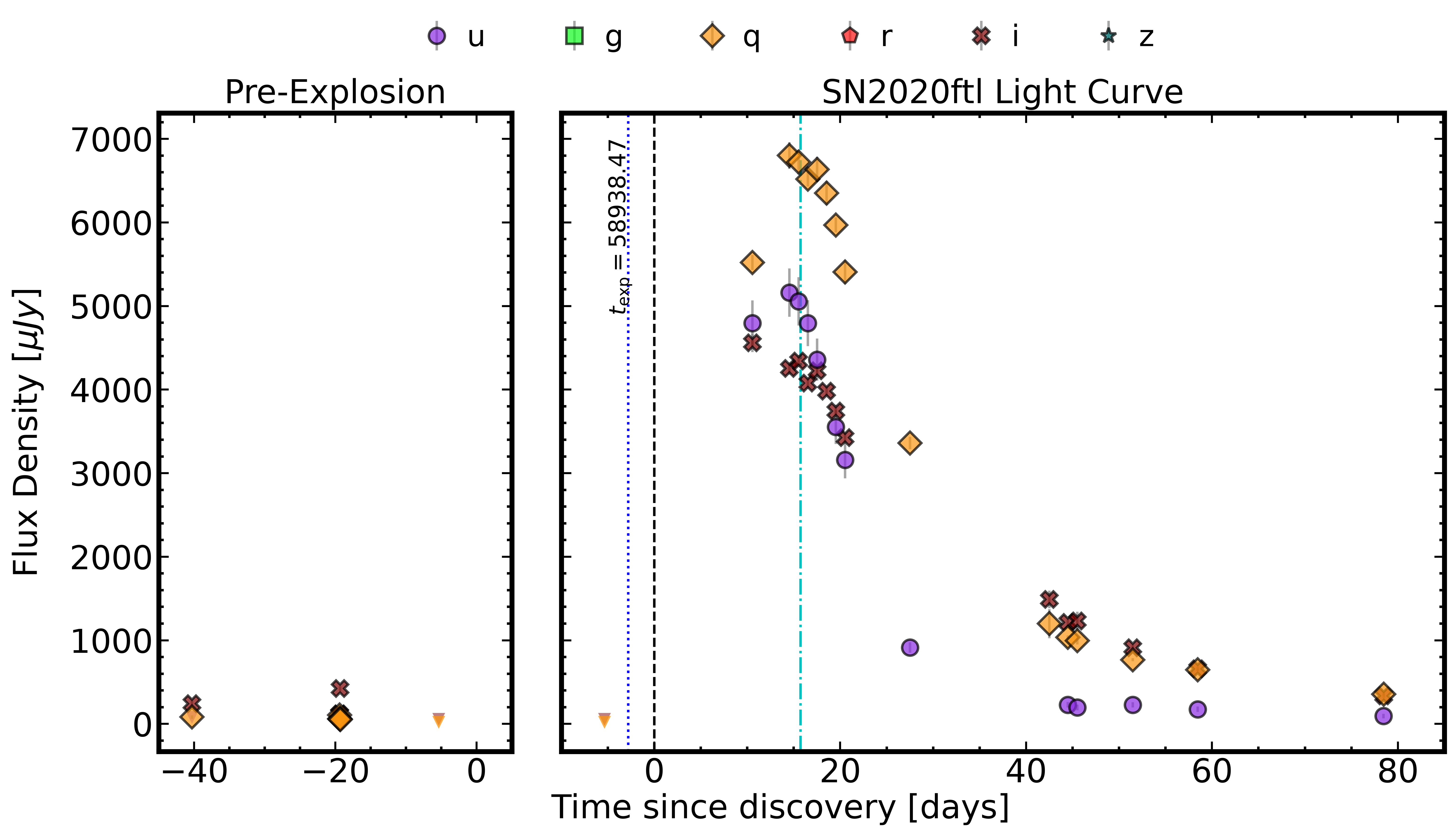}
\includegraphics[width=0.85\linewidth,trim={0.0cm 0.0cm 0.0cm 0.0cm},clip]{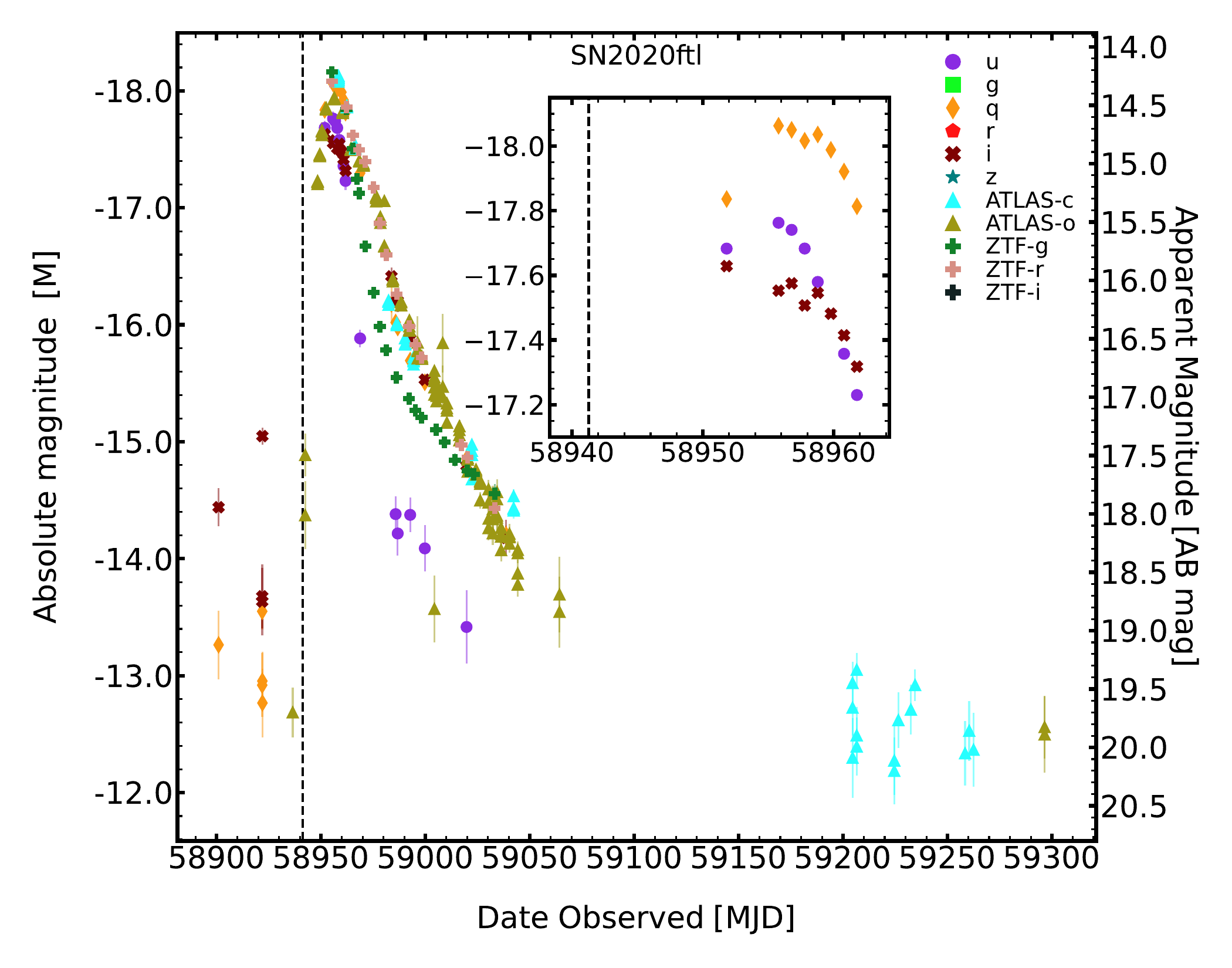}
\end{minipage}
\caption{
Pre- and post-explosion light curves of SN\,2020ftl from MeerLICHT. {\sl Top left}: Pre-explosion photometry up to 40 days before discovery. The shaded inverted triangles are upper limits, and all the other filled symbols are detections. 
{\sl Top right}: Multi-band optical light curves with the underlying background emission subtracted. The phases are relative to the time of discovery as reported on TNS, marked by the black vertical dashed line. The estimated explosion epoch and peak brightness epoch are designated with the dotted blue line and dash-dotted cyan line, respectively.
{\sl Bottom}: Absolute magnitude light curves from MeerLICHT, ATLAS, and ZTF. ATLAS magnitudes are represented by filled triangles, and ZTF corrected magnitudes are represented by filled 'plus' symbols. A zoom into the peak of SN\,2020ftl is plotted in the insert window, showing the  MeerLICHT observations of the steeply decaying phase of the SN.
}
\label{fig:SN2020ftl}
\end{figure}
%----------------------------------------------%
\begin{figure}
\centering
\begin{minipage}{0.491\textwidth}
\includegraphics[width=0.85\linewidth,trim={0.0cm 0.0cm 0.0cm 0.0cm},clip]{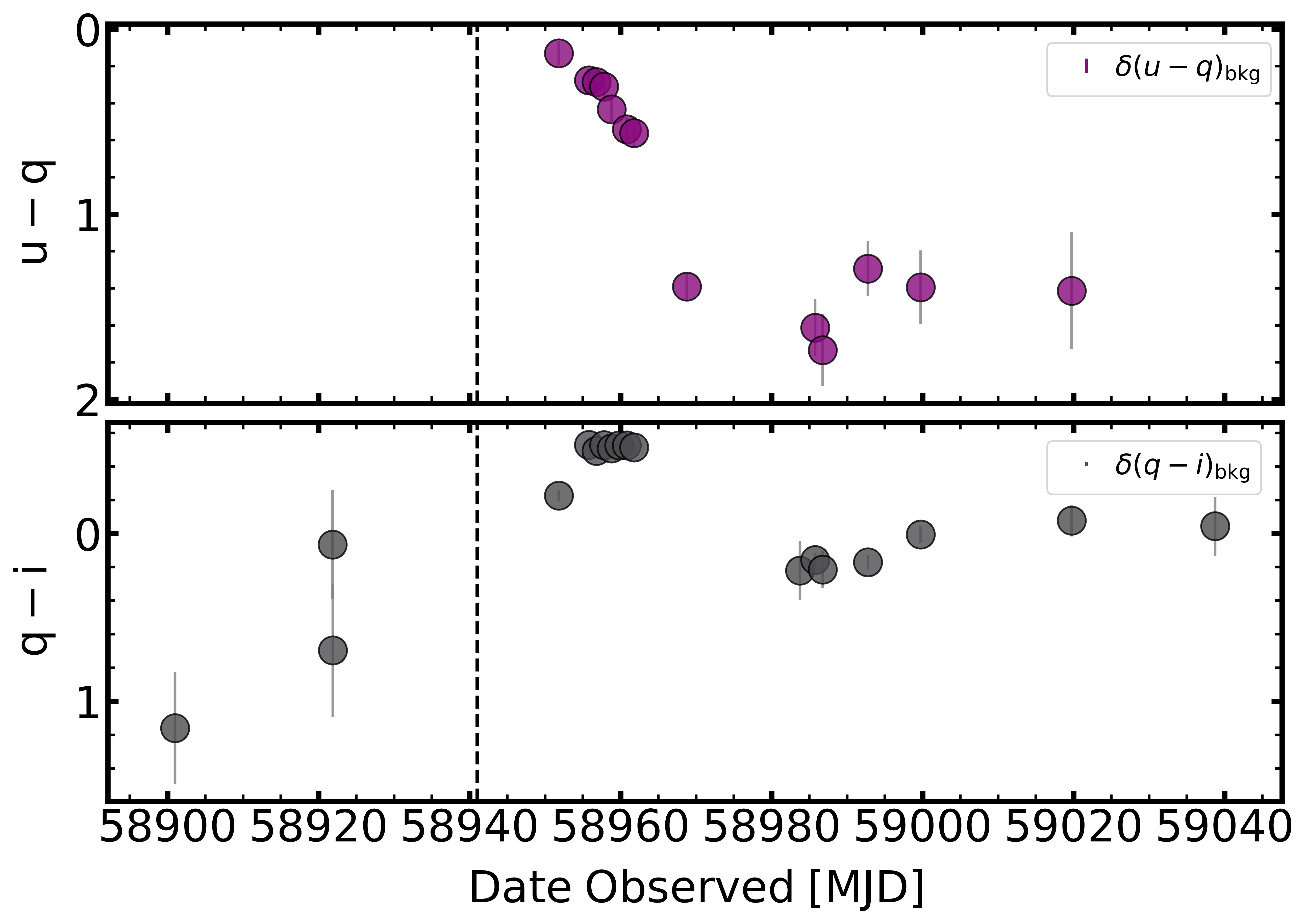}
\end{minipage}
\caption{
Extinction corrected ($A_{V,\mathrm{G}}=0.053$) color evolution of SN\,2020ftl from MeerLICHT observations. $\delta (color)_{\mathrm{bkg}}$ represents the magnitude of the systematic uncertainty in the computed color index.
}
\label{fig:SN2020ftl_color}
\end{figure}
%----------------------------------------------%
\begin{figure}
\centering
\includegraphics[width=0.35\columnwidth]{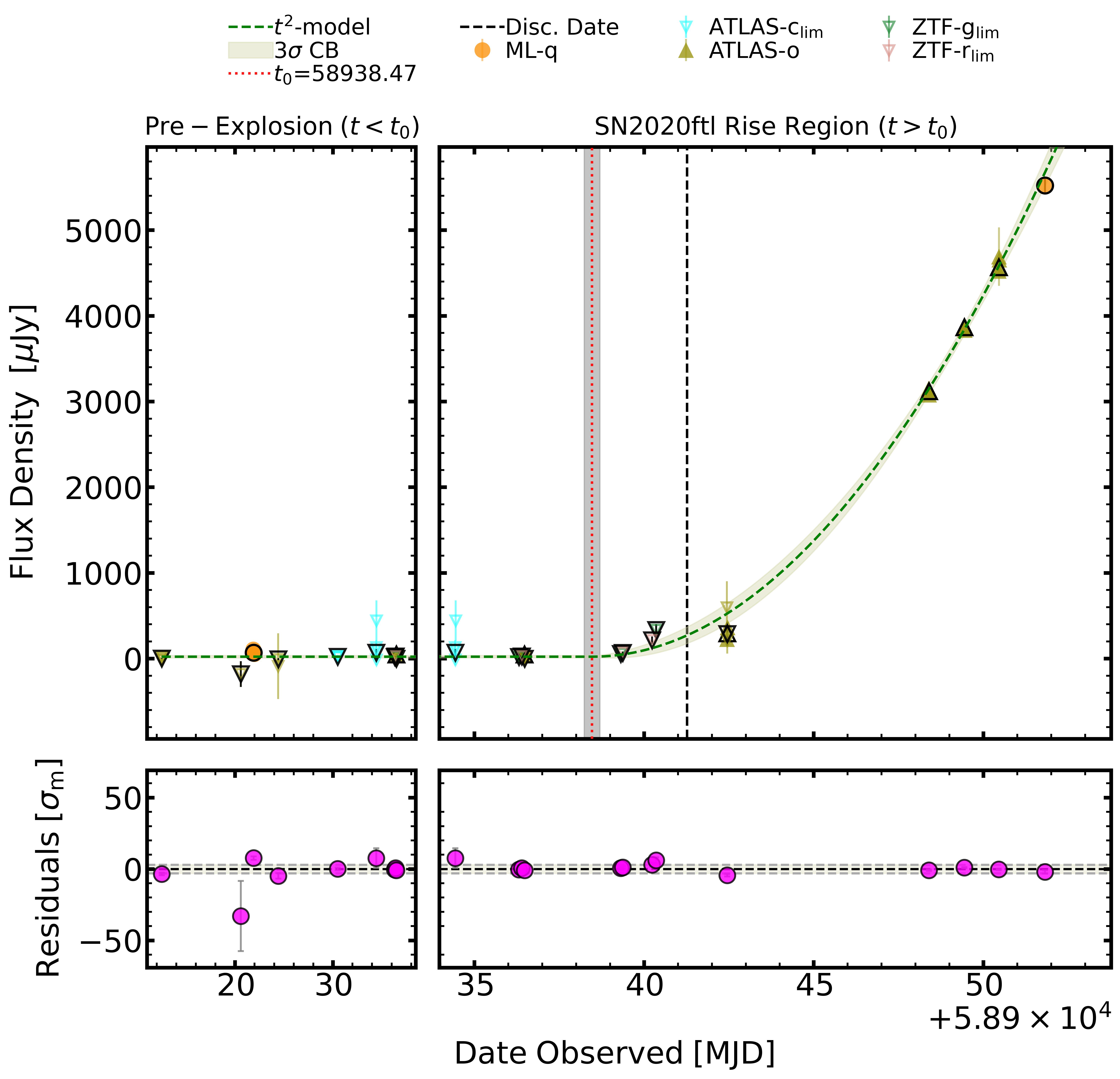}\\
\includegraphics[width=0.35\columnwidth]{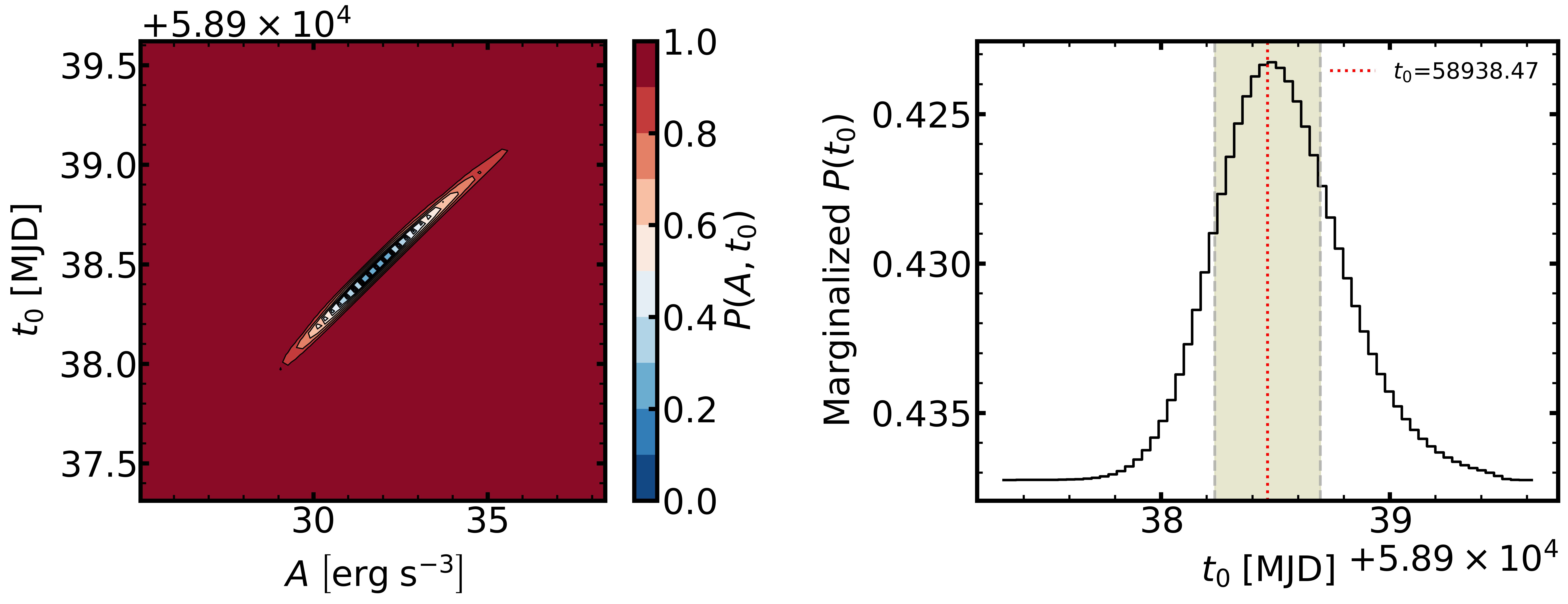}
\caption{Same as Figure\,\ref{fig:SN2020oi_t2model} for SN\,2020ftl.}
\label{fig:SN2020ftl_t2model}
\end{figure}
%%%%%%%%%%%%%%%%%%%%%%%%%%%%%%%%%%%%%%%%%%%%%%%
%\subsubsection{SN 2020jfo (IIP)}
%%%%%%%%%%%%%%%%%%%%%%%%%%%%%%%%%%%%%%%%%%%%%%%
%%%%%%%%%%%%%%%%%%%%%%%%%%%%%%%%%%%%%%%%%%%%%%%
\begin{figure}[!ht]
\centering
\begin{minipage}{0.491\textwidth}
\centering
\includegraphics[width=0.85\linewidth,trim={0.0cm 0.0cm 0.0cm 0.0cm},clip]{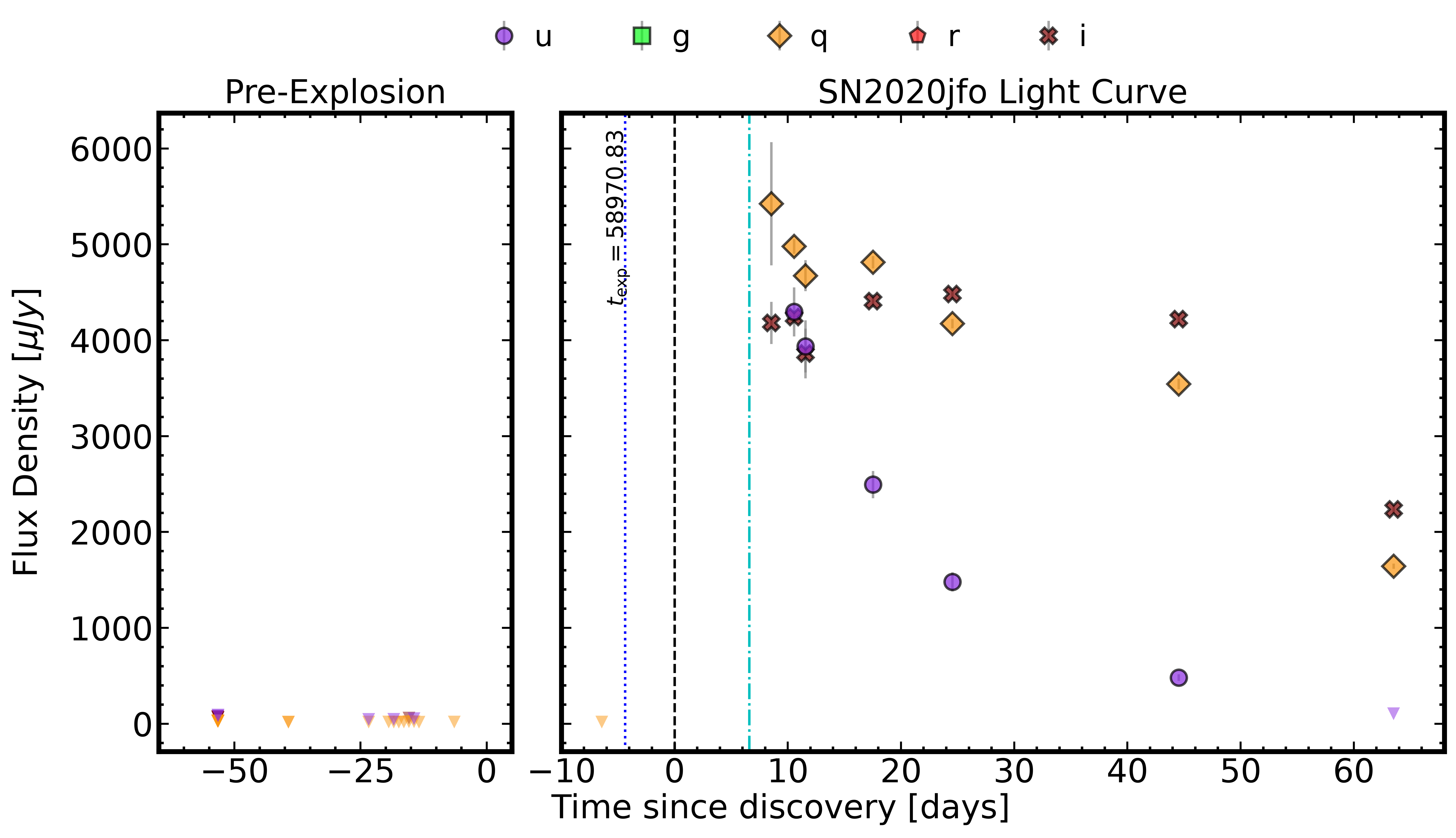}
\includegraphics[width=0.85\linewidth,trim={0.0cm 0.0cm 0.0cm 0.0cm},clip]{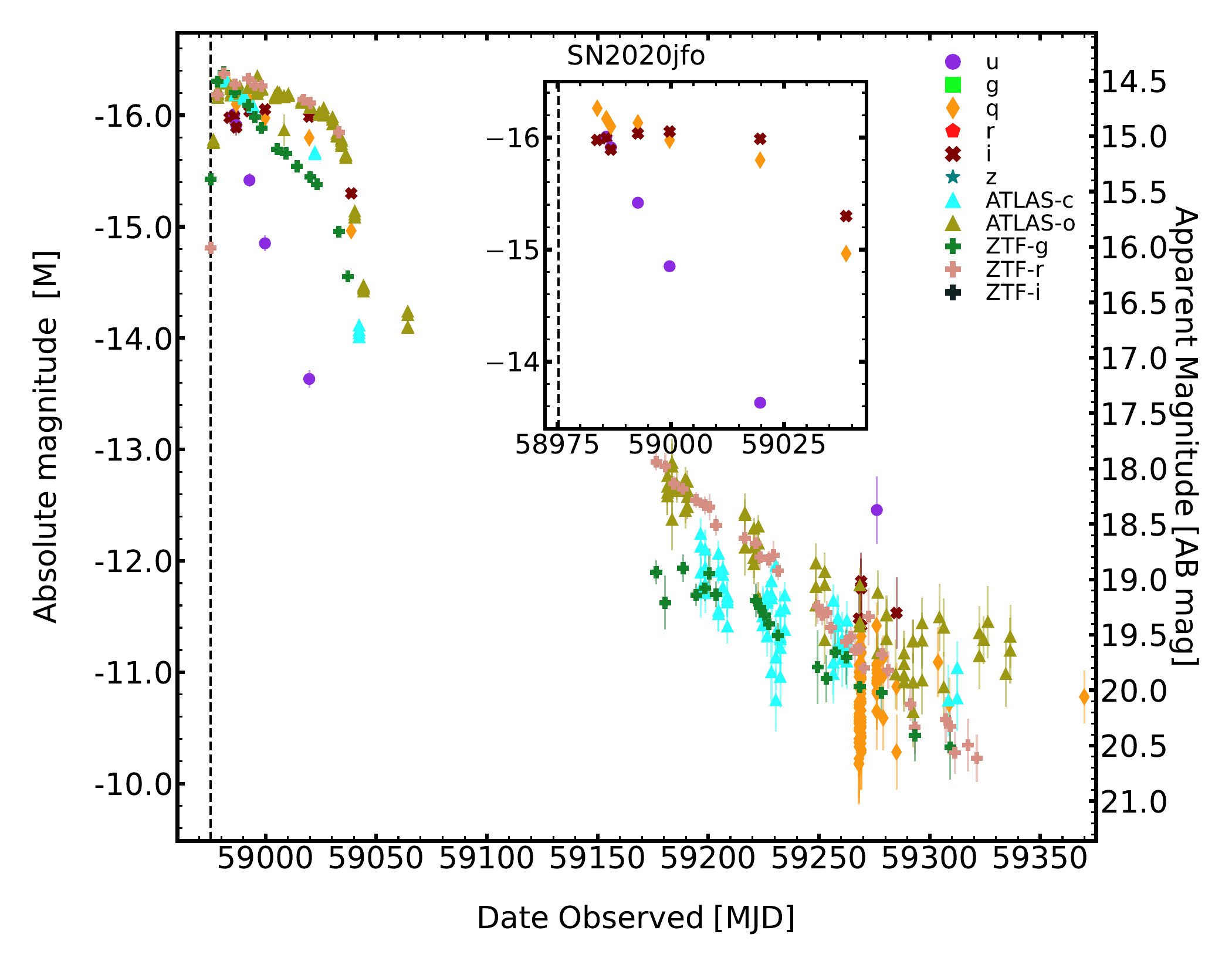}
\end{minipage}
\caption{
Pre- and post-explosion light curves of SN\,2020jfo from MeerLICHT. 
{\sl Top left}: Pre-explosion photometry up to 55 days before discovery. The shaded inverted triangles are upper limits, and all the other filled symbols are detections. 
{\sl Top right}: Multi-band optical light curves of SN\,2020jfo with the underlying background emission subtracted. The phases are relative to the time of discovery as reported on TNS, marked by the black vertical dashed line. The estimated explosion epoch and peak brightness epoch are designated with the dotted blue line and dash-dotted cyan line, respectively. 
{\sl Bottom}: Absolute magnitude light curves from MeerLICHT, ATLAS, and ZTF. ATLAS magnitudes are represented by filled triangles, and ZTF corrected magnitudes are represented by filled 'plus' symbols. A zoom into the peak of SN\,2020jfo is plotted in the insert window, showing the MeerLICHT observations of the short-plateau phase of the SN. 
}
\label{fig:SN2020jfo}
\end{figure}
%-----------------------------------------------%
\begin{figure}[!ht]
\centering
\begin{minipage}{0.491\textwidth}
\includegraphics[width=0.85\linewidth,trim={0.0cm 0.0cm 0.0cm 0.0cm},clip]{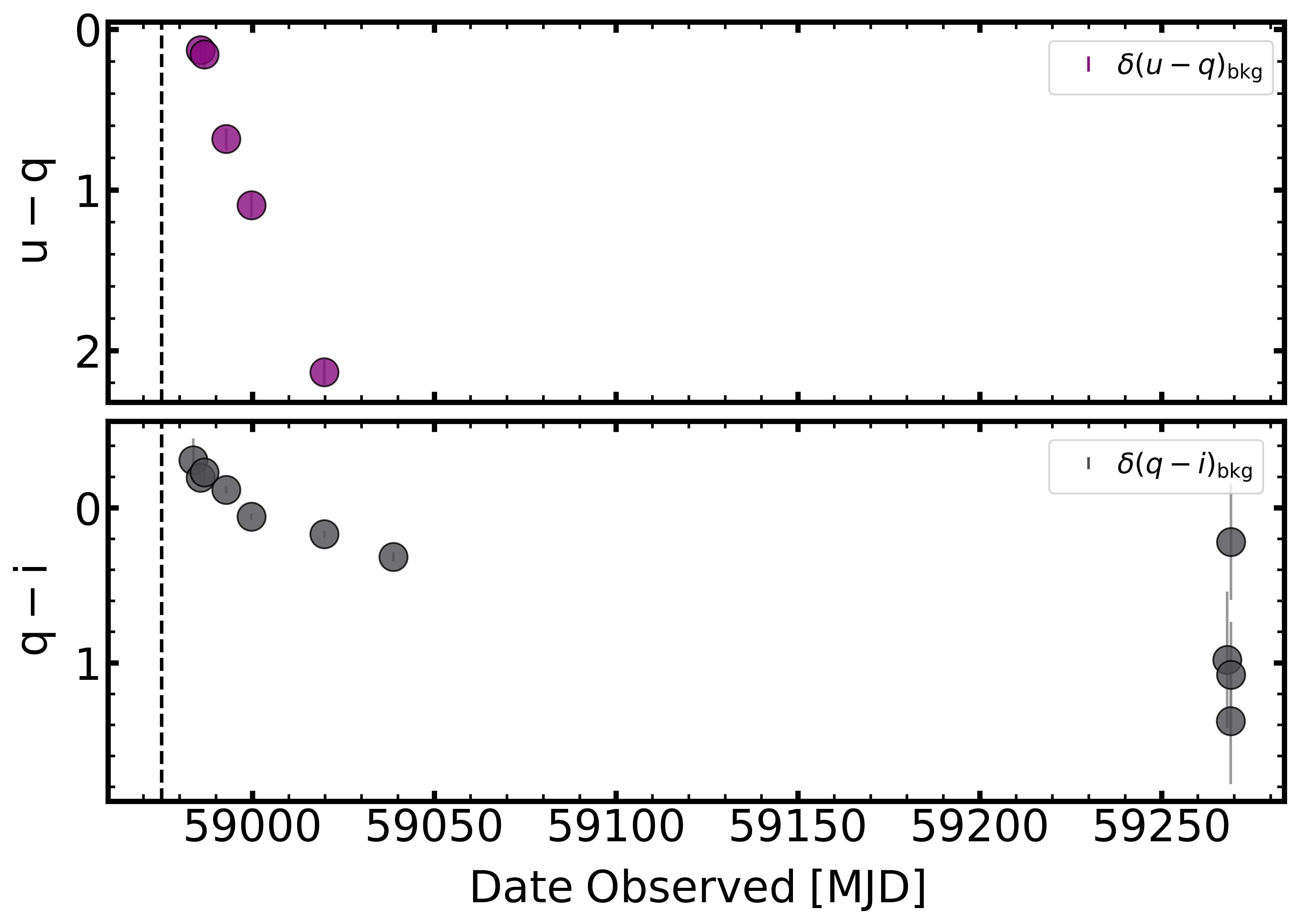}
\end{minipage}
\caption{
Extinction corrected ($A_{V,\mathrm{G}}=0.065$) color evolution of SN\,2020jfo from MeerLICHT observations. $\delta (color)_{\mathrm{bkg}}$ represents the magnitude of the systematic uncertainty in the computed color index.
} 
\label{fig:SN2020jfo_color}
\end{figure}
%-----------------------------------------------%
\begin{figure}[!ht]
\centering
\includegraphics[width=0.35\columnwidth]{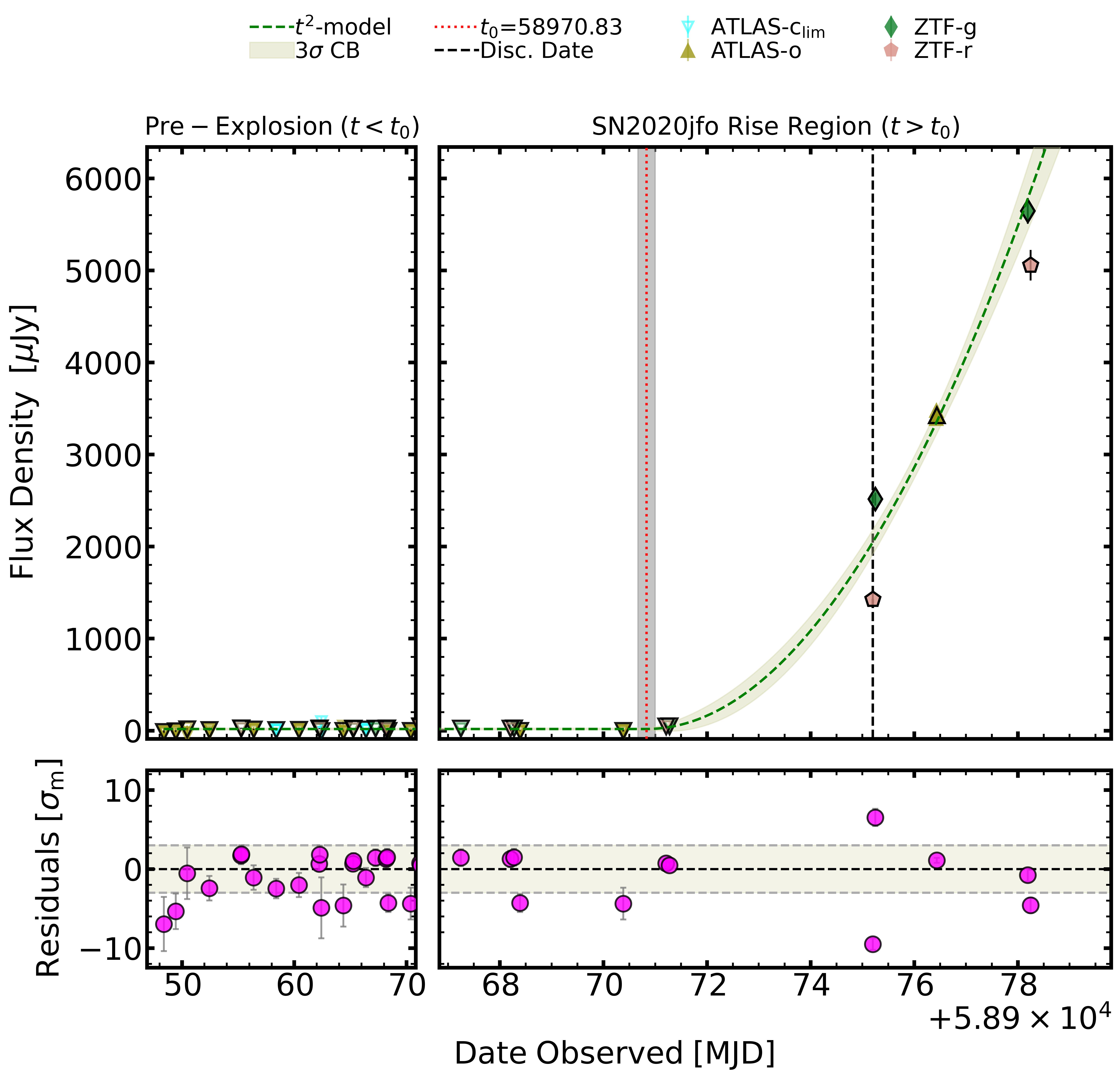}\\
\includegraphics[width=0.35\columnwidth]{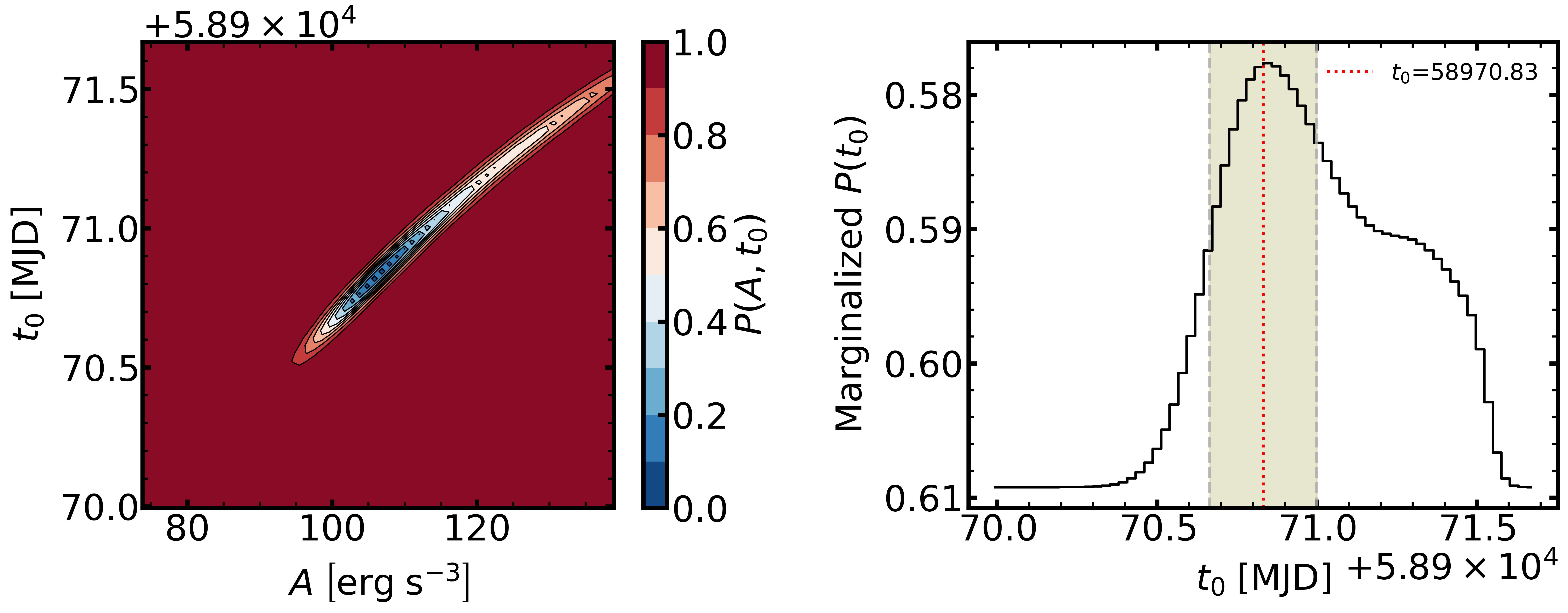}%
\caption{
Same as Figure\,\ref{fig:SN2020oi_t2model} for SN\,2020jfo. The uncertainty in the explosion time is large due to the sparse multi-survey coverage in the initial $\sim$$15$ days before peak brightness. As a result, there is an obvious degeneracy between the fitted parameters $A$ and $t_0$.
} 
\label{fig:SN2020jfo_t2model}
\end{figure}
%%%%%%%%%%%%%%%%%%%%%%%%%%%%%%%%%%%%%%%%%%%%%%%
%\subsubsection{SN 2021tkm (Ia)}
%%%%%%%%%%%%%%%%%%%%%%%%%%%%%%%%%%%%%%%%%%%%%%%
%%%%%%%%%%%%%%%%%%%%%%%%%%%%%%%%%%%%%%%%%%%%%%%
\begin{figure}[!ht]
\centering
\begin{minipage}{0.491\textwidth}
\centering
\includegraphics[width=0.85\linewidth,trim={0.0cm 0.0cm 0.0cm 0.0cm},clip]{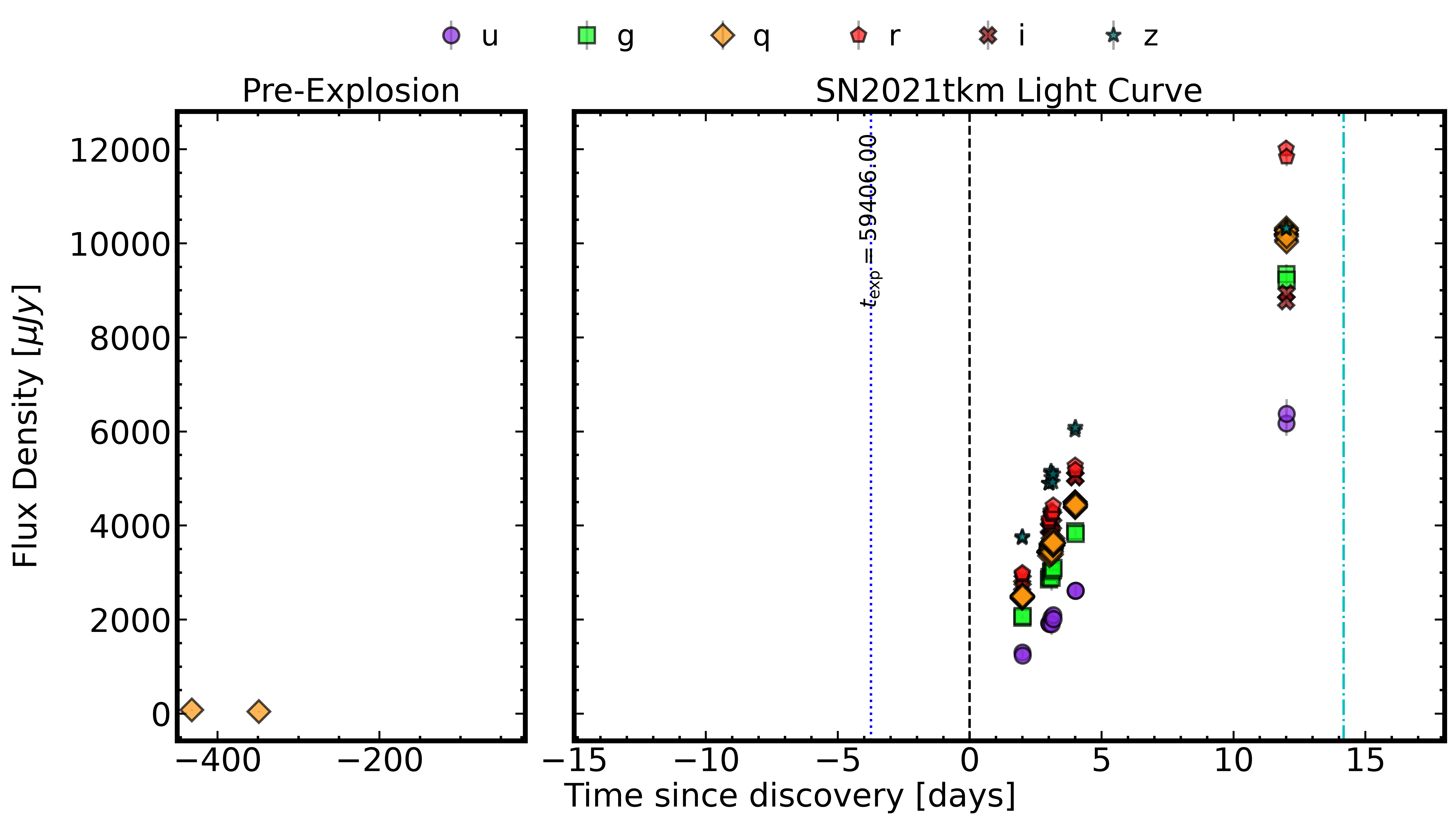}
\includegraphics[width=0.85\linewidth,trim={0.0cm 0.0cm 0.0cm 0.0cm},clip]{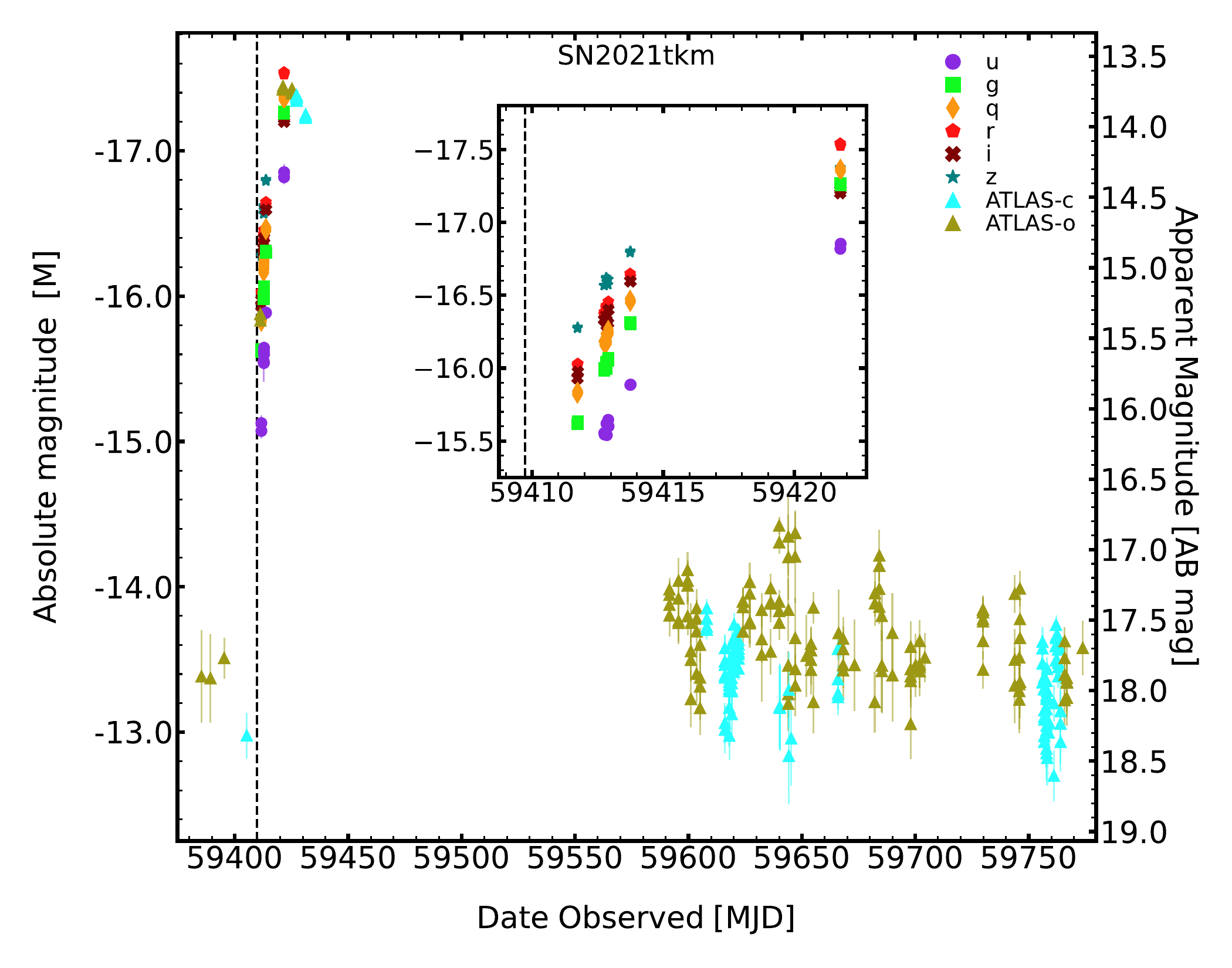}
\end{minipage}
\caption{
Pre- and post-explosion light curves of SN\,2021tkm from MeerLICHT.
{\sl Top left}: Pre-explosion photometry up to 450 days before discovery. The shaded inverted triangles are upper limits, and all the other filled symbols are detections. 
{\sl Top right}: Multi-band optical light curves of SN\,2021tkm without the underlying background emission subtracted. The phases are relative to the time of discovery as reported on TNS, marked by the black vertical dashed line. The estimated explosion epoch and peak brightness epoch are designated with the dotted blue line and dash-dotted cyan line, respectively. 
{\sl Bottom}: Absolute magnitude light curves from MeerLICHT and ATLAS. Filled triangles represent ATLAS magnitudes. A zoom into the early rise of SN\,2021tkm is plotted in the insert window, showing the MeerLICHT observations of the initial $15$ days post-discovery.
}
\label{fig:SN2021tkm}
\end{figure}
%---------------------------------------------%
\begin{figure}
\centering
\begin{minipage}{0.491\textwidth}
\includegraphics[width=0.85\linewidth,trim={0.0cm 0.0cm 0.0cm 0.0cm},clip]{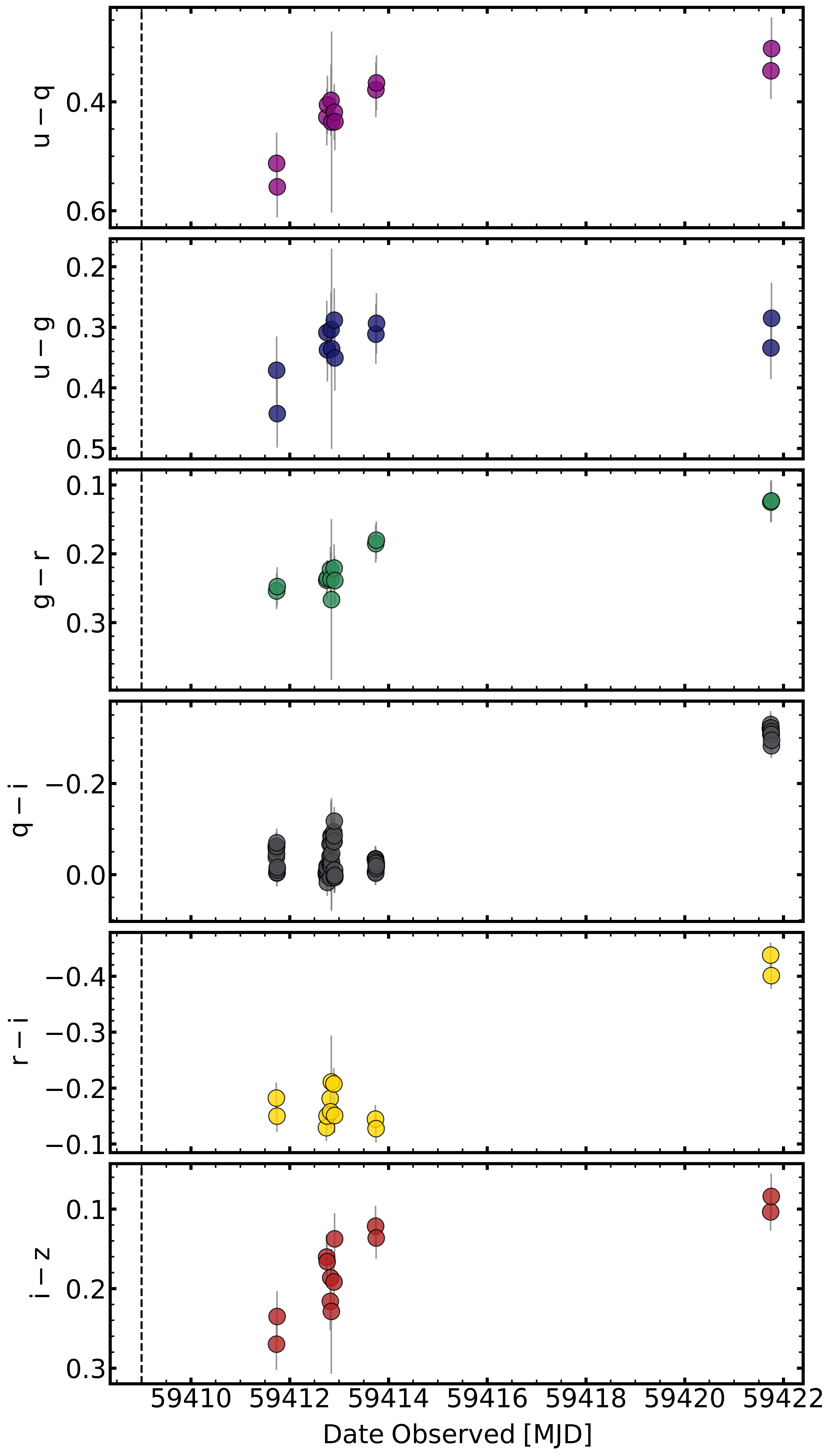}
\end{minipage}
\caption{
Extinction corrected ($A_{V,\mathrm{G}}=0.448$) color evolution of SN\,2021tkm from MeerLICHT observations. $\delta (color)_{\mathrm{bkg}}$ represents the magnitude of the systematic uncertainty in the computed color index.
} 
\label{fig:SN2021tkm_color}
\end{figure}
%---------------------------------------------%
\begin{figure}[!ht]
\centering
\includegraphics[width=0.35\columnwidth]{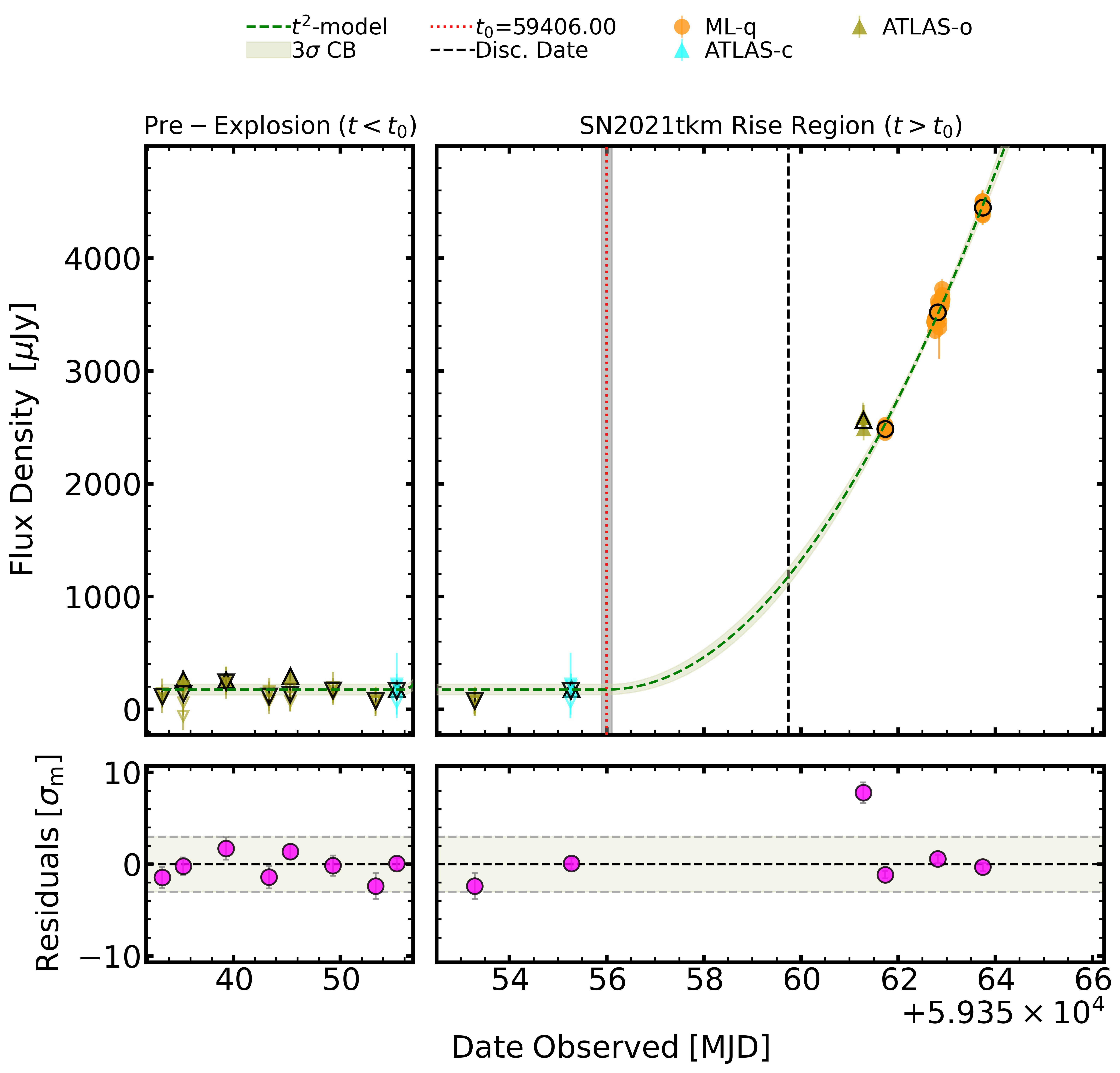}\\
\includegraphics[width=0.35\columnwidth]{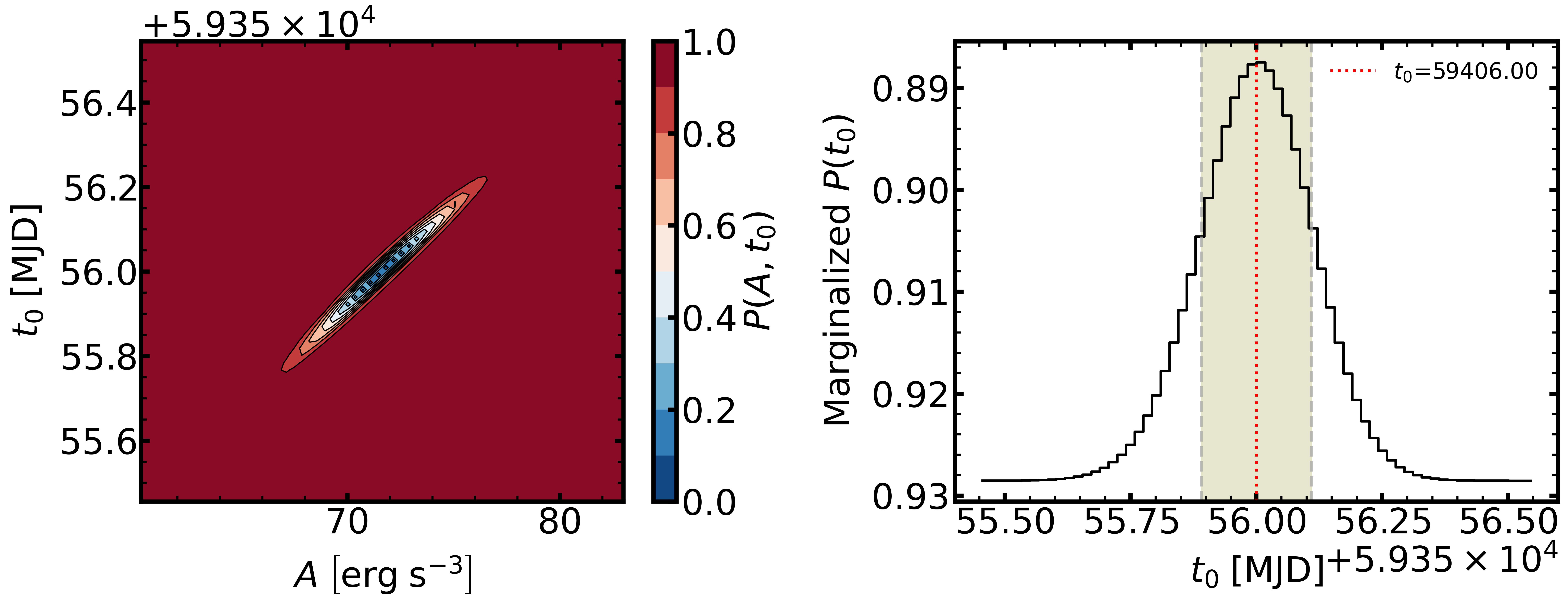}
\caption{
Same as Figure\,\ref{fig:SN2020oi_t2model} for SN\,2021tkm. The very narrow confidence region implies that this combination of $t_0$ and $A$ is well constrained. However, it also highlights one of the individual parameters having relatively high uncertainty.
}
\label{fig:SN2021tkm_t2model} 
\end{figure}
%%%%%%%%%%%%%%%%%%%%%%%%%%%%%%%%%%%%%%%%%%%%%%%%%%
%\subsubsection{SN 2021aele (Ia)}
%\label{sec:SN2021aele}
%%%%%%%%%%%%%%%%%%%%%%%%%%%%%%%%%%%%%%%%%%%%%%%%%%
%%%%%%%%%%%%%%%%%%%%%%%%%%%%%%%%%%%%%%%%%%%%%%%%%%
\begin{figure}[!ht]
\centering
\begin{minipage}{0.491\textwidth}
\centering
\includegraphics[width=0.85\linewidth,trim={0.0cm 0.0cm 0.0cm 0.0cm},clip]{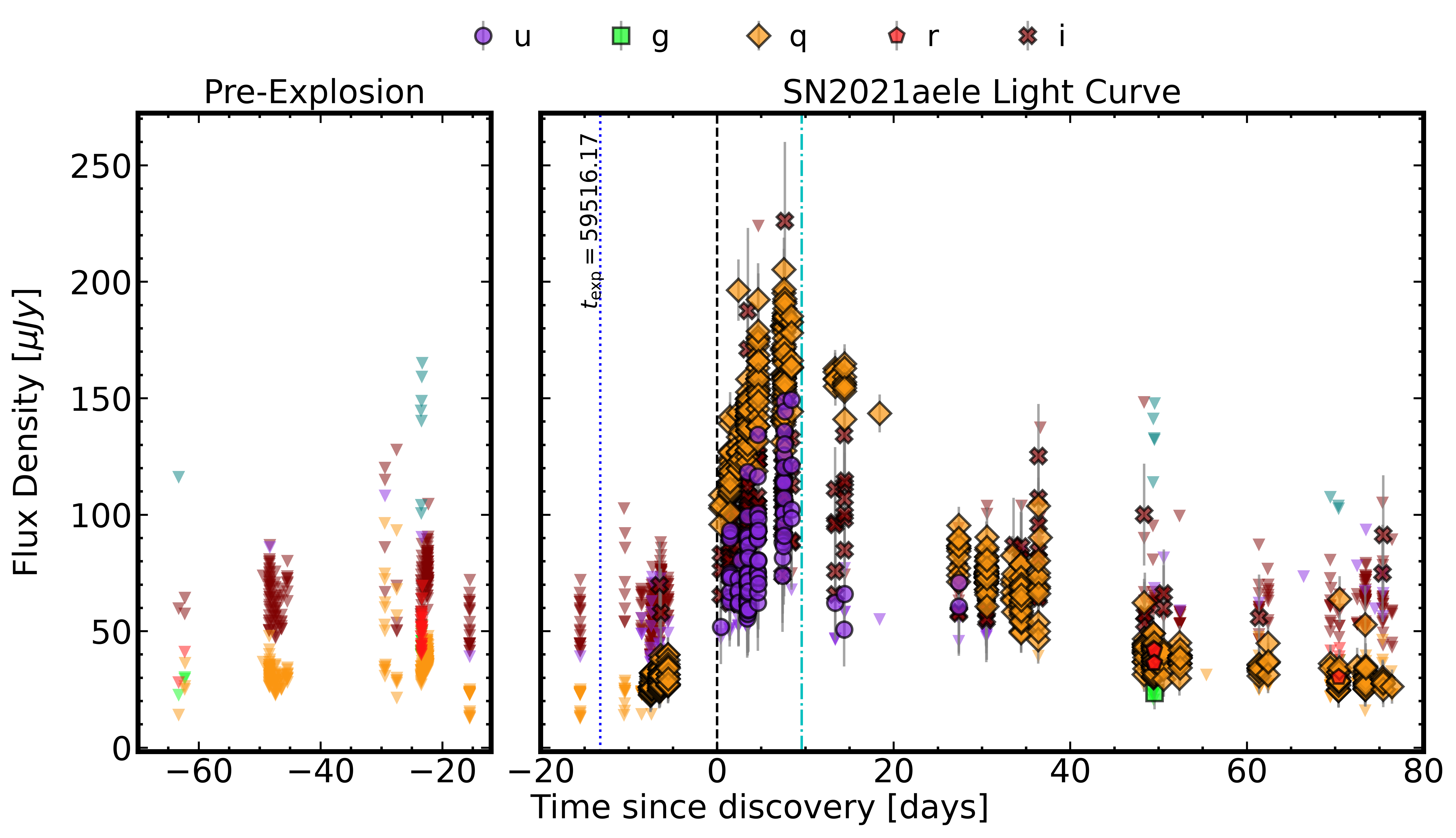}
\includegraphics[width=0.85\linewidth,trim={0.0cm 0.0cm 0.0cm 0.0cm},clip]{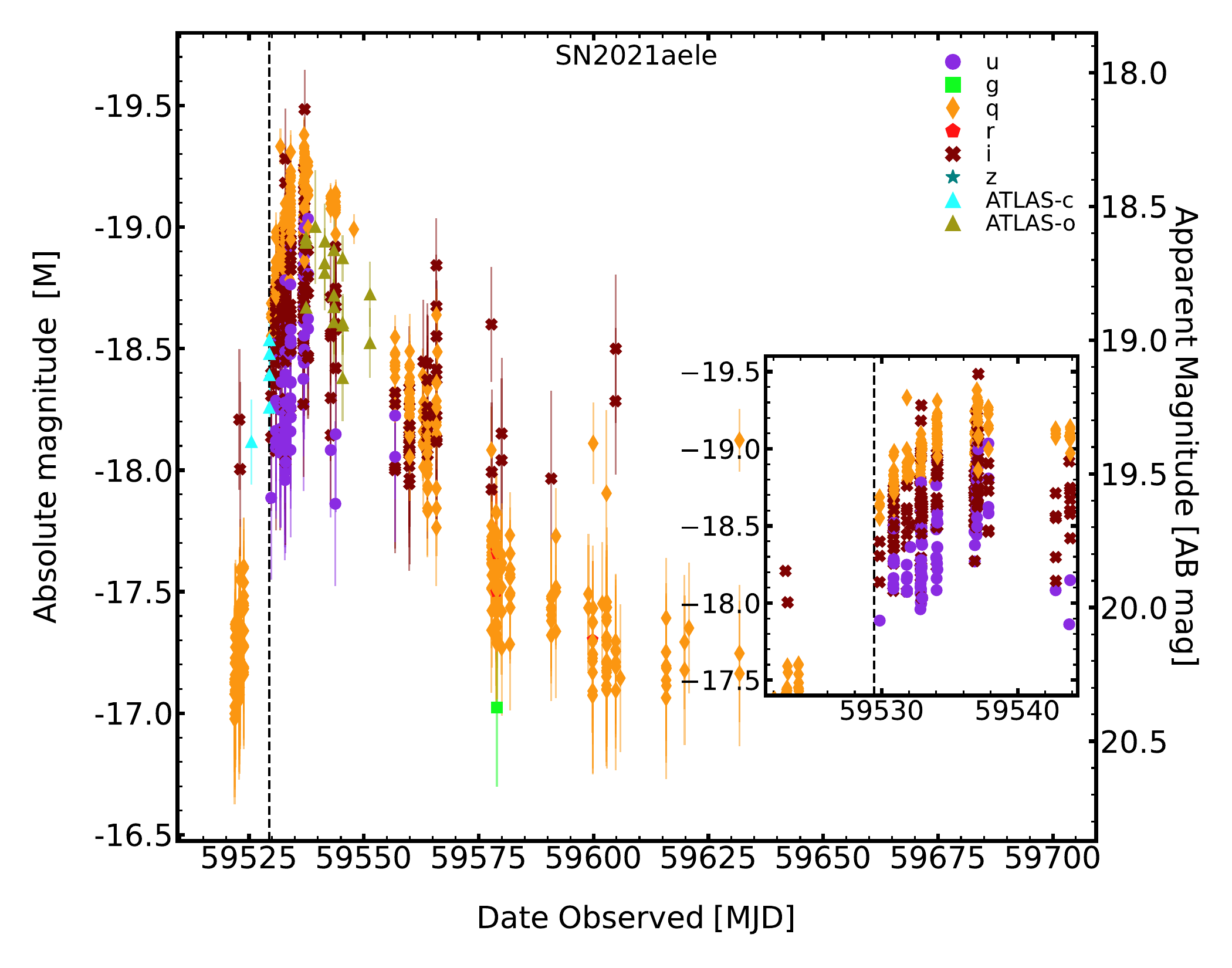}
\end{minipage}
\caption{ 
Pre- and post-explosion light curves of SN\,2021aele from MeerLICHT.
{\sl Top left}: Pre-explosion photometry up to 65 days before discovery. The shaded inverted triangles are upper limits, and all the other filled symbols are detections. 
{\sl Top right}: Multi-band optical light curves of SN\,2021aele with the underlying background emission subtracted. The phases are relative to the time of discovery as reported on TNS, marked by the black vertical dashed line. The estimated explosion epoch and peak brightness epoch are designated with the dotted blue line and dash-dotted cyan line, respectively. 
{\sl Bottom}: Absolute magnitude light curves from MeerLICHT and ATLAS. ATLAS magnitudes are represented by filled triangles. A zoom into the rise of SN\,2021aele is shown in the inset window, highlighting the high-cadence MeerLICHT observations around maximum brightness.
}
\label{fig:SN2021aele} 
\end{figure}
%---------------------------------------------%
\begin{figure}[!ht]
\centering
\begin{minipage}{0.491\textwidth}
\includegraphics[width=0.85\linewidth,trim={0.0cm 0.0cm 0.0cm 0.0cm},clip]{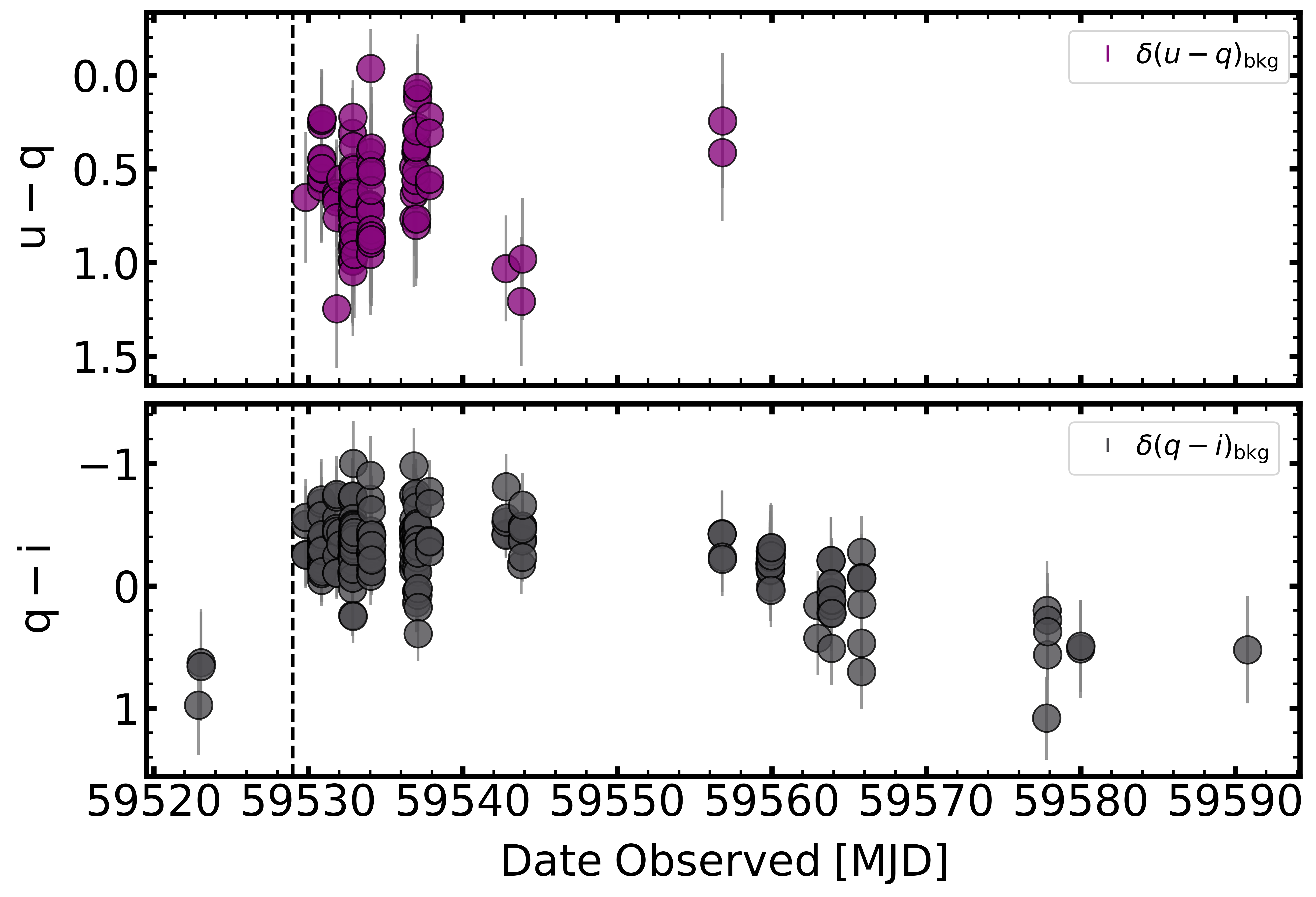}
\end{minipage}
\caption{
Extinction corrected ($A_{V,\mathrm{G}}=0.028$) color evolution of SN\,2021aele from MeerLICHT observations. $\delta (color)_{\mathrm{bkg}}$ represents the magnitude of the systematic uncertainty in the computed color index.
}
\label{fig:SN2021aele_color}
\end{figure}
%-----------------------------------------------%
\begin{figure}[!ht]
\centering
\includegraphics[width=0.35\columnwidth]{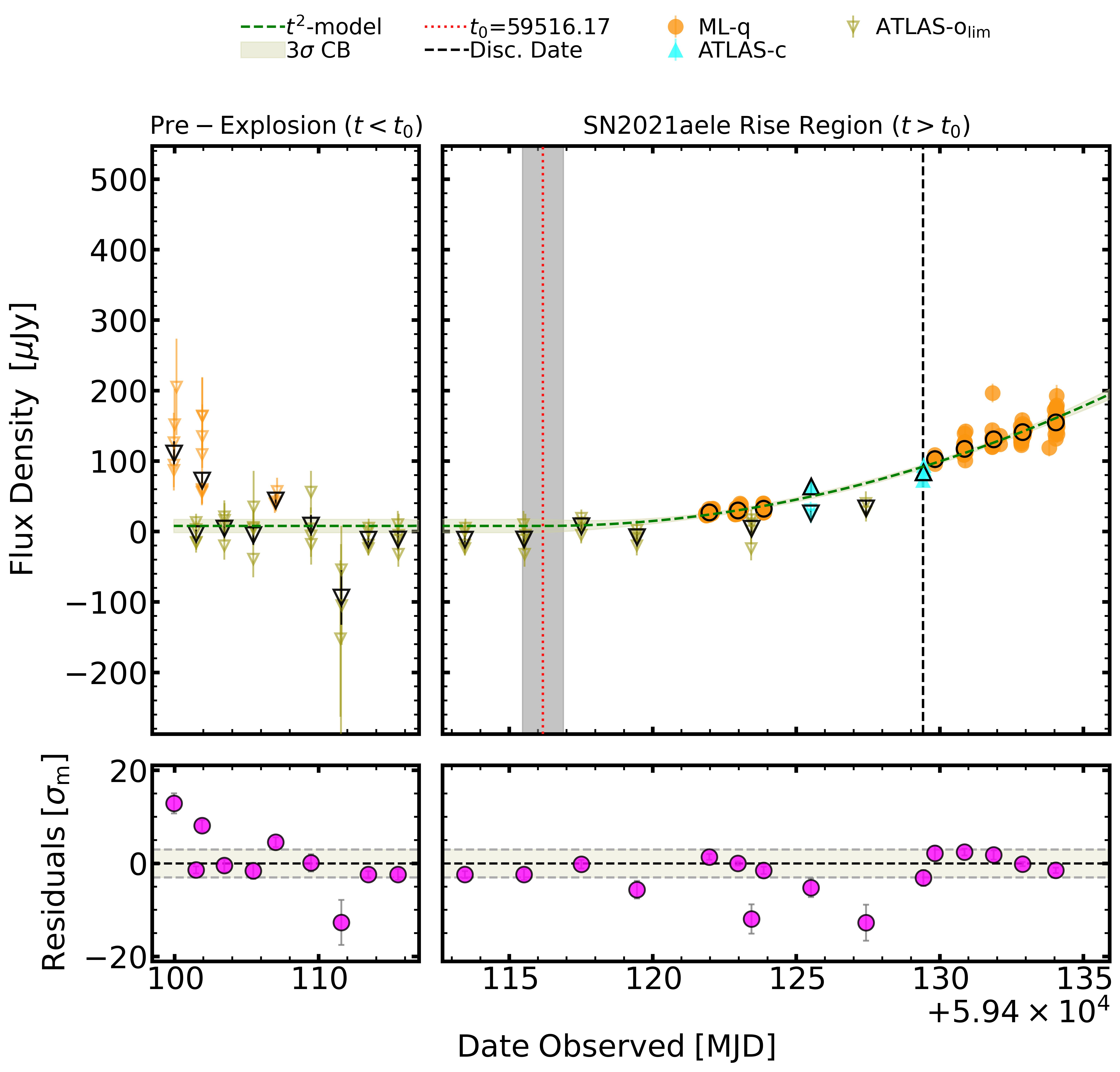}\\
\includegraphics[width=0.35\columnwidth]{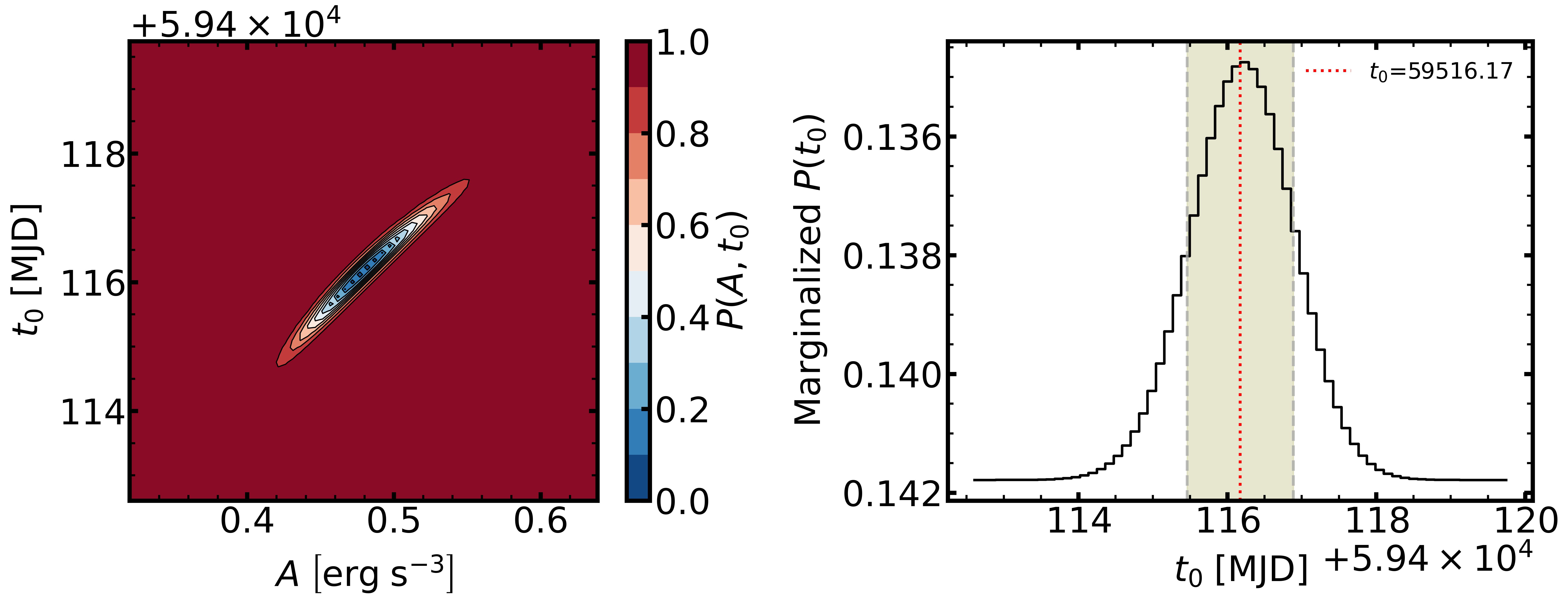}
\caption{
Same as Figure\,\ref{fig:SN2020oi_t2model} for SN\,2021aele
}
\label{fig:SN2021aele_t2model}
\end{figure}
%%%%%%%%%%%%%%%%%%%%%%%%%%%%%%%%%%%%%%%%%%%%%%%
%\subsubsection{SN 2022ame (II)}
%%%%%%%%%%%%%%%%%%%%%%%%%%%%%%%%%%%%%%%%%%%%%%%
%%%%%%%%%%%%%%%%%%%%%%%%%%%%%%%%%%%%%%%%%%%%%%%
\begin{figure}[!ht]
\centering
\begin{minipage}{0.491\textwidth}
\centering
\includegraphics[width=0.85\linewidth,trim={0.0cm 0.0cm 0.0cm 0.0cm},clip]{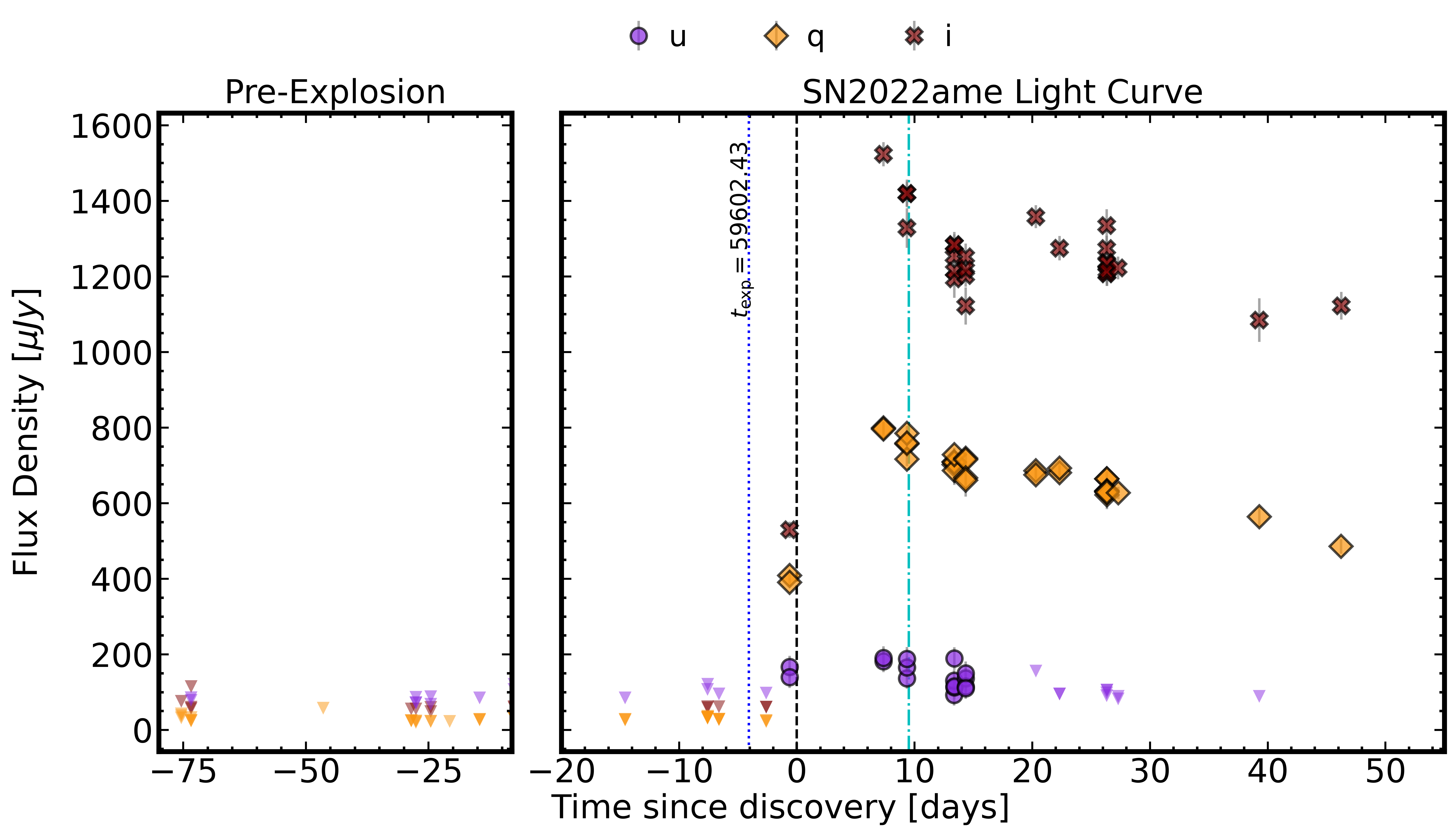}
\includegraphics[width=0.85\linewidth,trim={0.0cm 0.0cm 0.0cm 0.0cm},clip]{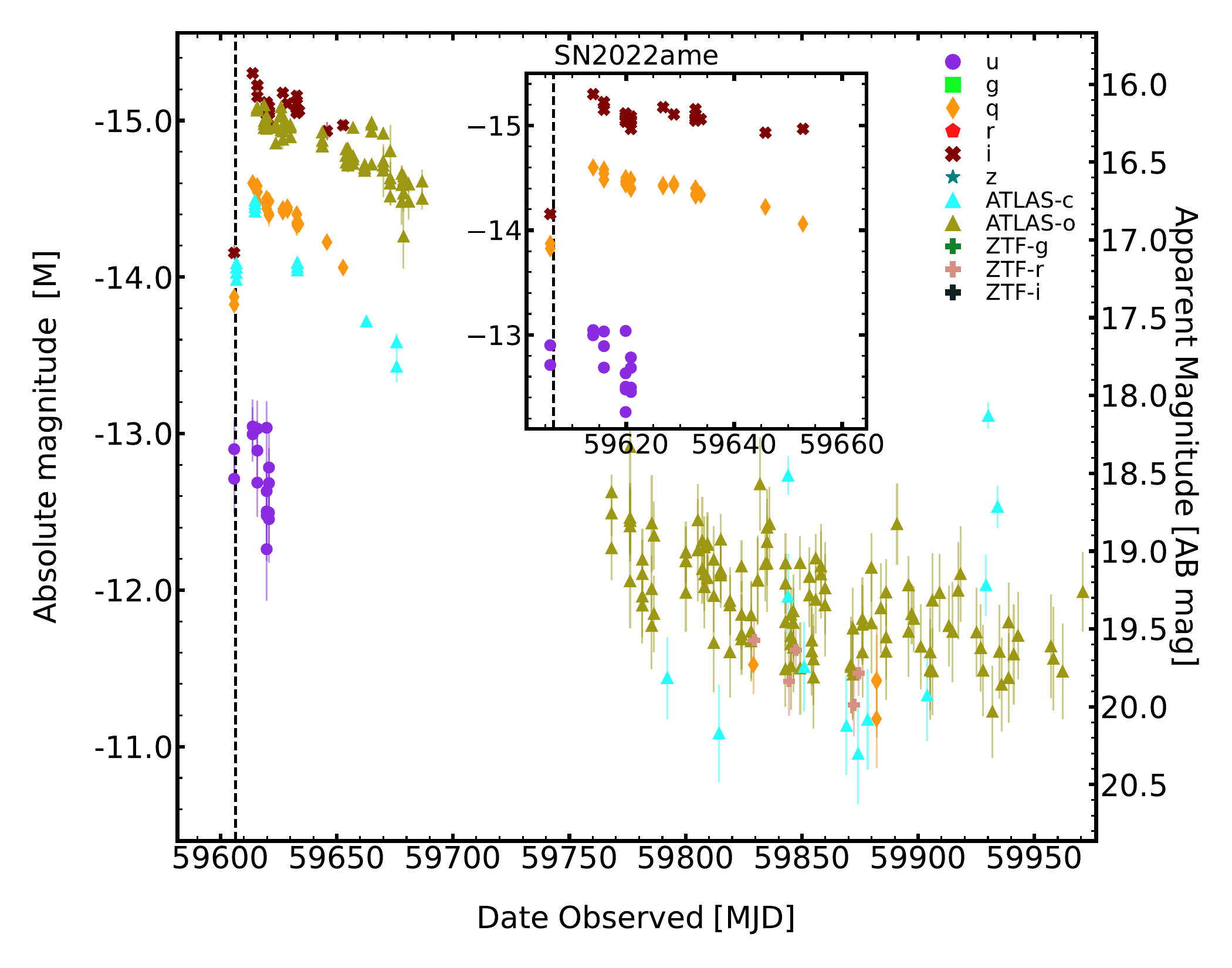}
\end{minipage}
\caption{
Pre- and post-explosion light curves of SN\,2022ame from MeerLICHT.
{\sl Top left}: Pre-explosion photometry up to 75 days before discovery. The shaded inverted triangles are upper limits, and all the other filled symbols are detections.
{\sl Top right}: Multi-band optical light curves of SN\,2022ame with the underlying background emission subtracted. The phases are relative to the time of discovery as reported on TNS, marked by the black vertical dashed line. The estimated explosion epoch and peak brightness epoch are designated with the dotted blue line and dash-dotted cyan line, respectively.
{\sl Bottom}: Absolute magnitude light curves from MeerLICHT, ATLAS, and ZTF. Filled triangles represent ATLAS magnitudes, and ZTF difference magnitudes are represented by filled 'plus' symbols. A zoom into the peak of SN\,2022ame is plotted in the insert window, showing the MeerLICHT observations during the slow decay phase of the SN in the first $60$ days after discovery.
}
\label{fig:SN2022ame}
\end{figure}
%--------------------------------------------%
\begin{figure}[!ht]
\centering
\begin{minipage}{0.491\textwidth}
\includegraphics[width=0.85\linewidth,trim={0.0cm 0.0cm 0.0cm 0.0cm},clip]{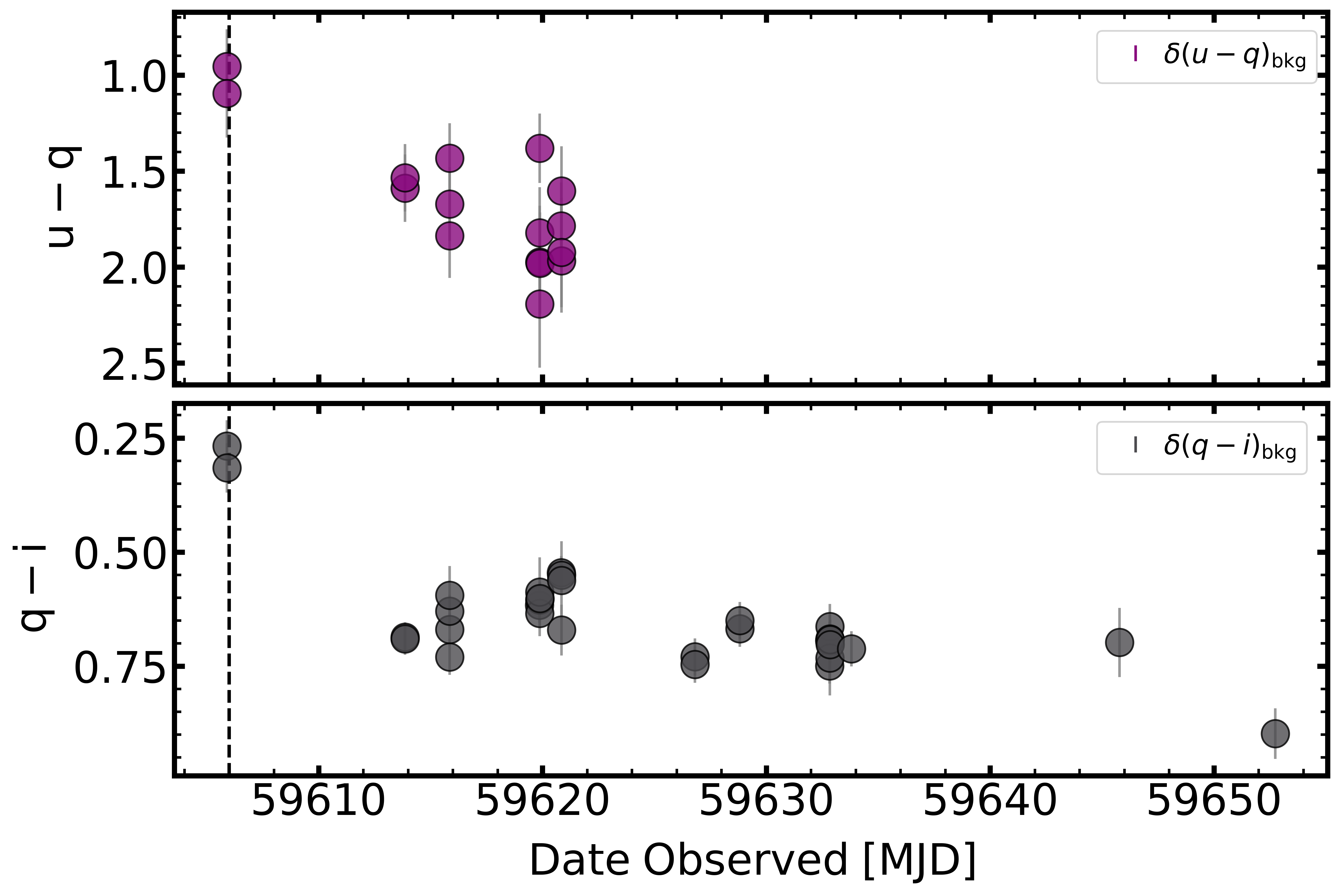}
\end{minipage}
\caption{
Extinction corrected ($A_{V,\mathrm{G}}=0.039$) color evolution of SN\,2022ame from MeerLICHT observations. $\delta (color)_{\mathrm{bkg}}$ represents the magnitude of the systematic uncertainty in the computed color index.
}
\label{fig:SN2022ame_color}
\end{figure}
%-----------------------------------------------%
\begin{figure}
\centering
\includegraphics[width=0.35\columnwidth]{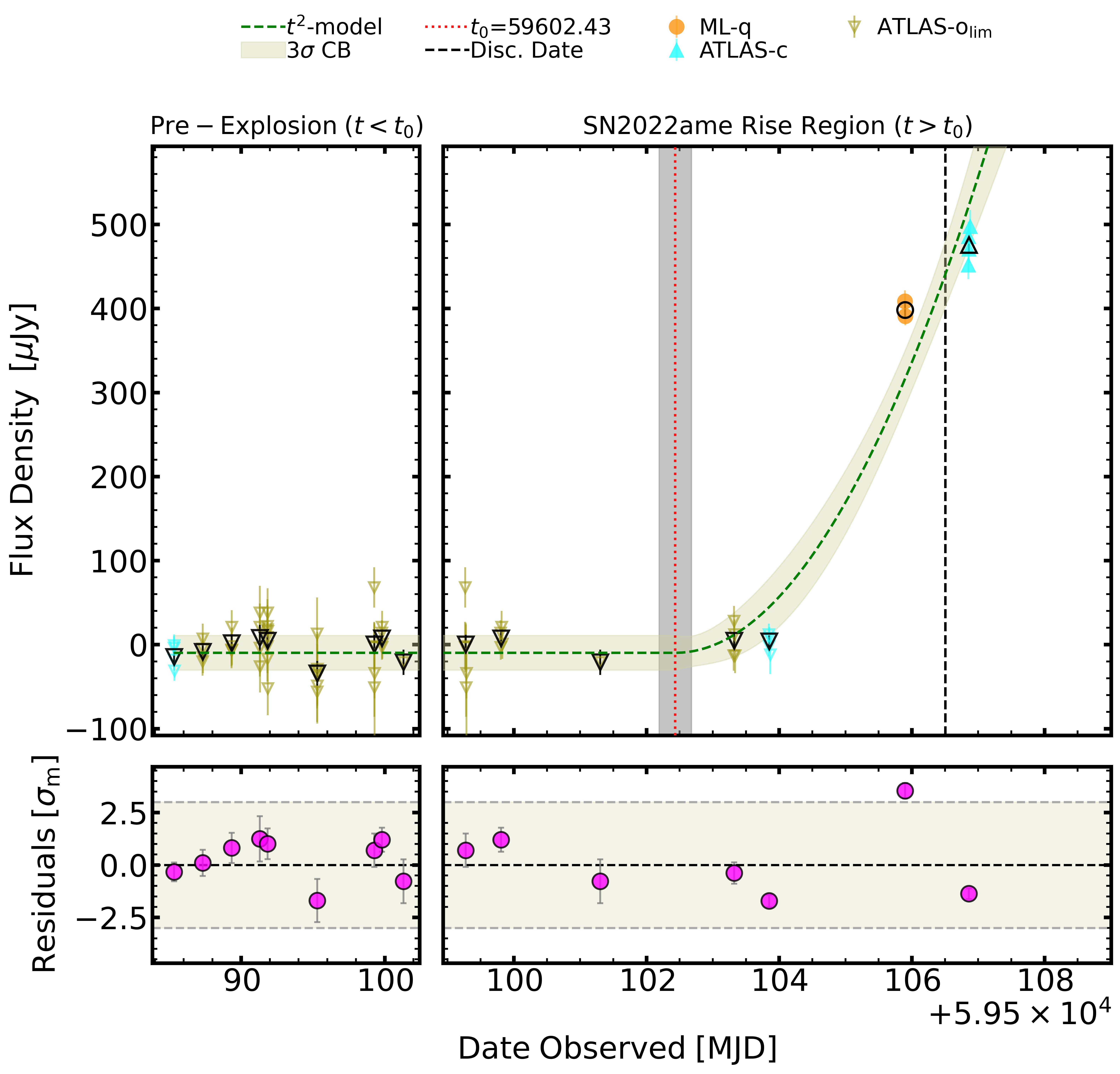}\\
\includegraphics[width=0.35\columnwidth]{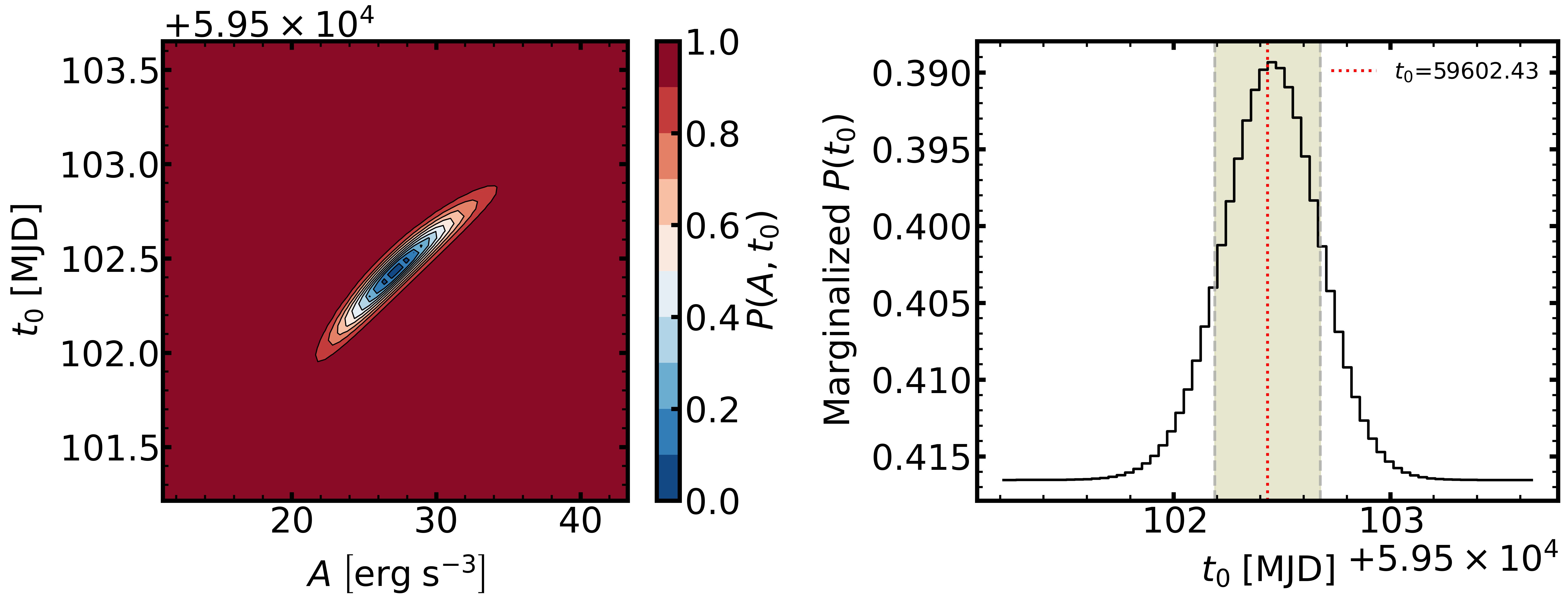}
\caption{
Same as Figure\,\ref{fig:SN2020oi_t2model} for SN\,2022ame
}
\label{fig:SN2022ame_t2model}
\end{figure}
%%%%%%%%%%%%%%%%%%%%%%%%%%%%%%%%%%%%%%%%%%%%%%%
%\subsubsection{SN 2022hrs (Ia)}
%%%%%%%%%%%%%%%%%%%%%%%%%%%%%%%%%%%%%%%%%%%%%%%
%%%%%%%%%%%%%%%%%%%%%%%%%%%%%%%%%%%%%%%%%%%%%%%
\begin{figure}[!ht]
\centering
\begin{minipage}{0.491\textwidth}
\centering
\includegraphics[width=0.85\linewidth,trim={0.0cm 0.0cm 0.0cm 0.0cm},clip]{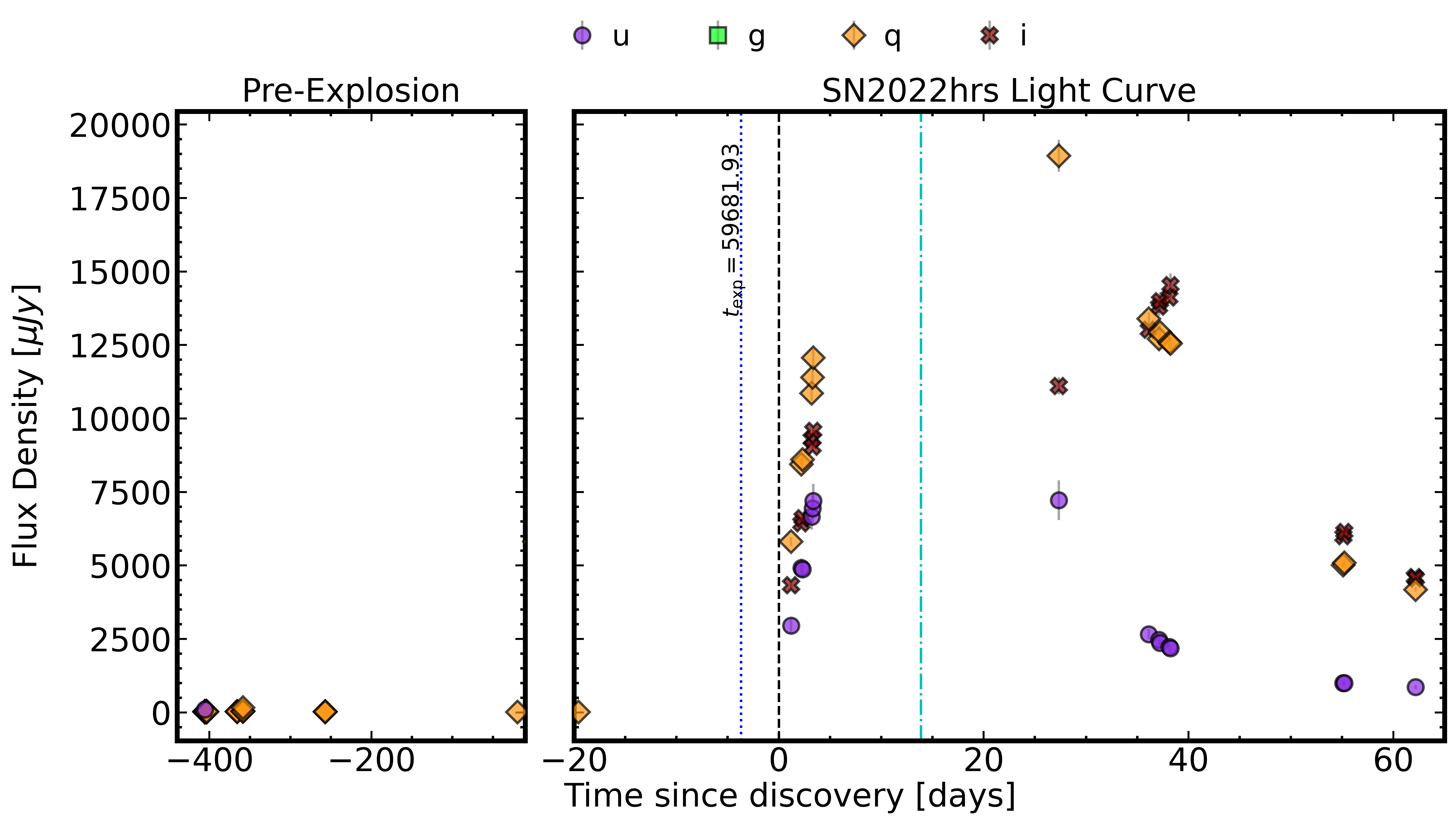}
\includegraphics[width=0.85\linewidth,trim={0.0cm 0.0cm 0.0cm 0.0cm},clip]{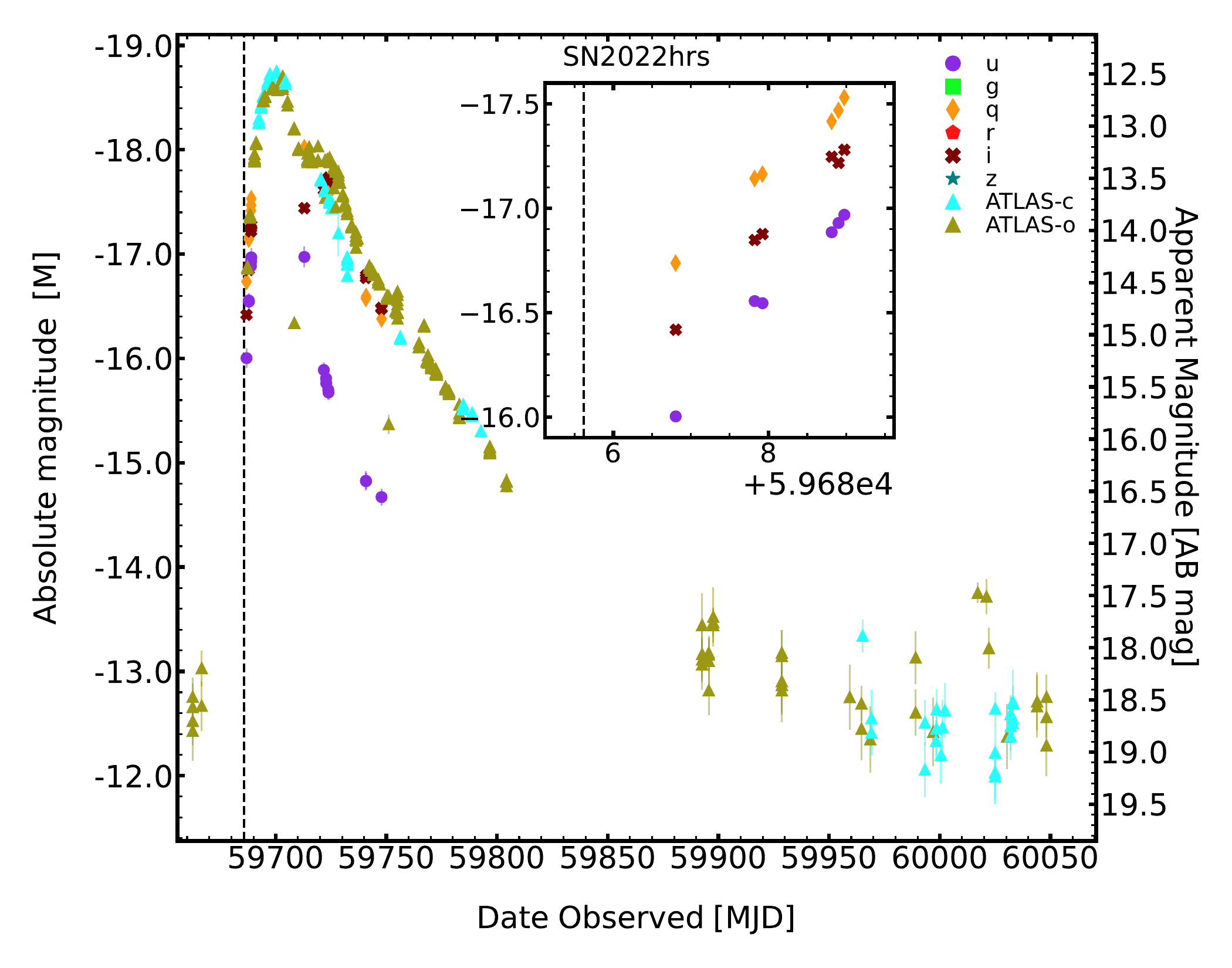}
\end{minipage}
\caption{
Pre- and post-explosion light curves of SN\,2022hrs from MeerLICHT. 
{\sl Top left}: Pre-explosion photometry up to 400 days before discovery. The shaded inverted triangles are upper limits, and all the other filled symbols are detections. 
{\sl Top right}: Multi-band optical light curves of SN\,2022hrs without subtracting the underlying background emission. The phases are relative to the time of discovery as reported on TNS, marked by the black vertical dashed line. The estimated explosion epoch and peak brightness epoch are indicated by the dotted blue line and the dash-dotted cyan line, respectively. 
{\sl Bottom}: Absolute magnitudes light curves from MeerLICHT and ATLAS. Filled triangles represent ATLAS magnitudes. A zoom into the rising phase of SN\,2022hrs is plotted in the insert window, showing the MeerLICHT observations in the first $10$ days after discovery.
}
\label{fig:SN2022hrs}
\end{figure}
%--------------------------------------------%
\begin{figure}[!ht]
\centering
\begin{minipage}{0.491\textwidth}
\includegraphics[width=0.85\linewidth,trim={0.0cm 0.0cm 0.0cm 0.0cm},clip]{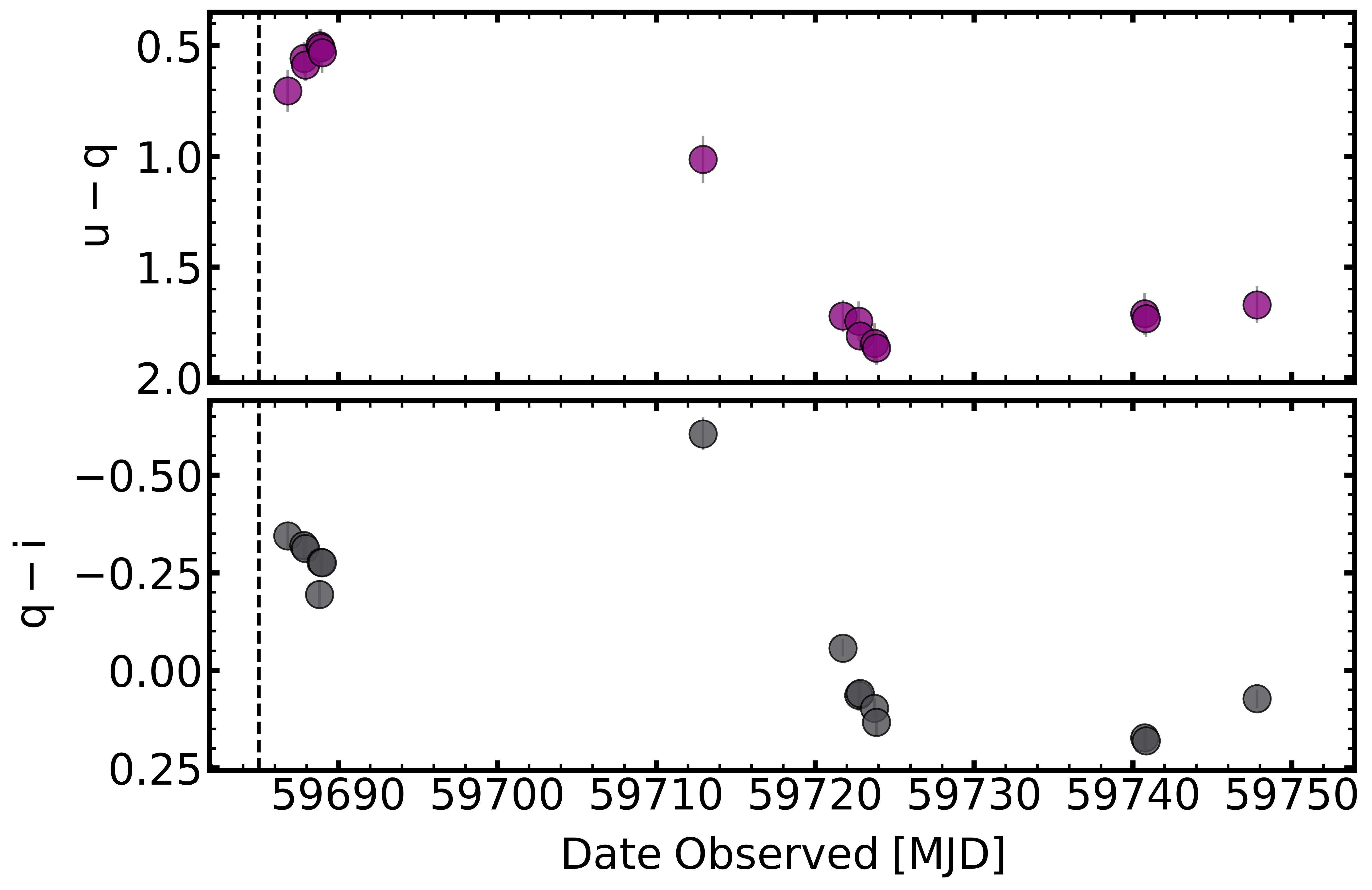}
\end{minipage}
\caption{
Extinction corrected ($A_{V,\mathrm{G}}=0.074$) color evolution of SN\,2022hrs from MeerLICHT observations. $\delta (color)_{\mathrm{bkg}}$ represents the magnitude of the systematic uncertainty in the computed color index.
}
\label{fig:SN2022hrs_color}
\end{figure}
%--------------------------------------------%
\begin{figure}[!ht]
\centering
\includegraphics[width=0.35\columnwidth]{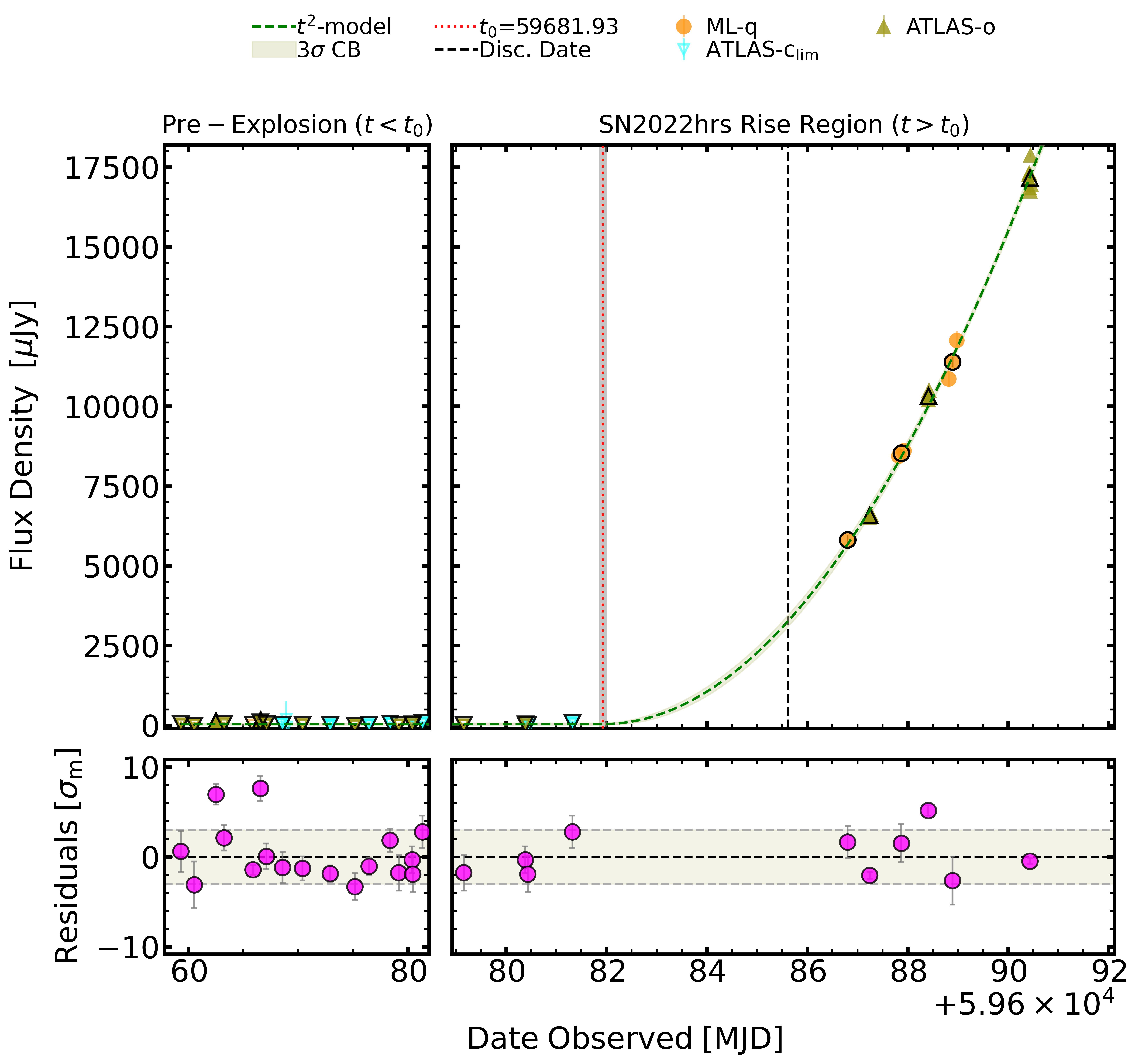}\\
\includegraphics[width=0.35\columnwidth]{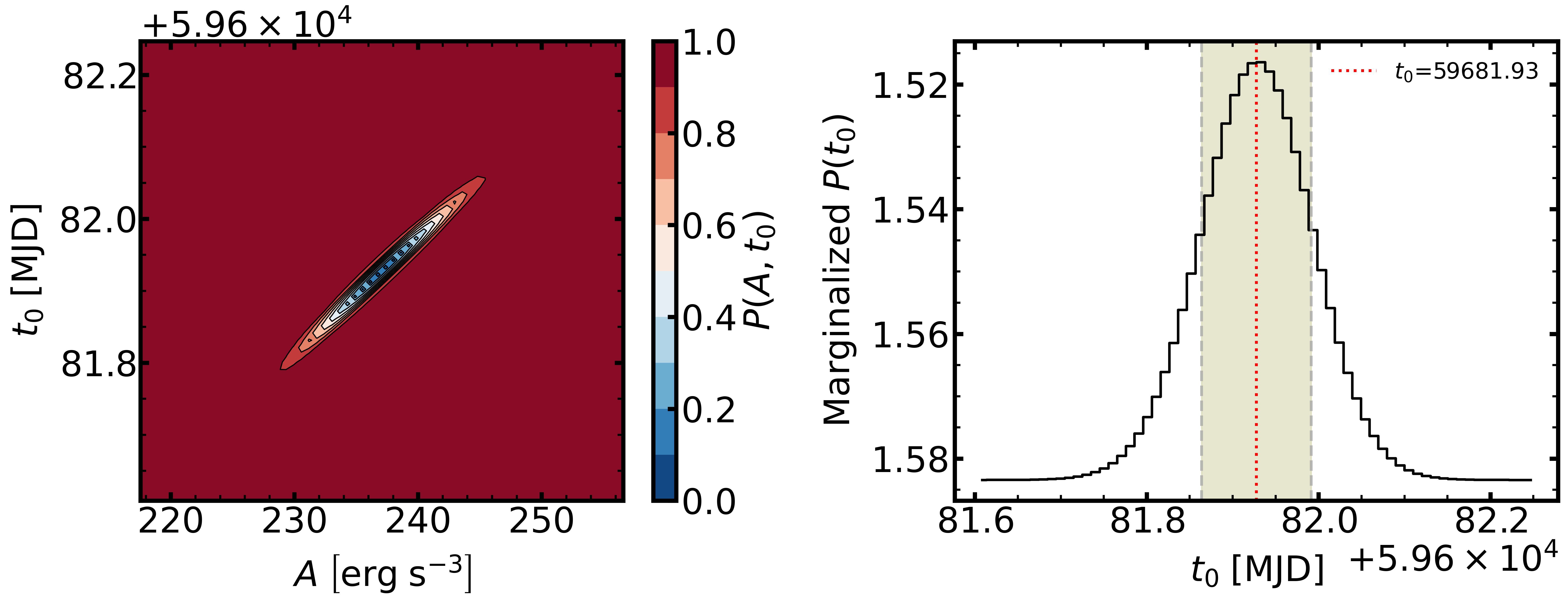}
\caption{
Same as Figure\,\ref{fig:SN2020oi_t2model} for SN\,2022hrs
}
\label{fig:SN2022hrs_t2model}
\end{figure}
%%%%%%%%%%%%%%%%%%%%%%%%%%%%%%%%%%%%%%%%%%%%%%%
%\subsubsection{SN 2022phj (II)}
%%%%%%%%%%%%%%%%%%%%%%%%%%%%%%%%%%%%%%%%%%%%%%%
%%%%%%%%%%%%%%%%%%%%%%%%%%%%%%%%%%%%%%%%%%%%%%%
\begin{figure}[!ht]
\centering
\begin{minipage}{0.491\textwidth}
\centering
\includegraphics[width=0.85\linewidth,trim={0.0cm 0.0cm 0.0cm 0.0cm},clip]{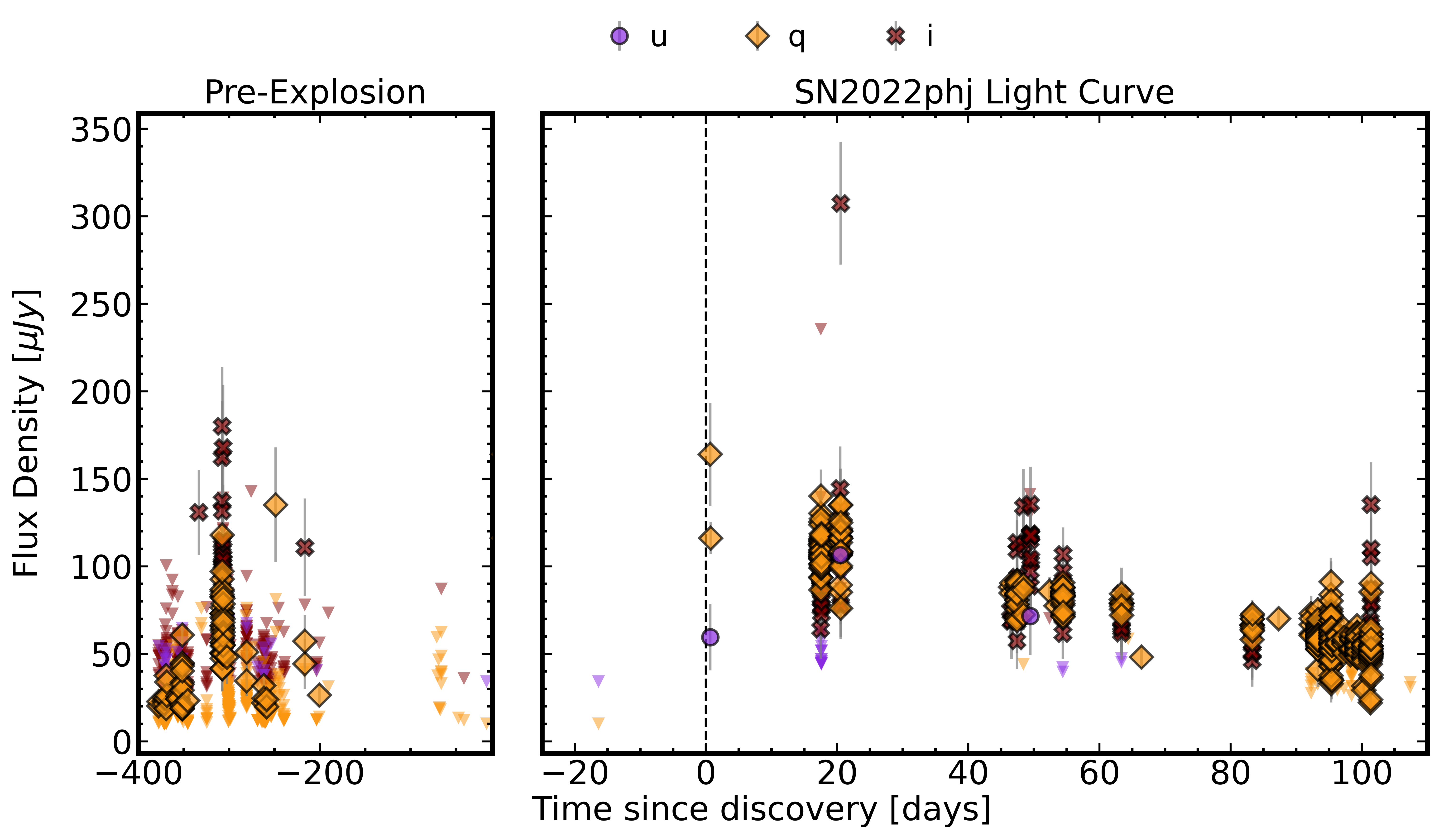}
\includegraphics[width=0.85\linewidth,trim={0.0cm 0.0cm 0.0cm 0.0cm},clip]{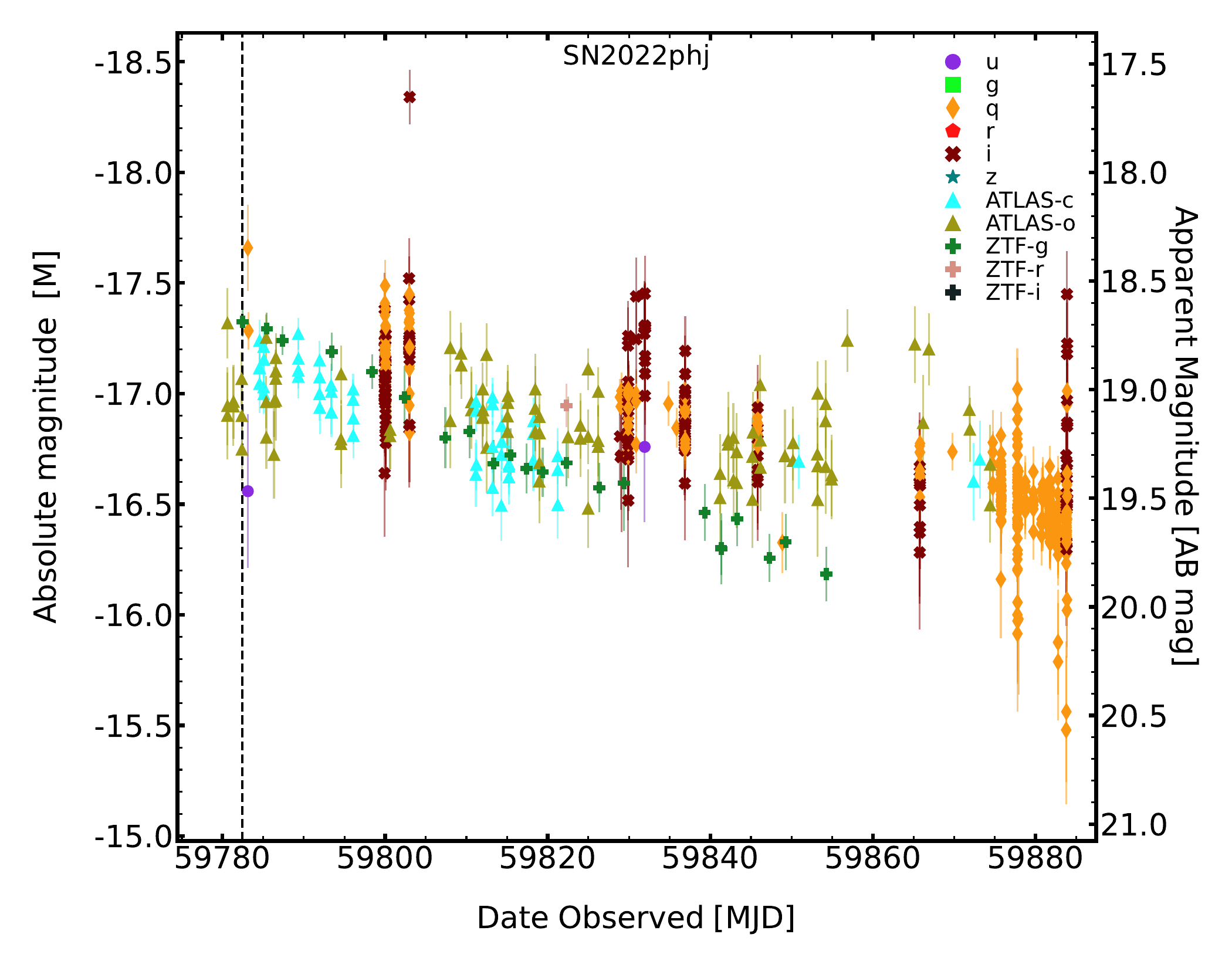}
\end{minipage}
\caption{
Pre- and post-explosion light curves of SN\,2022phj from MeerLICHT.
{\sl Top left}: Pre-explosion photometry up to 420 days before discovery. The shaded inverted triangles are upper limits, and all the other filled symbols are detections.
{\sl Top right}: Multi-band optical light curves of SN\,2022phj with the underlying background emission subtracted. The phases are relative to the time of discovery as reported on TNS, marked by the black vertical dashed line.
{\sl Bottom}: Absolute magnitudes light curves from MeerLICHT, ATLAS, and ZTF. ATLAS magnitudes are represented by filled triangles, and corrected ZTF magnitudes are represented by filled 'plus' symbols.
}
\label{fig:SN2022phj}
\end{figure}
%--------------------------------------------%
\begin{figure}[!ht]
\centering
\begin{minipage}{0.491\textwidth}
\includegraphics[width=0.85\linewidth,trim={0.0cm 0.0cm 0.0cm 0.0cm},clip]{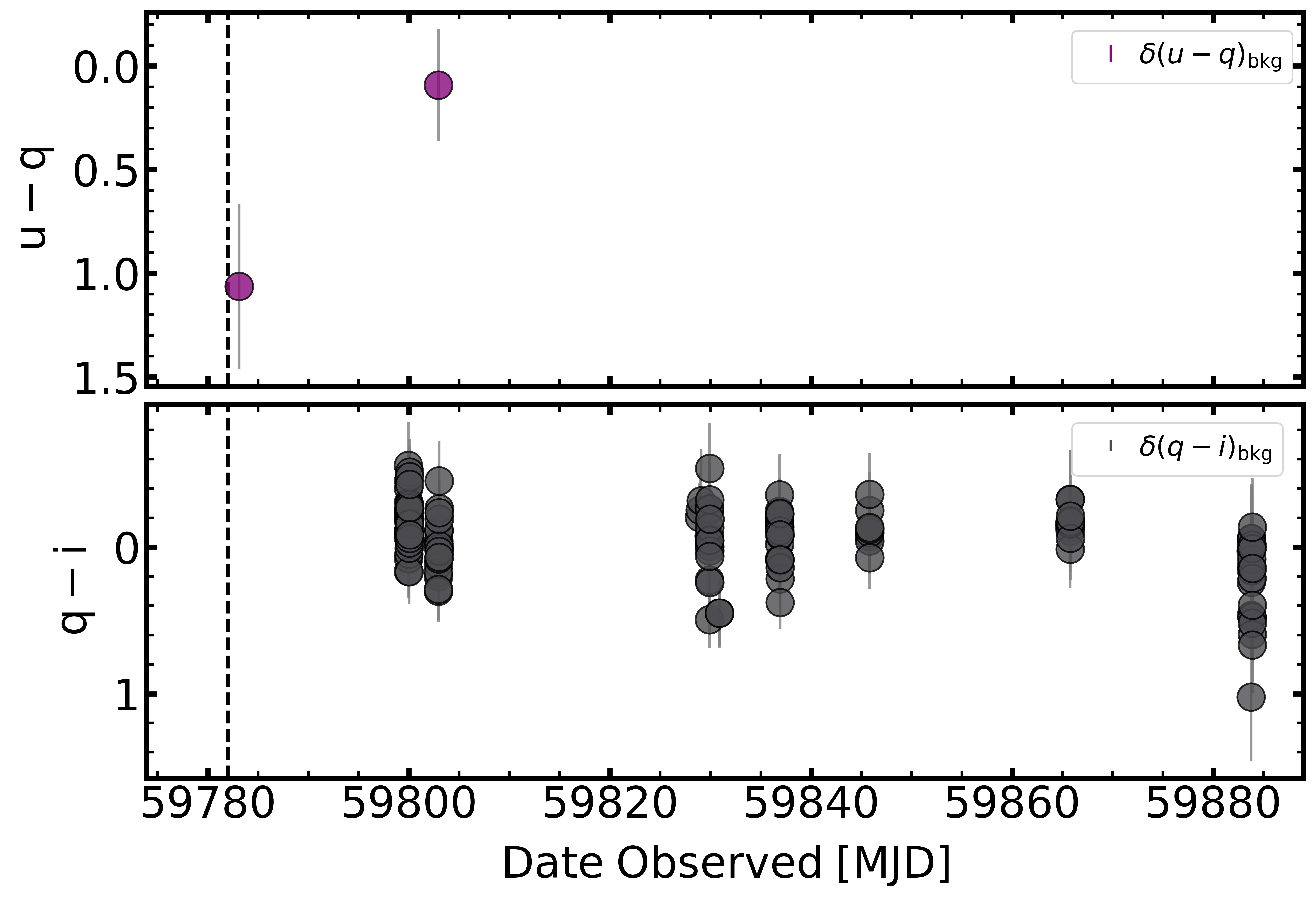}
\end{minipage}
\caption{
Extinction corrected ($A_{V,\mathrm{G}}=0.084$) color evolution of SN\,2022phj from MeerLICHT observations. $\delta (color)_{\mathrm{bkg}}$ represents the magnitude of the systematic uncertainty in the computed color index.
}
\label{fig:SN2022phj_color}
\end{figure}
%%%%%%%%%%%%%%%%%%%%%%%%%%%%%%%%%%%%%%%%%%%%%%%
%\subsubsection{SN 2019ano (Ia)}
%%%%%%%%%%%%%%%%%%%%%%%%%%%%%%%%%%%%%%%%%%%%%%%
%%%%%%%%%%%%%%%%%%%%%%%%%%%%%%%%%%%%%%%%%%%%%%%
\begin{figure}[!ht]
\centering
\begin{minipage}{0.491\textwidth}
\centering
\includegraphics[width=0.85\linewidth,trim={0.0cm 0.0cm 0.0cm 0.0cm},clip]{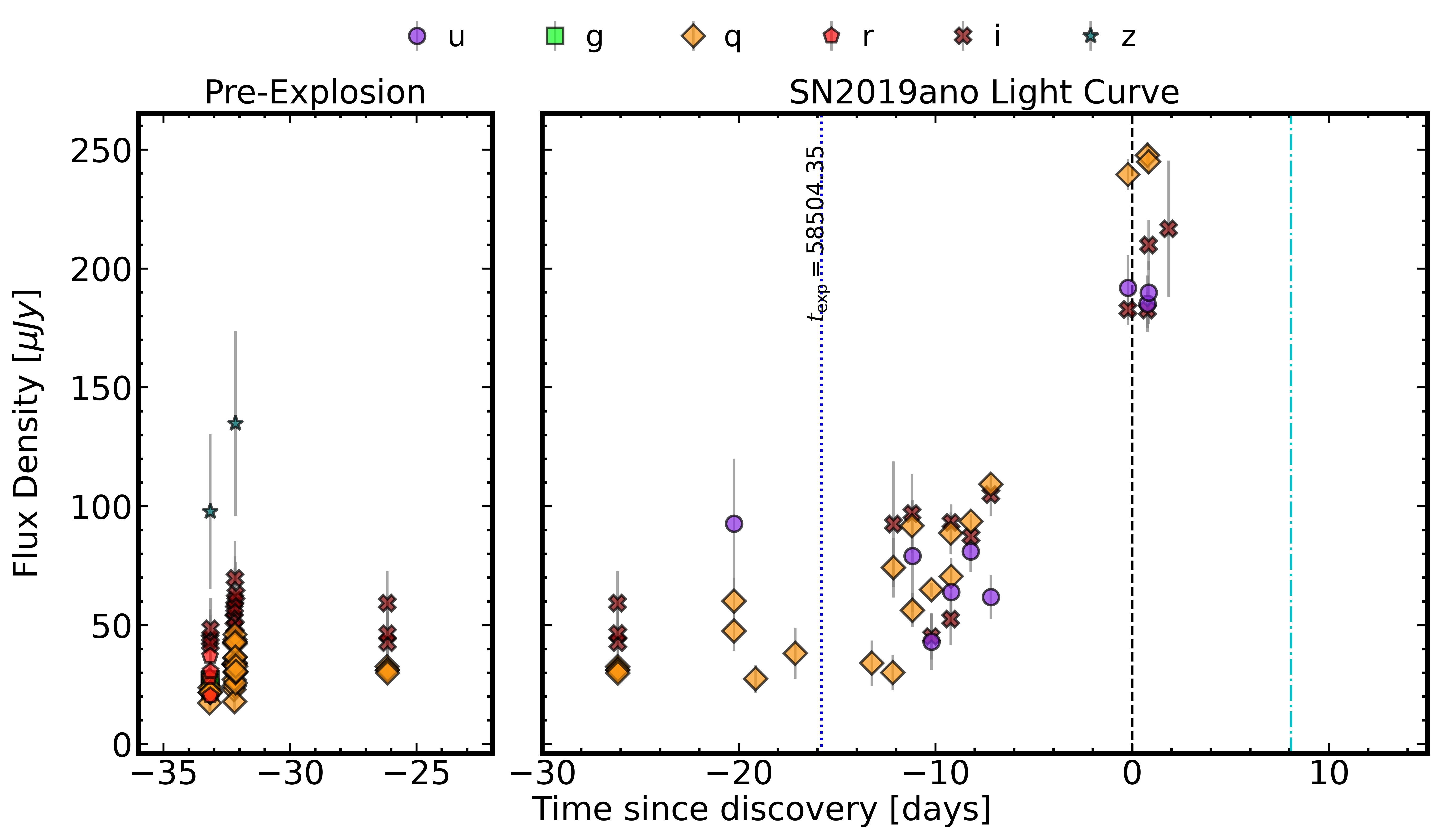}
\includegraphics[width=0.85\linewidth,trim={0.0cm 0.0cm 0.0cm 0.0cm},clip]{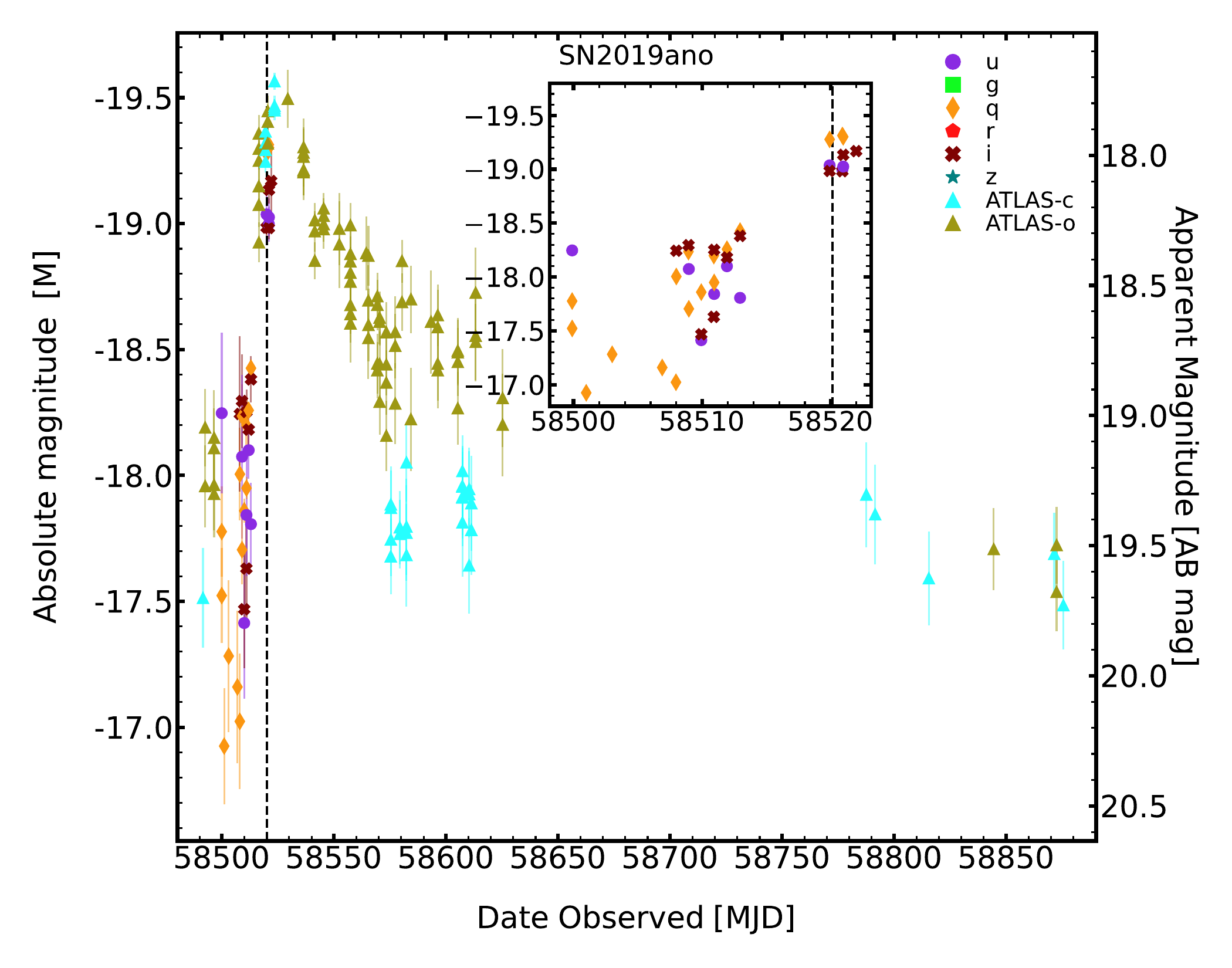}
\end{minipage}
\caption{
Pre- and post-explosion light curves of SN\,2019ano from MeerLICHT.
{\sl Top left}: Pre-explosion photometry up to 35 days before discovery. The shaded inverted triangles are upper limits, and all the other filled symbols are detections.
{\sl Top right}: Multi-band optical light curves of SN\,2019ano without subtracting the underlying background emission. The phases are relative to the time of discovery as reported on TNS, marked by the black vertical dashed line. The estimated explosion epoch and peak brightness epoch are indicated by the dotted blue line and the dash-dotted cyan line, respectively.
{\sl Bottom}: Absolute magnitude light curves from MeerLICHT and ATLAS. Filled triangles represent ATLAS magnitudes. A zoom into the rising phase of SN\,2019ano is plotted in the insert window, showing the MeerLICHT observations of the SN in the first $20$ days after discovery.
}
\label{fig:SN2019ano}
\end{figure}
%--------------------------------------------%
\begin{figure}[!ht]
\centering
\begin{minipage}{0.491\textwidth}
\includegraphics[width=0.85\linewidth,trim={0.0cm 0.0cm 0.0cm 0.0cm},clip]{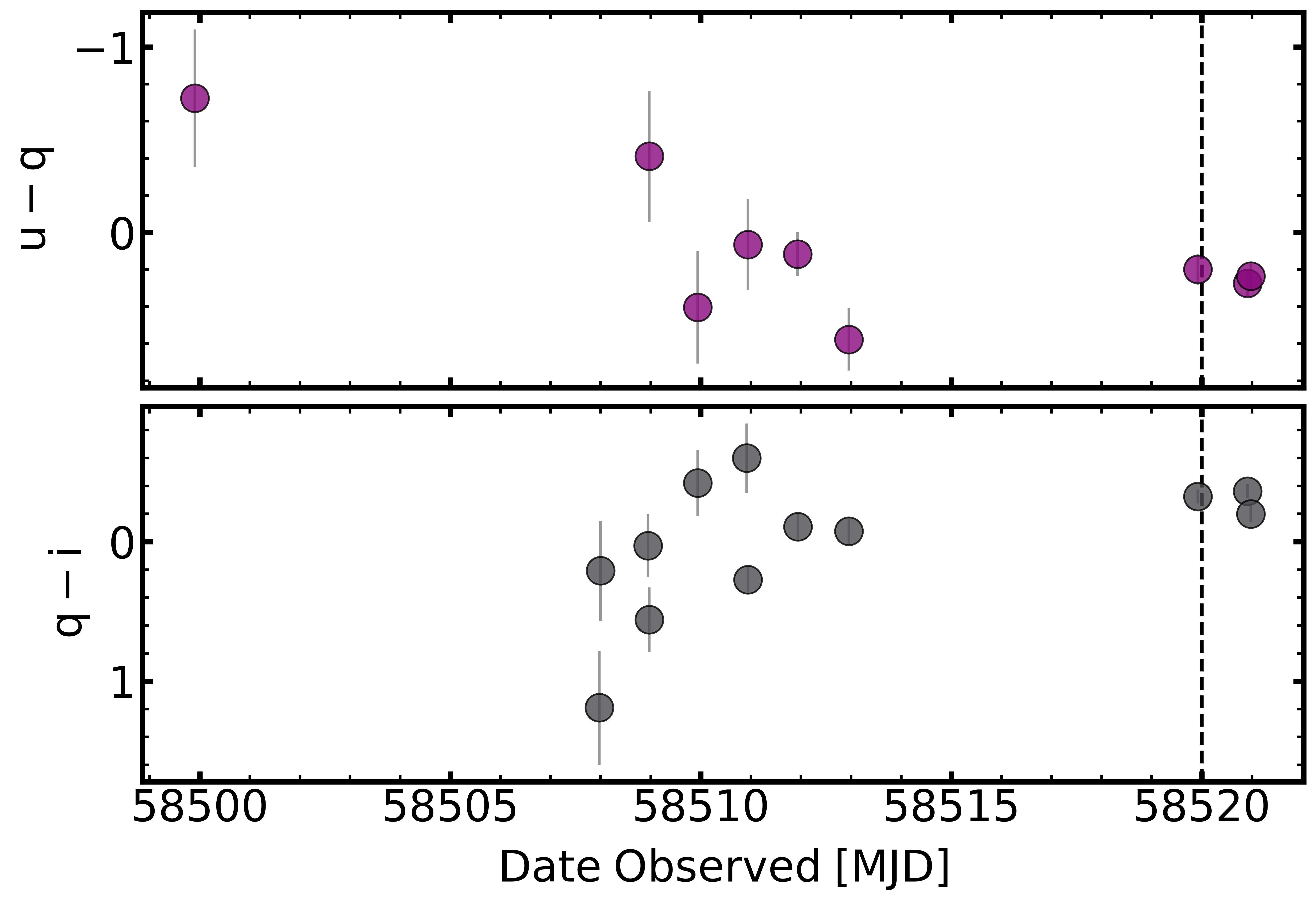}
\end{minipage}
\caption{
Extinction corrected ($A_{V,\mathrm{G}}=0.089$) color evolution of SN\,2019ano from MeerLICHT observations.
}
\label{fig:SN2019ano_color}
\end{figure}
%--------------------------------------------%
\begin{figure}[!ht]
\centering
\includegraphics[width=0.35\columnwidth]{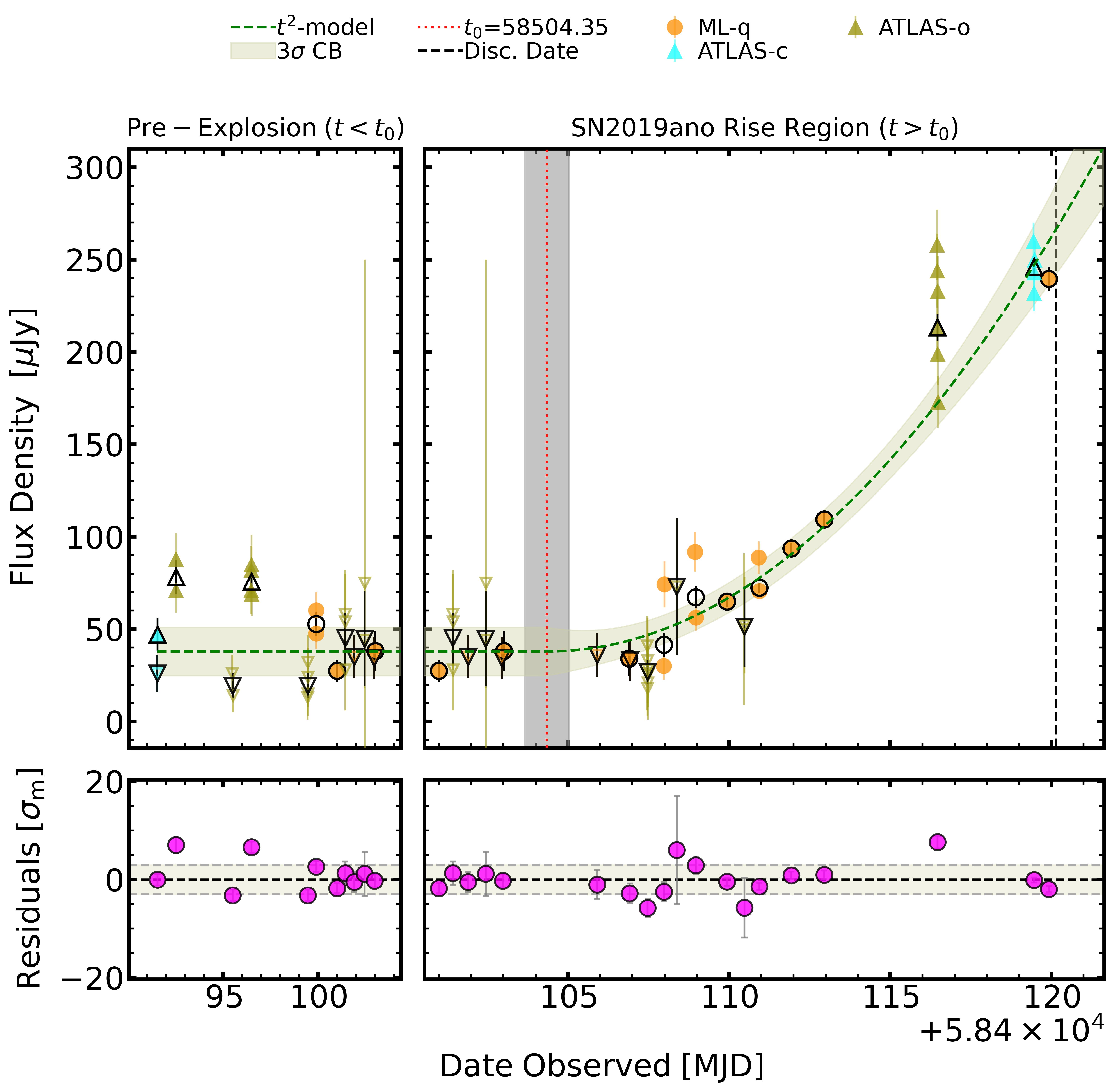}\\
\includegraphics[width=0.35\columnwidth]{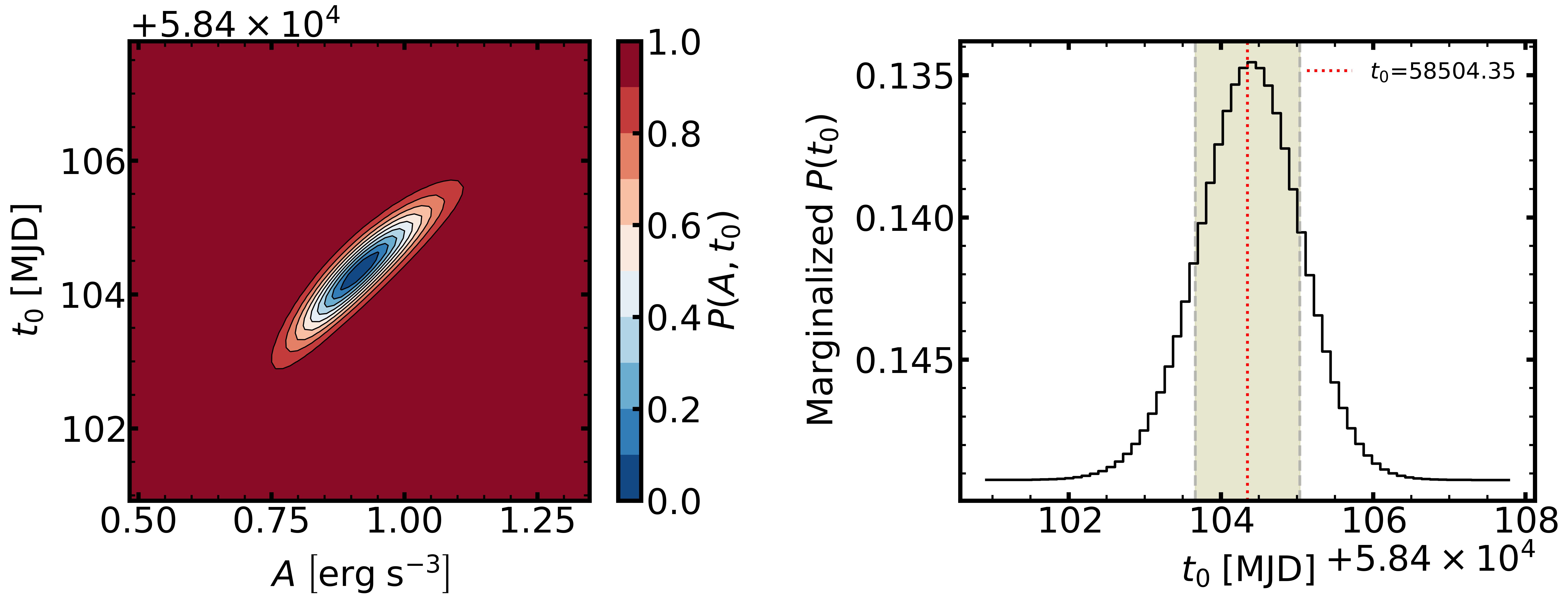}
\caption{
Same as Figure\,\ref{fig:SN2020oi_t2model} for SN\,2019ano
}
\label{fig:SN2019ano_t2model}
\end{figure}
%%%%%%%%%%%%%%%%%%%%%%%%%%%%%%%%%%%%%%%%%%%%%%%
%\subsubsection{SN 2020noz (IIn)} 
%%%%%%%%%%%%%%%%%%%%%%%%%%%%%%%%%%%%%%%%%%%%%%%
%%%%%%%%%%%%%%%%%%%%%%%%%%%%%%%%%%%%%%%%%%%%%%%
\begin{figure}[!ht]
\centering
\begin{minipage}{0.491\textwidth}
\centering
\includegraphics[width=0.85\linewidth,trim={0.0cm 0.0cm 0.0cm 0.0cm},clip]{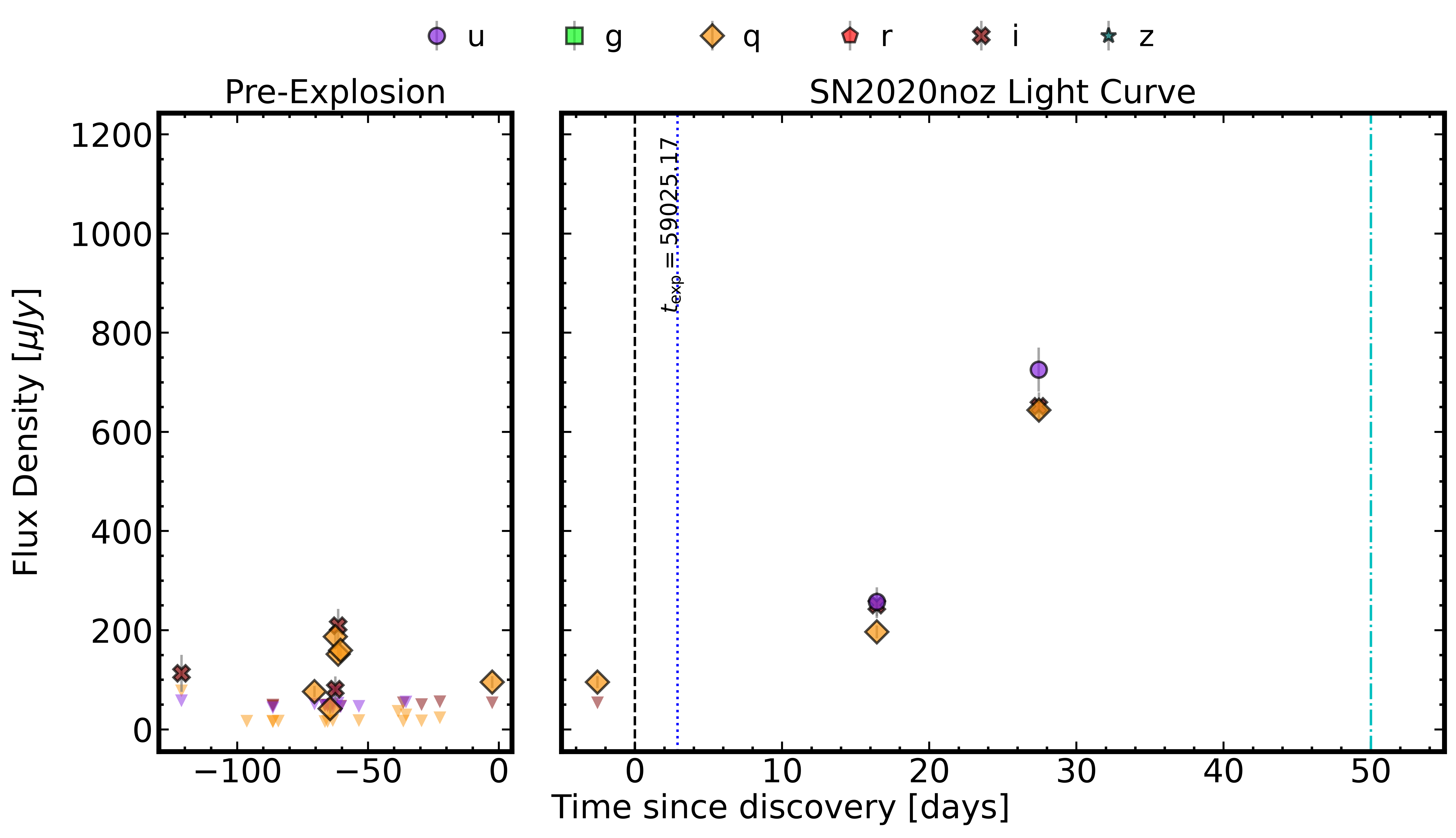}
\includegraphics[width=0.85\linewidth,trim={0.0cm 0.0cm 0.0cm 0.0cm},clip]{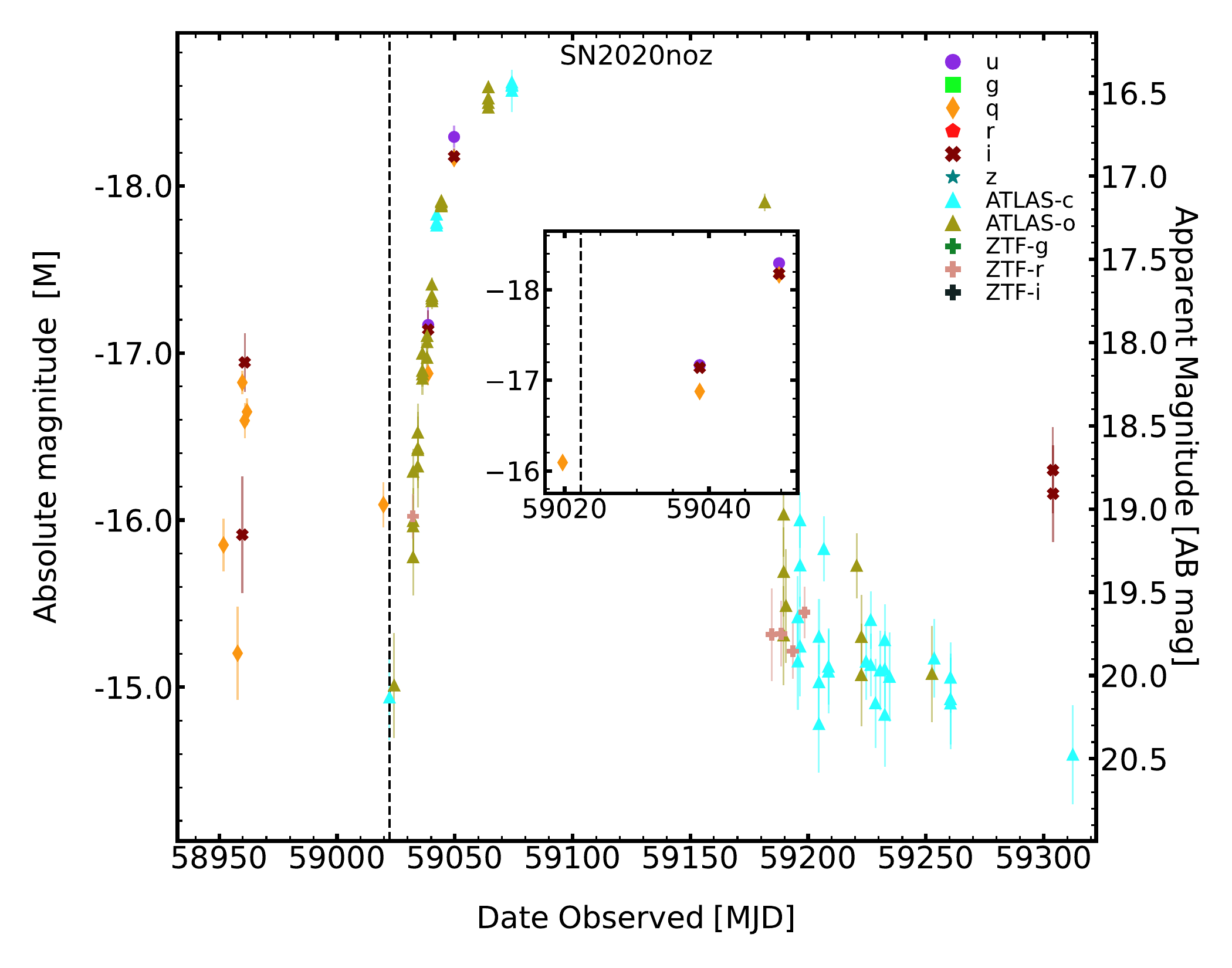}
\end{minipage}%
\hspace{1.5ex}
\caption{
Pre- and post-explosion light curves of SN\,2020noz from MeerLICHT
{\sl Top left}: Pre-explosion photometry up to 120 days before discovery. The shaded inverted triangles are upper limits, and all the other filled symbols are detections.
{\sl Top right}: Multi-band optical light curves of SN\,2020noz with the underlying background emission subtracted. The phases are relative to the time of discovery as reported on TNS, marked by the black vertical dashed line. The estimated explosion epoch and peak brightness epoch are indicated by the dotted blue line and the dash-dotted cyan line, respectively.
{\sl Bottom}: Absolute magnitude light curves from MeerLICHT, ATLAS, and ZTF. Filled triangles represent ATLAS magnitudes, and ZTF difference magnitudes are represented by filled 'plus' symbols. A zoom into the rise-to-peak region of SN\,2020noz is plotted in the insert window, showing the MeerLICHT observations of the SN in the first $35$ days after discovery.
}
\label{fig:SN2020noz}
\end{figure}
%--------------------------------------------%
\begin{figure}[!ht]
\centering
\begin{minipage}{0.491\textwidth}
\includegraphics[width=0.85\linewidth,trim={0.0cm 0.0cm 0.0cm 0.0cm},clip]{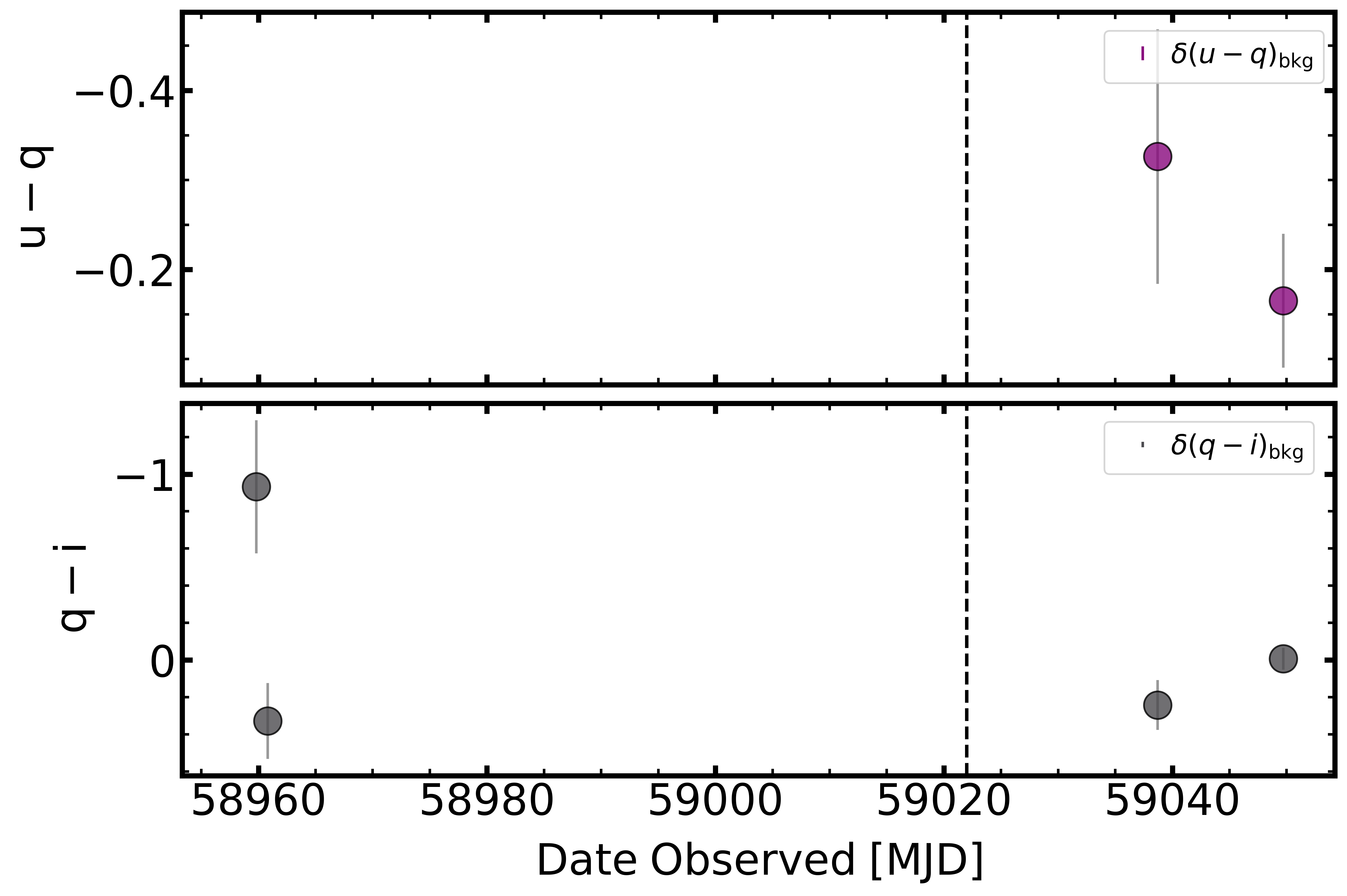}
\end{minipage}
\caption{
Extinction corrected ($A_{V,\mathrm{G}}=0.064$) color evolution of SN\,2020noz from MeerLICHT observations.
}
\label{fig:SN2020noz_color}
\end{figure}
%--------------------------------------------%
\begin{figure}[!ht]
\centering
\includegraphics[width=0.35\columnwidth]{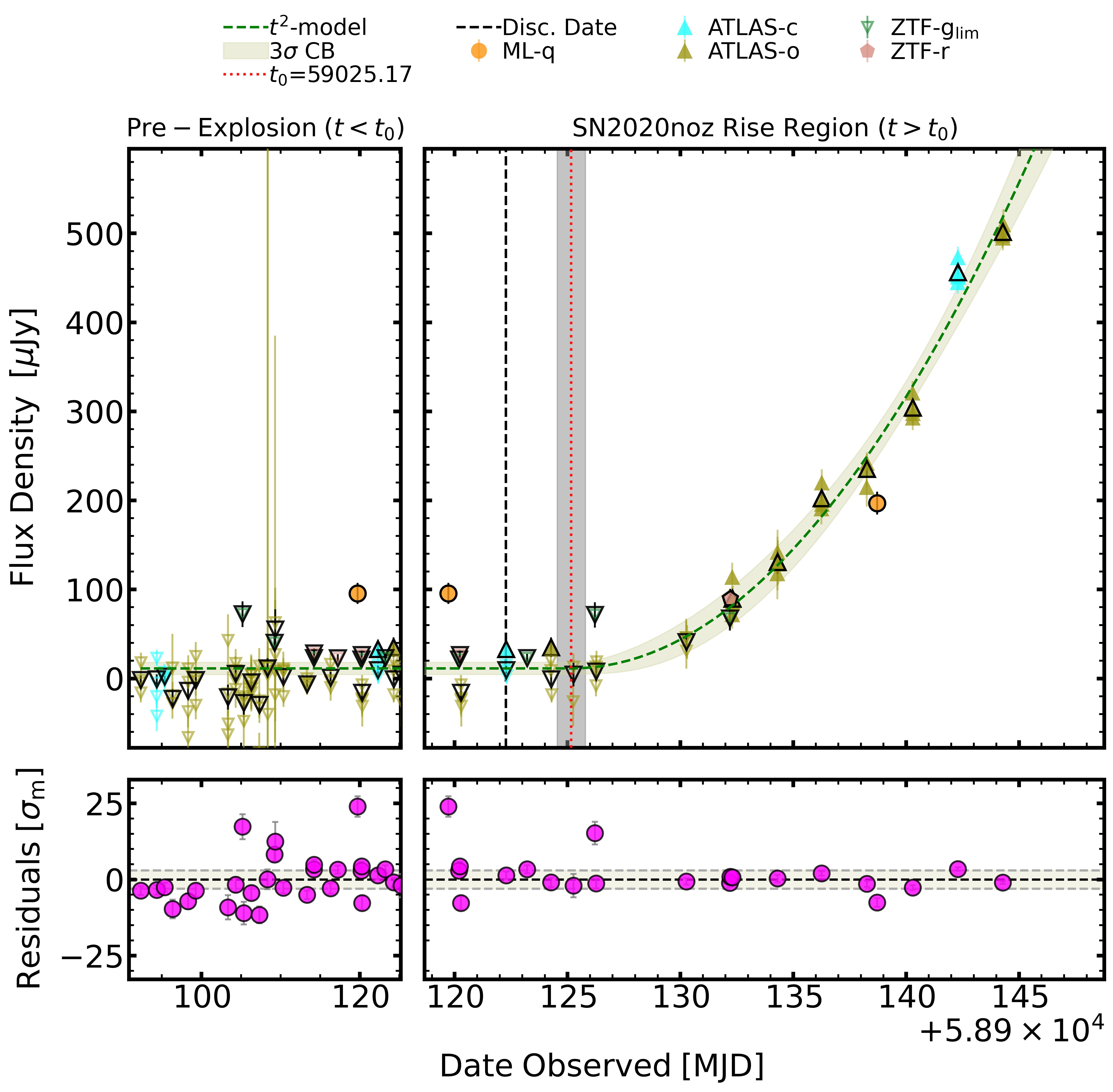}\\
\includegraphics[width=0.35\columnwidth]{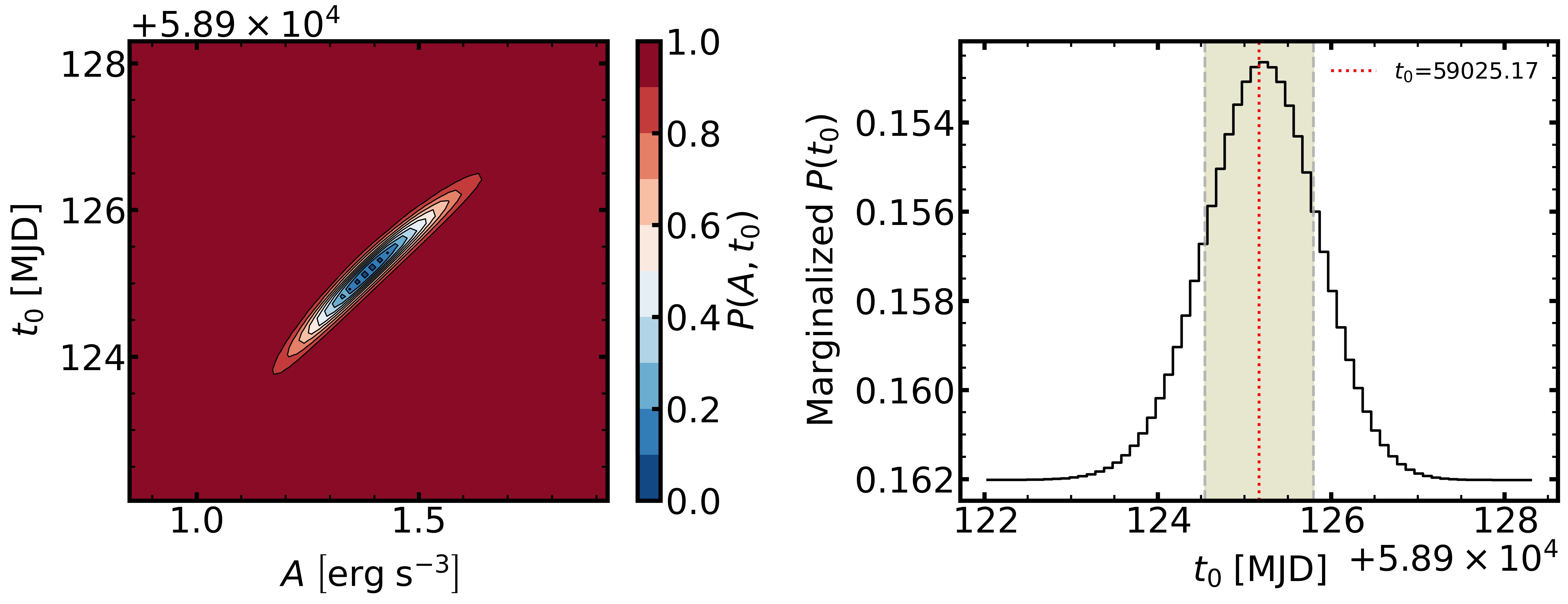}
\caption{Same as Figure\,\ref{fig:SN2020oi_t2model} for SN\,2020noz}
\label{fig:SN2020noz_t2model}
\end{figure}
%%%%%%%%%%%%%%%%%%%%%%%%%%%%%%%%%%%%%%%%%%%%%%%
%\subsubsection{SN 2021koq (Ia)}
%%%%%%%%%%%%%%%%%%%%%%%%%%%%%%%%%%%%%%%%%%%%%%%
%%%%%%%%%%%%%%%%%%%%%%%%%%%%%%%%%%%%%%%%%%%%%%%
\begin{figure}[!ht]
\centering
\begin{minipage}{0.491\textwidth}
\centering
\includegraphics[width=0.85\linewidth,trim={0.0cm 0.0cm 0.0cm 0.0cm},clip]{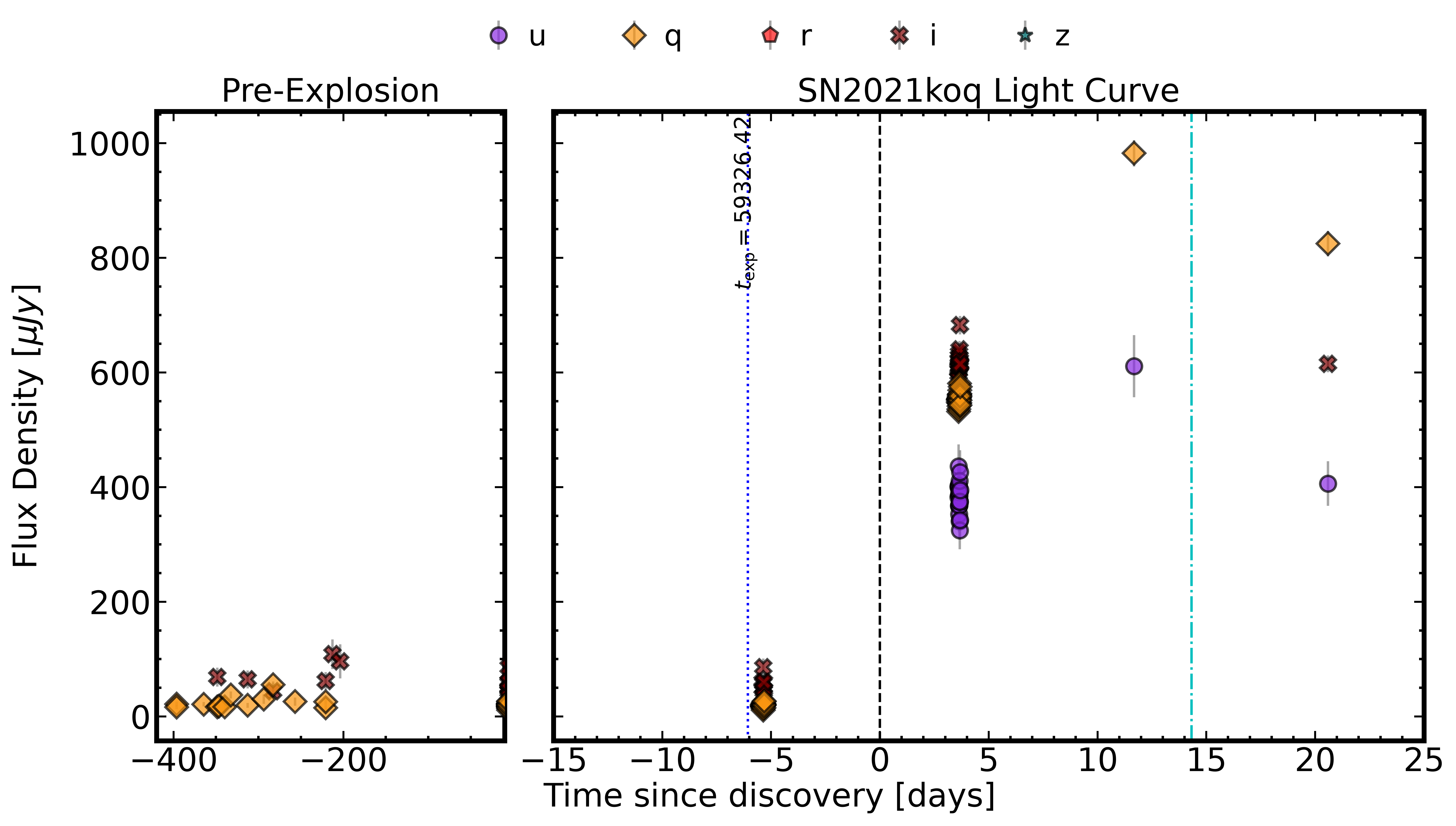}
\includegraphics[width=0.85\linewidth,trim={0.0cm 0.0cm 0.0cm 0.0cm},clip]{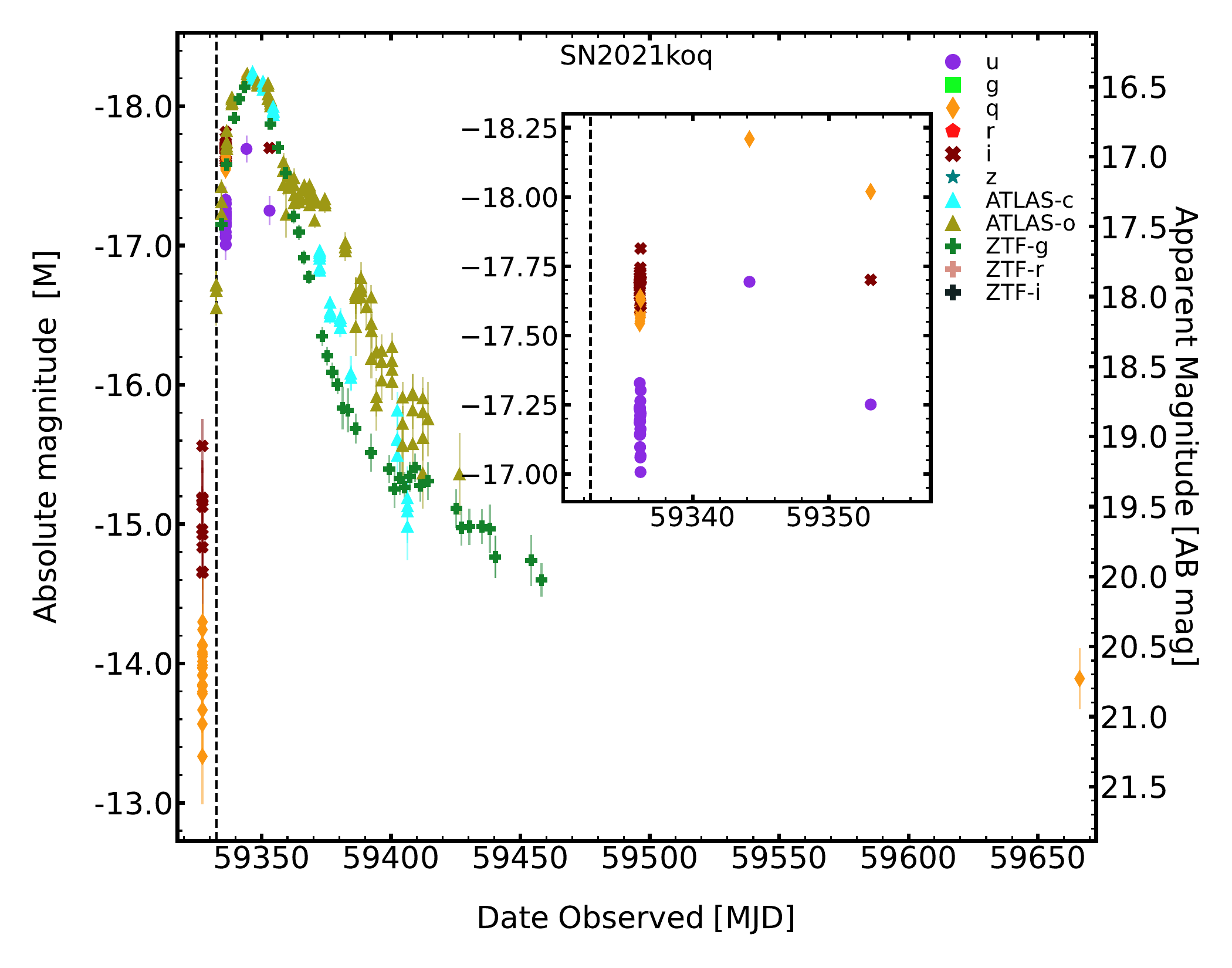}
\end{minipage}
\caption{
Pre- and post-explosion light curves of SN\,2021koq from MeerLICHT
{\sl Top left}: Pre-explosion photometry up to 400 days before discovery. The shaded inverted triangles are upper limits, and all the other filled symbols are detections.
{\sl Top right}: Multi-band optical light curves of SN\,2021koq without subtracting the underlying background emission. The phases are relative to the time of discovery as reported on TNS, marked by the black vertical dashed line. The estimated explosion epoch and peak brightness epoch are indicated by the dotted blue line and the dash-dotted cyan line, respectively. 
{\sl Bottom}: Absolute magnitude light curves from MeerLICHT, ATLAS, and ZTF. Filled triangles represent ATLAS magnitudes, and correct ZTF magnitudes are represented by 'plus' symbols. A zoom into the peak brightness of SN\,2021koq is plotted in the insert window, showing MeerLICHT observations spanning $\sim$17 days after discovery.
}
\label{fig:SN2021koq}
\end{figure}
%--------------------------------------------%
\begin{figure}[!ht]
\centering
\begin{minipage}{0.491\textwidth}
\includegraphics[width=0.85\linewidth,trim={0.0cm 0.0cm 0.0cm 0.0cm},clip]{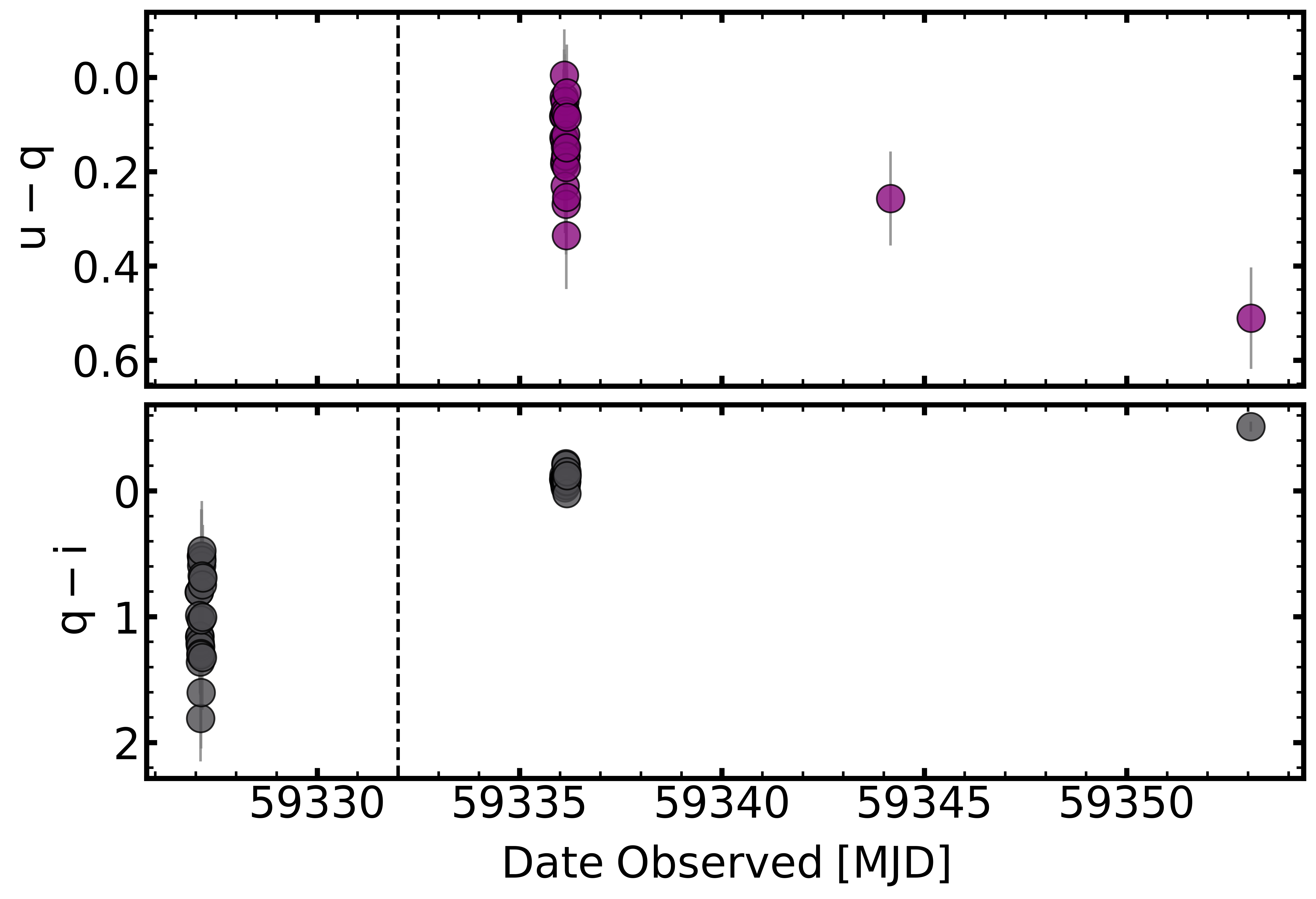}
\end{minipage}
\caption{
Extinction corrected ($A_{V,\mathrm{G}}=0.565$) color evolution of SN\,2021koq from MeerLICHT observations.
}
\label{fig:SN2021koq_color}
\end{figure}
%--------------------------------------------%
\begin{figure}[!ht]
\centering
\includegraphics[width=0.35\columnwidth]{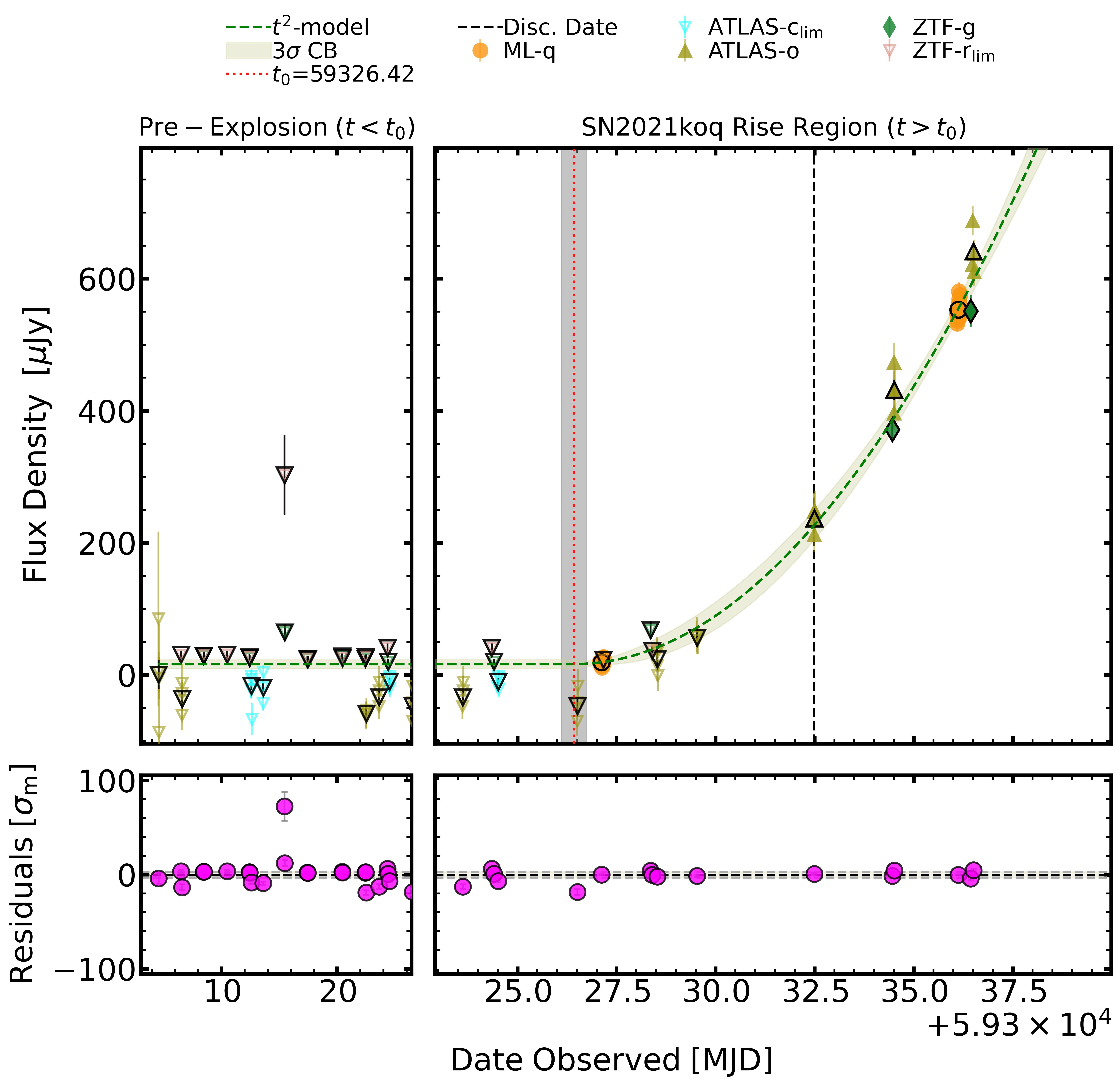}\\
\includegraphics[width=0.35\columnwidth]{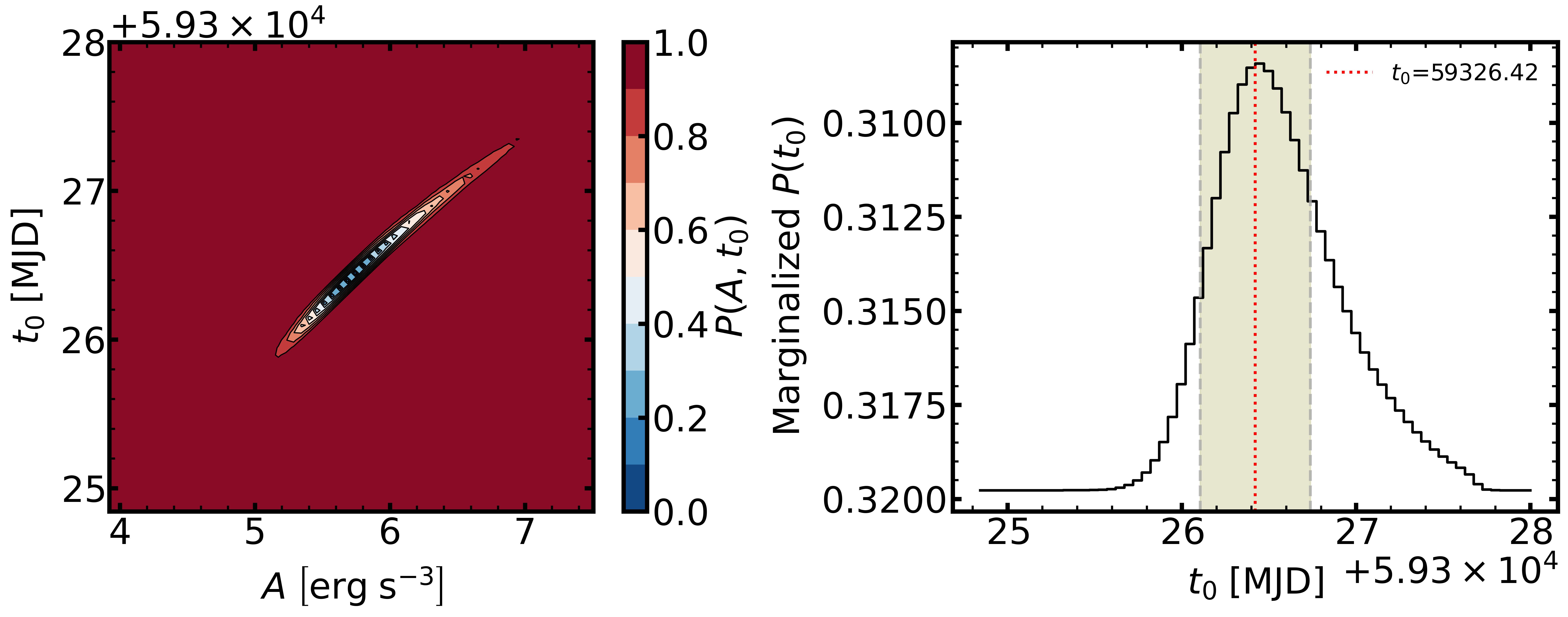}
\caption{Same as Figure\,\ref{fig:SN2021aele_t2model} for SN\,2021koq}
\label{fig:SN2021koq_t2model}
\end{figure}
%%%%%%%%%%%%%%%%%%%%%%%%%%%%%%%%%%%%%%%%%%%%%%%
%\subsubsection{SN 2022bll (IIP)} 
%%%%%%%%%%%%%%%%%%%%%%%%%%%%%%%%%%%%%%%%%%%%%%%
%%%%%%%%%%%%%%%%%%%%%%%%%%%%%%%%%%%%%%%%%%%%%%%
\begin{figure}[!ht]
\centering
\begin{minipage}{0.491\textwidth}
\centering
\includegraphics[width=0.85\linewidth,trim={0.0cm 0.0cm 0.0cm 0.0cm},clip]{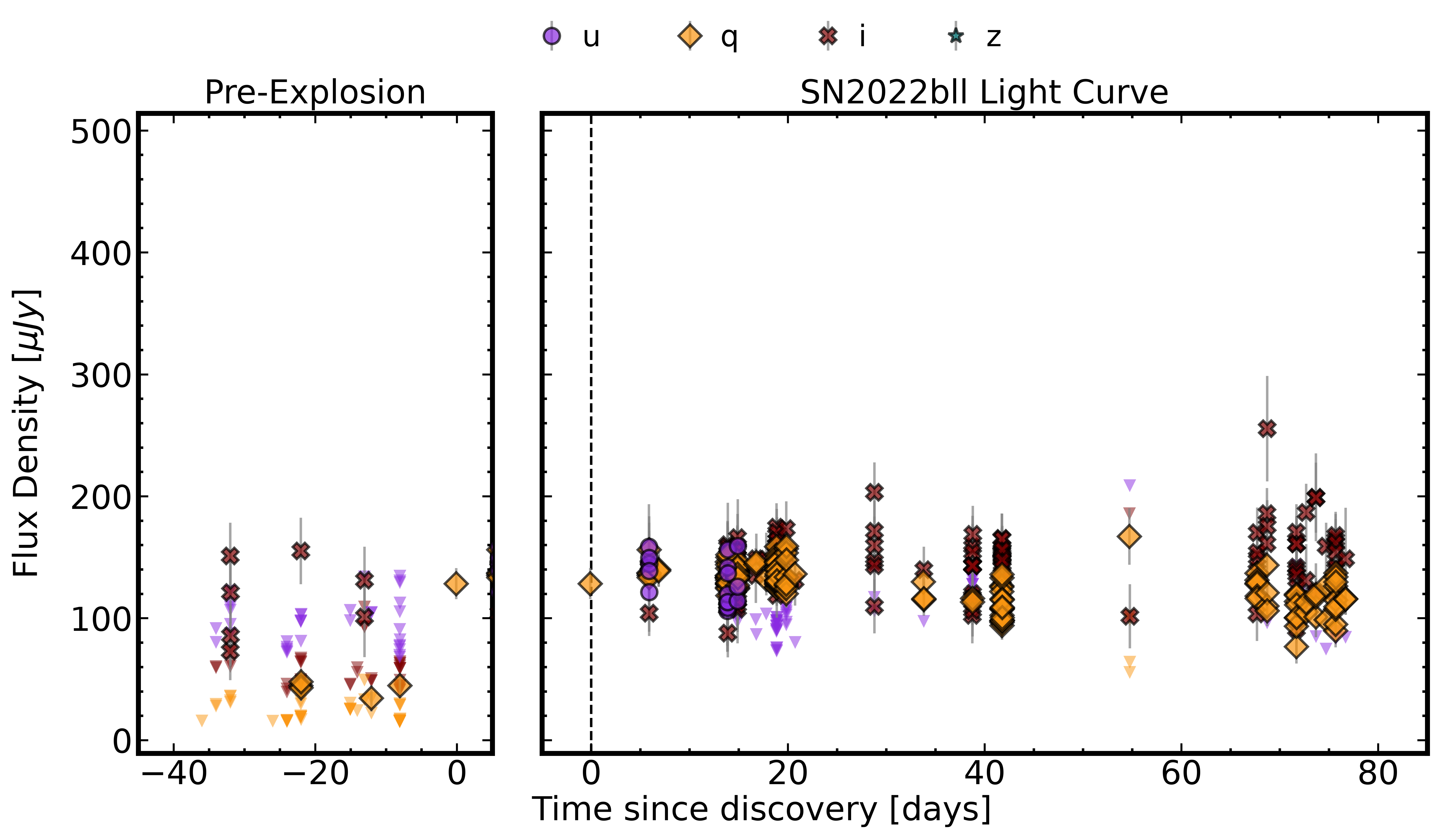}
\includegraphics[width=0.85\linewidth,trim={0.0cm 0.0cm 0.0cm 0.0cm},clip]{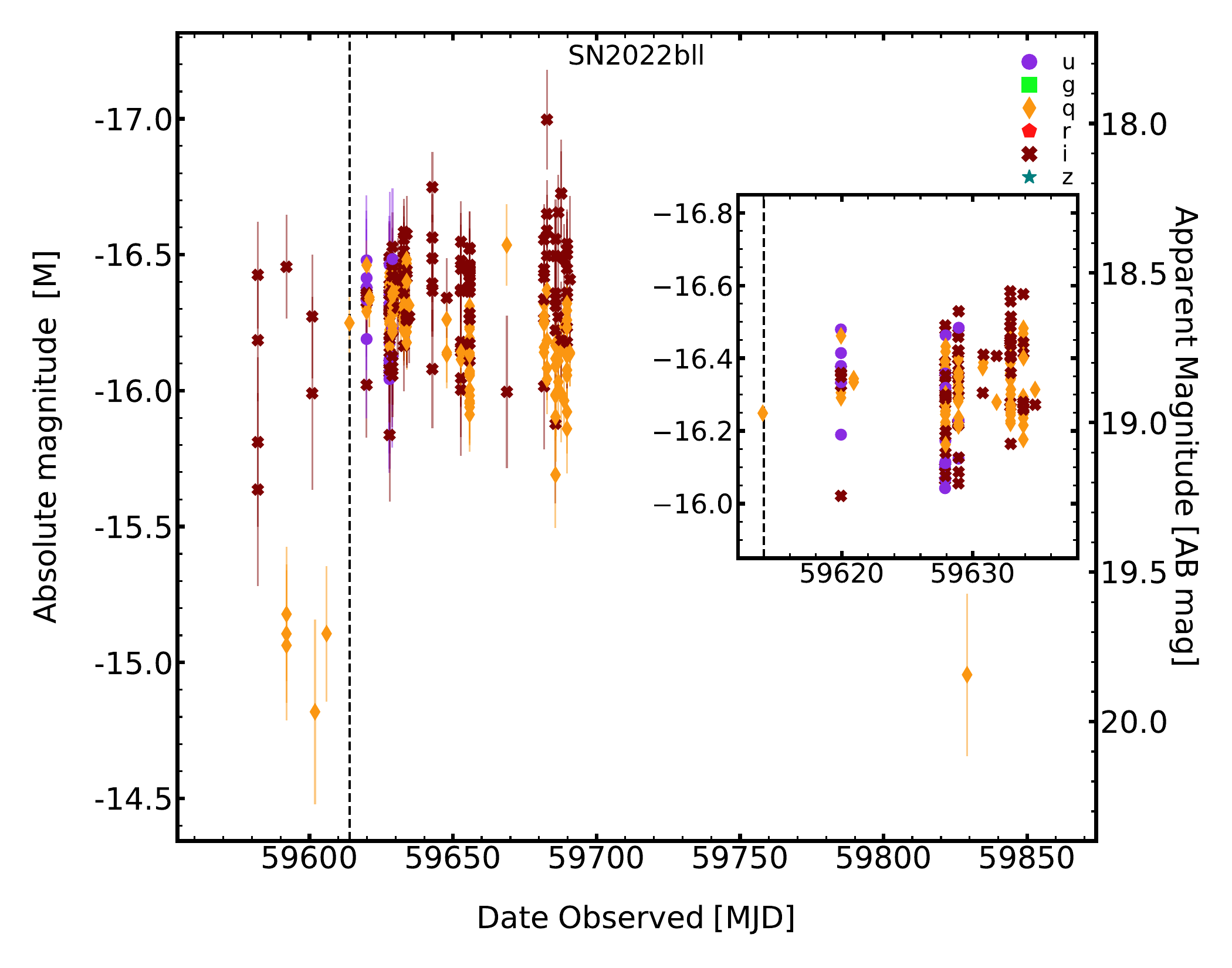}
\end{minipage}
\caption{
Pre- and post-explosion light curves of SN\,2022bll from MeerLICHT.
{\sl Top right}: Pre-explosion photometry up to 40 days before discovery. The shaded inverted triangles are upper limits, and all the other filled symbols are detections.
{\sl Top right}: Multi-band optical light curves of SN\,2022bll with the underlying background emission subtracted. The phases are relative to the time of discovery as reported on TNS, marked by the black vertical dashed line.
{\sl Bottom}: Absolute magnitude light curves from MeerLICHT. A zoom into the assumed peak of SN\,2022bll is plotted in the insert window, showing the high cadence MeerLICHT observations during the plateau phase of the highly reddened SN.
}
\label{fig:SN2022bll} 
\end{figure}
%---------------------------------------------%
\begin{figure}[!ht]
\centering
\begin{minipage}{0.491\textwidth}
\includegraphics[width=0.85\linewidth,trim={0.0cm 0.0cm 0.0cm 0.0cm},clip]{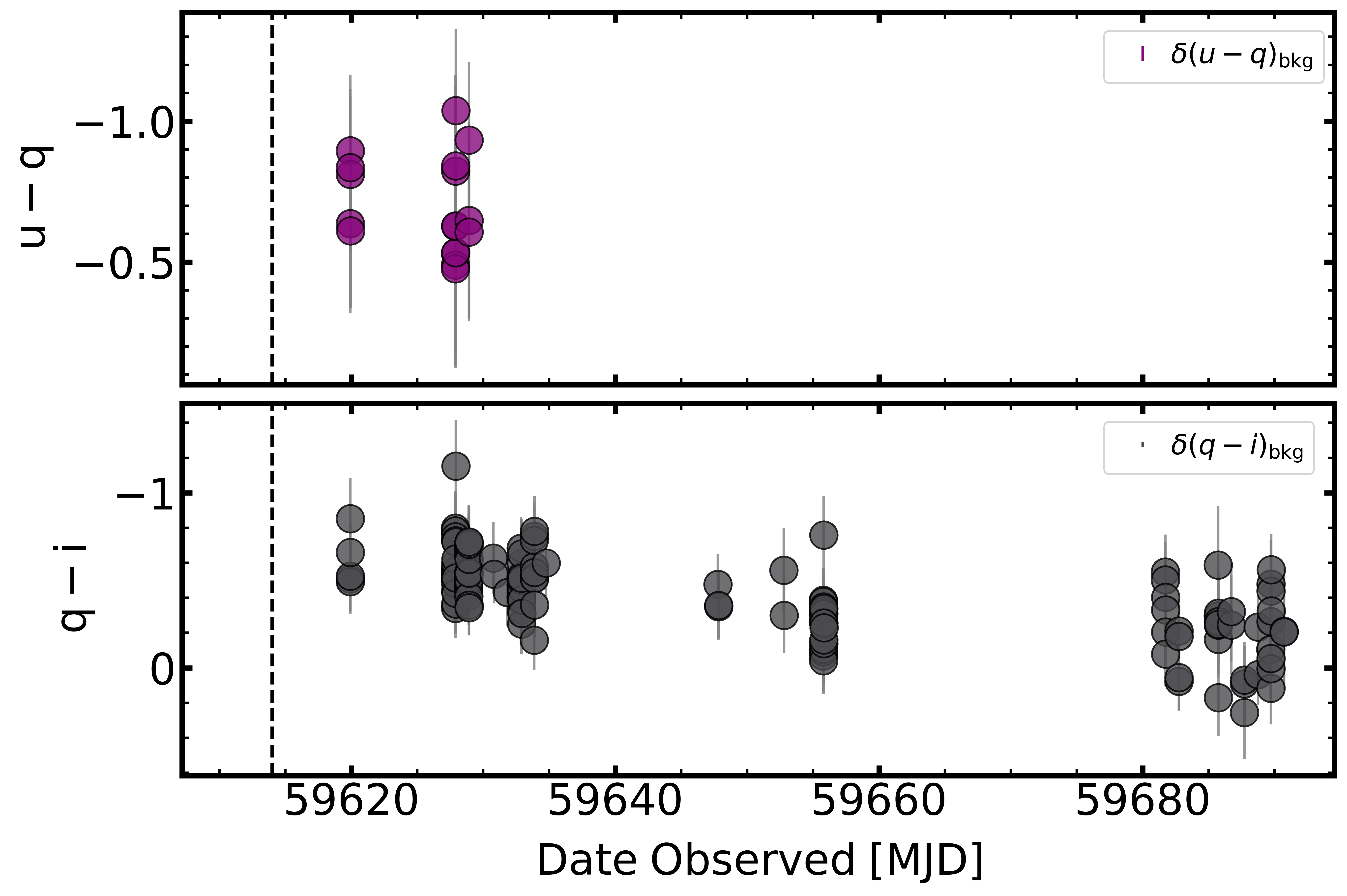}
\end{minipage}
\caption{
Extinction corrected ($A_{V,\mathrm{G}}=1.579$) color evolution of SN\,2022bll from MeerLICHT observations. $\delta (color)_{\mathrm{bkg}}$ represents the magnitude of the systematic uncertainty in the computed color index.
}
\label{fig:SN2022bll_color}
\end{figure}
%%%%%%%%%%%%%%%%%%%%%%%%%%%%%%%%%%%%%%%%%%%%%%%
%\subsubsection{SN 2019lub (IIn)}
%%%%%%%%%%%%%%%%%%%%%%%%%%%%%%%%%%%%%%%%%%%%%%%
%%%%%%%%%%%%%%%%%%%%%%%%%%%%%%%%%%%%%%%%%%%%%%%
\begin{figure}[!ht]
\centering
\begin{minipage}{0.491\textwidth}
\centering
\includegraphics[width=0.85\linewidth,trim={0.0cm 0.0cm 0.0cm 0.0cm},clip]{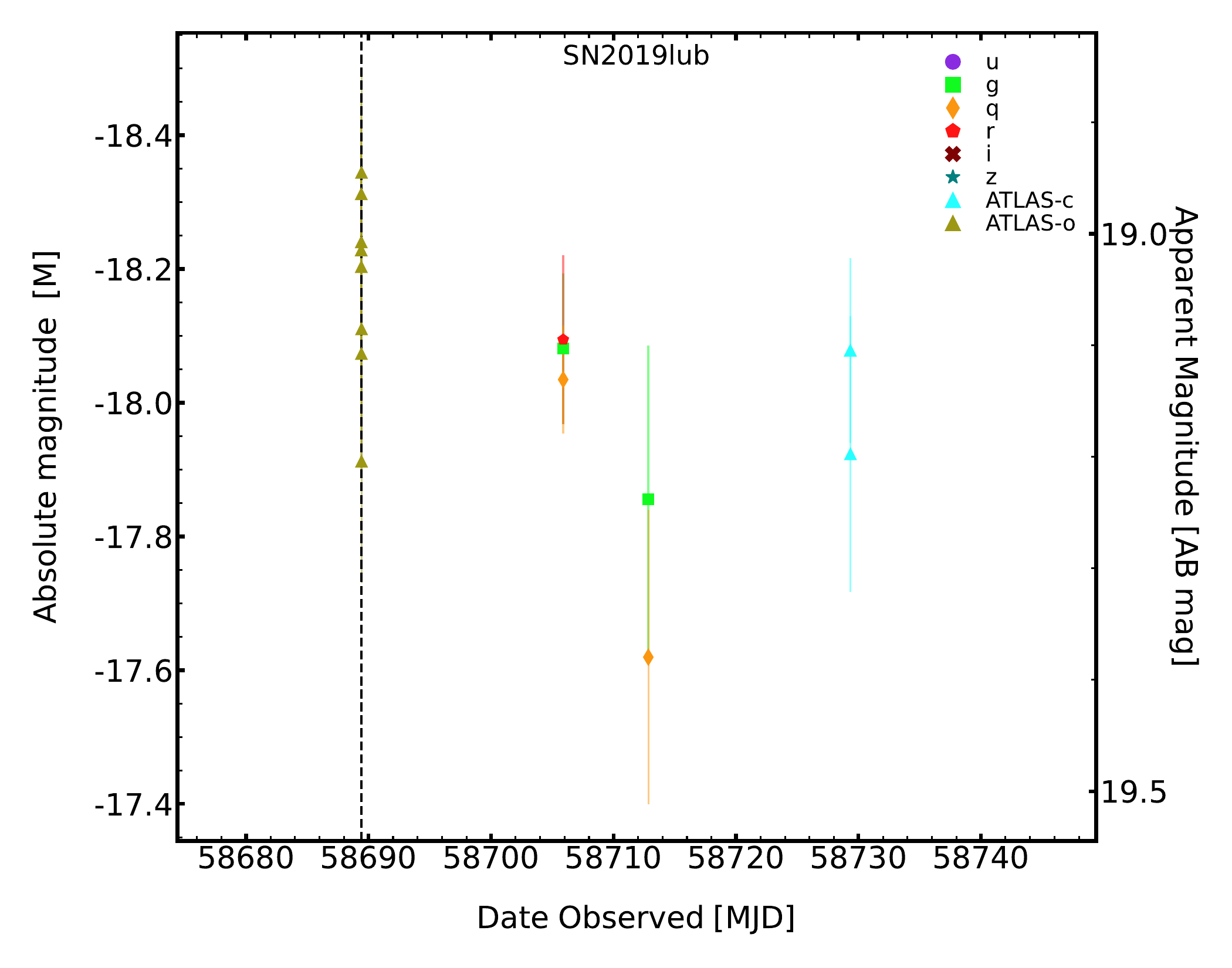}
\end{minipage}
\caption{
Sparse multi-band optical light curves of SN\,2019lub without subtracting the underlying background emission. Filled triangles represent ATLAS-derived absolute magnitudes, whereas the other symbols represent MeerLICHT-derived absolute magnitudes. The vertical dashed black line marks the discovery date as reported on TNS. 
}
\label{fig:SN2019lub}
\end{figure}
%%%%%%%%%%%%%%%%%%%%%%%%%%%%%%%%%%%%%%%%%%%%%%%
%\subsubsection{SN 2022vrr (Ia)}
%%%%%%%%%%%%%%%%%%%%%%%%%%%%%%%%%%%%%%%%%%%%%%%
%%%%%%%%%%%%%%%%%%%%%%%%%%%%%%%%%%%%%%%%%%%%%%%
\begin{figure}[!ht]
\centering
\begin{minipage}{0.491\textwidth}
\centering
\includegraphics[width=0.85\linewidth,trim={0.0cm 0.0cm 0.0cm 0.0cm},clip]{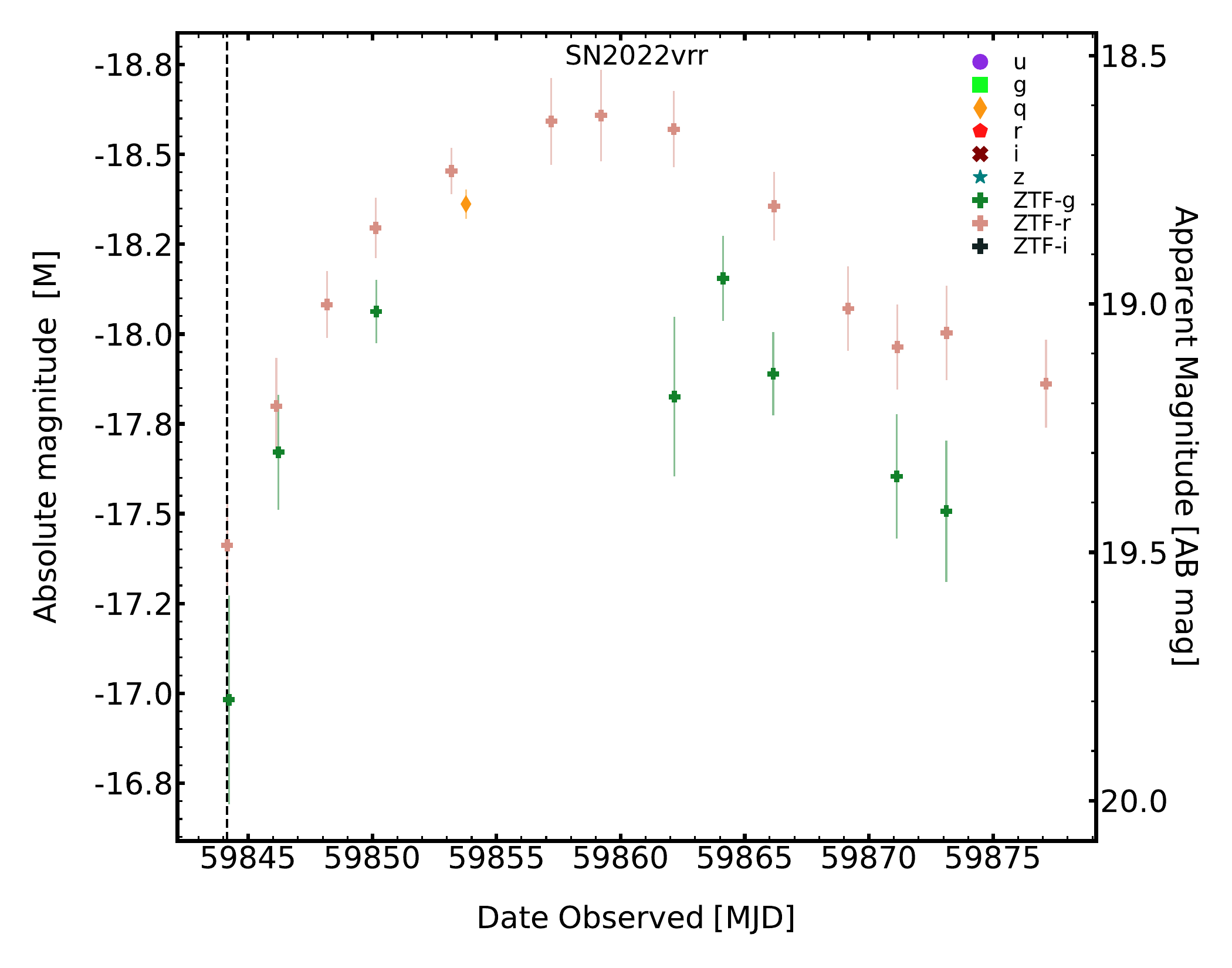}
\end{minipage}
\caption{Multi-band optical light curves of SN\,2022vrr without subtracting the underlying background emission. Filled 'plus' symbols represent ZTF-derived absolute magnitudes, whereas the other symbols are MeerLICHT-derived absolute magnitudes. The vertical dashed black line marks the discovery date as reported on TNS. 
\label{fig:SN2022vrr}}
\end{figure}
%---------------------------------------------%
\begin{figure}[!ht]
\centering
\includegraphics[width=0.35\columnwidth]{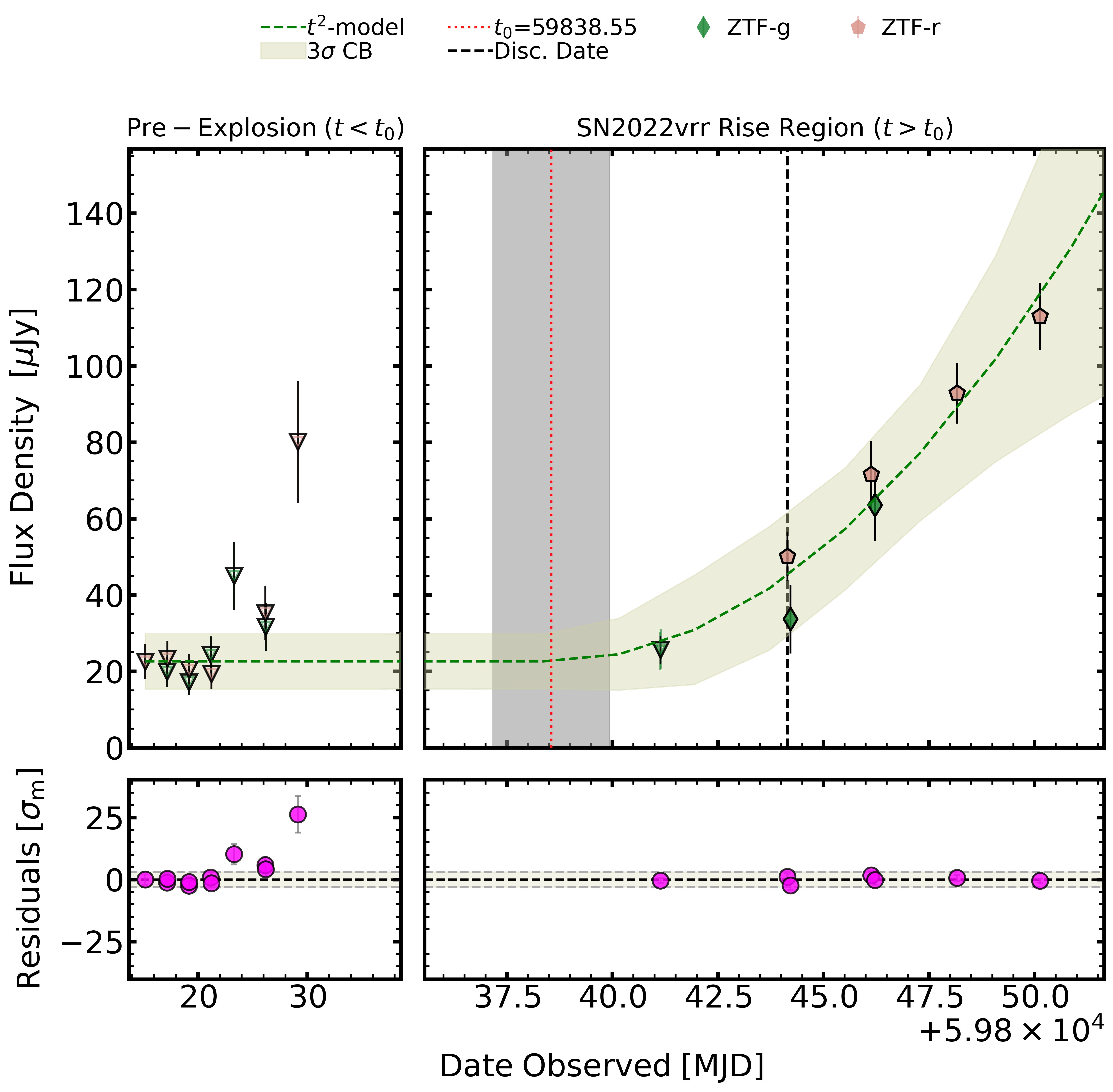}\\
\includegraphics[width=0.35\columnwidth]{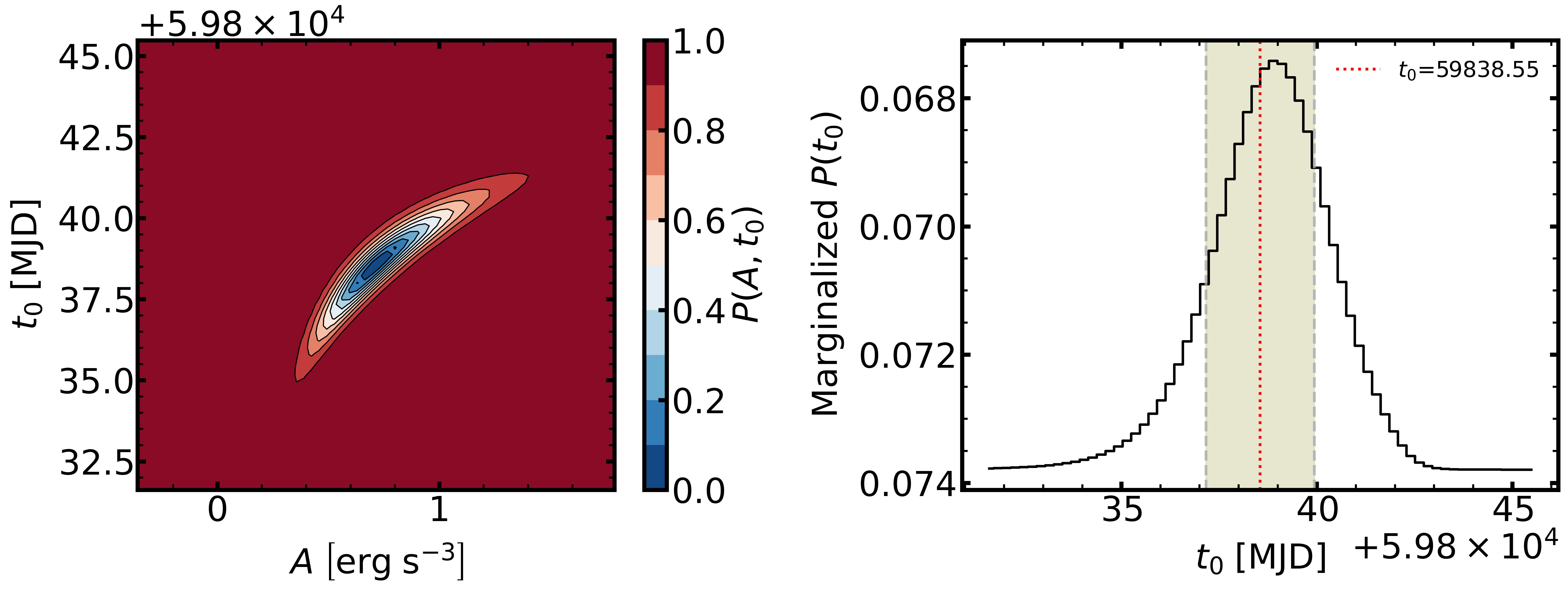}
\caption{Same as Figure\,\ref{fig:SN2020oi_t2model} for SN\,2022vrr}
\label{fig:SN2022vrr_t2model}
\end{figure}
%---------------------------------------------%

%-------------------------------------------------%
%% For this sample we use BibTeX plus aasjournalv7.bst to generate the
%% the bibliography. The sample7.bib file was populated from ADS. To
%% get the citations to show in the compiled file do the following:
%%
%% pdflatex sample7.tex
%% bibtext sample7
%% pdflatex sample7.tex
%% pdflatex sample7.tex

%\newpage
%\clearpage
%\pagestyle{empty}
\bibliography{reference}{}
\bibliographystyle{aasjournalv7}

%% This command is needed to show the entire author+affiliation list when
%% the collaboration and author truncation commands are used.  It has to
%% go at the end of the manuscript.
%\allauthors

%% Include this line if you are using the \added, \replaced, \deleted
%% commands to see a summary list of all changes at the end of the article.
%\listofchanges

\end{document}